\documentclass[11pt,a4paper]{article}

\usepackage[colorlinks=true, linkcolor=black!50!blue, urlcolor=blue, citecolor=blue, anchorcolor=blue]{hyperref}
\usepackage[font=small,labelfont=bf,margin=0mm,labelsep=period,tableposition=top]{caption}
\usepackage[a4paper,top=3cm,bottom=2.5cm,left=2.5cm,right=2.5cm,bindingoffset=0mm]{geometry}

\usepackage{graphicx,placeins}
\usepackage{float}
\usepackage{afterpage}
\usepackage{epsfig,cite}
\usepackage{amssymb}
\usepackage{amsmath}
\usepackage{dsfont}
\usepackage{multirow}
\usepackage{url}
\usepackage{xcolor}
\usepackage{float}
\usepackage{afterpage}

\usepackage{url}
\usepackage{hyperref}

\usepackage{booktabs}

\usepackage{tikz}
\usetikzlibrary{arrows}
\usepackage{tikz-3dplot}
\usepackage{enumitem}
\usepackage{hyperref}
\usepackage{cite}

\def\smallfrac#1#2{\hbox{$\frac{#1}{#2}$}}
\def\half{\smallfrac{1}{2}}

\newcommand{\be}{\begin{equation}}
\newcommand{\ee}{\end{equation}}
\newcommand{\bea}{\begin{eqnarray}}
\newcommand{\eea}{\end{eqnarray}}
\newcommand{\bi}{\begin{itemize}}
\newcommand{\ei}{\end{itemize}}
\newcommand{\ben}{\begin{enumerate}}
\newcommand{\een}{\end{enumerate}}
\newcommand{\la}{\left\langle}
\newcommand{\ra}{\right\rangle}
\newcommand{\lc}{\left[}
\newcommand{\rc}{\right]}
\newcommand{\lp}{\left(}
\newcommand{\rp}{\right)}
\newcommand{\as}{\alpha_s}

\def\frac#1#2{{{#1}\over {#2}}}
\def\gsim{\mathrel{\rlap{\lower4pt\hbox{\hskip1pt$\sim$}}
    \raise1pt\hbox{$>$}}}         
\def\lsim{\mathrel{\rlap{\lower4pt\hbox{\hskip1pt$\sim$}}
    \raise1pt\hbox{$<$}}}         

\newcommand{\draft}[1]{}

\def\beq{\begin{equation}}
\def\eeq{\end{equation}}

\def\lapprox{\lower .7ex\hbox{$\;\stackrel{\textstyle <}{\sim}\;$}}
\def\gapprox{\lower .7ex\hbox{$\;\stackrel{\textstyle >}{\sim}\;$}}
\def\half{\smallfrac{1}{2}}

\newcommand{\pmz}{{\pm}\hspace{-.9em}{\bigcirc}} 

\numberwithin{equation}{section}
\numberwithin{figure}{section}
\numberwithin{table}{section}

\usepackage{tabularx}
\newcolumntype{C}[1]{>{\centering\arraybackslash}p{#1}}

\begin{document}
\newgeometry{top=1.5cm,bottom=1.5cm,left=2.5cm,right=2.5cm,bindingoffset=0mm}
\begin{figure}[h]
  \includegraphics[width=0.32\textwidth]{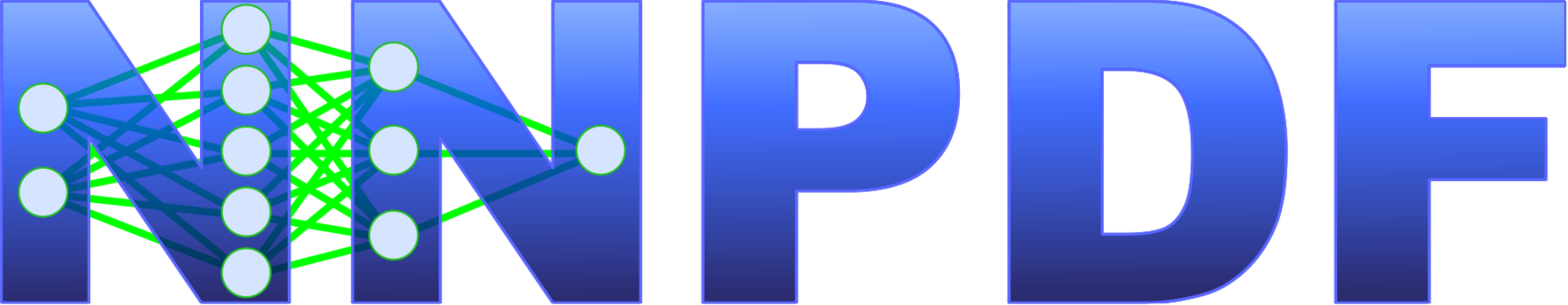}
\end{figure}
\vspace{-2.0cm}
\begin{flushright}
Edinburgh 2019/9\\
Nikhef/2019-014\\
TIF-UNIMI-2019-9\\
DAMTP-2019-24\\
CAVENDISH-HEP-19-11\\
\end{flushright}
\vspace{0.3cm}

\begin{center}
  {\Large \bf Parton Distributions with Theory
    Uncertainties:\\[0.2cm]
{\Large     General Formalism and First Phenomenological Studies}}
\vspace{1.1cm}

 {\small
  {\bf  The NNPDF Collaboration:} \\[0.2cm]
Rabah Abdul Khalek,$^{1,2}$
Richard D. Ball,$^{3}$
Stefano Carrazza,$^{4}$
Stefano Forte,$^{4}$
Tommaso Giani,$^{3}$\\
Zahari Kassabov,$^{5}$
Rosalyn L. Pearson,$^{3}$
Emanuele R. Nocera,$^{2}$
Juan Rojo,$^{1,2}$
Luca Rottoli,$^{6}$\\
Maria Ubiali,$^{7}$
Cameron Voisey,$^{5}$ and
Michael Wilson$^{3}$
}\\

 \vspace{0.7cm}
 
       {\it \small ~$^1$Department of Physics and Astronomy, VU University, NL-1081 HV Amsterdam,\\
~$^2$Nikhef Theory Group, Science Park 105, 1098 XG Amsterdam, The Netherlands\\[0.1cm]

         ~$^3$The Higgs Centre for Theoretical Physics, University of Edinburgh,\\
  JCMB, KB, Mayfield Rd, Edinburgh EH9 3JZ, Scotland\\[0.1cm]
  ~$^4$Tif Lab, Dipartimento di Fisica, Universit\`a di Milano and\\
INFN, Sezione di Milano, Via Celoria 16, I-20133 Milano, Italy\\[0.1cm]
~$^5$Cavendish Laboratory, University of Cambridge, Cambridge CB3
0HE, United Kingdom\\[0.1cm]
~$^6$Dipartimento di Fisica G. Occhialini, U2, Universit\`{a} degli Studi di Milano-Bicocca,
Piazza della Scienza, 3, 20126 Milano, Italy and \\
INFN, Sezione di Milano-Bicocca, 20126, Milano, Italy\\[0.1cm]
~$^7$DAMTP, University of Cambridge, Wilberforce Road, \\ Cambridge, CB3 0WA, United Kingdom
}

\vspace{1.0cm}

{\bf \large Abstract}

\end{center}
We formulate a general approach to the inclusion of theoretical uncertainties,
specifically those related to the missing higher order uncertainty (MHOU),
in the determination of parton distribution functions (PDFs).
We demonstrate how, under quite generic assumptions, theory
uncertainties can be included as an extra contribution to the covariance
matrix when determining PDFs from data.
We then review, clarify, and systematize the
use of renormalization and factorization scale variations as a 
means to estimate MHOUs consistently in deep inelastic and hadronic processes.
We define a set of prescriptions for constructing a theory covariance matrix 
using scale variations, which can be used in global fits of data from a wide 
range of different processes, based on choosing a set of
independent scale variations suitably correlated within and across processes.
We set up an algebraic framework for the choice and validation of an 
optimal prescription by comparing the estimate of MHOU encoded in the 
next-to-leading order (NLO) theory covariance
matrix to the observed shifts between  NLO and NNLO predictions.
We perform a NLO PDF determination which includes the MHOU,
assess the impact of the inclusion of MHOUs on the PDF central values
and uncertainties, and
validate the results by comparison to the known shift between NLO and
NNLO PDFs. 
We finally study the impact
of the inclusion of MHOUs in a global PDF determination on LHC
cross-sections, and provide guidelines for their use in
precision phenomenology.
In addition, we also compare the results based on the theory covariance matrix
formalism to those obtained by performing PDF
determinations based on different scale choices.

\clearpage

\tableofcontents

\clearpage

\section{Introduction}
\label{sec:introduction}

An accurate estimate of the uncertainty in Standard Model (SM) predictions
is a crucial ingredient for precision phenomenology at the Large Hadron 
Collider (LHC).
Now, and for several years to come~\cite{Cepeda:2019klc,Azzi:2019yne},
theoretical 
uncertainties for hadron collider processes are
dominated by the missing higher order uncertainty (MHOU) in perturbative 
QCD calculations, usually estimated by scale 
variation, and by parton distribution function (PDF) uncertainties.
Of course, PDFs 
summarize the information on the nucleon structure extracted from other
SM processes~\cite{Gao:2017yyd}: effectively, PDFs provide a way of obtaining
a prediction for a given process in terms of other processes.
This way of thinking
about PDFs immediately shows that MHOUs are present 
not only in the perturbative prediction for a particular process, but 
also in the underlying processes used for the PDF determination.

Current PDF uncertainties essentially only
include the propagated uncertainty arising from statistical and
systematic uncertainties in the experimental data used in their 
determination.
Methodological uncertainties related for example to
the choice of functional form for the PDFs, or the fitting methodology employed, can be kept under control using closure tests~\cite{Ball:2014uwa}, and with care can be made negligible in the data region.
Parametric uncertainties, such as those related to the value of
the strong coupling $\alpha_s(m_Z)$ or the charm mass $m_c$ can be included by 
performing fits for a range of parameters.
However up until now MHOUs have never been included in a PDF fit: what is 
usually called the ``PDF uncertainty'' does not include 
the MHOU in the theoretical calculations used for PDF determination, and, 
more generally, does not typically include any source of theory 
uncertainty. 

Historically,
this is related to the fact that MHOUs
have always been
considered as likely to be small in comparison to other
PDF uncertainties, especially since NNLO PDFs have become the default
standard.
However, it is clear that as PDF uncertainties become
smaller and smaller, at some point MHOUs will become significant. 
In the most recent NNPDF set, NNPDF3.1~\cite{Ball:2017nwa}, PDF
uncertainties at the electroweak scale can be as low as 1\%.
Given that the
typical size of MHOU on NNLO QCD processes is at the percent level (see
e.g.~\cite{Campbell:2017hsr}) their neglect seems difficult to justify a priori.

Besides contributing to the overall size of PDF uncertainty, more subtly the MHOU might affect the relative weights of different datasets 
included in the fit: a dataset which is accurately described by NNLO theory because it has small MHOU should in principle carry more weight than one which is poorly described because it has large MHOU. The neglect of MHOUs might thus be biasing current global PDF fits.

It is the purpose of this paper to set up a general formalism for the
inclusion of theoretical uncertainties, specifically MHOUs, in PDF
determinations, and then to perform a first phenomenological
exploration of their impact on LHC phenomenology.
The development of this treatment of MHOUs will involve three main ingredients.
The first
is the formulation of a general theory for the inclusion  in PDF fits of
generic theoretical uncertainties, of which MHOUs are a particular
case.
The second is the choice of a specific method for estimating the 
MHOU in each of the cross-sections that enter the PDF fit.
The third is the construction of a set of tools for the
validation of this methodology, to check that the MHOU is being correctly 
estimated.

The first ingredient in our approach is common to any kind of theory 
uncertainty: theory uncertainties include not only MHOUs, but also any other aspect in which the theory used in order to obtain
predictions for the physical processes that enter the PDF fit is 
incompletely known. These
include higher twists (see Refs.\cite{Alekhin:2000ch,Ball:2013gsa} and
Ref. therein) and other power-suppressed corrections, nuclear
corrections when nuclear targets are involved (see Refs.\cite{Paukkunen:2018kmm,Ball:2018twp} and
Ref. therein), final state corrections for non-inclusive processes, and so forth.
All of these uncertainties are only meaningful in a
Bayesian sense: there is only one correct value of the
next-order perturbative correction, not a distribution of values.
They thus necessarily involve a process of informed estimation
or guesswork: the only way to actually know the size of, say, a missing
higher order correction, is to calculate it.

We will show by adopting a Bayesian
point of view, and assigning a Gaussian probability distribution to the
expected true value of the theory calculation, that the impact of 
any missing theoretical contribution can be encoded as an 
additive contribution to the experimental covariance matrix used in the 
PDF fit \cite{Ball:2018odr}. The combination is additive because experimental and theoretical uncertainties are by their nature independent, and are thus combined in quadrature. In a global fit, theoretical uncertainties 
can be strongly correlated not only across data points within a given 
experiment, but also between different experiments, and even different 
processes, so we need a theoretical covariance matrix which includes all these correlations across all the datasets included in the fit.

This then immediately raises the issue of choosing a meaningful way to
estimate the MHOU, which in particular incorporates these correlations.
The standard way of estimating MHOUs in perturbative QCD calculations is 
to perform a variation of the
renormalization and factorization scales, denoted as $\mu_r$ and $\mu_f$
respectively, with various choices for the range and combination of variations existing.
While the shortcomings of this method are well known, and various alternatives have been discussed~\cite{Cacciari:2011ze,David:2013gaa,Bagnaschi:2014wea},
this remains the default and most widely used option.
In the present context, its main advantage is its universality (it can be applied in the same way to any
of the processes used in the fit), and the way in which it implicitly incorporates correlations (for example predictions for data points in the same process which are kinematically close will be automatically correlated), even across different processes (through the PDFs, which are the same in every process).
Thus while in principle our covariance matrix formalism allows for the
inclusion of any method for estimating MHOUs in a PDF determination, here 
we will specifically use scale variation.

In order to do this, we need to examine systematically the underpinnings 
of scale variation as a means to estimate theory uncertainties, since 
different definitions of scale variation have been used in different contexts.
Indeed, the standard definitions of renormalization and
factorization scale typically used for deep-inelastic scattering and 
hadronic collisions are not the same.
Because  PDF fits include both types of processes,
it is important to understand in detail how these
definitions relate to each other, in order to be
able to correlate the scale variations in a meaningful
way.
Specifically, we will show that one may estimate the MHOU for any process 
by combining two independent scale variations: one
to estimate the MHOU in the perturbative evolution of the PDFs
(missing higher orders in the DGLAP splitting functions), and the other 
to estimate the MHOU in the perturbative calculation of the 
partonic cross-sections (missing
higher orders in the hard-scattering matrix elements).

Once the scales to be varied are understood, the remaining task
is to choose a particular prescription to be used to construct the 
theoretical covariance matrix.
In estimating MHOUs for a given process, the most commonly adopted option 
is the so-called seven-point envelope prescription, in which 
$\mu_r$ and $\mu_f$ are independently varied by a factor of two about
the central choice 
while ensuring that $1/2 \le \mu_r/\mu_f\le 2$, and the MHOU is then
taken as the envelope of the results. 
For our purposes this is insufficient: rather than taking an envelope,
we wish to contruct a covariance matrix out of the scale variations.
In particular, because theoretical uncertainties are correlated across processes (through the evolution of the PDFs), we need a prescription for determining the entries of the covariance matrix both within a single process and 
across pairs of processes. 

We will discuss in detail a variety of options to achieve this, based on a general ``$n$-point prescription''.
These  options will differ from each other in the choice
of the number of independent variations, the directions of such
variations in the $(\mu_r,\mu_f)$ plane, and the way the variations
are correlated (or not) across different processes.

The validation of these point prescriptions, and the choice of the
optimal one to be used for PDF determinations is a nontrivial
problem, which however admits an elegant solution.
The validation can be performed at NLO, by comparing the estimate of the MHOU encoded in the theory covariance matrix to the known next (NNLO)
order correction. The problem is then to compare the probability
distribution of expected higher-order results to the unique answer given by the NNLO calculation.
The solution to this problem is to view the set
of shifts between the NLO and NNLO computations for all the processes under
consideration as a vector, with one component for each 
of the data points. The theory
covariance matrix corresponding to each prescription then defines a
one-sigma ellipsoid in a subspace of this space.
The validation is performed by projecting the shift vector into the ellipsoid: if the theory covariance matrix gives a sensible estimate of the MHOU at NLO, the shift vector will lie almost entirely within the ellipsoid.
Using this strategy, we will validate a variety of scale variation prescriptions
on a similar dataset to that of the global NNPDF3.1 analysis.
Since the dimension of the space of datapoints is typically two orders of magnitude higher than the dimension of the subspace of the ellipsoid, this is a highly nontrivial test.

Once a prescription has been selected and used to construct the theory
covariance matrix, it is possible to perform a PDF fit based
on it.
Within the NNPDF methodology, an ensemble of PDF replicas is fitted to data
replicas.
Data replicas are generated in a way which reflects the
uncertainties and correlations of the underlying data, as encoded in
their covariance matrix.
The best-fit PDF replica for each data
replica is then determined by minimizing a figure of merit ($\chi^2$) which
is computed using the covariance matrix.
As mentioned, and as we shall show in
Sect.~\ref{sec:thcovmat}, the theory contribution
appears as an independent contribution to the total covariance matrix,
uncorrelated with the experimental one and simply added to
it.
Therefore,  once the
covariance matrix is supplemented by an extra theory contribution coming 
from MHOUs, this should be treated on the same footing as any other
contribution, and it will thus affect both the data replica generation, and
the fitting of PDF replicas to data replicas.

Qualitatively, one may expect the inclusion of the MHOU in the data replica
generation to increase the spread of the data replicas, and thus lead in 
itself to an increase in overall PDF uncertainties. On the other hand 
the inclusion of the MHOU in the fitting might also  reduce 
tensions within the fit due to the imperfection of the theory and, since 
these are highly correlated, result in significant shifts in central values, 
and overall a better fit with reduced uncertainties. The combined effect
of including the MHOU in both the data generation and the fitting is thus not
at all obvious.

We will investigate these effects by performing PDF determinations in
which MHOUs are included in either, or both, the replica
generation and the PDF replica fitting. Once again, results can be 
validated at NLO by comparing NLO PDFs determined with the theory 
covariance matrix to NNLO PDFs.
A successful validation should show that the best-fit NLO PDF moves
towards the central NNLO result upon inclusion of the theory covariance 
matrix in both replica generation and fitting, due to a relaxation of 
tensions in the NLO fit, and that the NNLO PDF differs from the NLO PDF 
by an amount which is correctly estimated by the NLO uncertainty band.
As we shall see, this is indeed the case, and in fact it will
turn out that often the uncertainty band does not increase or even decreases 
upon inclusion of the theory covariance matrix.

Having determined PDFs which now account for the MHOU associated
to the processes that enter the fit, the natural
questions which then arise are what is their impact, and more
generallly
how they should be used for precision
LHC phenomenology.
In order to address the first question, we
will compute predictions with MHOUs for typical LHC standard
candle processes, both with and without including the MHOU in the PDF, and
provide a first phenomenological exploration and assessment of the
impact of these uncertainties.

The second  question is not entirely trivial
and we will address it in detail.
Indeed, scale variation is
routinely performed in order to estimate the MHOU in theoretical predictions for
hadron collider processes. Clearly, when obtaining a prediction, we
should avoid double counting a
MHOU which has already been included in the PDF.
Instances in which
this might happen include not only the trivial situation in which a
prediction is obtained for a process which has already been used for
PDF determination, but also the somewhat more subtle situation in
which the MHOU in the PDF and the observable which is being predicted
are correlated through perturbative
evolution~\cite{Harland-Lang:2018bxd}.
We will discuss this situation, and provide guidelines for the usage of PDFs with MHOUs. 

This paper is broadly divided into two main parts.
In the  first part, we
construct a general formalism for the inclusion of theory uncertainties
and specifically MHOUs in PDF determination, and show how to construct and 
validate a theory covariance matrix.
In the second part, we
perform a first investigation of the phenomenological implications of
these theory uncertainties.
The structure of the paper is the following: in
Sect.~\ref{sec:thcovmat} we show, using a Bayesian approach,
that under certain assumptions any type of theory uncertainty can be included as a contribution
to the covariance matrix.
In Sect.~\ref{sec:scalevarn} we summarize
the theory of scale variation and use it to review,
compare  and systematize different definitions which have been used
in the literature.
In Sect.~\ref{sec:prescriptions} we then formulate
a number of ``point prescriptions'' for the theory covariance matrix, both for a single process, and also to account for 
correlations between a pair of processes.
In Sect.~\ref{sec:results} we compute the theory covariance matrix for
a variety of prescriptions, we test them against known higher
order corrections, and use this comparison to select an optimal prescription.

We then move to the second, more phenomenological, part of the paper.
The centerpiece of this section is the determination of NLO PDF sets
with MHOU, presented  in Sect.~\ref{sec:fitstherr}.
We first only include deep-inelastic
scattering data (DIS-only fit), and then adopt a global data set, which is
compared to PDFs without MHOU, and validated against NNLO PDFs.
In Sect.~\ref{sec:pheno} we present initial studies of the phenomenological impact of the inclusion of MHOUs
in PDFs for representative LHC processes.
Finally in Sect.~\ref{sec:usage} we provide guidelines for the usage of PDFs
with MHOU, in particular concerning the combination of the PDF uncertainties
with the MHOU on the hard matrix element, and present the delivery of the
PDF sets produced in this work.

Two appendices contain further studies and technical details.
In Appendix~\ref{sec:diagonalization} we provide
additional details concerning the procedure adopted
to diagonalise  the theory covariance matrix.
Then in Appendix~\ref{sec:fitsscalesvar} we study another possible validation
of the results of Sect.~\ref{sec:fitstherr}, by comparing PDFs with MHOUs to
the PDFs obtained by adopting different choices of 
renormalization and factorization scales in the PDF determination.
Families of fits which
only differ in choices of scale have never been carried
out before and will be presented here for the first time.
Whereas they do  not necessarily give a fair estimate of the
MHOU on PDFs, they surely do provide an indication of the expected
impact of scale variation on PDFs, and the pattern of MHOU correlations.

A concise discussion of the main results of this work was
presented in Ref.~\cite{AbdulKhalek:2019bux}, of
which this paper represents the extended companion.

\section{A theoretical covariance matrix}
\label{sec:thcovmat}

\newcommand\True{{\cal T}}

Parton distribution functions are determined from a set of $N_{\rm dat}$
experimental data points, which we represent by an $N_{\rm dat}$-dimensional
vector $D_i$, $i=1,\ldots,N_{\rm dat}$.
These data points have experimental
uncertainties that may be correlated with each other, and this
information is encoded
in an experimental covariance matrix $C_{ij}$.
This covariance matrix may be
block-diagonal if some sets of data are uncorrelated.
Each experimental data point has associated with it a
``true'' value $\True_i$ --- the value given by Nature ---  whose
determination is the goal of the experiment. Since the 
experimental measurements are imperfect, they cannot determine $\True$ 
exactly, but they can be used to estimate the Bayesian probability 
of a given hypothesis for $\True$. Assuming that the
 experimental results are Gaussianly distributed about this hypothetical true
 value, the conditional probability for the true values $\True$ given
 the measured cross-sections
$D$ is
\begin{equation}\label{eq:PDT}
P(\True|D)=P(D|\True) \propto \exp\big(-\half (\True_i-D_i)C_{ij}^{-1}(\True_j-D_j)\big),
\end{equation}
up to an overall normalization constant. Note that this tacitly assumes equal priors for both $D$ and $\True$.

Of course the true values $\True_i$ are unknown.
However we can
calculate theoretical predictions for each data point $D_i$, which we
denote by $T_i$.
These predictions are computed using a theory framework which is
generally incomplete: for example because it is based on the
fixed-order truncation of a perturbative expansion, or because it excludes 
higher-twist effects, or nuclear effects, or some other effect that 
is difficult to calculate precisely.
Furthermore, these theory predictions $T_i$ depend on PDFs,
evolved to a suitable scale also using incomplete theory.
While the theory predictions
may correspond to a variety of different observables and processes, 
they all depend on the same underlying (universal) PDFs.

We now
assume, in the same spirit as when estimating experimental
systematics, that the true values $\True_i$ are centered on the theory
predictions $T_i$, and Gaussianly distributed
about the theory predictions, with which they would coincide 
if the theory were exact and the PDFs were known with certainty.
The conditional probability for the
true values $\True$ given theoretical predictions $T$  is then
\begin{equation}\label{eq:PT}
P(\True|T)=P(T|\True) \propto \exp\big(-\half (\True_i-T_i)S_{ij}^{-1}(\True_j-T_j)\big),
\end{equation}
again up to a normalization constant, where $S_{ij}$ is a ``theory
covariance matrix'', to be estimated in due course. 

PDFs are determined by  maximizing the probability of the theory given
the data $P(T|D)$, marginalised over the true values $\True$ which of
course remain unknown.
Now using Bayes' theorem 
\begin{equation}
  \label{eq:Bayes}
P(\True|DT)P(D|T)=P(D|\True T)P(\True|T) \, .
\end{equation}
Moreover, since the experimental data do not depend on 
the theorists' calculations $T$, but only on the `truth' $\True$,              
\begin{equation}\label{eq:indy}
P(D|\True T)=P(D|\True).
\end{equation} 
Then because by construction $\int \!D^{N}\True\, P(\True|TD)=1$,
\begin{equation}\label{eq:margin}
P(D|T) = \int \!D^{N}\True\, P(\True|D)P(\True|T) \, ,
\end{equation}
where the $N$-dimensional integral is over all of the possible values
of $\True_i$. The probability of the experimental data $D$ is now conditional 
on the theory $T$ because we have marginalised over the underlying `truth' 
$\True$, which is common to both.

Writing the difference between the true $\True_i$ and the actual $T_i$ values
of the theory prediction as
\be
\label{eq:shiftsTheory}
\Delta_i \equiv {\cal T}_i-T_i \, ,
\ee
we can change variables of integration
to convert the integral over $\True_i$ into an
integral over the shifts $\Delta_i$: using the Gaussian hypotheses 
Eqns.~(\ref{eq:PDT}) and~(\ref{eq:PT}), Eq.~(\ref{eq:margin}) becomes 
 that
\begin{equation}\label{eq:gauss}
  P(D|T) \propto \int \!D^{N}\Delta\, \exp\big(-\half  \lp D_i-T_i-\Delta_i\rp C_{ij}^{-1} \lp
  D_j-T_j-\Delta_j\rp -\half\Delta_i S_{ij}^{-1}\Delta_j\big).
\end{equation}
The Gaussian integrals can now be performed explicitly.
Adopting a vector notation in order to make the algebra more
transparent, we rewrite the exponent as 
\bea
  &&  (D - T - \Delta)^T C^{-1} (D - T - \Delta) + \Delta^T S^{-1} \Delta \qquad \qquad \qquad \\  \nonumber &=& \Delta^T (C^{-1} + S^{-1})\Delta - \Delta^T C^{-1} (D - T) - (D - T)^T C^{-1} \Delta + (D - T)^T C^{-1} (D - T) \\ \nonumber &=& (\Delta - (C^{-1} + S^{-1})^{-1} C^{-1} (D - T))^T (C^{-1} + S^{-1}) (\Delta - (C^{-1} + S^{-1})^{-1} C^{-1} (D - T)) \\ \nonumber &-& (D - T)^T C^{-1} (C^{-1} + S^{-1})^{-1} C^{-1} (D - T) + (D - T)^T C^{-1} (D - T),
\label{eq:square}
\eea
where we used the fact that both $C$ and $S$ are symmetric matrices, and in
the last line we completed the square. Integrating over $\Delta$,
ignoring the normalization, Eq.~(\ref{eq:gauss}) then becomes 
\begin{equation}
	P(T|D)=P(D|T) \propto \exp\big(- \half (D - T)^T (C^{-1} - C^{-1} (C^{-1} + S^{-1})^{-1} C^{-1}) (D - T)\big)\,.
	\label{eq:PDx}
\end{equation}
However
\begin{equation}
    (C^{-1} + S^{-1})^{-1} = (C^{-1} (C + S) S^{-1})^{-1} = S (C + S)^{-1} C,
\end{equation}
so that
\begin{align}
\begin{split}
    C^{-1} &- C^{-1} (C^{-1} + S^{-1})^{-1} C^{-1} = C^{-1} - C^{-1} S (C + S)^{-1} \\ &= (C^{-1} (C + S) - C^{-1} S) (C + S)^{-1} = (C + S)^{-1}.
\end{split}
\end{align}
Restoring the indices, we thus find the simple result
\begin{equation}
	P(T|D) \propto \exp\big(- \half (D_i - T_i) (C + S)_{ij}^{-1} (D_j - T_j)\big).
\label{eq:PDCS}
\end{equation}

Comparison of Eq.~(\ref{eq:PDCS}) with Eq.~(\ref{eq:PDT}) indicates that
when replacing the true $\True_i$ by the theoretical predictions
$T_i$ in the expression of the  $\chi^2$ of the data, the
theoretical covariance matrix $S_{ij}$ should simply be added
to the experimental covariance matrix $C_{ij}$ \cite{Ball:2018odr}.
In effect this implies that, at least within this
Gaussian approximation, when determining PDFs theoretical uncertainties 
can be treated simply as
another form of experimental systematic: it is an additional
uncertainty to be taken into account when trying to find the truth
from the data on the basis of a specific theoretical prediction.
The experimental and theoretical uncertainties are added in quadrature
because they are in principle uncorrelated. 

In the case for which theoretical uncertainties can be neglected, 
i.e. if $S_{ij}\to0$,
then $P(\True|T)$ in Eq.~(\ref{eq:PT}) becomes proportional
to $\delta^N(\True_i-T_i)$.
As  a result, in this case Eq.~(\ref{eq:PDCS}) reduces
 to Eq.~(\ref{eq:PDT}) with $\True_i$ replaced by the
predictions $T_i$. This shows that Eq.~(\ref{eq:PDCS})
remains true even if $S_{ij}$ has zero eigenvalues and is thus not
invertible.
Note however
that by construction $C_{ij}$ is positive definite, since any
experimental measurement always
has uncorrelated statistical uncertainties due to the finite number of events,
so $(C+S)_{ij}$ will always be invertible. 
 
The question remains of how to estimate the theory covariance matrix,
$S_{ij}$.
The Gaussian hypothesis Eq.~(\ref{eq:PT}) implies that
\begin{equation}
            S_{ij} = \big\langle(\True_i - T_i)(\True_j - T_j)\big\rangle=\big\langle\Delta_i\Delta_j\big\rangle,\label{eq:sdef}
      \end{equation}
where the average is taken over the true theory values $\True$ using the probability
distribution $P(\True|T)$, and 
$\langle\Delta_i\rangle=0$ consistent with the assumption that the probability 
distribution of the truth $\True$ is centred on the theoretical calculation $T$.
In practice however the formal definition
Eq.~(\ref{eq:sdef}) is not very helpful: we need some way to
estimate the shifts $\Delta_i$ --- `nuisance parameters', in the
language of systematic error determination --- in a way that takes into
account the theoretical correlations between different kinematic points
within the same dataset, between different datasets measuring
the same physical process, and between datasets corresponding to different
processes (with initial state hadrons).
Note that theory correlations will always be present even for entirely 
different processes, through the universal parton distributions: the only processes with truly independent theoretical uncertainties are those with only leptons in the initial state, which are of course irrelevant for PDF determination.

The most commonly used method of estimating the theory
corrections due to MHOUs, which can naturally incorporate all these
theoretical correlations, is scale variation.
This method is reviewed in Sect.~\ref{sec:scalevarn} in general terms and 
then used in Sect.~\ref{sec:prescriptions} in order to formulate
specific prescriptions for constructing the theory covariance matrix $S_{ij}$.
Other approaches which have been
discussed in the literature involve estimating MHOUs based on the
behaviour of the known perturbative
orders~\cite{Cacciari:2011ze,David:2013gaa,Bagnaschi:2014wea};
however, at least at present, these do not appear to provide a
formalism which is sufficiently well-established, and of appropriately
general applicability.
We emphasize however that the formalism presented in this section is independent of the specific method
adopted to estimate the correlated theory shifts $\Delta_i$ that enter Eq.~(\ref{eq:sdef}).

\section{MHOUs from scale variations}
\label{sec:scalevarn}

The variation of the renormalization and factorization scales is the most popular 
approach for estimating missing higher order uncertainties (MHOUs) in
QCD perturbative calculations.
It has a number of 
advantages: it naturally incorporates renormalization group (RG) invariance, 
thereby ensuring that as the perturbative order increases, estimates of MHOU 
decrease; the same procedure can be used for any perturbative process, since 
the scale dependence of the strong coupling $\as(\mu^2)$ and of PDFs
is universal; the 
estimates of MHOU it produces are smooth functions of the kinematics, and 
thereby correctly incorporate the strong correlations in nearby regions of 
phase space; and correlations between different processes due to universal ingredients 
such as PDFs can be easily incorporated.
Its drawbacks are also well known: 
there is no unique principle to determine the specific range of the scale variation (nor 
even the precise central scale to be adopted); and it misses uncertainties 
associated with new singularities or color structures present at higher orders but 
missing at lower orders.
The former problem may be dealt with, at least 
qualitatively, by validating a given range in situations where the next order 
corrections are known.
We will attempt such a validation in this paper.
The  latter problem is more challenging, requiring resummation in the case of 
unresummed logarithms, or other methods of estimating new types of
corrections, and it is unclear whether or not it admits a general solution.

While scale variation has been discussed many times in a variety of
contexts, there is no standard, commonly accepted formulation of it,
and specifically none that can be applied to both electroproduction
and hadroproduction processes, as we need to do if we wish to use
scale variation in the context of global PDF analyses.
In fact, it
turns out that the most commonly adopted approaches to scale variation
differ, typically according to the nature of the process which is
being considered, though also as a function of time, with different
prescriptions being favored in the past than those in common use 
at the present.
Moreover, even the terminology is not uniform: it has  evolved over time, 
resulting in the same names being used for what are essentially different 
scale variations.

To formulate prescriptions for the general use of scale
variation for MHOU estimation which can be applied to any process
included in present or future PDF determinations, it is thus necessary
to first review the underpinnings of scale variation, and to then use
them in order to set up a generally applicable formalism.
This
will be done in the current section, by specifically discussing the cases of
electroproduction and hadroproduction.
In particular, we will show that
for factorized processes MHOUs on  the partonic cross-sections and on
perturbative evolution are independent and can be
estimated through independent scale variations. We will then discuss 
how they can be combined, first with a single process and then for 
several processes, both correlated and uncorrelated. 

\subsection{Renormalization group invariance}
\label{rgi_sec}

The basic principle of scale variation is based
on the observation that scale-dependent
contributions to a perturbative prediction are fixed by
RG invariance, and therefore scale variation can be
used to generate higher order contributions, which are then taken as a
proxy for the whole missing higher orders.

More explicitly, consider
a generic theoretical prediction (typically a perturbative
cross-section) of the form $\overline{T}(\alpha_s(\mu^2), \mu^2/Q^2)$,
where $\mu^2$ is the renormalization scale and $Q^2$ is some physical
scale in the process. Thus $\overline{T}$ indicates the theory prediction 
$T$ when it is evaluated at some renormalization scale $\mu^2$ instead
of being evaluated at the physical scale $Q^2$: if we instead set 
$\mu^2=Q^2$, then
\begin{equation}
\label{xseccent}
T(Q^2) \equiv \overline{T} \lp \alpha_s(Q^2), 1 \rp \, .
\end{equation} 
The QCD running coupling $\alpha_s(\mu^2)$ satisfies the RG
equation 
\begin{equation} \label{1.1}
	\mu^2 \frac{d}{d \mu^2} \alpha_s(\mu^2) = \beta(\alpha_s(\mu^2)) \, ,
\end{equation}
where the QCD beta function 
has the following perturbative expansion:
\be
\label{eq:betafunctionQCD}
\beta(\alpha_s) = \beta_0 \alpha_s^2 + \beta_1 \alpha_s^3 
+ \beta_2 \alpha_s^4 + \ldots \, .
\ee 
RG invariance is the statement that
the all-order prediction is independent of the renormalization
scale: 
\begin{equation} \label{1.2}
  \mu^2 \frac{d}{d \mu^2} \overline{T} \lp
  \alpha_s(\mu^2), \mu^2/Q^2\rp  = 0 .
\end{equation}

It will be useful in what follows to define the variables
\begin{equation}\label{notn}
\mu^2 = k Q^2,\qquad t = \ln (Q^2 / \Lambda^2), \qquad \kappa = \ln k = \ln \mu^2/Q^2,
\end{equation}
so $\alpha_s(\mu^2)$ is a function of $\ln \mu^2/\Lambda^2 = t +
\kappa$.
We can then write  the RG equation (\ref{1.2})  as
\bea \label{1.3}
	0 & =& \frac{d}{d \kappa} \overline{T}(\alpha_s(t + \kappa), \kappa) \nonumber\\
	& =& \frac{d} {d \kappa} \alpha_s(t + \kappa) \frac{\partial}{\partial \alpha_s} \overline{T}(\alpha_s(t + \kappa), \kappa) \bigg|_\kappa + \frac{\partial}{\partial \kappa} \overline{T}(\alpha_s(t + \kappa), \kappa) \bigg|_{\alpha_s}\nonumber \\
	& =& \frac{\partial}{\partial t} \overline{T}(\alpha_s(t + \kappa), \kappa) \bigg|_\kappa + \frac{\partial}{\partial \kappa} \overline{T}(\alpha_s(t + \kappa), \kappa) \bigg|_{\alpha_s} \, ,
\eea
where in the second line we assume that $\overline{T}$ is analytic in
$\alpha_s$ and $\kappa$, and in the third we use 
\begin{equation} \label{1.4}
	\frac{d}{d \kappa} \alpha_s(t + \kappa) = \frac{d}{dt} \alpha_s(t + \kappa) = \beta(\alpha_s(t + \kappa) ) \, .
\end{equation}
Taylor expanding $\overline{T}(\alpha_s, \kappa)$ in $\kappa$ about $\kappa=0$ (i.e. $k=1$, $\mu^2=Q^2$) at fixed coupling $\alpha_s$,
\bea
  \label{1.5}
  \overline{T}(\alpha_s(t + \kappa), \kappa) &=& \overline{T}(\alpha_s(t + \kappa), 0)\nonumber\\&&\qquad\qquad +\kappa \frac{\partial}{\partial \kappa} \overline{T}(\alpha_s(t + \kappa), 0) \bigg|_{\alpha_s} + \half \kappa^2 \frac{\partial^2}{\partial \kappa^2} \overline{T}(\alpha_s(t + \kappa, 0)\bigg|_{\alpha_s} + \ldots \qquad \nonumber \\
	&=& \overline{T}(\alpha_s(t + \kappa), 0) - \kappa \frac{\partial}{\partial t} \overline{T}(\alpha_s(t + \kappa), 0) \bigg|_\kappa + \half \kappa^2  \frac{\partial^2}{\partial t^2} \overline{T}(\alpha_s(t + \kappa), 0)\bigg|_\kappa + \ldots\, ,
\eea
where in the second line we use the RG invariance condition,
Eq.~(\ref{1.3}), to replace $\frac{\partial}{\partial \kappa}$ with
$-\frac{\partial}{\partial t}$.
We can thus determine the $\kappa$
dependence of $\overline{T}(\alpha_s, \kappa)$ using the dependence of
$T(t)=\overline{T}(\alpha_s(t), 0)$ on $t$:
\begin{equation} \label{1.6}
	\overline{T}(\alpha_s(t + \kappa), \kappa) = T(t + \kappa) - \kappa \frac{d}{dt} T(t + \kappa) + \half \kappa^2  \frac{d^2}{dt^2} T(t + \kappa)+\ldots\>.
\end{equation}

Now since
\begin{equation} \label{1.7}
	\frac{d}{dt} T(t) = \frac{d \alpha_s(t)}{dt} \frac{\partial}{\partial \alpha_s} \overline{T}(\alpha_s(t), 0) = \beta(\alpha_s(t)) \frac{\partial}{\partial \alpha_s} \overline{T}(\alpha_s(t), 0),
\end{equation}
and $\beta(\alpha_s) = \mathcal{O}(\as^2)$, we see that 
$\frac{1}{T} \frac{dT}{dt}
= \mathcal{O}(\alpha_s)$, while $\frac{1}{T} \frac{d^2T}{dt^2} =
\mathcal{O}(\alpha_s^2)$ etc.: derivatives with respect to $t$ always
add one power of $\alpha_s$. It follows that in Eq.~(\ref{1.6}), the term
$\mathcal{O}(\kappa)$ is $\mathcal{O}(\alpha_s)$ with respect to the
leading term, and the term $\mathcal{O}(\kappa^2)$ is
$\mathcal{O}(\alpha_s^2)$ with respect to the leading term, and so
on.
We thus see explicitly that
the scale-dependent terms (those that depend on $\kappa$), 
at a given order in perturbation theory, are
determined by derivatives of the cross-section lower down the
perturbation series.  

This implies that if we know the cross-section $T(t)$ as a function of the central 
scale $Q^2$ to a given order in perturbation theory, we can then use 
Eq.~(\ref{1.6}) to determine the scale-dependent 
$\kappa$ terms directly from $T(t)$ at any given order,  by 
differentiating terms lower down the perturbative expansion.
For instance, truncating at LO, NLO, or NNLO, one has
\begin{align} \label{1.8}
\begin{split}
	\overline{T}_{\text{LO}}(\alpha_s(t + \kappa), \kappa) & = T_{\text{LO}}(t + \kappa), \\
	\overline{T}_{\text{NLO}}(\alpha_s(t + \kappa), \kappa) & = T_{\text{NLO}}(t + \kappa) - \kappa\smallfrac{d}{dt}{T}_{\text{LO}}(t + \kappa), \\
	\overline{T}_{\text{NNLO}}(\alpha_s(t + \kappa), \kappa) & = T_{\text{NNLO}}(t + \kappa) - \kappa\smallfrac{d}{dt}{T}_{\text{NLO}}(t + \kappa) + \half\kappa^2  \smallfrac{d^2}{dt^2}{T}_{\text{LO}}(t + \kappa).
\end{split}
\end{align}
The differentiation may be performed analytically,
which is trivial for a fixed 
order expansion, or numerically, which can be useful in a resummed expression
where the dependence on $\as(t)$ can be
nontrivial~\cite{Altarelli:2008aj}.
Note that when the renormalization scale coincides
with the physical scale of the process, $\mu^2=Q^2$, then $\kappa=0$
and $\overline{T}=T$ at every order in the perturbative expansion.

The MHOU can now be estimated as the difference between the scale
varied cross-section and the cross-section evaluated at the central
scale, namely
\begin{equation}\label{1.9}
\Delta(t,\kappa) = \overline{T}(\alpha_s(t + \kappa), \kappa) - T(t) \, .
\end{equation}
Thus at LO, NLO and NNLO we have, using Eq.~(\ref{1.8}), that the theory nuisance
parameters are given by
\begin{align}
  \label{1.10}
\begin{split}
\Delta_{\text{LO}}(t,\kappa) & = T_{\text{LO}}(t + \kappa)-T_{\text{LO}}(t), \\
{\Delta}_{\text{NLO}}(t,\kappa) & = (T_{\text{NLO}}(t + \kappa) - \kappa\smallfrac{d}{dt}{T}_{\text{LO}}(t + \kappa))-T_{\text{NLO}}(t), \\
{\Delta}_{\text{NNLO}}(t, \kappa) & = (T_{\text{NNLO}}(t + \kappa) - \kappa\smallfrac{d}{dt}{T}_{\text{NLO}}(t + \kappa) + \half\kappa^2  \smallfrac{d^2}{dt^2}{T}_{\text{LO}}(t + \kappa))-T_{\text{NNLO}}(t) \, .
\end{split}
\end{align}
One finds that while at LO the theory uncertainty is entirely due to the scale chosen
for $\as$, at NLO the dependence on scale is milder since the leading
dependence is subtracted off by the $O(\kappa)$ term.  At NNLO it
is milder still, since the $O(\kappa)$ term subtracts the leading
dependence in the first term, and the $O(\kappa^2)$ removes the
subleading dependence in the first two terms.
RG
invariance then guarantees that the terms generated by scale variation
are always subleading, so if the perturbation series is well
behaved, the theory shifts $\Delta$ becomes smaller and smaller as the order of the
expansion is increased. 

Clearly the size of the MHOU, estimated in this way, will depend on
the size of the scale variation, and thus on the value chosen for
$\kappa$. Typically one varies the renormalization scale by a factor of two in each
direction, i.e. $\kappa\in[-\ln 4,\ln4]$, since this range is
empirically found to yield sensible results for many processes.
However, in principle, one should treat $\kappa$ as a free parameter,
whose magnitude needs to be validated whenever possible by comparing
to known higher order results.

In the present work, we are specifically interested in the application of this
method to processes
with one or more hadrons in the initial state, i.e. to
cross-sections factorized into a  hard
cross-section convoluted with a PDF or a parton luminosity.
There are then two independent sources of MHOU: the perturbative
expansion of the hard partonic cross-section, and the perturbative expansion of
the anomalous dimensions that determine the perturbative evolution of the
parton distributions.
It is convenient to obtain each of these from an independent scale
variation, and this can be done by writing separate RG
equations for the hard cross-section and for the PDF, as we will
demonstrate below.
This approach is completely equivalent to the perhaps more familiar point
of view in which MHOUs on perturbative evolution are instead obtained
by varying the scale at which the PDF is evaluated in the factorized
expression, as we will also show.

We will begin
by considering the MHOU in the hard-scattering
partonic cross-sections; we will then turn to a discussion of
MHOUs in the PDF evolution, and show that the latter can be obtained
by several equivalent procedures.
We will then discuss how both scale variations can be obtained
from double scale variation of the hard cross-section, and how this
fact also offers the possibility of performing scale variation in
alternative ways whereby these two sources of MHOU are mixed.
We will discuss these for completeness, since in the past scale 
variations were often performed in this way.
Finally, we will address scale
variations and their correlations when several processes are considered at once.

\subsection{Scale variation for partonic cross-sections}
\label{hard_xsec_sec}

We start by considering scale variation in hard-scattering
partonic cross-sections, first in the case of electroproduction
(that is, for lepton-proton deep-inelastic scattering, DIS), and then for the case of
hadroproduction (proton-proton or proton-antiproton collisions).

\subsubsection{Electroproduction}

Consider first an electroproduction process, such as DIS, with an associated structure function given by
\begin{equation} 
    {F}(Q^2) = {C}(\as(Q^2))\otimes f(Q^2) \, ,
\label{2.4}
\end{equation}
where $\otimes$ is the convolution in the momentum fraction $x$ between the perturbative
coefficient function $C(x,\as)$ and the PDF $f(x,Q^2)$, and where the sum over parton
flavors is left implicit.
In Eq.~(\ref{2.4}) both $\as$ and the PDF
are evaluated at the physical scale of the process, so nothing depends
on unphysical renormalization or factorization scales.
We
can determine the MHOU associated with the structure function $F$ due to the truncation of the
perturbative expansion of the coefficient function by fixing the
factorization scheme and keeping fixed the scale
at which the PDF is evaluated (usually referred to as factorization
scale), but varying the renormalization scale used in the computation
of the coefficient function itself.

The scale-dependent structure function $\overline{F}$ will then be given by
\begin{equation}
    \overline{F}(Q^2, \mu^2) = \overline{C}(\alpha_s(\mu^2), \mu^2/Q^2)\otimes f(Q^2)\, ,
\label{2.5}
\end{equation}
where $\mu^2$ is the renormalization scale used in the computation of the coefficient function, or equivalently by 
\begin{equation}
    \overline{F}(t, \kappa) = \overline{C}(\alpha_s(t + \kappa), \kappa)\otimes f(t),
\label{2.5a}
\end{equation}
where as in Eq.~(\ref{notn}) we are using
the notation $t=\ln Q^2/\Lambda^2$ and $\kappa = \ln \mu^2/Q^2$.
Note that in Eq.~(\ref{2.5}) the structure function is written as a
function of $\mu^2$ in the sense of the RG equation~(\ref{1.2}): the
dependence on $\mu^2$ cancels order by order, and the residual
dependence can be used to estimate the MHOU.

In phenomenological applications, it
is more customary to  write $F(Q^2)$, i.e.  {\it not} to write the
dependence of $F$ on $\mu^2$, thereby emphasizing the renormalization
scale independence of the physical observable, and just to indicate
the scale dependence of the hard coefficient function
$\overline{C}(\alpha_s(\mu^2), \mu^2/Q^2)$. Here and in the sequel we
will stick to the  notation used in RG equations
since we wish to emphasize that, as the
scale is varied, we are
dealing with a one-parameter family of theory predictions for the
physical (RG invariant) observable, which all coincide to the accuracy
at which they are calculated but which differ by higher order terms.

Now, the RG invariance of physical cross-sections, and therefore
of the structure function $F$, requires RG
invariance of the coefficient function. This is because we are not varying the
factorization scheme, so the PDF is independent of the renormalization scale $\mu$.
It follows that, as in Eq.~(\ref{1.8}),
\begin{equation}
  \label{2.6}
	\overline{C}(\alpha_s(t + \kappa), \kappa) = C(t + \kappa) - \kappa \smallfrac{d}{dt} C(t + \kappa) + \half \kappa^2  \smallfrac{d^2}{dt^2} C(t + \kappa)+\ldots,
\end{equation}
where $C(t) = \overline{C}(\as(t),0)$ is the coefficient function evaluated at 
$\mu^2=Q^2$, and thus $\kappa=0$.
Then, given the perturbative expansion of the coefficient function,
\begin{equation}
C(t) = c_0 + \alpha_s(t) c_1 + \alpha_s^2(t) c_2 + \alpha_s^3(t) c_3 +\ldots, 
\label{2.7}
\end{equation}
its derivatives can be easily evaluated using the beta function expansion
Eq.~(\ref{eq:betafunctionQCD}), 
\begin{equation}
  \label{2.2}
\begin{split}
\smallfrac{d}{dt}{C}(t) & = \alpha_s^2(t) \beta_0 c_1+ \alpha_s^3(t) (\beta_1c_1+2\beta_0c_2) + \ldots,\\
\smallfrac{d^2}{dt^2}{C}(t) & = 2\alpha_s^3(t) \beta_0^2 c_1+ \ldots,
\end{split}
\end{equation}
and we find that the renormalization scale variation of the
coefficient function is 
\begin{equation} \label{2.8}
\begin{split}
\overline{C}(\alpha_s(t + \kappa), \kappa) = c_0 &+ \alpha_s(t + \kappa) c_1 + \alpha_s^2(t + \kappa) (c_2 - \kappa \beta_0 c_1)\\ 
&+\alpha_s^3(t + \kappa)\big(c_3-\kappa (\beta_1c_1+2\beta_0c_2)+ \kappa^2 \beta_0^2 c_1\big) + \ldots \, .
\end{split}
\end{equation}
Again, note that in the case where $\mu^2=Q^2$, and so $\kappa=0$, one recovers
the standard perturbative expansion Eq.~(\ref{2.7}).
We can now find the scale-dependent structure function,
\begin{equation}
 \label{2.9}
\begin{split}
\overline{F}(t, \kappa) = c_0\otimes f(t) &+ \alpha_s(t + \kappa) c_1\otimes f(t) + \alpha_s^2(t + \kappa)  \lp c_2 - \kappa \beta_0 c_1 \rp \otimes f(t)\\ 
&+\alpha_s^3(t + \kappa) \lp c_3-\kappa (\beta_1c_1+2\beta_0c_2)+ \kappa^2 \beta_0^2 c_1 \rp \otimes f(t) + \ldots\>. 
\end{split}
\end{equation}

Note that evaluating these expressions is numerically very
straightforward, in that the scale-varied expression
Eq.~(\ref{2.9}) has the same form, involving the same
convolutions of $c_i$ with $f$, as the
convolution with the PDFs to the given order at the central scale
Eqs.~(\ref{2.4}) and~(\ref{2.7}),  only with rescaled coefficients.
This means there is no need to recompute NNLO corrections, $K$-factors, etc.: all
that is necessary is to change the coefficients in the perturbative expansion at
the central scale according to Eq.~(\ref{2.9}).  

\subsubsection{Hadronic processes}

MHOUs in the partonic hard cross-sections of
hadronic processes can be computed  in the same way as for DIS.
The only additional
 complication is that the physical observable -- typically, a
cross-section $\Sigma$ -- now depends on the convolution of two PDFs:
\begin{equation}\label{H.1}
    \Sigma(t) = H(t)\otimes( {f}(t)\otimes  {f}(t)) \, ,
\end{equation}
 where again the physical scale is $t = \ln (Q^2 / \Lambda^2)$, $H(t)$
 is the partonic hard-scattering cross-section,  the PDFs are convoluted
 together into a parton luminosity $\mathcal{L}=f\otimes f$, and  the sum
 over parton flavors is left implicit.
 Then, varying the
 renormalization scale $\kappa = \ln \mu^2/Q^2$ in the hard
 cross-section, we have 
\begin{equation}\label{H.2}
  \overline{\Sigma}(t,\kappa) = \overline{H} (\as(t+\kappa), \kappa)\otimes(f(t)\otimes f(t)).
\end{equation}
where, just as for electroproduction, for PDFs evaluated at a fixed scale $T$, the
RG invariance tells us that $\overline{H} (\as(t),
\kappa)$ is given in terms of $H(t)$ by Eq.~(\ref{1.6}):  
\begin{equation} \label{1.6a}
	\overline{H}(\alpha_s(t), \kappa) = H(t) - \kappa \smallfrac{d}{dt} H(t) + \half \kappa^2  \smallfrac{d^2}{dt^2} H(t)+\ldots\>.
\end{equation}

If the partonic process begins at $O(\alpha_s^n)$,
with $n=0,1,2,\ldots$, then one can expand the hard cross-section as
follows
\begin{equation} \label{H.3} 
H(t) = \alpha_s^n(t)h_0 + \alpha_s^{n+1}(t)h_1 + \alpha_s^{n+2}(t)h_2+ \ldots\>.
\end{equation}
Then, as in the case of electroproduction, using Eq.~(\ref{eq:betafunctionQCD}) we
can readily evaluate these derivatives,
\begin{equation} \label{H.4}
\begin{split}
\smallfrac{d}{dt}{H}(t) & = n\alpha_s^{n-1}(t)\beta(\alpha_s) h_0 + (n+1)\alpha_s^n(t)\beta(\alpha_s) h_1 + \ldots \\
&= \alpha_s^{n+1} n \beta_0 h_0 + \alpha_s^{n+2} (n \beta_1 h_0 + (n+1) \beta_0 h_1) + \ldots\\
\smallfrac{d^2}{dt^2}{H}(t) & = \alpha_s^{n+2} n(n+1) \beta_0^2 h_0 + \ldots
\end{split}
\end{equation}
so that, putting everything together, the expression for the scale-varied
partonic cross-section to be used to evaluate the scale-varied
hadronic cross-section $\overline{\Sigma}$, Eq.~(\ref{H.2}), will be given by
\bea \label{H.5}
    \overline{H}(\alpha_s, \kappa) &=& \alpha_s^n h_0 + \alpha_s^{n+1} (h_1 - \kappa n \beta_0 h_0) \nonumber\\ &&\qquad+\alpha_s^{n+2} (h_2 - \kappa(n \beta_1 h_0 + (n+1) \beta_0 h_1) + \half \kappa^2 n(n+1) \beta_0^2 h_1) + \ldots .
\eea
This is rather more involved than Eq.~(\ref{2.9}), but
shares the same advantages: the convolutions to be evaluated in
Eq.~(\ref{H.2}) have the same structure as those in Eq.~(\ref{H.1}), so
all that is required to vary the renormalization scale is to modify
their coefficients. 

\subsection{Scale variation for PDF evolution}
\label{ren_pdfs_sec}

The renormalization scale variation described in the previous section
can be used to estimate the MHOU in any partonic cross-section of an
electroproduction or hadroproduction
process evaluated to a fixed order in perturbation theory.
However, when computing  factorized observables
of the form  Eqs.~(\ref{2.4}, \ref{H.1}), an entirely independent source
of MHOU arises from the truncation of the perturbative expansion of the
splitting functions
(or anomalous dimensions in Mellin space) that govern the PDF evolution equations.
We now show that this MHOU
can again be estimated by scale variation; we will also show that
this scale variation can be performed in different ways:
either at the level of the anomalous dimension; or at the level of the
PDFs themselves; or finally at the level of the
hard-scattering partonic coefficient functions, by exploiting the fact that physical
results cannot depend on the scale at which the PDF is evaluated, and
so one may trade the effect of scale variation between the PDF and
the hard coefficient function.

Consider a PDF $f(\mu^2)$, where $\mu$ is  the scale at which the
PDF is evaluated. For simplicity, in this section all the argument is 
presented implicitly assuming a Mellin space formalism, so that 
convolutions are replaced by ordinary products.
Also, indices labeling different PDFs are left implicit,
so our argument applies directly to the
nonsinglet case but can be straightforwardly generalized to the singlet
evolution and to other flavor combinations.

The scale dependence of $f(\mu^2)$ is fixed by the evolution equation
\begin{equation} \label{3.1}
	\mu^2 \frac{d}{d \mu^2} f(\mu^2) = \gamma(\alpha_s(\mu^2)) f(\mu^2)\, ,
\end{equation}
which applies also to the general singlet case assuming that a
sum over parton flavors is left implicit.
The anomalous dimension admits a perturbative expansion of the form
\begin{equation}\label{3.3a} 
\gamma(t) =  \alpha_s(t) \gamma_0 + \alpha_s^2(t) \gamma_1^2  + \alpha_s^3(t) \gamma_2^3 + \cdots .
\end{equation}
Eq.~(\ref{3.1}) can be integrated to give
\begin{equation} \label{3.2}
	f(\mu^2) = \text{exp}\bigg(\int^{\mu^2} \frac{d \mu'^{2}}{\mu'^{2}} \gamma(\alpha_s(\mu'^{2}))\bigg) f_0 \, ,
\end{equation}
where $f_0$ indicates the PDF at the initial scale $\mu_0$.
Of course, the left-hand
side of the equation  is
independent of this initial scale $\mu_0$, so  the dependence  can be left
implicit also on the right-hand side, by not specifying the lower
limit on the integral. In practice, if the PDF $f_0$ were extracted from
data, any change in this scale would be entirely reabsorbed by the fitting
procedure.

We now observe the well-known fact that the anomalous dimension in Eq.~(\ref{3.1})
is a RG invariant quantity, and
therefore the scale on which it depends is physical.
However, this
physical scale can in general be different from the renormalization
scale used to determine the anomalous dimension itself
(e.g. if it were determined through the renormalization 
of a twist-two operator).
We let $\mu^2 = k Q^2$, where as in the
general argument of Sect.~\ref{rgi_sec}, $\mu^2$ is an arbitrary
renormalization scale and $Q^2$ is a physical scale.
We can make $\gamma$ independent of the renormalization
scale order by order in perturbation theory if we define its scale-varied
counterpart in the same way as before
\begin{equation} \label{3.3}
	\overline{\gamma}(\alpha_s(t), \kappa) = \gamma(t) - \kappa
        \smallfrac{d}{dt}{\gamma}(t) + \half \kappa^2
        \smallfrac{d^2}{dt^2}{\gamma}(t) + \cdots ,
\end{equation}
with $\kappa$ given by Eq.~(\ref{notn}) and
$\gamma(t)= \overline{\gamma}(\alpha_s(t), 0)$, so that given
the perturbative expansion Eq.~(\ref{3.3a}) one has that
\bea
  \label{3.4}
    \overline{\gamma}(\alpha_s(t + \kappa), \kappa) &=& \alpha_s(t+\kappa) \gamma_0 + \alpha_s^2(t+\kappa) (\gamma_1 - \kappa \beta_0 \gamma_0) \nonumber\\&&\qquad+ \alpha_s^3(t+\kappa) (\gamma_2 - \kappa (\beta_1 \gamma_0 + 2 \beta_0 \gamma_1) + \kappa^2 \beta_0^2 \gamma_0) + \cdots
\eea
is independent of $\kappa$ up to higher orders terms, order by order.
Note that Eq.~(\ref{3.4}) has the same form as
Eqs.~(\ref{H.3}-\ref{H.5}) (with $n=1$).

We have shown that
variation of the scale on which the anomalous dimension depends
can be used, in the usual way, to generate higher order terms which estimate
MHOUs in the expansion of the anomalous dimension itself. We now show how the
same result can be obtained by scale variation at the PDF level.
Inserting the result Eq.~(\ref{3.4})  in the solution
of the evolution equations for the PDFs, Eq.~(\ref{3.2}), one finds that the evolution
factor can be expressed as
\bea 
  &&\exp\lp \int^{t} dt' \overline{\gamma}(\alpha_s(t' + \kappa), \kappa)\rp =
  \exp \lp \int^{t + \kappa} dt' \overline{\gamma}(\alpha_s(t'), \kappa)\rp \nonumber\\
  &=& \exp\lp \lc \int^{t + \kappa} dt' \gamma(t')\rc  - \kappa  \gamma(t + \kappa) + \half \kappa^2 \frac{d}{dt} {\gamma}(t + \kappa) + \ldots \rp \nonumber\\
 &=& \lc 1 - \kappa \gamma(t + \kappa) + \half \kappa^2
    (\gamma^2(t + \kappa)+\frac{d}{dt}{\gamma}(t + \kappa)) + \ldots
    \rc \exp\lp \int^{t + \kappa} dt' \gamma(t')\rp \ , \label{3.5}
\eea
where in the first line we changed integration variable (ignoring any change in the lower limit of integration), in the second
we used Eq.~(\ref{3.3}), and in the third we expanded the exponential 
perturbatively.
We can now use this result to
determine renormalization scale variation in the evolution directly
from the scale dependence of the PDF, as in
Ref.~\cite{Altarelli:2008aj}.
Defining a scale-varied PDF as
\begin{equation} \label{3.6}
	\overline{f}(\as(t + \kappa), \kappa) = \text{exp}\bigg(\int^t dt' \overline{\gamma}(\alpha_s(t' + \kappa), \kappa)\bigg) f_0 \, ,
\end{equation}
that is, as the PDF obtained by varying the renormalization scale in the
anomalous dimension, then $f(t) = \overline{f}(\as(t), 0)$, and using
Eq.~(\ref{3.5}) we find that
\begin{equation} \label{3.6a}
	\overline{f}(\as(t + \kappa), \kappa) = \lc 1 - \kappa \gamma(t + \kappa) + \half \kappa^2  (\gamma^2(t + \kappa)+\smallfrac{d}{dt}{\gamma}(t + \kappa)) + \ldots \rc\,f(t+\kappa)\, ,
\end{equation}
provided only that any variation of the initial scale $\mu_0$
due to changes in $\kappa$ has been
reabsorbed into the initial PDF $f_0$. 

Eq.~(\ref{3.6a}) is the same as the result obtained from 
varying the scale $\mu^2$ at which the PDF is evaluated about the 
physical scale $Q^2$: just as
in the derivation of Eq.~(\ref{1.6a}), this gives
\begin{equation} \label{3.7}
\begin{split}
	\overline{f}(\as(t + \kappa), \kappa) & = f(t + \kappa) - \kappa\smallfrac{d}{dt} {f}(t + \kappa) + \half \kappa^2  \smallfrac{d^2}{dt^2}{f}(t + \kappa) + ... \\
	& = f(t + \kappa) - \kappa \gamma f(t + \kappa) + \half \kappa^2  \big(\gamma^2 + \smallfrac{d}{dt}\gamma\big) f(t + \kappa) + ..., 
\end{split}
\end{equation}
where in the second line we used the PDF evolution equation,
Eq.~(\ref{3.1}).
Thus there is little point in varying the renormalization scale of the anomalous dimension and the scale at which the PDF is evaluated independently: provided we absorb changes in the initial scale in the initial PDF, and use the linearised solution of the evolution equation, the result (Eq.~(\ref{3.6a}) or Eq.~(\ref{3.7})) is precisely the same.
This is essentially because the PDF $f(t)$ depends on only a single scale. 

Equation~(\ref{3.6a}) indicates that the $\kappa$ dependence can be factorized
out of the PDF.
We can use this property to factor it into the hard-scattering
coefficient
function.
Consider for example electroproduction, whose factorized
structure function is given by Eq.~(\ref{2.4}):
\bea\label{3.7a}
    \widehat{F}(t, \kappa) &=& {C}(t) \overline{f}(\as(t+\kappa),\kappa)\nonumber\\
    &=& {C}(t) \lc
    1 - \kappa \gamma(t + \kappa) + \half \kappa^2  (\gamma^2(t + \kappa)+\smallfrac{d}{dt}{\gamma}(t + \kappa)) + \ldots \rc f(t+\kappa)\nonumber\\
    &\equiv& \widehat{C}(t, \kappa) f(t+\kappa) \, ,
\eea
where in the second line we used the expansion Eq.~(\ref{3.6a}),
and the third line
should be viewed as the  definition of
the scale-varied coefficient function $\widehat{C}(t+\kappa, \kappa)$.
Moreover, given the relation
\be
\frac{d}{dt}\gamma(\alpha_s) = \beta(\alpha_s) \frac{d\gamma}{d\alpha_s} \, ,
\ee
and then using the perturbative expansions of
the beta function $\beta$, the anomalous dimension $\gamma$, and the coefficient function
$C$, Eqs.~(\ref{eq:betafunctionQCD}),~(\ref{3.3a}), and~(\ref{2.7}),
respectively, one finds
\begin{equation}
  \label{3.7c}
	\widehat{C}(t, \kappa) 
	 = c_0 + \alpha_s(t) (c_1 -\kappa\gamma_0) 
   + \alpha_s^2(t)  \lp c_2- \kappa (\gamma_0 c_1 +\gamma_1 c_0)  + \half \kappa^2  \gamma_0(\gamma_0 + \beta_0)c_0)\rp + \ldots \, .
\end{equation}

Note that this result for $\widehat{C}(t, \kappa)$ is not the same as $\overline{C}(t + \kappa, \kappa)$,
Eq.~(\ref{2.8}).
The reason is that 
$\overline{C}(t + \kappa, \kappa)$ is obtained from the
variation of the renormalization scale of the hard coefficient
function, and can be used to estimate the MHOU in the perturbative
expansion of the coefficient function, while $\widehat{C}(t, \kappa)$
is obtained from the variation of the renormalization scale of the
anomalous dimension, and can be used to estimate the MHOU in the
perturbative evolution of the PDF.
We have obtained the former from
RG invariance of the hard cross-section, and the
latter from RG invariance of the anomalous dimension.
However,
Eq.~(\ref{3.7a}) can be equivalently viewed as expressing the fact
that the physically observable structure function cannot depend on the
scale at which the PDF is evaluated in the factorized expression,
usually referred to as factorization scale: provided we absorb changes
in the initial scale in 
the initial PDF, varying the scale of the anomalous dimension is
identical to
varying the scale of the  PDF.

It is customary to refer to the  scale variation which
estimates MHOU in the coefficient function as renormalization scale
variation: this corresponds to evaluating $\overline{C}(t +\kappa, \kappa)$ in
Eq.~(\ref{2.8}).
The scale variation
which estimates MHOU in the
anomalous dimension, and corresponds to $\widehat{C}(t +
\kappa, \kappa)$ in Eq.~(\ref{3.7c}), is usually called instead
factorization scale variation.
This terminology
is used for example by the Higgs Cross-Section working group~\cite{deFlorian:2016spz} and more generally within the context of LHC physics; in the older DIS literature
the same terminology has a somewhat different meaning, as we shall
discuss in Sect.~\ref{double_var_sec} below. 

The previous discussion entails that in practice there are
(at least) three different ways of estimating the MHOU associated
to the PDF evolution in terms of the anomalous dimension at fixed order in
perturbation theory by means of scale variations: 
\begin{description}
\item{(A)} The renormalization scale of the anomalous dimension can be
  varied directly, using Eq.~(\ref{3.4}).
  This approach works well provided that the
  initial PDF $f_0$ is refitted, but
  if it is held fixed care must be taken to absorb scale variations of
  the initial scale into the initial PDF.
  This method was used for DIS
  renormalization scale variations in many older papers, see
  e.g. Refs.~\cite{Martin:1990fq,Virchaux:1991jc,Ridolfi:1999vr}).
  It has the
  disadvantage that it requires refitting the PDF as the scale is varied,
  which is cumbersome for most applications.
  
\item{(B)} The scale at which the PDF is evaluated can be varied,
  either analytically or numerically, using Eq.~(\ref{3.7}). This is
  in many ways the simplest method, as the initial PDF remains
  unchanged, while only the PDF is involved so the result is
  manifestly universal.
  Furthermore it is easily adapted to a variable
  flavor number scheme (VFNS),
  since the MHOUs in the PDFs with different numbers of active
  flavors can each be estimated separately. The numerical method was
  employed in~\cite{Altarelli:2008aj}, in the context of small $x$
  resummation.
  It has the disadvantage that if one wishes to estimate
  the impact on a given physical observable one needs to first generate 
  the scale-varied PDF, before combining it with the hard coefficient function. 
\item{(C)} The scale at which the PDF is evaluated is varied, but the
  compensating scale-dependent terms are factorized into the
  coefficient function using for example Eq.~(\ref{3.7c}).
  This
  factorization scale variation is most commonly used when evaluating
  a new process using an established PDF set, e.g. in studies of LHC
  processes (as in Ref.~\cite{deFlorian:2016spz})
  since it has the
  advantage that it can be implemented directly using an external
  interpolated PDF set (such as provided by
  LHAPDF~\cite{Buckley:2014ana}).
  It has the conceptual disadvantage that the
  universality of the variation is obscured, since the scale dependent
  terms are mixed in the expansion of the coefficient function (this
  is particularly complicated  in a VFNS, where the coefficient functions
  also depend on heavy quark masses), and the practical disadvantage
  that it requires the evaluation of new contributions to the
  coefficient function involving additional convolutions. Also, it can be
  impractical in situations where higher order corrections are difficult 
  to evaluate precisely due to numerical issues.  
\end{description}

Note that whereas these methods are in principle completely
equivalent, they can differ by subleading terms according to the
convention used to truncate the
perturbation expansion.
Indeed, in method (A)
the expansion of the anomalous dimension is truncated, but higher
order terms in the exponentiation may be retained depending on the
form of the solution to the evolution equations adopted; in method (B)
the exponential has been expanded (see Eq.~(\ref{3.5})) so the result
is the same as would be obtained with a linearized solution of the
evolution equation; while in method (C) cross-terms between the
expansion of linearized evolution and coefficient function expansion
have also been dropped (compare Eq.~(\ref{3.7a}) with
Eq.~(\ref{3.7c})).
However, since the differences always involve higher
order terms, each method can be regarded as giving an equally valid
estimate of the MHOU in the perturbative evolution: differences
between methods should be viewed as the uncertainty on the MHOU itself
when estimated by scale variation.

\subsection{Double scale variations} \label{double_var_sec}

We now discuss the combination of the  two independent scale
variations of Sects.~\ref{hard_xsec_sec} and~\ref{ren_pdfs_sec},
respectively estimating MHOUs in the hard cross-section and in
perturbative evolution, thereby deriving master formulae for scale
variation up to NNLO which will then be used in the subsequent sections.
For completeness, we will also discuss
different options for scale variation which have been considered in
the literature, and clarify some terminological mismatches, especially
between the older studies of DIS and the more recent applications to
LHC processes.

\subsubsection{Electroproduction} \label{double_var_sec_DIS}

Consider first
the more general factorization of an electroproduction cross-section,
such as a DIS structure function:
\begin{equation}
  \overline{F}(Q^2, \mu_f^2, \mu_r^2) = \overline{C} \lp
  \alpha_s(\mu_r^2),\ \mu_r^2/Q^2 \rp \otimes \overline{f} \lp
  \alpha_s(\mu_f^2), \mu_f^2/Q^2\rp \, ,
    \label{5.1}
\end{equation}
where here and in the following we adopt the (standard)
terminology  that
we introduced in Sect.~\ref{ren_pdfs_sec}, and the viewpoint which
corresponds to option (B) of that section: $\mu_r$ denotes the
renormalization scale, whose dependence is entirely contained in the
hard coefficient function $\overline{C}$ (as in Eq.~(\ref{2.5})),
and whose variation estimates MHOUs in its expansion; while $\mu_f$ denotes the
factorization scale, whose dependence is entirely contained in the PDF
(as in Eq.~(\ref{3.6})), and whose variation estimates MHOUs in the
expansion of the anomalous dimension (or equivalently the
splitting functions).
In the following, as in Sect.~\ref{ren_pdfs_sec}, we will omit the
convolution as well as the parton indices.

Note that again, as in Eq.~(\ref{2.5}), and then in
Eqs.~(\ref{H.2}),~(\ref{3.3}), and~(\ref{3.7}), the dependence on the scales
$\mu_f$ and $\mu_r$ should be understood in the sense of the RG equation: the structure
function does not depend on them, but as the scales are varied there
remains a subleading dependence which estimates the MHOU.
As
already mentioned, this notation, while standard in the context of RG
equations, is somewhat unusual in the context of  factorization, where
instead it is more customary to omit the scale dependence of the
physical observable.

Given that the structure function $\overline{F}(Q^2, \mu_f^2, \mu_r^2)$ factorizes into
the hard coefficient function and the PDF,
the factorization and renormalization
scales $\mu_f$ and $\mu_r$ can be chosen completely independently; the
scale dependence will also factorize.
Explicitly, we define
\be\label{eq:scaledef} \mu_f^2 =
k_fQ^2 \, , \quad  \mu_r^2 = k_rQ^2\, , \quad {\rm with} \quad  t_f = t + \kappa_f \, , \quad
t_r = t +
\kappa_r \, ,
\ee
and then $\kappa_f=\ln k_f$, $\kappa_r=\ln k_r$.
In terms of these variables, the factorized structure function will be given by
\begin{equation}
    \overline{F}(t, \kappa_f, \kappa_r) = \overline{C}(t_r, \kappa_r)
    \overline{f}(t_f, \kappa_f), 
\end{equation}
where, as in Sects.~\ref{hard_xsec_sec} and \ref{ren_pdfs_sec}, the scale-varied PDF and coefficient
functions are
\begin{equation}
\begin{split}
    &\overline{f} (t_f, \kappa_f) = f(t_f) - \kappa_f
  \smallfrac{d}{dt}{f}(t_f) + \half \kappa_f^2
  \smallfrac{d^2}{dt^2}{f}(t_f) + ... \, , \\ 
    &\overline{C} (t_r, \kappa_r) = C(t_r) - \kappa_r
  \smallfrac{d}{dt}{C}(t_r) + \half \kappa_r^2
  \smallfrac{d^2}{dt^2}{C}(t_r) + ...   \, ,
\end{split}    
\label{5.4}
\end{equation}
where $f(t_f) \equiv \overline{f}(t_f, 0)$ and $C(t_r) \equiv
\overline{C}(t_r, 0)$ stand for the PDF and the coefficient function evaluated
at the central scale, $\mu_f^2=Q^2$ and $\mu_r^2=Q^2$, respectively.
Recalling that $\smallfrac{\partial}{\partial
  t} \sim \mathcal{O}(\alpha_s)$, the structure function is therefore given by
\bea
    \overline{F}(t, \kappa_f, \kappa_r) 
    &=& C(t_r)f(t_f) - \lp \kappa_r \smallfrac{d}{dt}{C}(t_r) f(t_f) +
    \kappa_f C(t_r) \smallfrac{d}{dt}{f} (t_f)\rp  
    + \half\Big( \kappa_r^2 \smallfrac{d^2}{dt^2}{C}(t_r)f(t_f) \nonumber\\
    &&\qquad+ 2\kappa_r \kappa_f
    \smallfrac{d}{dt}{C}(t_r)\smallfrac{d}{dt}{f}(t_f) + \kappa_f^2
    C(t_r) \smallfrac{d^2}{dt^2} f(t_f) \Big) +
    \mathcal{O}(\alpha_s^3) \, .
\label{5.5}    
\eea
From this expression, it follows that
scale variations with respect to $\kappa_f$ can be determined by
taking derivatives with respect to $t_f$ while holding $t_r$ fixed and vice-versa, so one has
\bea
    \overline{F}(t, \kappa_f, \kappa_r)  &=& F(t_f,t_r) - \bigg( \kappa_f\ \frac{\partial F}{\partial t_f}\bigg|_{t_r} + \kappa_r\ \frac{\partial F}{\partial t_r}\bigg|_{t_f} \bigg) \nonumber\\
    &&\qquad+ \half \bigg( \kappa_f^2 \frac{\partial^2 F}{\partial t_f^2}\bigg|_{t_r} +  2 \kappa_f \kappa_r \frac{\partial^2 F}{\partial t_f \partial t_r} +  \kappa_r^2 \frac{\partial^2 F}{\partial t_r^2}\bigg|_{t_f} \bigg) + \cdots \, .    
\eea
In other words, we can think of the two variations as being generated by $\kappa_f
\smallfrac{\partial}{\partial t_f}$ and $\kappa_r
\smallfrac{\partial}{\partial t_r}$ respectively.

We can equivalently treat the factorization scale variation using
method (C) of the previous subsection, and thus
factorize both scale variations into the coefficient function, as done in
Eq.~(\ref{3.7c}).
In the case of
electroproduction, inserting the expansions of Eq.~(\ref{2.7}) in Eq.~(\ref{5.5}) one obtains
\begin{equation}
    \overline{F}(t,\kappa_f, \kappa_r) = \widehat{\overline{C}}(\alpha_s(t_r), \kappa_f, \kappa_r) f(t_f) \, ,
    \label{5.7}
\end{equation}
with now all dependence on $\kappa_r$ and $\kappa_f$ encoded into a redefined coefficient function:
\bea
    \widehat{\overline{C}}(\alpha_s(t_r), \kappa_f, \kappa_r) &\equiv& c_0 +
    \alpha_s(t_r) c_1 - \alpha_s(t_f) \kappa_f\ c_0 \gamma_0  \nonumber\\
   &&\qquad+ \alpha_s(t_r)^2 (c_2 - \kappa_r \ \beta_0c_1) - \alpha_s(t_r)
    \alpha_s(t_f) \kappa_f \ c_1 \gamma_0 \nonumber\\ 
    &&\qquad + \alpha_s^2(t_f)(-\kappa_f \ c_0\gamma_1 + \half
    \kappa_f^2 c_0 \gamma_0(\beta_0 + \gamma_0)) + \cdots\nonumber\\ 
    &=& c_0 + \alpha_s(t_r)(c_1 -\kappa_f \ c_0 \gamma_0) +
    \alpha_s^2(t_r) \big(c_2 - \kappa_r \ \beta_0 c_1 -
    \kappa_f\ (c_1\gamma_0 + c_0\gamma_1) \nonumber\\ 
    &&\qquad + \half
    \kappa_f^2 c_0\gamma_0(\gamma_0 - \beta_0) + \kappa_f \kappa_r
    \beta_0 c_0 \gamma_0\big) + \cdots \label{5.8} 
\eea
up to terms of $\mathcal{O}(\alpha_s^3(t_r))$, given that one can change the scale
that enters the coupling using
\be
\alpha_s(t_f)=\alpha_s(t_r)
+(\kappa_f-\kappa_r)\beta_0\alpha_s^2(t_r)+\ldots \, .
\ee
Note that in the expression for $ \widehat{\overline{C}}$ the coupling constant
is always evaluated at the renormalization scale $\mu_r$, and that for
$\kappa_r=\kappa_f=0$ one gets back the original
perturbative expansion Eq.~(\ref{2.7}).

However, especially in the context of PDF determinations, as opposed to the
situation in which a pre-computed PDF set is being used, it is rather
more convenient
to use either of methods (A) or (B) from Sect.~\ref{ren_pdfs_sec}
when estimating the MHOU in the scale
dependence of the PDF, since this can be done without reference to
any particular process.
We can then determine the universal $\mu_f$
variation by varying the scale in the PDF evolution, as done
for instance in
Eq.~(\ref{3.4}) or Eq.~(\ref{3.7}), while instead the process-dependent $\mu_r$
variation is estimated by varying the renormalization scale in the coefficient
function, as done in Eq.~(\ref{2.8}), or Eq.~(\ref{H.5}) in the case of hadronic
processes. 

Note that since all scale-varied terms ultimately derive from the
scale dependence of the universal QCD coupling $\as(\mu^2)$, it is
reasonable to treat the
independent scale variations of $\mu_f$ and $\mu_r$ symmetrically,
e.g. by varying in the range $|\kappa_f|, |\kappa_r| 
\leq \ln 4$.
Indeed, this symmetry is an advantage  of the
method: we use the same variation for estimating all MHOUs.
Since
$\mu_f$ and $\mu_r$ can each be varied independently, a simple option
is to perform the double scale variations by considering the five  scale choices
$(\kappa_f,\kappa_r)= (0,0),(\pm 
\ln 4,0),(0,\pm \ln 4)$.
We will refer to this as 5-point scale
variation; alternative schemes will be considered in the
next section.

Note finally that if we set the renormalization and factorization
 scales in Eq.~(\ref{5.1}) to be equal to each other, $\mu_f^2=\mu_r^2
= \tilde{\mu}^2$, we have the factorization 
\begin{equation} 
    \widetilde{F}(Q^2, \tilde\mu^2) = \widetilde{C}(\alpha_s(\tilde\mu^2), \tilde\mu^2/Q^2)\ f(\tilde\mu^2) \, .
\label{4.1}
\end{equation}
In most of the earlier papers, mainly concerned
with DIS structure functions,
e.g. \cite{Altarelli:1988qr,Nason:1987xz,Close:1987ay,Martin:1990fq,Virchaux:1991jc},
the scale
$\tilde\mu^2$ was termed the factorization scale:
this  originates in the earliest papers on the OPE. However, in our current
terminology it corresponds to both renormalization and factorization
scales taken equal to each other.
Likewise, in the earlier papers what here we call the factorization scale 
$\mu_f$ was referred to as the renormalization scale.
Here, to avoid
 confusion, we will call $\tilde\mu^2$ in Eq.~(\ref{4.1})  the scale
 of the process.
 For clarity the different nomenclatures for the various scales used in 
the earlier papers, and in more modern work (and in this paper), are 
 summarized in Table~\ref{tab:scale_nomenclature}. 

\begin{table}[t!]
	\centering
	\renewcommand*{\arraystretch}{1.8}
	\begin{tabular}{|c|c|c|c|}
          \toprule
Scale & MHOU &`Traditional' name\cite{Altarelli:1988qr,Nason:1987xz,Close:1987ay,Martin:1990fq,Virchaux:1991jc} & `Modern' name \cite{vanNeerven:2000uj},[PDG]  \\
\midrule
    $\mu_r$& in hard xsec & --- & renormalization scale \\
    $\mu_f$ & in PDF evolution &renormalization scale & factorization scale  \\
      $\widetilde\mu$ & in physical xsec& factorization scale & scale of the process\\
\bottomrule
        \end{tabular}
        \vspace{0.3cm}
	\caption{\small Nomenclatures for the different scale variations used in 
some of the earlier papers (mainly in the context of DIS), and in more recent work (mainly in the context of hadronic processes), as discussed in detail in the text. The `modern' terminology is adopted throughout this paper.}
	\label{tab:scale_nomenclature}
\end{table}

Consider now the effect on the structure function of varying the scale of the process.
As before, we define $\tilde\kappa = \ln\tilde{\mu}^2/Q^2$ and write
\begin{equation}
    \widetilde{F}(t + \tilde\kappa, \tilde\kappa) = \widetilde{C}(\alpha_s(t + \tilde\kappa), \tilde\kappa)\ f(t+ \tilde\kappa) \, .
\label{4.2}
\end{equation}
Now the renormalization group invariance of the cross-section [i.e. Eq.~(\ref{1.2})] requires a cancellation between scale variations in the coefficient function and the PDF: with $F(t) \equiv \widetilde{F}(t, 0)$,
\begin{equation}
\begin{split}
     \widetilde{F}(t + \tilde\kappa, \tilde\kappa) &= F(t + \tilde\kappa) - \tilde\kappa \smallfrac{d}{dt}{F}(t+\tilde\kappa) +\half \tilde\kappa^2  \smallfrac{d^2}{dt^2}{F}(t + \tilde\kappa) + ... \\
     &= Cf - \tilde\kappa(\smallfrac{d}{dt} C + \gamma C)f + \half \tilde\kappa^2  \big(\smallfrac{d^2}{dt^2}{C} + 2\gamma\smallfrac{d}{dt}{C} + C\smallfrac{d}{dt}{\gamma} + C\gamma^2\big)f + ...
\end{split}
\label{4.3}
\end{equation}
where the first line is the same as Eq.~(5.8) in
Ref.~\cite{Altarelli:2008aj} while in the second line we used
Eq.~(\ref{3.7}) for scale variation of the PDF. Then, expanding in the
usual way, we find that 
\begin{equation}
\begin{split}
    \overline{C}(t + \tilde\kappa, \kappa) &= c_0 + \alpha_s(t+ \tilde\kappa)(c_1 -\tilde\kappa c_0 \gamma_0) \\
    &+ \alpha_s^2 ( t + \tilde\kappa)\big(c_2 - \tilde\kappa(\beta_0 c_1 + c_1 \gamma_0 + c_0 \gamma_1) + \half \tilde\kappa^2 \ c_0 \gamma_0 (\beta_0 + \gamma_0)\big)+\cdots
\end{split}
\label{4.6}
\end{equation}
which indeed coincides with the expression for what is referred to
as factorization scale  
variation in this earlier literature: see
e.g. Ref.~\cite{vanNeerven:2000uj}, Eq.~(2.17).
Therefore, varying  
the scale of the process mixes together the scale dependence in the 
coefficient function and the scale dependence in the PDF: indeed, if
in Eq.~(\ref{5.8}) we set
$\kappa_f=\kappa_r=\tilde\kappa$, it reduces to
Eq.~(\ref{4.6}). 

\begin{figure}[b!]
\centering
\begin{tikzpicture}
\draw[->,>=stealth,black] (0,0) -> (2,0);
\draw[->,>=stealth,black] (0,0) -> (-2,0);
\draw[->,>=stealth,black] (0,0) -> (0,2);
\draw[->,>=stealth,black] (0,0) -> (0,-2);
\draw[->,>=stealth, red] (0,0) -> (1.5,1.5);
\draw[->,>=stealth,red] (0,0) -> (-1.5,-1.5);
\node at (0.5,2) {$\kappa_r$};
\node at (2.4,0) {$\kappa_f$};
\node at (1.75,1.75) {$\tilde{\kappa}$};
\end{tikzpicture}
\caption{The two-dimensional space of scale variations for a single
  process: $\kappa_r$ is the renormalization scale (giving the MHOU in
  the hard cross-section), $\kappa_f$ is the factorization scale
  (giving the MHOU in the evolution of the PDF) and $\tilde\kappa$ is
  the variation of the scale of the process (called factorization
  scale variation in the earlier literature), obtained by setting
  $\kappa_f=\kappa_r$. \label{fig:DoubleScaleVar1}} 
\end{figure}
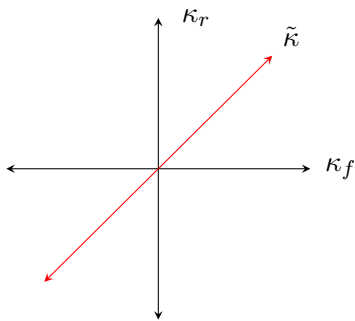

Clearly, variations of $\tilde\mu^2$ are not independent of the
variations of $\mu_f^2$ or $\mu_r^2$: rather they are generated by
$\tilde\kappa\ (\smallfrac{\partial}{\partial t_f} +
\smallfrac{\partial}{\partial t _r})$, so they correspond to directions along
the diagonal in the space of $\kappa_f$ and $\kappa_r$, see Fig.~\ref{fig:DoubleScaleVar1}.
In the earlier literature, MHOUs were estimated by combining
renormalization scale variation with this latter variation, namely by
varying $\tilde\mu^2$ and
$\mu_f^2$: see e.g.
Refs.~\cite{Martin:1990fq,Virchaux:1991jc}. This however has the
disadvantage of generating large scale ratios:
performing variations of
$\tilde\mu^2$ and $\mu_f^2$ sequentially we can obtain $\kappa_f = 2
\ln 4$, because
\be
\tilde\kappa
\lp \frac{\partial}{\partial t_f} + \frac{\partial}{\partial t
  _r} \rp + \kappa_f\ \frac{\partial}{\partial t_f} =
(\tilde\kappa+\kappa_f)\ \frac{\partial}{\partial t_f} +
\tilde\kappa \ \frac{\partial}{\partial t _r} \, .
\ee
A way of
avoiding these large ratios was constructed in
Ref.~\cite{vanNeerven:2000uj}: first do the  scale
variation of Eq.~(\ref{4.6}), but then substitute
\begin{equation}
    c_2  \to c_2 - (\kappa_r - \kappa_f) \beta_f c_1 = c_2 - (\ln \mu_f^2/\mu_r^2) \beta_0 c_1 \, ,
    \label{5.9}
 \end{equation}
where care must  be taken to use the correct argument of $\alpha_s$ in
each term.
Indeed, this procedure then agrees with Eq.~(\ref{5.7}) given that
\be
\kappa_f\ \frac{\partial}{\partial t_f} +
\kappa_r\ \frac{\partial}{\partial t _r} = \kappa_f
 \lp \frac{\partial}{\partial t_f} + \frac{\partial}{\partial t
  _r}\rp + (\kappa_r  -  \kappa_f)\frac{\partial}{\partial t _r} \, .
\ee

\subsubsection{Hadronic processes} \label{double_var_sec_hadr}

Consider now the case of  hadronic process as in  Eq.~(\ref{H.1}).
For these processes, the factorization  has the general form
\begin{equation}
	\overline{\Sigma}(t_f, t_r, \kappa_f, \kappa_r) = \overline{H}(\as(t_r), \kappa_r)\otimes  \lp \overline{f}(t_f, \kappa_f)\otimes\overline{f}(t_f, \kappa_f) \rp \, .
\end{equation}
The hard coefficient function will have the same expansion as
Eq.~(\ref{H.5}). Just as for electroproduction, it is 
possible to factorize variations of $\kappa_f$ into the hard
coefficient functions: then  
\begin{equation}
	\overline{\Sigma}(t_f, t_r, \kappa_f, \kappa_r) = \widehat{\overline{H}}(\as(t_r), \kappa_r,\kappa_f)\otimes ({f}(t_f)\otimes{f}(t_f)),
\end{equation}
where (using as above  Mellin space, to avoid the convolutions), one finds
\bea
    \widehat{\overline{H}} &=& ~\alpha_s^n(t_r) h_0 +
    \alpha_s^{n+1}(t_r)(h_1 - \kappa_r \ \beta_0 h_0) -
    2\alpha_s^n(t_r)\alpha_s(t_f) \kappa_0 \ h_0 \gamma_0 \nonumber\\ 
    &&\qquad + \alpha_s^{n+2}(t_r)\big(h_2 - \kappa_2(n\beta_1h_0 +
    (n+1)\beta_0 h_1) + \half \kappa_2^2 n(n+1) \beta_0^2 h_1\big) \nonumber\\ 
    &&\qquad - \alpha_s^{n+1}(t_r)\alpha_s(t_f)\big(\kappa_0 (h_1 -
    \kappa_2 \beta_0 h_0)2 \gamma_0\big)\nonumber\\ 
    &&\qquad +\alpha_s^n(t_r)\alpha_s^2(t_f)\big(- \kappa_0 h_0
    2\gamma_1 + \half \kappa_0^2h_02 \gamma_0 (\beta_0 +
    2\gamma_0)\big) + \ldots \, .    
\label{6.3}
\eea
However these expressions are even more cumbersome than in the case of
electroproduction,  thereby demonstrating the greater clarity of methods (A) or
(B) in determining the dependence on the scale $\mu_f$.
By adopting one of these two methods,
we can determine the MHOU in a hadronic process
through independent variations of the factorization scale $\mu_f$ and
the renormalization scale $\mu_r$ in just the same way as we estimated
the MHOU in the deep inelastic structure function in the previous section.

\subsection{Multiple scale variations} \label{multiple_var_sec}

We finally consider simultaneous scale variation in a pair of
processes: for instance the electroproduction
process of Sect.~\ref{double_var_sec_DIS} and a hadronic
process as in Sect.~\ref{double_var_sec_hadr}.
Clearly, the
PDF is universal, but the coefficient functions are
process-dependent.
It follows that while the scale variations of 
$\kappa_r$ in the two coefficient functions will be totally
independent, the 
scale variation $\kappa_f$ of the
PDF will be  correlated between these two processes.

The degree of
this correlation is somewhat subtle: indeed, $\kappa_f$ generates MHO
terms in anomalous dimensions, but the anomalous dimension matrix has
several independent eigenvalues (two singlet and one nonsinglet which
at NLO and beyond further splits into C-even and C-odd). Hence in
principle one should introduce an independent factorization
scale variation for each of
these components, which is then fully correlated across all
processes. For the time being, we will perform fully correlated
variations of the factorization scale. This is an approximation, which may not be accurate particularly for processes which depend
on PDFs whose evolution is controlled by different anomalous
dimensions (such as, say, the singlet and the isospin triplet). We
will comment further on this approximation in the sequel.

Now, considering both processes together, we have three
independent scales to vary, $\mu_f$, $\mu_{r_1}$, and $\mu_{r_2}$, where
$\mu_{r_1}$ is the renormalization scale for the deep inelastic process,
and $\mu_{r_2}$ is the renormalization scale for the hadronic process. The
relation of the
factorization scale $\mu_f$ to the physical scale of each process
(whatever that is) is the same for both processes, since the PDFs
are universal.
Thus if we vary all scales independently
by a factor two about their central value we end up with seven
scale choices.
We can think of the additional renormalization scale as an
extra dimension in the space of possible scale variations.

By trivial
generalization for $p$ independent processes $\pi_a$, $a=1,\ldots,p$,
we will have $p+1$ independent scale parameters
$\mu_f,\mu_{r_1},\ldots\mu_{r_p}$ corresponding to a total of 3+2$p$ scale variations.
Writing
$\kappa_{r_a} = \ln \mu_{r_a}^2/Q^2$ with $a= 1,\ldots,p$, the traditional range
of variation of $\kappa_f, \kappa_{r_1}, ..., \kappa_{r_p}$ would then be  defined by
\[
|\kappa_f| \leq \ln 4,\qquad |\kappa_{r_a}| \leq \ln 4,\qquad a=1,\ldots p \, .\label{eq:range}
\] 
Clearly all prescriptions constructed in this way will be symmetrical
in the different scales. 

We now see why, for the determination of MHOUs in PDFs, it is advantageous 
to work with 
the independent scales $\kappa_f$, $\kappa_{r_a}$, $a = 1,\ldots,p$ rather
than with  the traditional factorization scales $\tilde\kappa$ used in the
older treatments of scale variation: while the scale $\kappa_f$ used
to estimate MHOUs in the PDF evolution is universal, the scales
$\kappa_{r_a}$ used to estimate MHOUs in the hard cross-sections are instead
process-dependent.
We can therefore only define process scales $\tilde\kappa$ by
either introducing artificial correlations between the scales of the
hard cross-sections for different processes (which would result in
underestimated MHOU in the hard cross-sections), or else by sacrificing
universality of the PDFs, with uncorrelated evolution uncertainties
for different processes (which would result in overestimated MHOU from
PDF evolution). Neither of these options is very satisfactory, though
we consider the latter briefly in Sect.~\ref{sec_asym}
below, where it gives rise
to asymmetric scale-variation prescriptions.

\section{Scale variation prescriptions for the theory covariance matrix}
\label{sec:prescriptions}

Having set out a general formalism for the inclusion of MHOUs
through a theory covariance matrix, based on assuming a
distribution of shifts between a  theory calculation at finite
perturbative order and the true all--order value (Sect.~\ref{sec:thcovmat}), and having discussed
how scale variation can be used to produce estimates for such shifts (Sect.~\ref{sec:scalevarn}), we
now provide an explicit prescription for the construction of a theory covariance
matrix from scale variation.
Because of the intrinsic arbitrariness involved in
the procedure, we actually propose several alternative prescriptions,
which will be then validated in the next section
by studying cases in which the next
perturbative order is in fact known.
We will also assess the impact at the PDF fit level of varying the prescription
used for constructing the theory covariance matrix.

We consider a situation in which we have $p$ different types of
processes $\pi_a =\{i_a\}$, where $i_a$ labels the data points
belonging to the $a$-th process and $a = 1,\ldots,p$.
Each of the $p$ processes is characterized  by a factorization scale
$\mu_f$ (associated with the PDFs) and a renormalization 
scale $\mu_{r_a}$ (associated with the hard coefficient functions), to be
understood in the sense of the `modern'
terminology in
Table~\ref{tab:scale_nomenclature}.
We will perform scale variation of both scales
following Sect.~\ref{double_var_sec}, by taking them as independent, as
discussed in that section.
When considering a pair of different processes, as
explained in Sect.~\ref{multiple_var_sec}, we assume the variations of
$\mu_{r_a}$ to be uncorrelated among them, while those of $\mu_f$ are taken to be fully
correlated.

The theory covariance matrix is then constructed by averaging outer products of the shifts
with respect to the central scales, given for the $a$-th process as
\begin{equation} \label{P.1}
  \Delta_{i_a} (\kappa_f, \kappa_{r_a} ) \equiv
  T_{i_a}(\kappa_f, \kappa_{r_a}) - T_{i_a}(0,0) \, ,
\end{equation}
over points in the space of scales.
Here, as before, we have defined 
$\kappa_{r_a}=\ln k_{r_a} = \ln \mu_{r_a}^2/Q^2$
and
$\kappa_{f}=\ln k_{f} = \ln \mu_{f}^2/Q^2$.
In Eq.~(\ref{P.1}),
$T_{i_a}(\kappa_f, \kappa_{r_a})$ indicates
the theoretical prediction evaluated at these scales with
$T_{i_a}(0,0)$ being the central theory prediction, and the index
$i_a$ running over all data points corresponding to process $a$.

We assume here that all scale variations correspond to the same range
\[
|\kappa_f| \leq w,\qquad |\kappa_{r_a}| \leq w,\qquad a=1,\ldots, p, \label{eq:rangew}
\] 
for some $w$ (typically $w = \ln 4$, as in Eq.~(\ref{eq:range})).
In practice, in each  prescription  the three points
$\kappa= 0, \pm w$ are sampled for each scale.
The theory covariance matrix is then
\begin{equation} \label{P.2}
  S_{ij} = n_m \sum_{V_m} \Delta_{i_a} (\kappa_f, \kappa_r) \Delta_{i_b} (\kappa_f, \kappa_s)
\end{equation}
where $i_a\ \in\ \pi_a$ and $i_b\ \in\ \pi_b$ indicate two data points,
possibly corresponding to different processes $\pi_a$ and $\pi_b$,
$m$ labels the
prescription, $V_m$ is the set of scale points to be summed over in
the given prescription, and $n_m$ is a normalization factor, both to
be determined.
Different prescriptions to construct the theory covariance matrix $S_{ij}$
vary in the set of  combination of scales which
are summed over in Eq.~(\ref{P.2}), as we will discuss below.

Because Eq.~(\ref{P.2}) is a sum of outer products of shifts, the theory
covariance matrix $S_{ij}$ is positive semi-definite by construction.
To demonstrate this, consider a real vector $v_i$: then it follows that 
 \begin{equation} \label{P.2pos}
 \sum_{ij} v_iS_{ij}v_j = N_m \sum_{V_m}  \lp \sum_i v_i\Delta_{i}\rp^2 \geq 0.
\end{equation}
Note however that because the number of elements of $V_m$ is finite,
$S_{ij}$ will generally be singular, since for any vector $z_j$ which
is orthogonal to the space $S$ spanned by the set of vectors
$\{\Delta_{i_a}(\kappa_f, \kappa_{r_a}): \kappa_f, \kappa_{r_a} \in V_m\}$,
$S_{ij}z_j=0$.
This property will be important when we come to validate the
covariance matrix in the following section, by constructing the set of
orthonormal eigenvectors $e_i^\alpha$ which span the space $S$.

It is interesting to  note that the diagonalization of ${\widehat
S}_{ij}$ can be rephrased in terms of nuisance parameters of the
systematic uncertainties associated with the MHOU. For example,
following the
notation of Appendix A.2 of Ref.~\cite{Ball:2012wy},
the
absolute correlated 
uncertainties $\beta_{i,\alpha}$ may be expressed in terms of the
eigenvalues and eigenvectors of the normalised covariance matrix
${\widehat S}_{ij} = \sum_{\alpha=1}^{N_{\rm sub}} (s^\alpha)^2
e_i^\alpha e_j^{\alpha}$ as
\begin{equation}\label{eq:nuisance}
\beta_{i,\alpha} = T_i^{\rm NLO} s^{\alpha} e_i^\alpha,
\end{equation}
for $\alpha=1,\ldots,N_{\rm sub}$. An algorithm for constructing the
eigenvectors $e_i^\alpha$ from the shifts induced by scale variation
is given in Appendix~\ref{sec:diagonalization}. This way of looking at the theory covariance
matrix might be useful in that the nuisance parameters can be
interpreted in terms of missing higher order contributions. For
instance, the values of the nuisance parameters which optimize the
agreement bwetween data and theory are the most likely guess for MHO
terms which is favored by the data, everything else being equal.

We now consider various prescriptions.
Because $S_{ij}$ will in general span the full set of data points, we must
consider both the case in which points $i,\>j$ in Eq.~(\ref{P.2})
belong to the same process (``single process prescription'') and the
case in which they belong to two different processes  (``multiple
process prescription'').
We first discuss the case of symmetric scale
variation, in which the two scales are varied independently, and then
the case in  which the two scales are varied in a correlated way, the latter scenario being equivalent to varying the ``scale of the process" (in the sense of Table~\ref{tab:scale_nomenclature}), thereby
leading to asymmetric prescriptions as already mentioned in Sect.~\ref{multiple_var_sec}.
 
\subsection{Symmetric prescriptions for individual processes}
\label{sec:sympres}

We  consider first the prescriptions for when there is just a
single process, that is, $p=1$.
In this case, there are at most two
independent scales, the factorization and renormalization scales $\kappa_f$ and $\kappa_r$. The theory covariance matrix is then constructed as
\begin{equation}
  \label{P.3}
  S_{ij} = n_{m} \sum_{v_{m}} \Delta_{i} (\kappa_f, \kappa_r)\Delta_{j} (\kappa_f, \kappa_r)\, ,
\end{equation}
where again $v_m$ represents the set of points to be summed over in the given
prescription, limited here to points in the space of the two scales
$\kappa_f$ and $\kappa_r$, and $n_m$ is the normalization factor.
Let $s$ be
the number of independent scales being varied (so $s=1$ or $s=2$), and
$m$ be the number of points in the variation (so $m$ is the number of
elements of $v_m$): a given scheme is then usually described as an
`$(m+1)$-point scheme'.
Note that we do not include in $v_m$ trivial points
for which $\Delta_{i}$ vanishes (which in practice means the single
point $\kappa_f=\kappa_r=0$), since these do not contribute to the
sum. 

The normalization factor $n_m$ in Eq.~(\ref{P.3})
is determined by averaging over the number of
points associated with the variation of each scale,  and adding the
contributions from variation of independent scales. This means that  
\begin{equation}
  \label{P.4}
n_m = s/m.
\end{equation}

\begin{figure}[t]
\centering
{\begin{tikzpicture}
\draw[->] (-1.5,0) -- (1.5,0);
\draw[->] (0,-1.5) -- (0,1.5);
\filldraw[black] (0,0) circle (2pt);
\filldraw[black] (-1,0) circle (2pt);
\filldraw[black] (0,-1) circle (2pt);
\filldraw[black] (1,0) circle (2pt);
\filldraw[black] (0,1) circle (2pt);
\node at (0.5,1.5) {$\kappa_r$};
\node at (1.9,0) {$\kappa_f$};
\end{tikzpicture}}\qquad
{\begin{tikzpicture}
\draw[->] (-1.5,0) -- (1.5,0);
\draw[->] (0,-1.5) -- (0,1.5);
\filldraw[black] (0,0) circle (2pt);
\filldraw[black] (-1,-1) circle (2pt);
\filldraw[black] (1,1) circle (2pt);
\filldraw[black] (1,-1) circle (2pt);
\filldraw[black] (-1,1) circle (2pt);
\node at (0.5,1.5) {$\kappa_r$};
\node at (1.9,0) {$\kappa_f$};
\end{tikzpicture}}\qquad
{\begin{tikzpicture}
\draw[->] (-1.5,0) -- (1.5,0);
\draw[->] (0,-1.5) -- (0,1.5);
\filldraw[black] (0,0) circle (2pt);
\filldraw[black] (-1,0) circle (2pt);
\filldraw[black] (0,-1) circle (2pt);
\filldraw[black] (1,0) circle (2pt);
\filldraw[black] (0,1) circle (2pt);
\filldraw[black] (-1,-1) circle (2pt);
\filldraw[black] (1,-1) circle (2pt);
\filldraw[black] (-1,1) circle (2pt);
\filldraw[black] (1,1) circle (2pt);
\node at (0.5,1.5) {$\kappa_r$};
\node at (1.9,0) {$\kappa_f$};
\end{tikzpicture}}
\begin{caption}{Symmetric prescriptions for a single process, indicating
    the sampled values for the factorization scale $\kappa_f$ and 
    renormalization scale $\kappa_r$ in each case.
    The origin of coordinates corresponds to the
    central scales $\kappa_f=\kappa_r= 0$.
    We show the three prescriptions 
    $5$-point (left), $\bar{5}$-point (center) and $9$-point (right).
\label{fig:symmetricPrescriptions}
  }
\end{caption}
\end{figure}

We consider three different prescriptions, represented schematically
in Fig.~\ref{fig:symmetricPrescriptions}.
\begin{itemize}
\item \textbf{5-point}: we vary $\kappa_f$ keeping $\kappa_r = 0$ and vice versa, so 
$v_4 = \{(\pm;0), (0; \pm) \}$, where the pairs denote the values of
  the two independent scales $(\kappa_f; \kappa_r)$. Then $s=2$,
  $m=4$, and the normalisation is $n_4 = 1/2$.
  This definition implies that we can average over the two nontrivial values
  of the each scale in turn, and add the results: 
\begin{equation}\label{5S}
    S^{\rm (5pt)}_{ij} = \smallfrac{1}{2}\big\{ \Delta_i^{+0}\Delta_j^{+0} + \Delta_i^{-0} \Delta_j^{-0} + \Delta_i^{0+} \Delta_j^{0+} + \Delta_i^{0-} \Delta_j^{0-}  \big\} \, ,
\end{equation}
where 
we have adopted the abbreviated notation $\Delta_i^{+0}=\Delta_i(+w,0)$, $\Delta_i^{0-}=\Delta_i(0,-w)$, etc. for the shifts. 

Note that the variations of $\kappa_f$ and $\kappa_r$ are added in quadrature since they are
  independent: this is why it is important to make sure that the
  variations are indeed independent, as is the case for
  renormalization and factorization scales, as discussed in Sect.~\ref{double_var_sec}. 

\item \textbf{$\overline{5}$-point}: this is an alternative 5-point
  prescription, which is basically the complement of 5-point:
  $\overline{v}_4 = \{(\pm; \pm) \}$, where $(\pm;\pm)$ are assumed
  uncorrelated, i.e. 4 independent points.
  The counting is the same
  as for 5-point: $s=2$, $m=4$ and again $\overline{n}_4 = 1/2$:  
\begin{equation}\label{5bS}
    S^{(\rm \overline{5}pt)}_{ij} = \smallfrac{1}{2}\big\{ \Delta_i^{++}\Delta_j^{++} + \Delta_i^{--}\Delta_j^{--} + \Delta_i^{+-}\Delta_j^{+-} + \Delta_i^{-+} \Delta_j^{-+}\big\} \, .
\end{equation}
As before, the two scales are varied in a manifestly independent way.

\item \textbf{9-point}: here we vary $\kappa_f$ and $\kappa_r$
  completely independently, giving the union of the 5-point and
  $\overline{5}$-point prescriptions: $v_8=v_4\oplus \overline{v}_4$.  Now $s=2$, $m=8$ and $n_8=1/4$, and
  the theory covariance matrix is given by
\begin{equation}\label{9S}
\begin{split}
    S^{(\rm 9pt)}_{ij} = \smallfrac{1}{4}\big\{ &\Delta_i^{+0} \Delta_j^{+0} + \Delta_i^{-0}\Delta_j^{-0}
                            + \Delta_i^{0+} \Delta_j^{0+} + \Delta_i^{0-}\Delta_j^{0-} \\
                            + &\Delta_i^{++}\Delta_j^{++} + \Delta_i^{+-} \Delta_j^{+-}
                            + \Delta_i^{-+}\Delta_j^{-+} + \Delta_i^{--} \Delta_j^{--} \big\} \, .
\end{split}                            
\end{equation}
\end{itemize}

\subsection{Symmetric prescriptions for multiple processes}
\label{sec:sympresmul}

Now we consider multiple processes, i.e. $p>1$, with scale variations
either uncorrelated or partially correlated.
In Eq.~(\ref{P.2}), the set $V_m$ now involves possible variations
of the $p+1$ scales $\kappa_f$, $\kappa_{r_1},\ldots\kappa_{r_p}$, where $\kappa_{r_a}$ indicates
the renormalization
scale for process $a=1,\ldots,p$.
This implies that now
  $V_m$ is a much bigger set than $v_m$.
  However any given element of $S_{ij}$
in Eq.~(\ref{P.2}) can involve at most two different processes, $\pi_a$
and $\pi_b$, so to compute this element we can simply ignore the other
processes.
Consequently, it is sufficient to consider $p=2$, since
generalization to $p>2$ will then be straightforward.  

For a given pair of processes, say $\pi_1$ and $ \pi_2$, the
covariance matrix has
diagonal elements $S_{i_1j_1}, S_{i_2j_2}$ and off-diagonals
$S_{i_1j_2} = S_{j_2i_1}$, where as above the extra subscript indicates the
process: $i_1,j_1\in\pi_1$, $i_2,j_2\in\pi_2$.
Thus one can write
\begin{equation}
  \label{P.2.2}
  S_{ij} = \left(\begin{array}{cc}
S_{i_1j_1}&S_{i_1j_2}\\ 
S_{i_2j_1}&S_{i_2j_2}\end{array}\right)\, .
\end{equation} 
Consider first the diagonal blocks $S_{i_1j_1}$ and
$S_{i_2j_2}$.
Adding process $\pi_2$ cannot change the theoretical
uncertainty in process $\pi_1$, although the two uncertainties may be
correlated.
Consequently $S_{i_1j_1}$ and $S_{i_2j_2}$ are each given
by the same expression as in the single-process case, Eq.~(\ref{P.3}), so we must have 
 \begin{equation} \label{P.2.3}
S_{i_1j_1}= N_m \sum_{V_m} \Delta_{i_1} (\kappa_f, \kappa_{r_1}) \Delta_{j_1} (\kappa_f, \kappa_{r_1})=
n_{m} \sum_{v_{m}} \Delta_{i_1} (\kappa_f, \kappa_{r_1})\Delta_{j_1} (\kappa_f, \kappa_{r_1}) \, .
\end{equation}
 This can only be true if the set of points $v_m$ in
 Eq.~(\ref{P.3}) is a
subset of the set $V_m$ in Eq.~(\ref{P.2}): so when for example computing
$S_{i_1j_1}$, $\Delta_{i_1}$ and $\Delta_{j_1}$ depend only on the
scales $\kappa_f$ and $\kappa_{r_1}$ associated with $\pi_1$, and are
independent of the scale $\kappa_{r_2}$ associated with $\pi_2$.
Consequently, when we sum over $V_m$ in Eq.(\ref{P.2}), performing the
trivial sum over $\kappa_{r_2}$ must reduce $V_m$ to its subset $v_m$, up
to a degeneracy factor $d_m$ which counts the number of copies of
elements of $v_m$ contained in $V_m$.
This fixes the overall
normalization factor $N_m$: 
\begin{equation} \label{P.5}
N_m = n_m/d_m \, .
\end{equation}

\begin{figure}[t]
\centering
\includegraphics[scale=0.39]{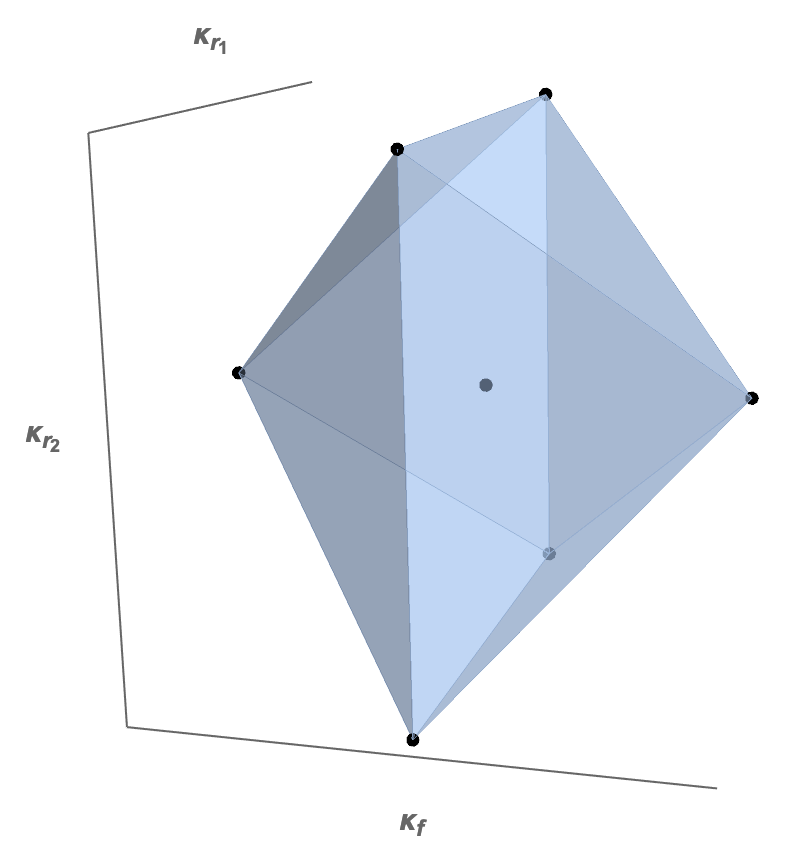}
\includegraphics[scale=0.39]{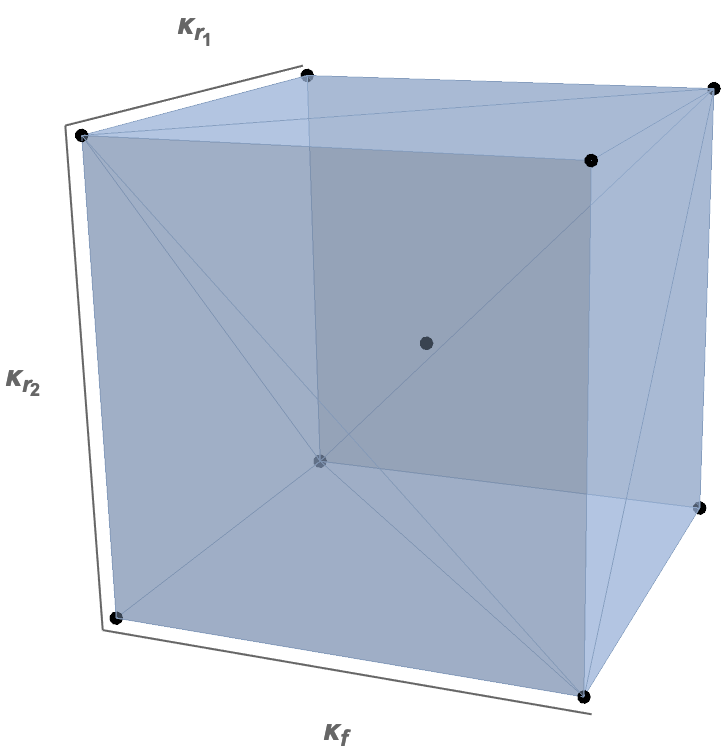}
\includegraphics[scale=0.39]{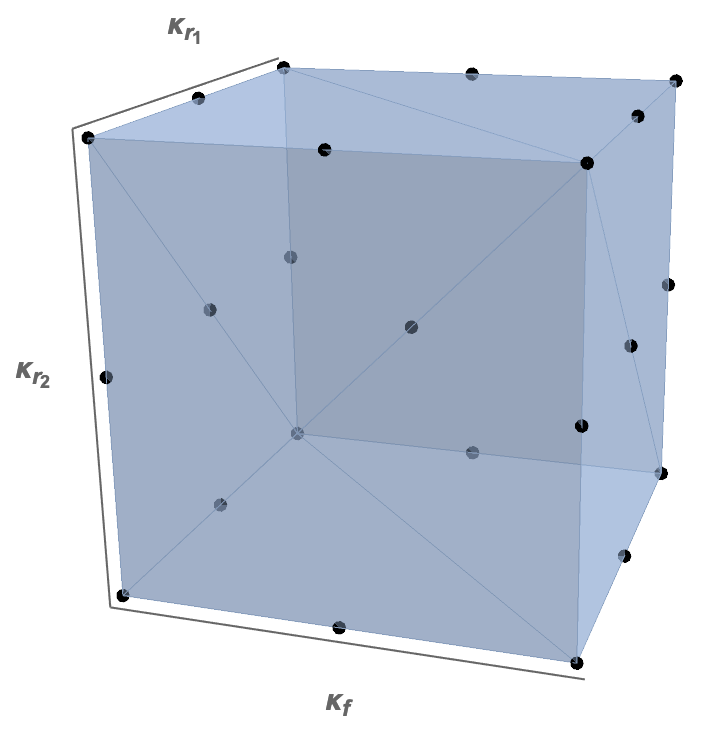}
\begin{caption}{\small Same as Fig.~\ref{fig:symmetricPrescriptions}, now
    for the case of two different processes $\pi_1$ and $\pi_2$
    with a common factorization scale $\kappa_f$ and different renormalization scales $\kappa_{r_1}$
    and $\kappa_{r_2}$, so the diagrams are now in three dimensions. The origin 
of coordinates is associated to the central scale, $\kappa_f=\kappa_{r_1}=\kappa_{r_2}=0$
    We again show the three prescriptions $5$-point (left), $\bar{5}$-point (center) and $9$-point (right).}
  \label{fig:symmetricPrescriptions2proc}
\end{caption}
\end{figure}

It remains to determine $V_m$ for a given $(m+1)$-point prescription. It is
easy to see that in each case we obtain a unique result, which is
in a sense a direct product of $p$ copies of $v_m$, taking into account the common scale $\kappa_f$.
The points in the $(\kappa_f,\kappa_{r_1},\kappa_{r_2})$ space that are being
sampled in each prescription when there are two processes are shown in Fig.~\ref{fig:symmetricPrescriptions2proc} (corresponding to the single-process prescriptions shown in 
Fig.~\ref{fig:symmetricPrescriptions}). 

To show how this works, we consider each prescription 
in turn, starting with the $\overline{5}$-point prescription which is easier to construct than $5$-point.
%

\begin{itemize}

\item \textbf{$\overline{5}$-point}: for two processes, $\pi_1$ and 
$\pi_2$ say, we now have 
three scales, $\kappa_f, \kappa_{r_1}, \kappa_{r_2}$ which can each be varied
independently.
For the $\overline{5}$-point prescription we only consider variations in
which none of the scales is at the central value: $\overline{v}_4 =
\{(\pm;\pm)\}$,  where 
the $\pm$ variations are performed independently.
It follows that  
$\overline{V}_4 = \{(\pm;\pm,\pm)\}$, where the triples denote the
three independent scales, $(\kappa_f; \kappa_{r_1},\kappa_{r_2})$, varied
independently.

The set $\overline{V}_4$ thus has eight points in total.
For each element of $\overline{v}_4$, there are two elements of
$\overline{V}_4$, so $\overline{d}_4=2$, and since
$\overline{n}_4=1/2$, $\overline{N}_4=1/4$.
The result for the
off-diagonal blocks of the theory covariance matrix in this prescription is thus
given by
\begin{equation}
  \label{5bM}
    S^{(\rm \overline{5}pt)}_{i_1j_2} = \smallfrac{1}{4}\big\{ \big(\Delta_{i_1}^{++} + \Delta_{i_1}^{+-}\big) \big(\Delta_{j_2}^{++} + \Delta_{j_2}^{+-} \big) 
    + \big(\Delta_{i_1}^{-+} + \Delta_{i_1}^{--}\big) \big(\Delta_{j_2}^{-+} + \Delta_{j_2}^{--} \big) \big\} \, .
\end{equation}
From this expression it is clear that while the scale $\kappa_f$ is
varied coherently between the two processes, the scales $\kappa_{r_1}$ and
$\kappa_{r_2}$ are varied incoherently, as required. 

It is straightforward to generalize this procedure to three processes:
then $\overline{V}_4 = \{(\pm;\pm,\pm,\pm)\}$, so $\overline{d}_4=4$,
and since $\overline{n}_4=1/2$, $\overline{N}_4=1/8$.  However
Eq.~(\ref{5bM}) remains unchanged, in the sense that it can be used to
evaluate all three off-diagonal blocks $s_{i_1j_2}$, $s_{i_2j_3}$,
$s_{i_3j_1}$: this must always be the case, since each term in the sum
Eq.~(\ref{P.2}) involves at most three scales.
For $p$ processes, it is
easy to see that the number of distinct elements of $V_4$ is
$2^{p+1}$. 

\item \textbf{5-point}: again for two processes we have three
  scales, but now one varies each holding the other fixed to its central 
value: $v_4 =  \{(\pm;0),(0;\pm)\}$, so  
$V_4 = \{2(\pm;0,0), (0;\pm,\pm)\}$, where the two in front of the
  first element indicates that there are two copies of it, so $V_4$
  has eight elements in total.
  Then for each element of $v_4$, there are
  precisely two elements of $V_4$, so $d_4=2$, and since $n_4=1/2$,
  $N_4=1/4$.
  The result for the off-diagonal entries of the theory covariance matrix
  in this prescription is thus  given by
\begin{equation}\label{5M}
    S^{(\rm 5pt)}_{i_1j_2} = \smallfrac{1}{4}\big\{ 2\Delta_{i_1}^{+0}\Delta_{j_2}^{+0} + 2\Delta_{i_1}^{-0}\Delta_{j_2}^{-0}  
            + \big(\Delta_{i_1}^{0+} + \Delta_{i_1}^{0-} \big)\big(\Delta_{j_2}^{0+} + \Delta_{j_2}^{0-}\big)\big\}.
\end{equation}
Note that also in this expression the variations of  $\kappa_f$ are
manifestly correlated between the two processes, whereas the
variations of $\kappa_{r_1}, \kappa_{r_2}$ are not.  

When there are three processes, it is easy to see that 
$V_4 = \{4(\pm;0,0,0), (0;\pm,\pm,\pm)\}$, i.e. it has 16 elements,
though only 10 are distinct: the other six are simply copies,
necessary to obtain the correct coefficients in Eq.~(\ref{5S}) and
Eq.~(\ref{5M}).
There are now four elements of $V_4$ for each element of
$v_4$, so now $d_4=4$, and $N_4=1/8$.
Again Eq.~(\ref{5M}) can be used
to calculate all three off-diagonal blocks. For $p$ processes, it is
easy to see that $V_4$ has $2^{p+1}$ elements, but that only  
$2+2^p$ of these are actually distinct.   

\item \textbf{9-point}: here we vary the three scales completely
  independently: $v_8 = v_4 \oplus \overline{v}_4$. Constructing
  $V_{8}$ is now somewhat more involved, since while terms with $\kappa_f=0$
  have degeneracy 2, terms where $\kappa_f=0$ is varied have
  degeneracy 3, so we need three copies of the former and two of the
  latter to take the overall degeneracy to 6.
  The solution is thus 
$V_8= \{3(0;\pm,\pm), 2(\pm;\pmz,\pmz)\}$, where  
$\pmz$ means either $+$, $-$ or $0$. Thus $V_8$ has $48$ elements, of 
  which only $22$ are actually distinct.
  Since the first term of $V_8$
has a degeneracy of $2$, while the last has a degeneracy of $3$, the
overall degeneracy is $d_8=6$, and since $n_8=1/4$, $N_8=1/24$.
It follows that the off-diagonal blocks of the theory covariance matrix
in this prescription are
\begin{equation}\label{9M}
\begin{split}
    S^{(\rm 9pt)}_{i_1j_2} =
    \smallfrac{1}{24}\big\{&2\big(\Delta_{i_1}^{+0}+\Delta_{i_1}^{++}
    + \Delta_{i_1}^{+-}\big) \big(\Delta_{j_2}^{+0} +
    \Delta_{j_2}^{++} + \Delta_{j_2}^{+-} \big) \\ 
            + &2\big(\Delta_{i_1}^{-0} + \Delta_{i_1}^{-+} +
            \Delta_{i_1}^{--}\big)\big(\Delta_{j_2}^{-0} +
            \Delta_{j_2}^{-+} + \Delta_{j_2}^{--} \big) \big\}\\ 
            + &3\big(\Delta_{i_1}^{0+}+ \Delta_{i_1}^{0-}\big)\big(\Delta_{j_2}^{0+} + \Delta_{j_2}^{0-} \big)\big\}.
\end{split}            
\end{equation}
The pattern of correlations in the variation of the three scales in this expression should be clear from the way it is written.

When there are  three processes, $V_8= \{9(0;\pm,\pm,\pm),
4(\pm;\pmz,\pmz,\pmz)\}$, whence $d_8=36$, and since $n_8=1/4$,
$N_8=1/144$.
Again, Eq.~(\ref{9M}) can be used to calculate all three
off-diagonal blocks. $V_8$ now has $288$ elements, of which $62$ are
distinct. For $p$ processes, there are $2^p+2\cdot3^p$ distinct elements.  

\end{itemize}

\subsection{Asymmetric prescriptions}\label{sec_asym}
\label{sec:asympres}

It is sometimes argued that since only the cross-section is actually physical, 
a single process has only one scale, namely the scale of the process in the
sense of Table~\ref{tab:scale_nomenclature} and
Eq.~(\ref{4.1}).
Therefore,  in order to estimate the MHOUs, only this single scale should be varied.
Alternatively, one may consider the
variation of the scale of the process {\it on top} of the variation of
the renormalization and factorization scales considered
previously.

The logic of the first alternative (variation of the scale of the
process only) is that after all
there is only one scale in the factorised expressions, for example those given by the Wilson expansion applied to DIS.
The logic of the second alternative (variation of the scale of the process, the
renormalization scale, and the factorization scale) is that each of
these estimates a different source of MHOU:
varying the scale of the
process generates terms related to missing higher order contributions to the 
hard cross-section which are proportional to collinear logarithms, the
renormalization scales to missing higher order contributions to  the hard
cross-section which are proportional to the beta function, and finally
the factorization scale to missing higher order contributions to the anomalous
dimension.

On the other hand, both alternatives might be criticized on the grounds that they suppress correlations between the uncertainties in PDF evolution across
different processes, and thus seriously overestimate uncertainties (the first worse than the second). Ultimately, however, they can 
be considered as possible options to be tested in a situation
in which the true answer is known. Such a validation will be performed in the next section.

\begin{figure}[t]
\centering
{\begin{tikzpicture}
\draw[->] (-1.5,0) -- (1.5,0);
\draw[->] (0,-1.5) -- (0,1.5);
\filldraw[black] (0,0) circle (2pt);
\filldraw[black] (-1,-1) circle (2pt);
\filldraw[black] (1,1) circle (2pt);
\node at (0.5,1.5) {$\kappa_{r_1}$};
\node at (1.9,0) {$\kappa_f$};
\end{tikzpicture}}\qquad
{\begin{tikzpicture}
\draw[->] (-1.5,0) -- (1.5,0);
\draw[->] (0,-1.5) -- (0,1.5);
\filldraw[black] (0,0) circle (2pt);
\filldraw[black] (-1,0) circle (2pt);
\filldraw[black] (0,-1) circle (2pt);
\filldraw[black] (1,0) circle (2pt);
\filldraw[black] (0,1) circle (2pt);
\filldraw[black] (-1,-1) circle (2pt);
\filldraw[black] (1,1) circle (2pt);
\node at (0.5,1.5) {$\kappa_{r_1}$};
\node at (1.9,0) {$\kappa_f$};
\end{tikzpicture}}
\begin{caption}{Same as Fig.~\ref{fig:symmetricPrescriptions},
    now in the case of the asymmetric prescriptions for a single process
    with factorization scale $\kappa_f$ and renormalization scale $\kappa_r$.
    We display the 
    $3$-point (left) and $7$-point (right) prescriptions, defined in the text.}
  \label{fig:AsymmetricPrescriptions1proc}
\end{caption}
\end{figure}

We now consider these two options in turn, both for the single-process
case, which is represented schematically
in Fig.~\ref{fig:AsymmetricPrescriptions1proc}, and for multiple-processes.

\begin{itemize}

\item \textbf{3-point}: For a single process,
  we set $\kappa_f = \kappa_r$  and only vary the single resulting scale. Then $v_2 = \{\pm\}$
  in an obvious notation, and $s =1$, $m=2$ and $n_2 = 1/2$, i.e. we
  simply average over the two nontrivial values of the single
  scale.
  For a single process we thus find that
\begin{equation}\label{3S}
    S^{(\rm 3pt)}_{ij} = \smallfrac{1}{2}\big\{ \Delta_i^{++}\Delta_j^{++}  + \Delta_i^{--}\Delta_j^{--}\big\} \, ,
\end{equation}
whenever $i,j\in \pi$.

Likewise, for two different processes $\pi_1$ and $\pi_2$, we set $\kappa_f = \kappa_{r_1}$ 
for $\pi_1$, set $\kappa_f = \kappa_{r_2}$ for $\pi_2$, and then vary
$\kappa_{r_1}$ and $\kappa_{r_2}$ independently.
This procedure necessarily
ignores the correlations in  
the variation of $\kappa_f$ between $\pi_1$ and $\pi_2$. Since $v_2 =
\{\pm\}$, $V_2 = v_2 \otimes v_2 = \{\pm, \pm \}$, where the ordered
pairs denote the two independent scales $(\kappa_{r_1},\kappa_{r_2})$.
Clearly,
for each element of $v_2$ there are two elements of $V_2$, so $d_2 =2$, 
Eq.~(\ref{P.5}) gives $N_2 = 1/4$, and the off-diagonal
entries of the theory covariance matrix are
\begin{equation}
    S^{(\rm 3pt)}_{i_1j_2} = \smallfrac{1}{4}\big\{\big(\Delta_{i_1}^{++} + \Delta_{i_1}^{--} \big) \big(\Delta_{j_2}^{++} + \Delta_{j_2}^{--} \big) \big\}\, .
\end{equation}
It can be seen from this factorised expression that the
variations for each process are entirely uncorrelated. Generalization
to more than two processes is straightforward: for $p$ processes $V_2$
has $2^p$ elements, all of them distinct. 

Because in this prescription we ignore correlations in the PDF evolution uncertainties, we expect this prescription to significantly overestimate the MHOUs.
Note that a fully correlated 3-point prescription in which we set
$\kappa_f = \kappa_{r_1} = \kappa_{r_2}$ would instead significantly underestimate
the MHOUs, which is why we do not consider it.

\item \textbf{7-point}:  We now  combine the variation of the scale
  of the process to the variation of renormalization and factorization scales.
As we saw in Sect.~\ref{double_var_sec}, a change in the scale
of the process is generated by $\tilde\kappa (\partial_{t_r} +
\partial_{t_f})$, so it moves diagonally in the
$(\kappa_f,\kappa_{r})$ plane.
Thus for a single process, varying the
scale of the process just corresponds to a new point-prescription,
symmetric only about the line $\kappa_f=\kappa_r$, but
asymmetric about the $\kappa_f$ and $\kappa_r$ axes.
However, because
variations of the scale of the process are assumed uncorrelated across
different processes, while $\mu_f$ variations are correlated, such a
scheme can give reduced correlations when there several processes. 

For a single process, variation of the scale of the process just gives
two extra points $(+;+),(-;-)$ (in the same notation as before,
i.e. variations in the $\kappa_f=\kappa_r$ plane), so $v_4 =\{(\pm;0),(0;\pm)\}$ becomes $v_6 = \{(\pm;0),(0;\pm),(+;+),(-;-)\} =\{(\pm;0),(0;\pm),(\overline{\pm;\pm})\}$, where $(\overline{\pm;\pm})$ simply means that the variation is fully correlated (so there are only 2 terms, not 4).

We then have $v_6 = \{(\pm;0),(0;\pm),(\overline{\pm;\pm})\}$, $s=2$ (note we still have only two independent scales), $m=6$ and $n_6 = 1/3$, and thus for a single process
\begin{equation}\label{7S}
    S^{(\rm 7pt)}_{ij} = \smallfrac{1}{3}\big\{ \Delta_i^{+0}\Delta_j^{+0} + \Delta_i^{-0}\Delta_j^{-0} + \Delta_i^{0+}\Delta_j^{0+}  + \Delta_i^{0-}\Delta_j^{0-}  
        + \Delta_i^{++}\Delta_j^{++} + \Delta_i^{--}\Delta_j^{--}  \big\} \, .
\end{equation}

When there is more than one process, we have to remember that
variations of the scale of the process are uncorrelated between
different processes, so they can decorrelate the allowed variations of
$\mu_f$.
This means the allowed variations for two processes are in a space
of four dimensions rather than three: call these say
($\kappa_{f_1},\kappa_{r_1};\kappa_{f_2},\kappa_{r_2})$. The extension of $v_6$ is then  
$V_6=\{2(+,0;+,0),2(-,0;-,0),(0,\pm;0,\pm),(\overline{\pm,\pm};\overline{\pm,\pm})\}$, where $(\overline{\pm,\pm};\overline{\pm,\pm})=\{(+,+;+,+),(+,+;-,-),(-,-;+,+),(-,-;-,-)\}$, and thus $d_6=2$, so $N_6=1/6$, and the off-diagonal theory covariance matrix reads
\begin{equation}\label{7M}
  \begin{split}
    S^{(\rm 7pt)}_{i_1j_2} =& \smallfrac{1}{6}\big\{ 2\Delta_{i_1}^{+0} \Delta_{j_2}^{+0}  + 2\Delta_{i_1}^{-0} \Delta_{j_2}^{-0} 
             + \big(\Delta_{i_1}^{0+}+\Delta_{i_1}^{0-}\big)\big(\Delta_{j_2}^{0+} + \Delta_{j_2}^{0-} \big)
            \\&+\big(\Delta_{i_1}^{++}+\Delta_{i_1}^{--}\big) \big(\Delta_{j_2}^{++} + \Delta_{j_2}^{--} \big) \big\} \, .
            \end{split}
\end{equation} 
This prescription gives smaller correlations than the 
symmetric prescriptions
since the variation of the two
factorization scales $\mu_{f_1}$ and $\mu_{f_2}$ is now entirely
uncorrelated. 

Generalization to $p$ processes is again straightforward: since the
variations of the scale of the process are in effect independent of
the separate variations of $\mu_f$ and $\mu_r$, $V_6=V_4\oplus V_2$
for any number of processes, so there are in total $2+2^{p+1}$
distinct elements. 

\end{itemize}

\section{Validation of the theory covariance matrix}
\label{sec:results}

In this section we determine the theory covariance matrix $S_{ij}$ at NLO
using the different prescriptions formulated in
Sect.~\ref{sec:prescriptions}, we introduce a method for the
validation of the theory covariance matrix when the next-order result
is known, and we use it to validate the theory covariance matrices
that we computed against the known NNLO results.
This validation is performed on a global dataset based on the same processes
as those used in the NNPDF3.1 PDF determination.
This dataset will then be used to produce fits incorporating MHOUs using 
the theory covariance matrix (Sect.~\ref{sec:fitstherr}), and also,
for comparison,  
fits using scale-varied theories (Appendix~\ref{sec:fitsscalesvar}).
 
\subsection{Input data and process categorization}
\label{sec:inputdata}

The validation of the theory covariance matrix and the PDF
determination to be discussed in the next section are performed using
a set of theory predictions for a dataset which is very similar to 
that used in the NNPDF3.1 PDF determination~\cite{Ball:2017nwa}, but
differs from it in some details, as we now discuss.

The input dataset used here
includes fixed-target~\cite{Arneodo:1996kd,Arneodo:1996qe,
Whitlow:1991uw,bcdms1,bcdms2,Goncharov:2001qe,MasonPhD,Onengut:2005kv} 
and HERA~\cite{Abramowicz:2015mha} deep-inelastic inclusive structure functions;
charm cross-sections from HERA~\cite{Abramowicz:1900rp};
gauge boson production from the Tevatron~\cite{Aaltonen:2010zza,Abazov:2007jy,
D0:2014kma,Abazov:2013rja}; and electroweak boson production, 
inclusive jet, $Z$ $p_T$ distributions, and $t\bar{t}$ total and differential
cross-sections from ATLAS~\cite{Aad:2011dm,Aaboud:2016btc,Aad:2014qja,
Aad:2013iua,Aad:2015auj,Aad:2011fc,Aad:2014kva,Aaboud:2016pbd,Aad:2015mbv},
CMS~\cite{Chatrchyan:2013tia,Chatrchyan:2012xt,Chatrchyan:2013mza,
Khachatryan:2016pev,Khachatryan:2015oaa,Chatrchyan:2012bja,Khachatryan:2016mqs, 
Khachatryan:2015uqb,Khachatryan:2015oqa} 
and LHCb~\cite{Aaij:2012vn,Aaij:2012mda,Aaij:2015gna,Aaij:2015zlq} 
at $\sqrt{s}=7$ and 8 TeV (two data points for the ATLAS and 
CMS total $t\bar{t}$ cross-sections are at 13 TeV).
\begin{table}[htbp!]
  \centering
\begin{center}
  \renewcommand*{\arraystretch}{1.50}
  \small
\begin{tabular}{|c|l|c|c|c|}
\toprule
  Process Type & Dataset   &  Reference    &  $N_{\rm dat}$     & $N_{\rm dat}$ (total)      \\
\midrule
\multirow{5}{*}{DIS NC}  & NMC    &  \cite{Arneodo:1996kd,Arneodo:1996qe}  & 134  &
\multirow{5}{*}{1593} \\
        & SLAC     & \cite{Whitlow:1991uw}     & 12  &    \\
        & BCDMS   & \cite{bcdms1,bcdms2}      & 530  & \\
        & HERA $\sigma^p_{NC}$     & \cite{Abramowicz:2015mha} & 886  & \\
        & HERA $\sigma^c_{NC}$     & \cite{Abramowicz:1900rp}  & 31  & \\
\midrule
\multirow{3}{*}{DIS CC}  & NuTeV dimuon  &\cite{Goncharov:2001qe,MasonPhD}
& 41  &
\multirow{3}{*}{552}\\
         & CHORUS  & \cite{Onengut:2005kv} & 430  &\\
        & HERA $\sigma^p_{CC}$   & \cite{Abramowicz:2015mha}& 81  & \\
\midrule
\multirow{19}{*}{DY}      &
  ATLAS $W,Z$, 7 TeV 2010  & \cite{Aad:2011dm} & 30  & \multirow{19}{*}{484}  \\
& ATLAS $W,Z$, 7 TeV 2011  &  \cite{Aaboud:2016btc} & 34  & \\
& ATLAS low-mass DY 2011  & \cite{Aad:2014qja} & 4  &\\
& ATLAS high-mass DY 2011  & \cite{Aad:2013iua} & 5  &\\
& ATLAS $Z$ $p_T$ 8 TeV ($p_T^{ll}, M_{ll}$) & \cite{Aad:2015auj} & 44  &\\
& ATLAS $Z$ $p_T$ 8 TeV ($p_T^{ll}, y_Z$) & \cite{Aad:2015auj} &  48   &\\
& CMS Drell-Yan 2D 2011    &  \cite{Chatrchyan:2013tia}    & 88  &   \\
& CMS $W$ asy 840 pb   & \cite{Chatrchyan:2012xt}    & 11   &   \\
& CMS $W$ asy 4.7 pb       & \cite{Chatrchyan:2013mza}  &11   & \\
& CMS $W$ rap 8 TeV   &  \cite{Khachatryan:2016pev}     &  22  &     \\
& CMS $Z$ $p_T$ 8 TeV ($p_T^{ll},M_{ll}$)      &  \cite{Khachatryan:2015oaa}    & 28  &  \\
& LHCb $Z$ 940 pb      &  \cite{Aaij:2012vn}    & 9  &  \\
& LHCb $Z \to ee$ 2 fb   & \cite{Aaij:2012mda}    & 17  &      \\
& LHCb $W, Z \to \mu$ 7 TeV  &  \cite{Aaij:2015gna}    & 29  &   \\
& LHCb $W, Z \to \mu$ 8 TeV &  \cite{Aaij:2015zlq}     & 30  &      \\
& CDF $Z$ rap    &  \cite{Aaltonen:2010zza}       & 29   &     \\
& D0 $Z$ rap    &  \cite{Abazov:2007jy}        & 28  &     \\
& D0 $W\to e\nu$ asy  & \cite{D0:2014kma}         & 8  &     \\
& D0 $W\to\mu\nu$ asy     &  \cite{Abazov:2013rja}        &  9 &   \\
\midrule
\multirow{2}{*}{JET}   & ATLAS jets 2011 7 TeV    & \cite{Aad:2011fc}  &  31 &  \multirow{2}{*}{164}    \\
& CMS jets 7 TeV 2011       & \cite{Chatrchyan:2012bja}  &  133  &  \\
\midrule
\multirow{4}{*}{TOP}     &
ATLAS $\sigma_{tt}^{\rm top}$       & \cite{Aad:2014kva, Aaboud:2016pbd}  & 3  & \multirow{4}{*}{26}  \\
& ATLAS $t\bar{t}$ rap &  \cite{Aad:2015mbv} & 10   &\\
& CMS $\sigma_{tt}^{\rm top}$   & \cite{Khachatryan:2016mqs, Khachatryan:2015uqb}       &  3 &   \\
& CMS $t\bar{t}$ rap   & \cite{Khachatryan:2015oqa}& 10   &   \\
\bottomrule
Total    &         &        & 2819  &  2819  \\
\bottomrule
\end{tabular}
\end{center}

  \caption{\small The categorization of the input datasets into different processes
    adopted
    in this work.
    Each dataset is assigned to one of five categories:
    neutral-current DIS (DIS NC),
    charged-current DIS (DIS CC), Drell-Yan (DY), jet production (JET) and top quark
    pair production (TOP).
    For each dataset, we also provide the corresponding publication reference
    and the number of data points after cuts.
We also show the total number of points in each of the five
categories of process.
    \label{tab:datasets_process_categorisation}}
\end{table}

This input dataset differs in many small respects from that used in the
NNPDF3.1 baseline.
Firstly, the fixed-target Drell-Yan (DY)
cross-sections~\cite{Webb:2003ps,Webb:2003bj,Towell:2001nh,Moreno:1990sf} are
excluded from the fit  since {\tt APFEL} currently does
not allow the calculation of scale-varied fixed-target DY cross-sections.
Secondly, the value of the lower kinematic
cut has been increased from
$Q_{\rm min}^2=2.69$~GeV$^2$ to
$13.96$~GeV$^2$ in order to ensure the validity of the
perturbative QCD expansion when scales 
are varied downwards.
Thirdly, we include only jet data for which the exact NNLO calculations
are available, as discussed in~\cite{Ball:2018iqk}, namely the
ATLAS and CMS inclusive jet cross-sections at 7 TeV
from the 2011 dataset.
Finally, we exclude the  bottom structure
function $F_2^b$ measurements, for which the implementation
of scale variations is complicated by the crossing of the heavy quark
thresholds.

Also, in original NNPDF3.1 determination
somewhat different cuts were applied to data at
NLO and NNLO (essentially in order to remove from the NLO fit data
which are subject to large NNLO corrections). Here we wish to have
exactly the same dataset at NLO and NNLO, in order to make sure that 
the differences between NLO and NNLO are due purely to differences in 
the theoretical calculations, and not in the input datasets. Therefore, the
baseline kinematic cuts of NNPDF3.1 have been slightly
modified so that the data points excluded at NLO are also excluded at NNLO
and vice-versa.

Taking into account all these modifications,  in total
the input dataset includes $N_{\rm dat}=2819$ datapoints.
The fact that the dataset differs somewhat from that of
Ref.~\cite{Ball:2017nwa} must be kept in mind when assessing the
impact of theory uncertainties, and indeed to this purpose in
Sect.~\ref{sec:globmhou} we will construct a new baseline PDF set
which only differs from that of Ref.~\cite{Ball:2017nwa} in that it is
based on the dataset we present here. Specifically, the loss of
Drell-Yan data will lead to an increased uncertainty in the $\bar
u-\bar d$ combination, and the higher $Q^2$ cutoff to somewhat larger
uncertainties in the small-$x$ region where the low $Q^2$ data are
concentrated. Here our main goal is to assess the impact of
theory uncertainties, not to construct the most competitive,
state-of-the art PDF set, which will be the subject of future work. 

Because the prescriptions in
Sect.~\ref{sec:prescriptions} assume that
renormalization scale variation is fully correlated within a given
process, but uncorrelated between different processes, it is necessary
to define what it is meant by ``process'', i.e., to classify
datasets into processes. This requires an educated
guess as to which theory computations share the same higher order
corrections. For example, it is necessary to decide whether
charged-current (CC) and neutral-current (NC) DIS are the
same process or not, and whether the transverse momentum
and rapidity distributions for one observable (such as, say, $Z$
production) should be grouped together.
Our categorization is summarized in Table~\ref{tab:datasets_process_categorisation}.

Specifically, we group the data into five distinct categories: DIS NC, DIS CC, Drell-Yan (DY), inclusive jet
production (JET), and top quark pair production (TOP). More refined
categorizations will be considered elsewhere, but we consider this to
be sufficient for a first study. The logic underlying this choice is
that we group together processes that are likely to share the same MHO
terms. Thus for instance the predictions for
all DY processes are obtained by integrating
the same underlying fully differential distributions, and thus have a
similar perturbative structure. Because different
distributions impact different PDF combinations -- so e.g. the $Z$ $P_t$
distribution mostly impacts the gluon, while the $W$ rapidity
distributions mostly impact flavor separation -- this will induce
nontrivial correlations in the PDF fitting.

All calculations are performed using 
the same settings as in~\cite{Ball:2017nwa}:
PDF evolution and the calculation of DIS structure functions up to NNLO
are carried out using the {\tt APFEL}~\cite{Bertone:2013vaa} program;
heavy quark mass effects are included by means of the FONLL general-mass variable
flavor number scheme~\cite{Forte:2010ta,Ball:2015dpa,Ball:2015tna}; 
the charm PDF is fitted alongside the light quark PDFs~\cite{Ball:2016neh}, 
rather than being generated from perturbative evolution of light 
partons; the charm quark pole mass is taken to be $m_c=1.51$~GeV,
and the strong coupling constant is fixed to be $\as(m_Z) = 0.118$, consistent
with the latest PDG average~\cite{Olive:2016xmw}.

In order to evaluate the theory covariance matrix $S_{ij}$, it is necessary
to be able to evaluate both DIS structure functions and hadronic
cross-sections for a range of values of the factorization
and renormalization scales, i.e., in the notation of Eq.~(\ref{eq:scaledef}), for $\kappa_f\ne 0$ and $\kappa_r\ne 0$.
In this case, the entries of the NLO theory covariance matrix have been 
constructed
by means of the {\tt ReportEngine} software~\cite{zahari_kassabov_2019_2571601}
taking the
scale-varied NLO theory cross-sections $T_i(k_f,k_r)$  as input.
These are provided 
by {\tt APFEL}~\cite{Bertone:2013vaa} for the DIS structure functions
and by {\tt APFELgrid}~\cite{Bertone:2016lga} combined with
{\tt APPLgrid}~\cite{Carli:2010rw} for the hadronic
cross-sections.
The evaluation of these scale-varied cross-sections has been validated
by means of independent programs, in particular with {\tt HOPPET}~\cite{Salam:2008qg}
and {\tt OpenQCDrad}~\cite{Alekhin:2012ig}
for the DIS structure functions, and with the built-in scale variation
functionalities of {\tt APPLgrid}. All these NLO cross-sections are evaluated using the central NLO PDF obtained by performing a NLO fit to the same dataset, for consistency.

\subsection{The theory covariance matrices at NLO}
\label{sec:nlothcov}

We now present results for the
theory covariance matrices, constructed using NLO calculations
and evaluated according to the prescriptions
introduced in Sect.~\ref{sec:prescriptions}, and discuss some of their
qualitative features.

\begin{figure}[t!]
  \begin{center}
    \includegraphics[width=1.0\linewidth]{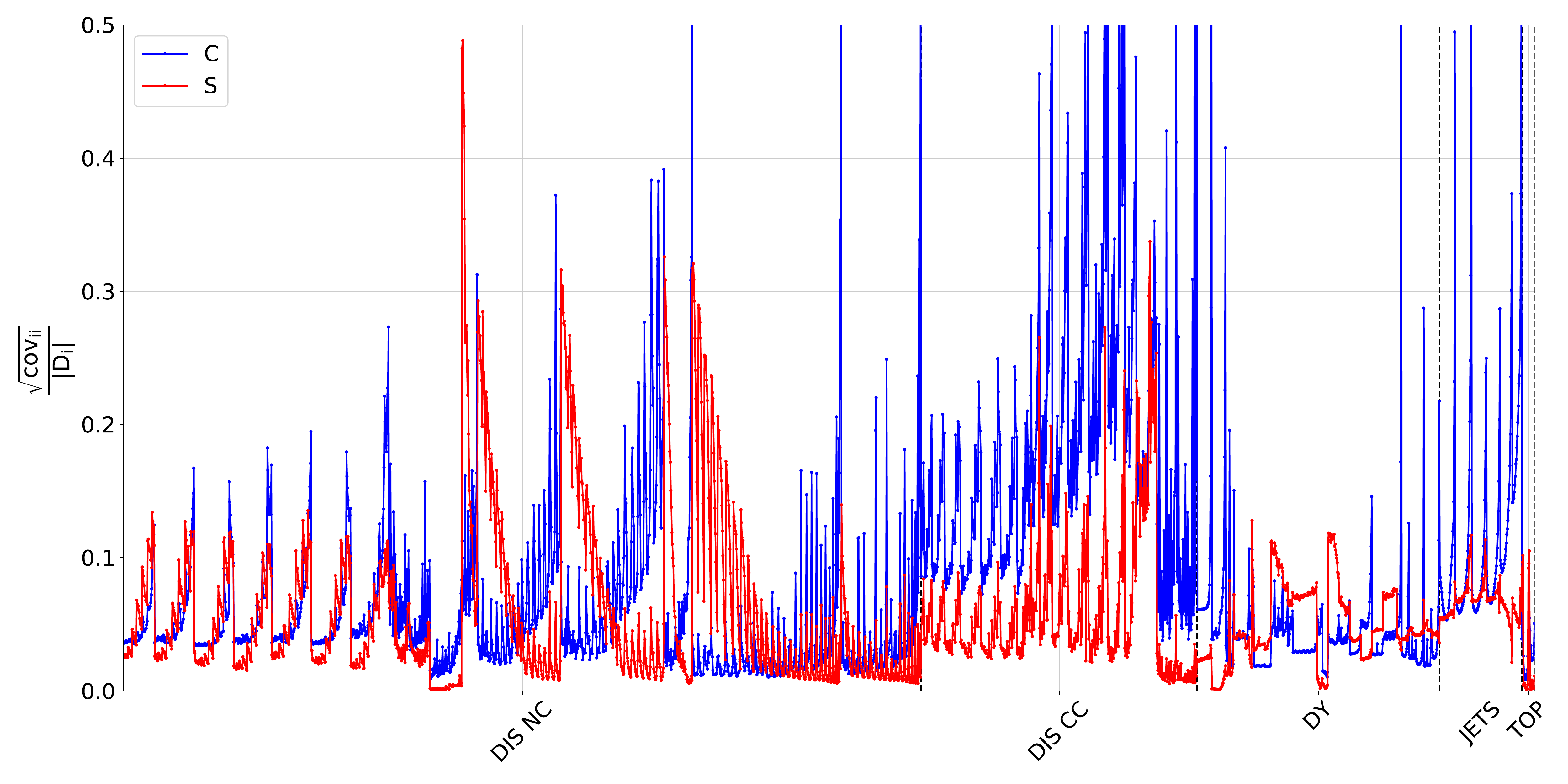}
    \caption{\small Comparison of the diagonal experimental uncertainties 
    (blue) and the diagonal theoretical uncertainties  evaluated using
      the 9-point       prescription 
      (red),      all normalized to the central experimental value.
  The data are grouped by process and, within a process, by experiment, following
  Table~\ref{tab:datasets_process_categorisation} 
        \label{fig:diag_covmats} }
  \end{center}
  \begin{center}
    \includegraphics[width=0.49\linewidth]{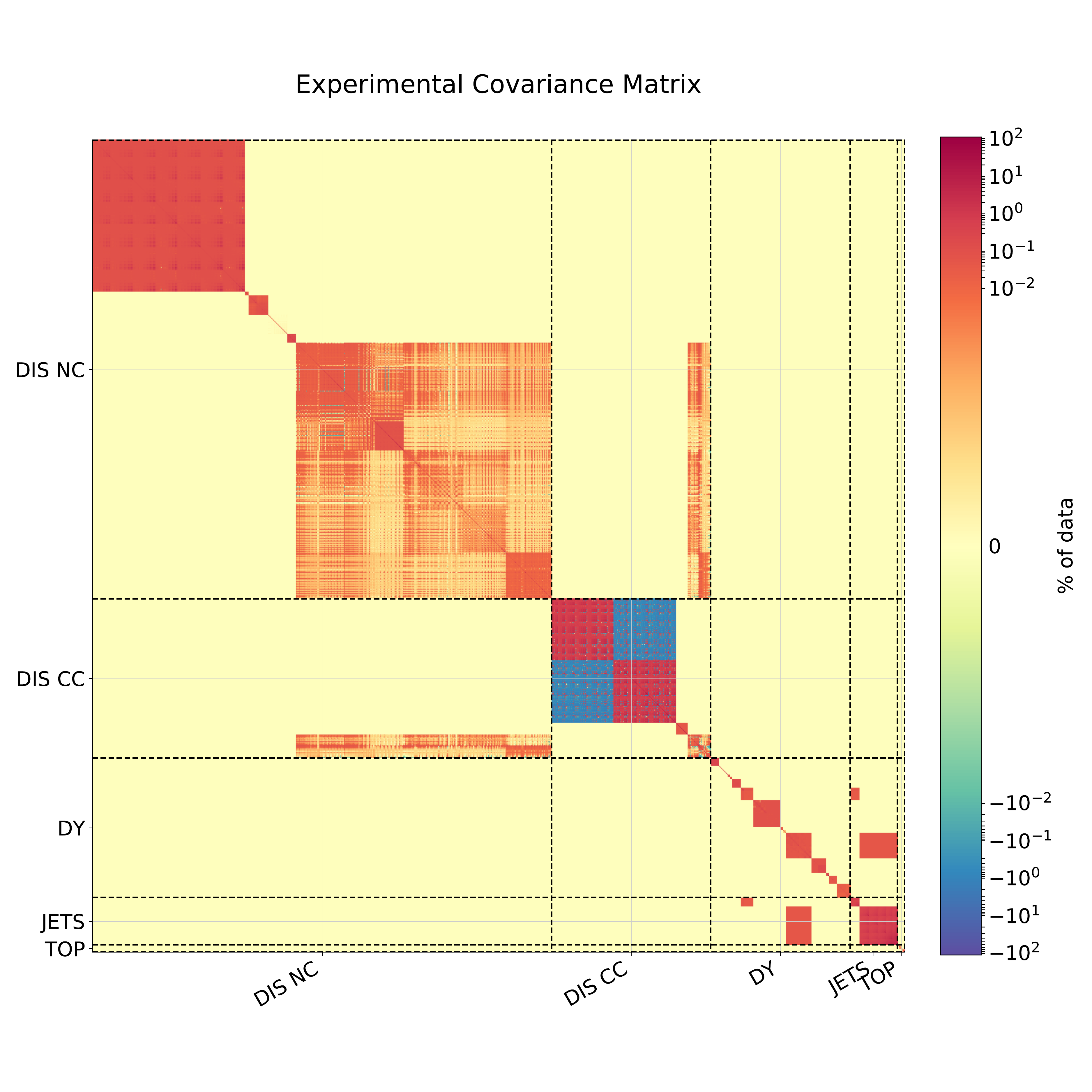}
    \includegraphics[width=0.49\linewidth]{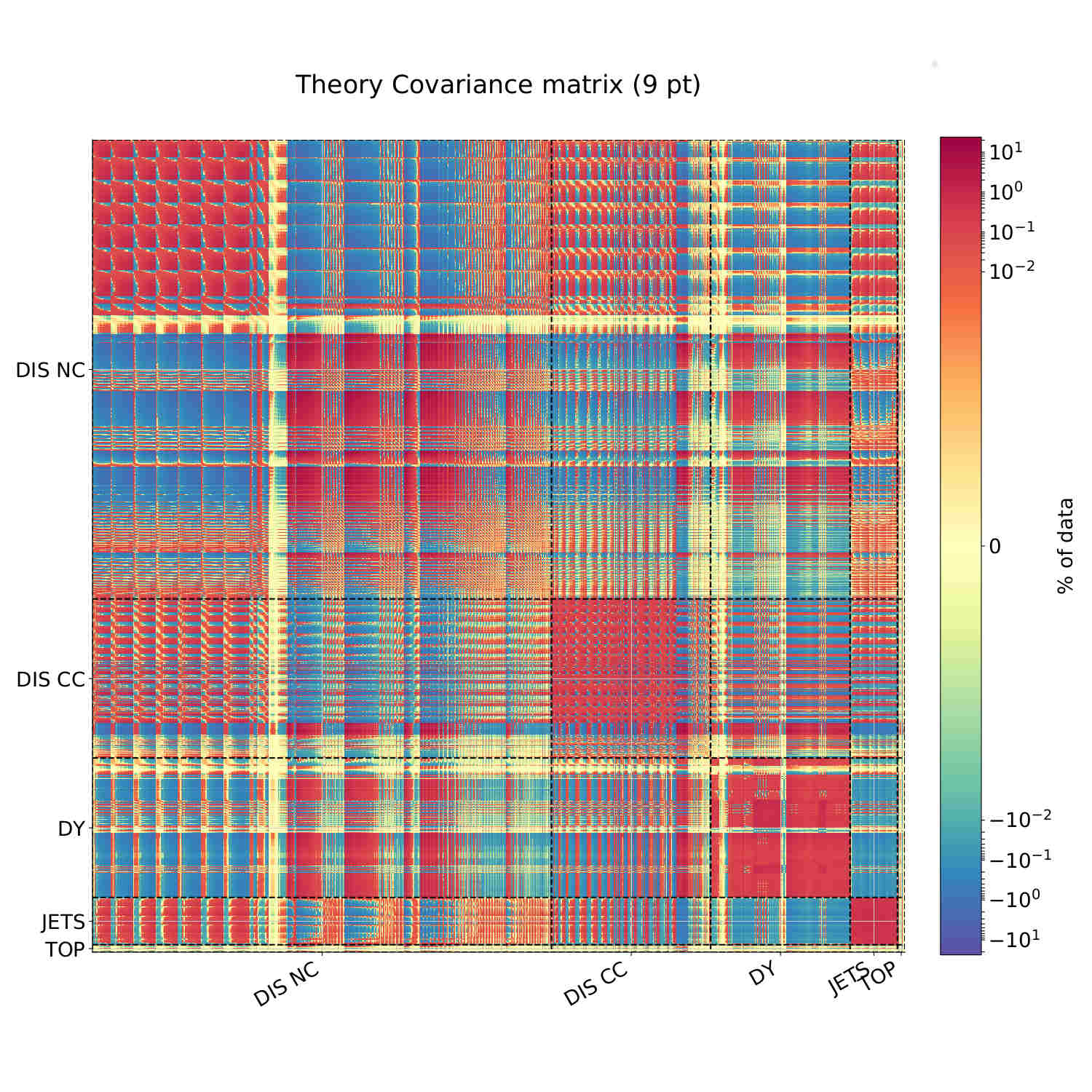}
    \caption{\small Comparison of the  experimental $C_{ij}$ (left)
      and the theoretical  $S_{ij}$  
      (right) covariance matrices, the latter evaluated using the 9-point
      prescription.
      All entries are normalized to the central  experimental value.
  The data are grouped by process and, within a process, by experiment, following
  Table~\ref{tab:datasets_process_categorisation} 
        \label{fig:covmats} }
  \end{center}
\end{figure}

In Fig.~\ref{fig:diag_covmats} we show the diagonal elements of 
the experimental and theory covariance matrices, or more specifically 
the experimental uncertainty normalized to the data, $(C_{ii})^{1/2}/D_i$, and 
the MHOU normalized to the data, $(S_{ii})^{1/2}/D_i$, for
$i=1,\ldots,N_{rm dat}$, where $D_i$ is the $i$-th datapoint. Here and
henceforth, the experimental covariance matrix $C_{ij}$ includes all
uncorrelated statistical uncertainties as well as correlated
systematic uncertainties, as published by the respective
experiments, and used to assess fit quality as e.g. in Sect.~3.2 of
Ref.~\cite{Ball:2018iqk}. Note that this differs from the
covariance matrix $C_{ij}$ used for PDF minimization in the treatment
of multiplicative uncertainties (such as normalization or luminosity
uncertainties) in that the latter must be treated using the so-called
$t_0$ method of Ref.~\cite{Ball:2009qv} in order to avoid bias. As in
all previous NNPDF determinations, the
$t_0$ covariance matrix is  used for PDF minimization while the
experimental covariance matrix is used in order to assess fit quality,
in order to ensure reproducibility of results.

The datapoints are grouped by process and, within a 
process by experiment, following 
Table~\ref{tab:datasets_process_categorisation}. The theory covariance matrix $S_{ij}$ is computed using the 9-point prescription (the one with the largest number of independent variations; recall
Sect.~\ref{sec:prescriptions}). Broadly speaking, the estimated NLO MHOU is roughly comparable to experimental uncertainties, as expected. 
However for some datapoints the 
experimental uncertainty is dominant (and thus the theory uncertainty
will have only a small effect), while for others the MHOU is
dominant. These latter datapoints will carry less weight
in a PDF fit with MHOU included, depending also on the underlying
correlation pattern. Some datasets have
datapoints in both these categories: the HERA NC DIS are particularly
striking, since at high $Q^2$ (where statistics are low) the dominant
uncertainty is experimental, while at low $Q^2$ (and thus small $x$,
where perturbation theory is less reliable) the dominant uncertainty
is due to MHO.  

In Fig.~\ref{fig:covmats} we compare the 
complete experimental covariance matrix $C_{ij}$ to the theory covariance
matrix $S_{ij}$, again computed using the 9-point prescription. Both covariance matrices are displayed
as heat maps, with each entry expressed as a fraction with respect to the
corresponding experimental central value; i.e. $C_{ij}/D_iD_j$ and
$S_{ij}/D_iD_j$. It is clear from Fig.~\ref{fig:covmats} that
the theory covariance matrix has, as expected, a 
richer structure of correlations than
its experimental counterpart: for example data from the same
process (such as DIS) are correlated even when the corresponding
experimental measurements are completely uncorrelated (such as HERA and
fixed target). Furthermore, correlation of the factorization scale variation 
between disparate processes, such as DIS processes and hadronic processes, 
results in nonzero entries in the theory covariance matrix even  
in these regions.
\begin{figure}[htb!]
  \begin{center}
    \includegraphics[width=0.49\linewidth]{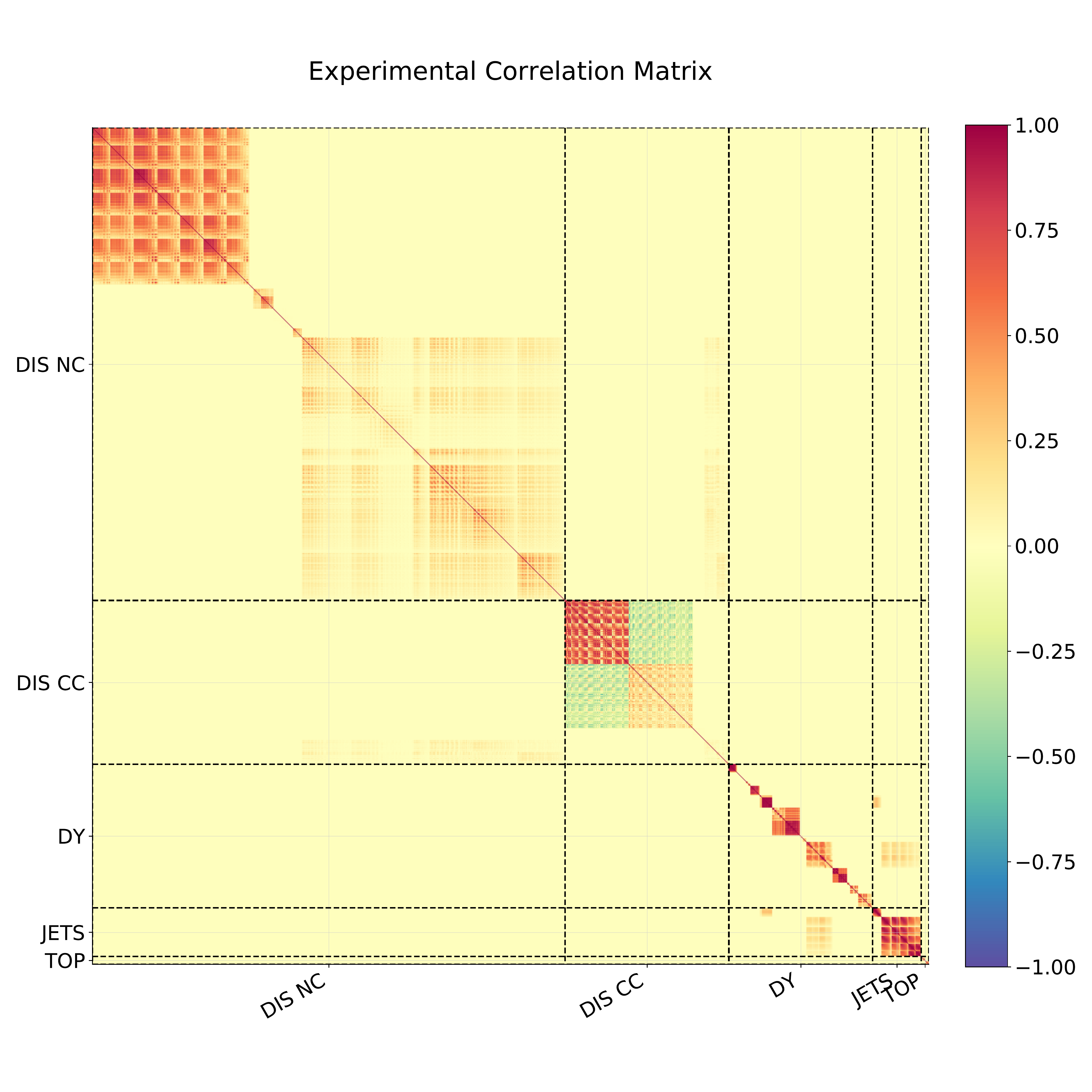}
    \includegraphics[width=0.49\linewidth]{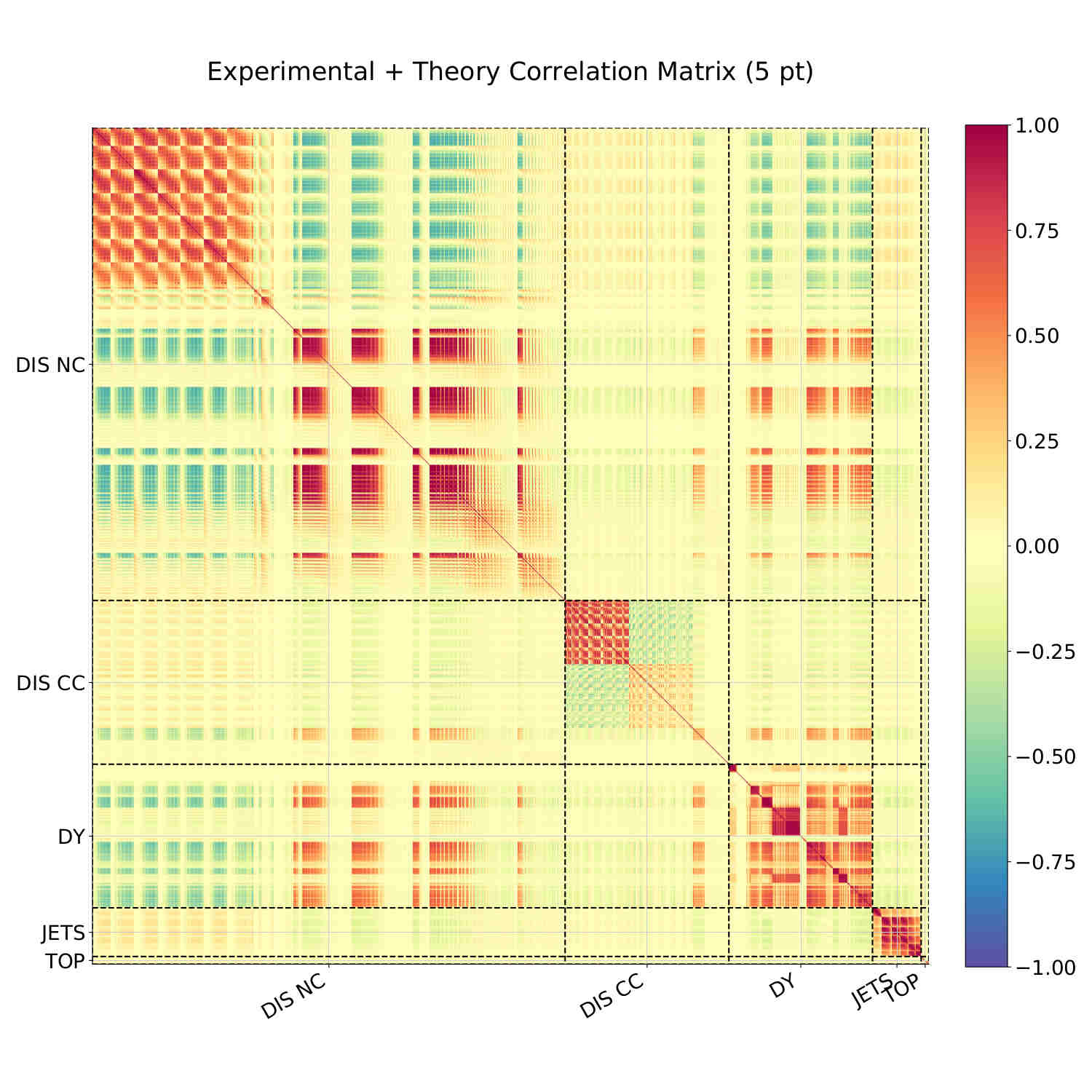}
\vskip-0.5cm
    \includegraphics[width=0.49\linewidth]{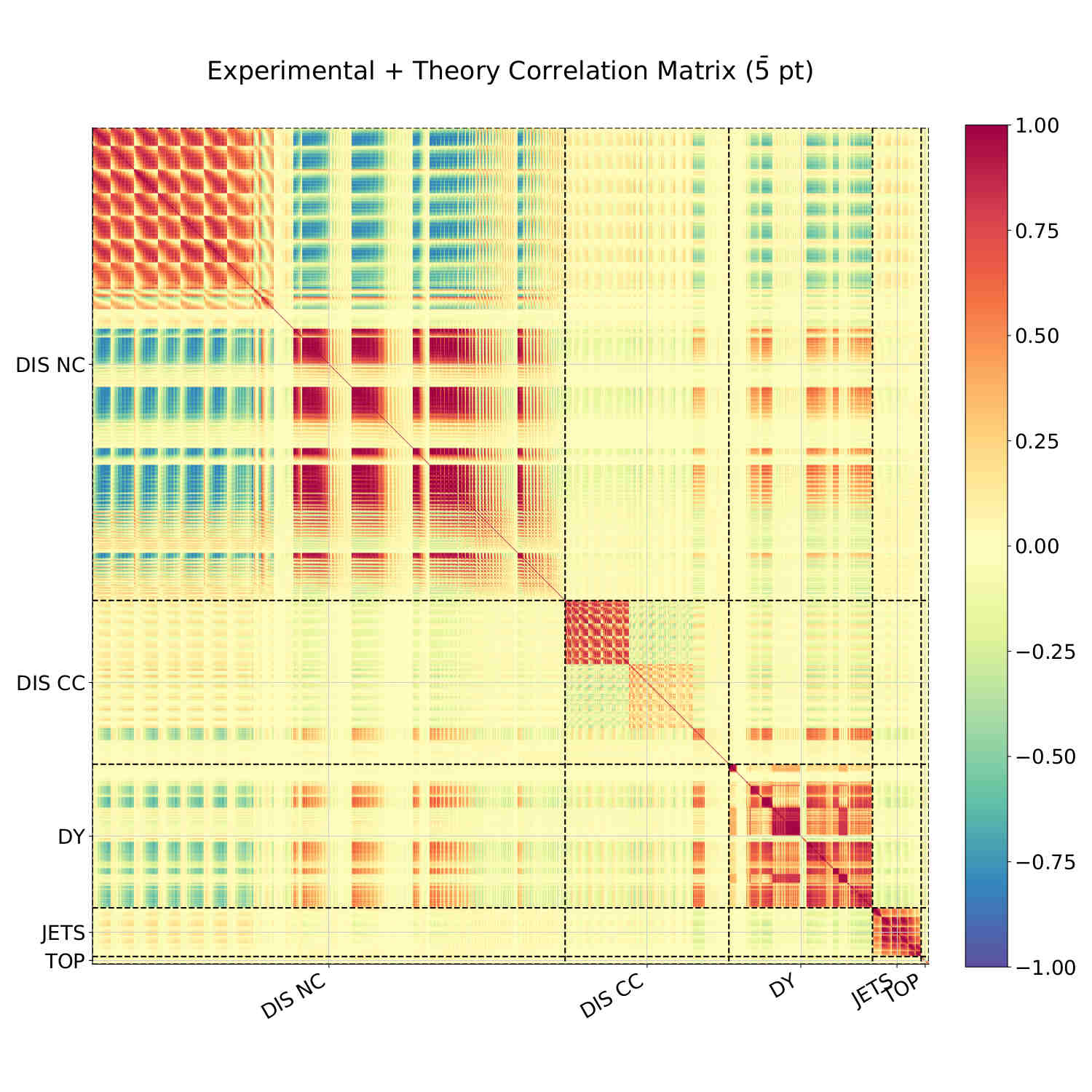}
    \includegraphics[width=0.49\linewidth]{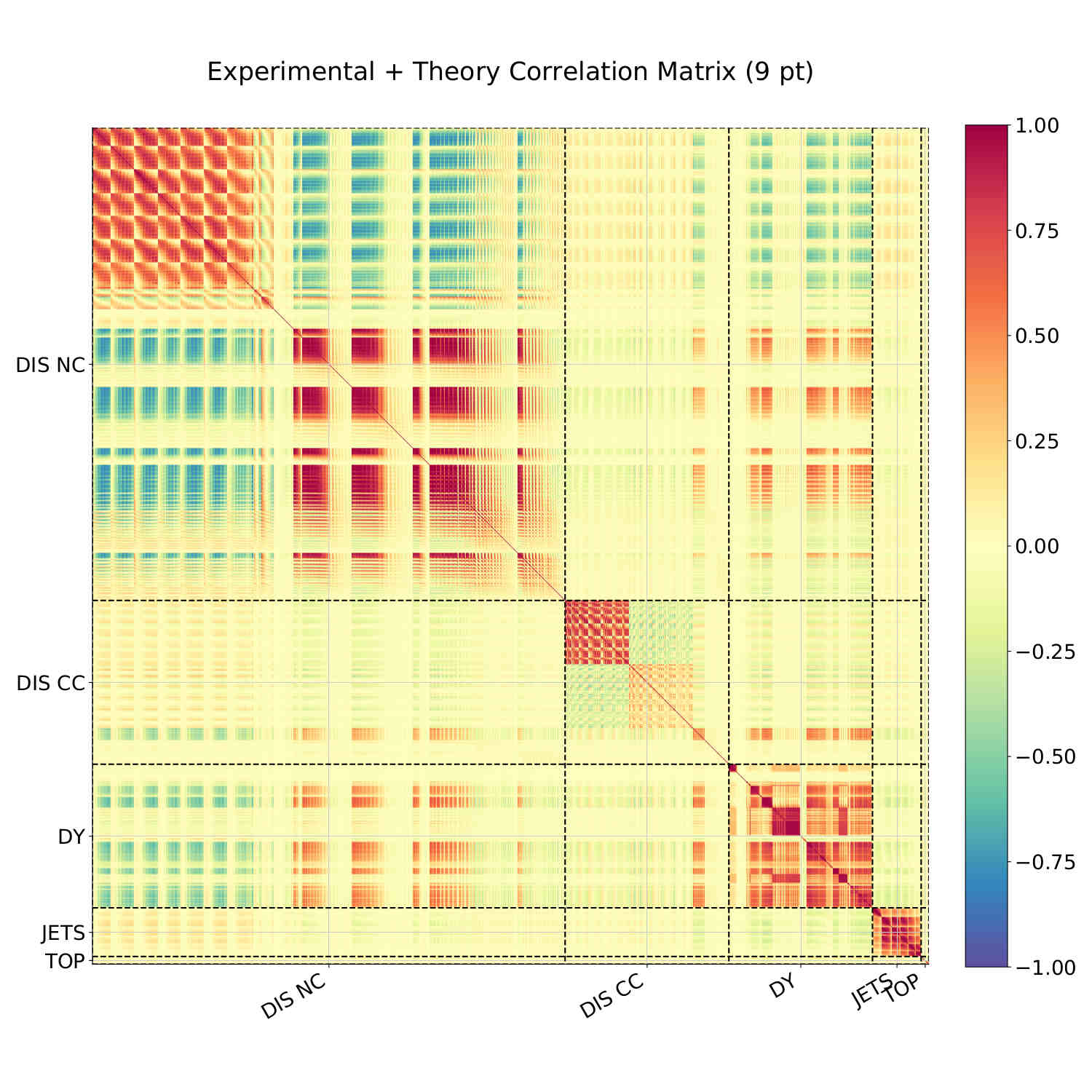}
\vskip-0.5cm
\includegraphics[width=0.49\linewidth]{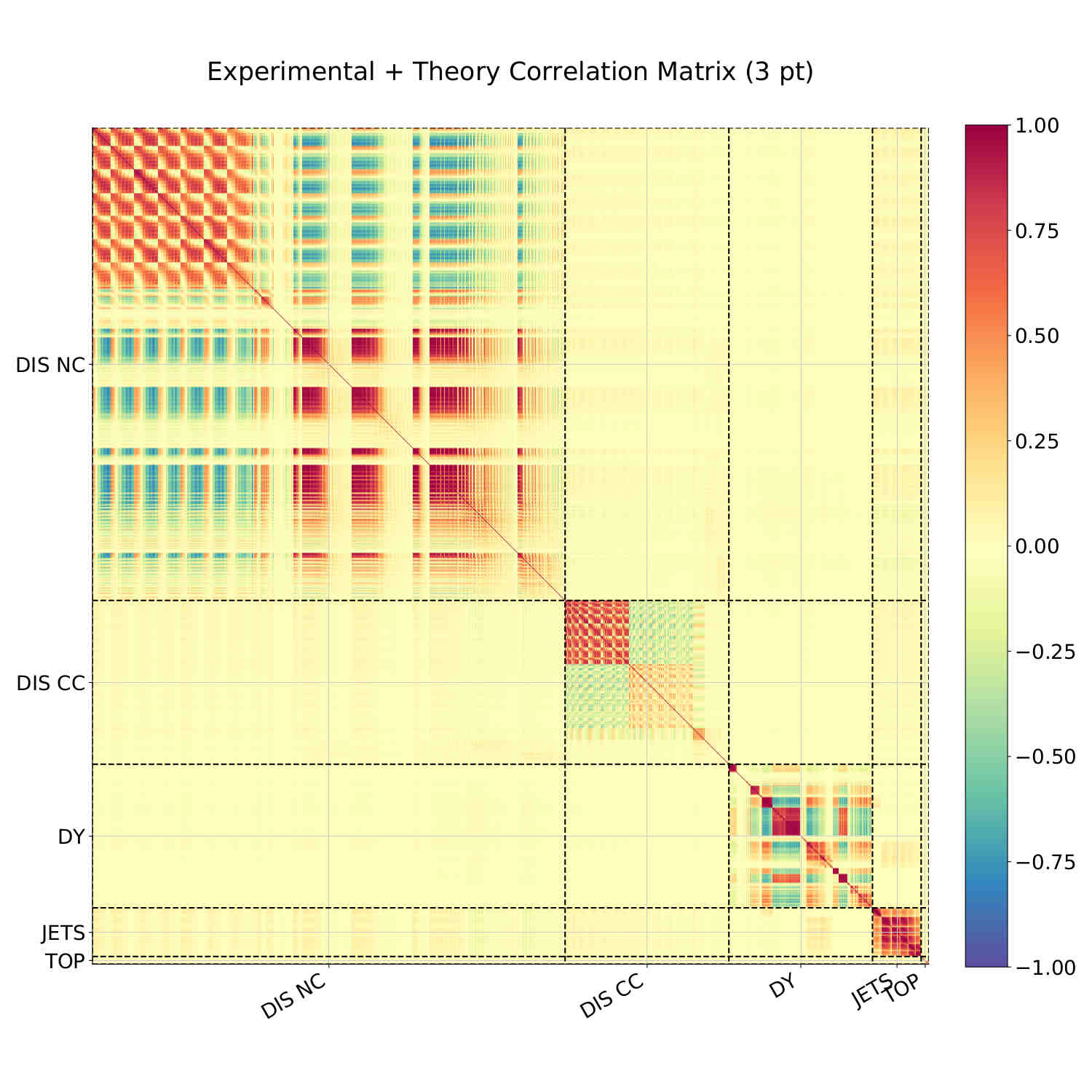}
    \includegraphics[width=0.49\linewidth]{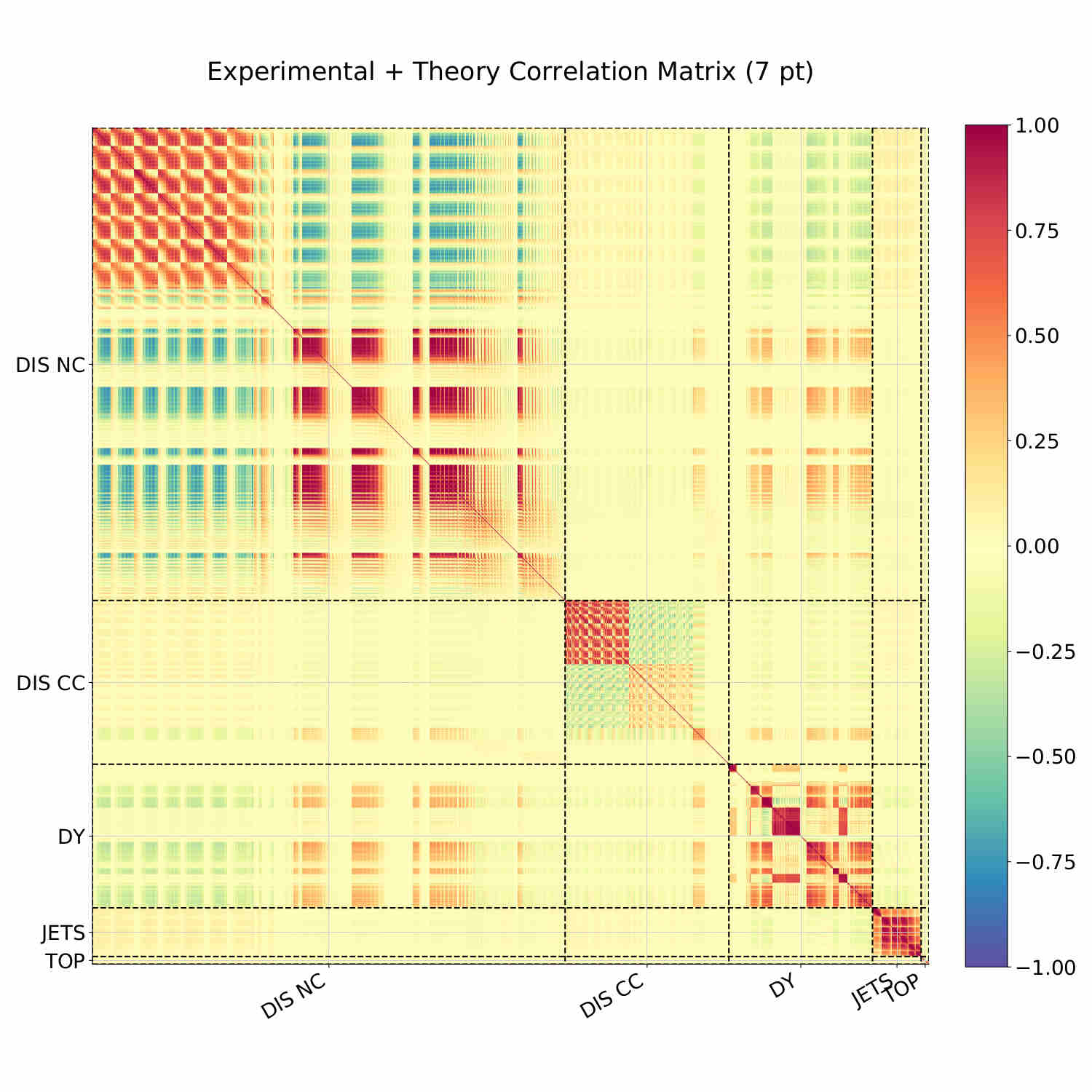}
    \caption{\small Comparison of the experimental correlation matrix
      Eq.~(\ref{eq:expcorrmat}) (top left) and the
      the combined experimental and theoretical correlation matrices
      Eq.~(\ref{eq:expthcorrmat}) computed using the prescriptions
      described in Sect.~\ref{sec:prescriptions}: the symmetric
      prescriptions (5-pt top right, $\overline{5}$-pt center left,
      9-pt center right), and asymmetric prescriptions (3-pt bottom
      left, 7-pt bottom right). The data are grouped by process and
      within a process by experiment, as in Fig.~\ref{fig:covmats}. 
  \label{fig:corrmats} }
  \end{center}
\end{figure}

The precise structure of these theory-induced correlations depends
on the 
choice of prescription adopted.
To illustrate this, 
Fig.~\ref{fig:corrmats} compares the experimental correlation matrix, given by
\be
\label{eq:expcorrmat}
\rho^{(C)}_{ij} = \frac{C_{ij}}{  \sqrt{C_{ii}}\sqrt{C_{jj}} } \, ,
\ee
with the corresponding combined experimental and theoretical correlation matrices, defined by
\be
\label{eq:expthcorrmat}
\rho^{(C+S)}_{ij} = \frac{( C+S)_{ij} }{ \sqrt{(C+S)_{ii}} \sqrt{(C+S)_{jj}}  } \, ,
\ee
for all the prescriptions defined in Sect.~\ref{sec:prescriptions}.
Specifically, from top left to bottom
right we have the experimental correlations $\rho^{(C)}$ followed by $\rho^{(C+S)}$ for the symmetric 5, $\bar{5}$, 9 point prescriptions, and the asymmetric 3 and 7 point prescriptions.
As in Fig.~\ref{fig:covmats},
the cross-sections are grouped by process type and, within that, by experiment.

Some qualitative features of the theory-induced correlations are apparent.
There are clearly large positive correlations within individual
experiments along the diagonal blocks, this being particularly evident for 
DIS NC and DY data, which have large numbers of points which are 
relatively close kinematically.
Off the diagonal, but still within the same process, there are 
large correlations between experiments for the DY, jets and top data points, and
large anticorrelations for the DIS NC data points (these mostly between fixed 
target and HERA). Correlations and
anticorrelations between different processes are also often present but are
somewhat weaker: for example the DY data points (from LHC) are quite 
correlated with the HERA NC DIS data points, but anticorrelated with 
fixed target NC DIS data points.  

When comparing different prescriptions, it is clear that the
3-point prescription leads to especially small correlations
between processes, which is expected because with this prescription the
factorization scale and renormalization scale variations are uncorrelated between processes. The correlations between processes are also weaker in
7-point than in 5-point, due to the fact that (as discussed in
Sect.~\ref{sec:asympres}) the correlated variation of the
factorization scale is combined with the uncorrelated variations of
the scale of the process for the pair of processes involved. It is worth
noting, however, that the pattern of correlations is similar 
for all the symmetric prescriptions.

In order to decide which prescriptions are best, and
more generally whether or not they produce a reliable estimate
of MHOUs, we must proceed to their validation.

\subsection{Construction of validation tests}
\label{sec:validconstruction}

We wish to construct a validation test for the NLO theory covariance
matrix, by comparing it to the known NNLO theoretical result. We do so by
viewing the set of experimental data as a vector with components
$D_i$, where $i=1, \ldots, N_{\rm dat}$. The vector lives in an
$N_{\rm dat}$-dimensional ``data" space D, on which the theory
covariance matrix $S_{ij}$ acts as a linear operator. The matrix
$S_{ij}$ is symmetric and positive semi-definite, meaning that all of its non-zero eigenvalues are positive.
In a PDF fit, $S_{ij}$ always enters as an additive contribution to the experimental covariance matrix $C_{ij}$, and thus 
their sum is always invertible, owing to the non-zero statistical uncertainties on the data, which bound the eigenvalues from below.

The matrix
$S_{ij}$ defines ellipsoids $E$ corresponding to a given confidence level 
in the data space, centered on the NLO theoretical 
prediction, $T^{\rm NLO}_i\equiv T^{\rm NLO}_i(0,0)$ evaluated using the central scale
choice.
In the context of MHOUs, we can take $T^{\rm NLO}_i$ to be the predictions
at NLO, with the one-sigma ellipsoid $E_{1\sigma}$
estimating a 68\% confidence level for the MHO correction.
We can validate whether $S_{ij}$
correctly predicts both the size and the correlation pattern of the
MHO terms by testing the extent to which the shift vector $\delta_i \sim T^{\rm NNLO}_i-T^{\rm NLO}_i$, i.e. the difference between the NNLO and NLO predictions 
for $T_i$, falls within a given ellipsoid $E$. These predictions
should be taken with a fixed underlying PDF (which could indeed be 
any standard reference PDF): it is the change in prediction due to the
change of perturbative evolution and hard matrix element which are
relevant here. 
Note that the dimensionality of the subspace spanned by
the ellipsoid $E$ is much smaller than that of the data space $D$: in a 
global fit the data space has dimension
$\mathcal{O}(3000)$ (Table~\ref{tab:datasets_process_categorisation}), while 
even the most complex prescriptions in
Sect.~\ref{sec:prescriptions} have at $\mathcal{O}(30)$ independent
variations, not all of which correspond to independent eigenvectors, as
we will see shortly. So 
$E$ actually lives in a subspace $S$ of dimension $N_{\rm sub}$ of
the full space  $D$: $E \in S \in D$. For a single process we expect
$N_{\rm sub}$ to be of order a dozen or so at most. 
In fact, even for a single process
(see Table~\ref{tab:datasets_process_categorisation}) we always 
have $N_{\rm sub} \ll N_{\rm dat}$.
Hence, a nontrivial validation of the theory covariance matrix is if the component of the shift vector $\delta_i$ lying outside $E$ is small, i.e. if the angle
between $\delta_i$ and the projection of $\delta_i$ onto $S$ is small. 

Furthermore we expect the component of $\delta_i$ along each axis of the
ellipsoid $E$ to be of the same order as the typical one-sigma variation.
The physical interpretation of such a successful validation is that
the eigenvectors of $S_{ij}$ correctly estimate
the independent directions of uncertainty in theory space, with the
size of the shift estimated by the corresponding eigenvalue. The null
subspace of $E$, i.e. the directions of vanishing eigenvalues, would
then correspond to directions in $D$ for which the theory uncertainty
is so small that it cannot  be reliably estimated and so can be
safely neglected. These are highly nontrivial tests, given the huge 
discrepancy between the dimensionality of the space $D$, and the 
dimensionality of $S$.

Let us now see how this works in detail. First, we
need to identify
the spaces $E$ and $S$.
To do this, we normalize the NLO theory covariance matrix $S_{ij}$ to
the central NLO theory prediction $T_i$, so that all its elements 
are dimensionless, allowing a meaningful comparison: we define
\begin{equation}
\widehat{S}_{ij} = S_{ij}/(T^{\rm NLO}_iT^{\rm NLO}_j)\, .\label{eq:Snorm}
\end{equation}
Likewise, we define a normalized shift vector with components 
\begin{equation}
{\delta}_i = (T^{\rm NNLO}_i-T^{\rm NLO}_i)/T^{\rm NLO}_i \, .\label{eq:normshift}
\end{equation}
The NNLO prediction $T^{\rm NNLO}_i$ is computed using NNLO matrix elements 
and parton evolution, but with the same NLO PDF set used in the computation of 
$T^{\rm NLO}_i$ and $S_{ij}$. In this way the shift $\delta_i$ only takes 
account of the perturbative effects due to NNLO corrections, which are estimated by $S_{ij}$, and not the additional effect of refitting.

We then diagonalize $\widehat{S}_{ij}$, to give
eigenvectors, $e_i^\alpha$ (chosen to be orthonormal, i.e. $\sum_i
e_i^\alpha e_i^\beta = \delta^{\alpha\beta}$), with corresponding
non-zero eigenvalues, $\lambda^\alpha=(s^\alpha)^2$; $\alpha = 1,
\ldots, N_{\rm sub}$. All these eigenvalues are real and positive, see
Eq.~(\ref{P.2pos}). The eigenvectors span the subspace $S$.
There are also  $N_{\rm dat}-N_{\rm sub}$ zero eigenvalues. These are
degenerate, and their eigenvectors span the space $D/S$. Because of
the zero eigenvalues, the diagonalization of $\widehat{S}_{ij}$ is in
practice rather difficult: the procedure we use to identify the
subspace $S$ and its dimensionality $N_{\rm sub}$, and then
diagonalize the projection of $\widehat{S}_{ij}$ into $S$,  is 
described in some detail in Appendix~\ref{sec:diagonalization}.  

Next we project the shift vector $\delta_i$ onto the eigenvectors,
\begin{equation}\label{eq:deltaproj}
\delta^\alpha = \sum_{i=1}^{N_{\rm dat}} \delta_i e_i^\alpha \, .
\end{equation} 
These projections $\delta^\alpha$ should be of the same order as the size 
of the ellipse in this direction, i.e. the $s^\alpha$: more specifically in an ideal world 68\% of the $\delta^\alpha/s^\alpha$ would be less than one.
This is all the meaningful statistical information that is contained
in $\widehat{S}_{ij}$.

\begin{figure}[t]
  \begin{center}
    \includegraphics[scale=0.27]{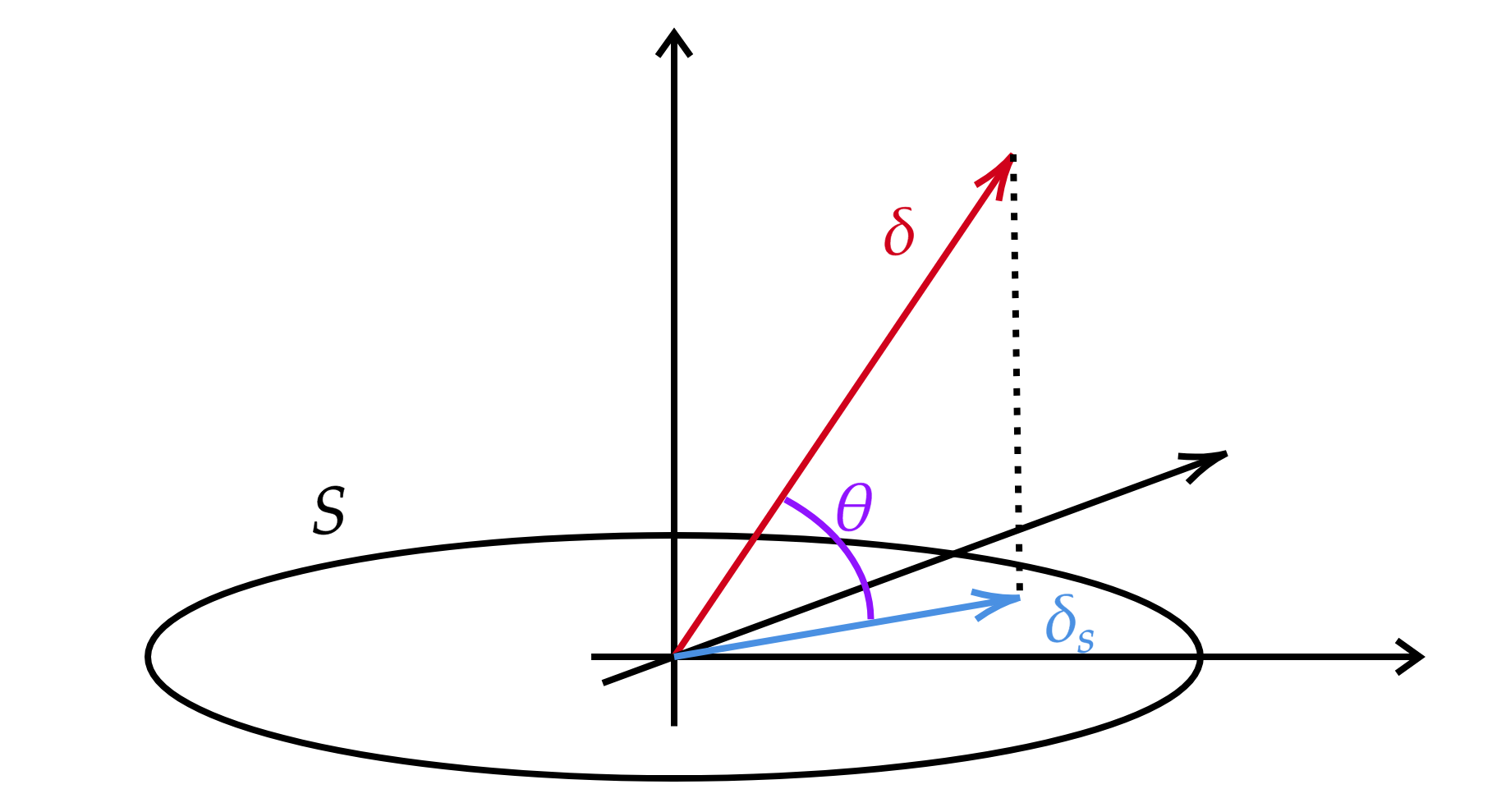}
    \caption{\small Schematic representation of the geometric relation
      between the shift vector $\delta\in D$ (here drawn as a three dimensional space), and
      the component $\delta^S$ of the shift vector which lies in the 
subspace $S$ (here drawn as a two dimensional space, containing the ellipse E defined by the theory covariance matrix). The angle $\theta$ between $\delta$ and $\delta^S$ is also shown: the dotted line shows the other side of the triangle, $\delta^{\rm miss}\in D/S$.
    \label{fig:subspace_diagram} }
  \end{center}
\end{figure}

Finally, we can now resolve the shift vector $\delta_i$ into its component 
lying within $S$
\begin{equation}\label{eq:deltaS}
\delta_i^{S} = \sum_{\alpha=1,\ldots, N_{\rm sub}} \delta^\alpha e_i^\alpha, 
\end{equation}
and the complementary component within the remaining space $D/S$,
$\delta_i^{\rm miss} = \delta_i-\delta^S_i$.
For a successful test, we expect most of $\delta$ to lie within $S$,
so $|\delta_i^{\rm miss}|\ll |\delta_i|$, or equivalently
$|\delta_i^{S}|\approx |\delta_i|$.
By construction $\delta_i^S$ and $\delta_i^{\rm miss}$ are orthogonal
(since the subspaces $S$ and $D/S$ are orthogonal spaces), 
thus the three vectors $\delta_i^S$, $\delta_i^{\rm miss}$ and $\delta_i$ 
form a right-angled triangle, with $\delta_i$ being its hypotenuse.  The  
geometrical relation between the shift vector $\delta_i$, and
the component of the shift vector which lies in the subspace
$S$, $\delta^S_i$ is illustrated in Fig.~\ref{fig:subspace_diagram}.

With these definitions, the theory covariance matrix $S_{ij}$
provide a reasonable estimate of the MHOU if the angle
\begin{equation}
\theta = \arccos  \lp \frac{|\delta^S_i|}{|\delta_i|} \rp = \arcsin  \lp \frac{|\delta^{\rm miss}_i|}{|\delta_i|}\rp \, 
\label{eq:theta}
\end{equation}
between the shift $\delta_i$ and its component in the subspace
$S$, $\delta_i^S$ is reasonably small.
As mentioned above, for a global PDF fit
the typical situation that one encounters is that
$N_{\rm dat}\gg N_{\rm sub}$ (in the present case $N_{\rm dat}\sim\mathcal{O}(3000)$, while $N_{\rm sub}\sim\mathcal{O}(30)$).
So this validation test is highly nontrivial, since finding the
relatively small subspace $S$ in the huge space $D$ is rather hard:
for a random symmetric matrix $S_{ij}$, components of $\delta_i$ in
$D/S$ will generally be as large as those in $S$, and thus
$|\delta_i^{S}|\ll |\delta_i|$, and $\theta$ will be very close to a
right angle.  
%

\subsection{Results of validation tests}
\label{sec:validresults}

\begin{figure}[t!]
  \begin{center}
    \includegraphics[width=14cm, height=4.4cm]{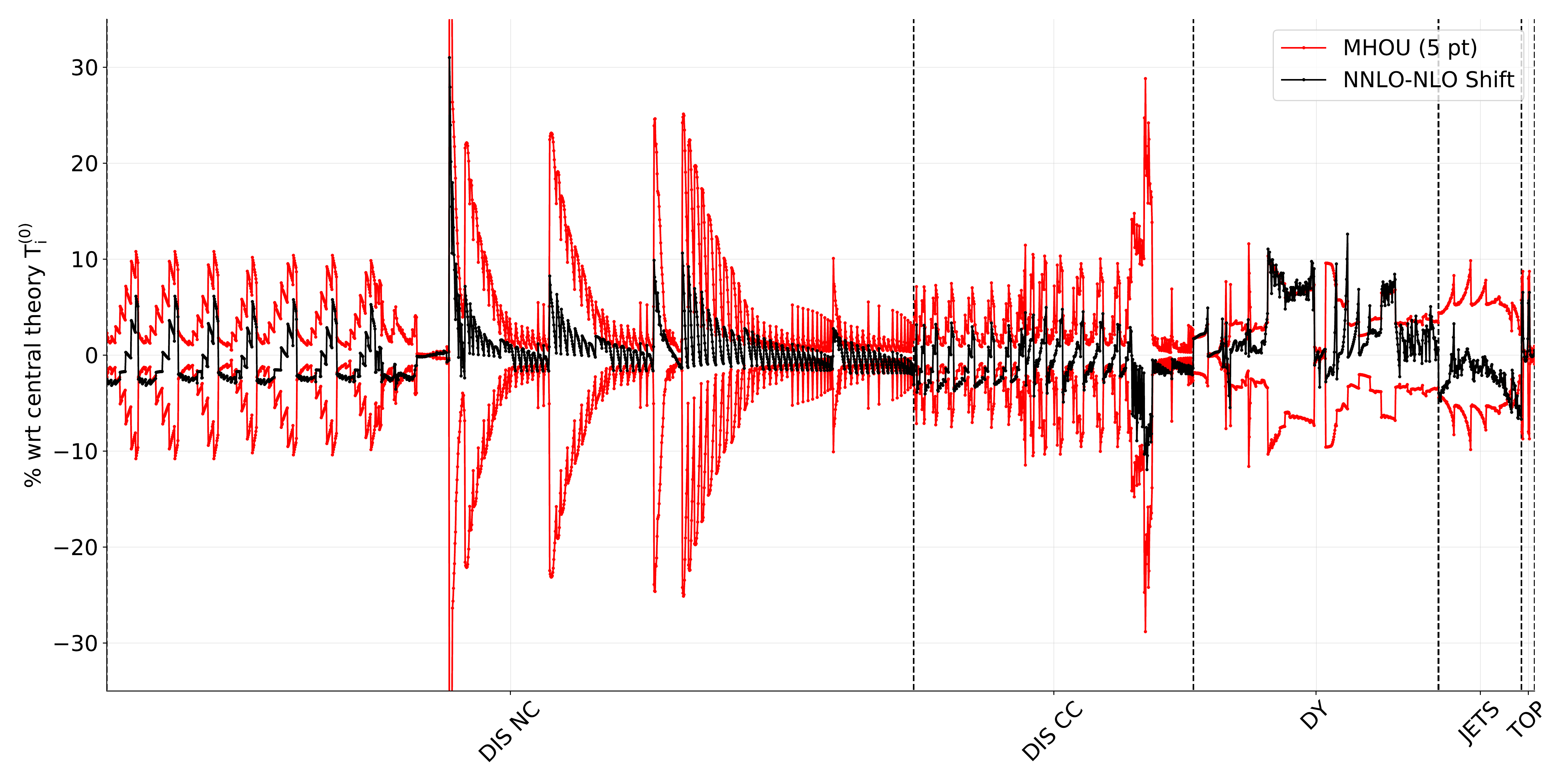}
    \includegraphics[width=14cm, height=4.4cm]{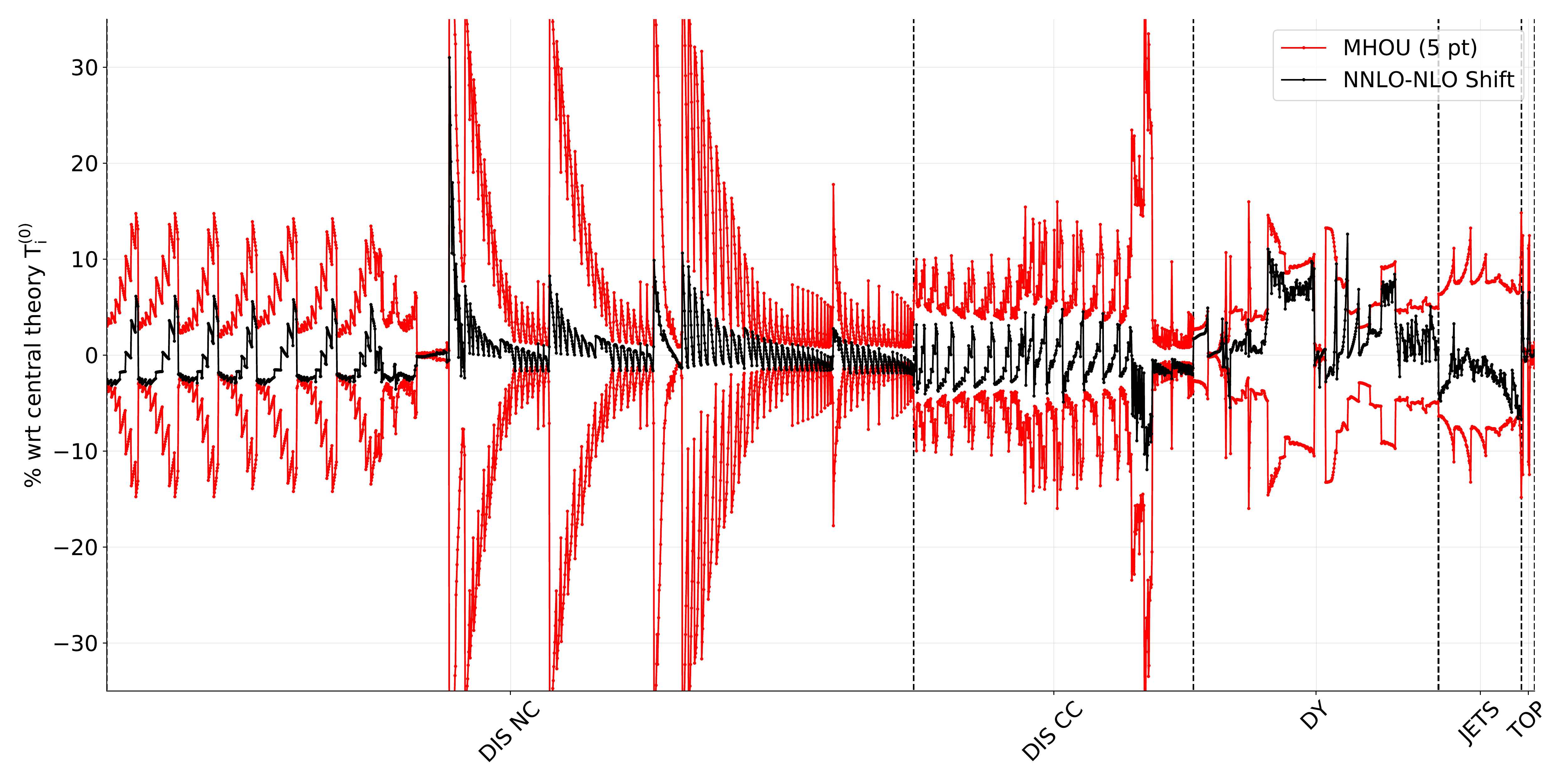}
    \includegraphics[width=14cm, height=4.4cm]{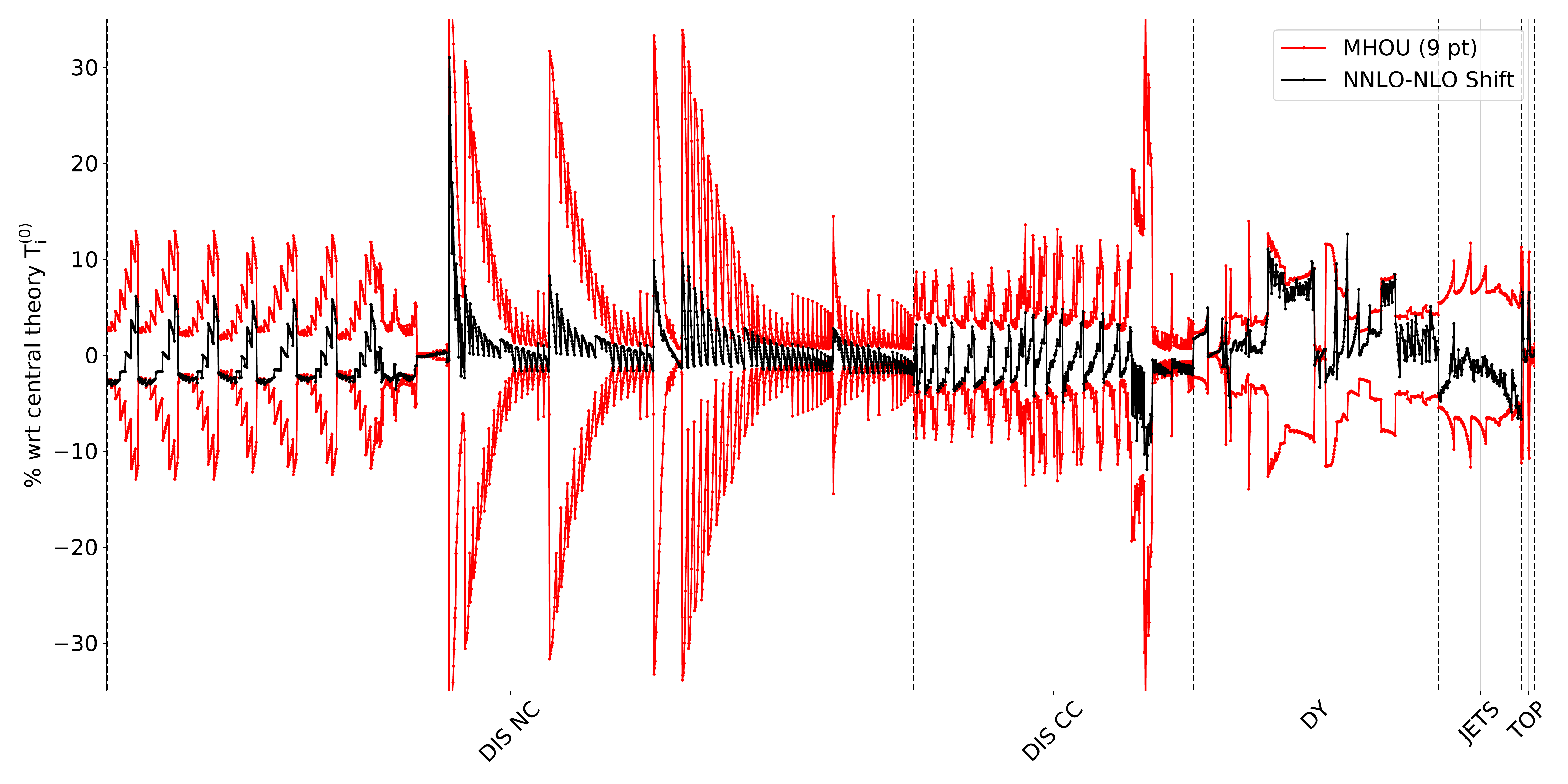}
    \caption{\small The diagonal uncertainties  $\sigma_i$ (red)
      symmetrized about zero,
      compared to the shift $\delta_i$ for each
      datapoint (black), for the symmetric prescriptions:  5-point (top),
      $\overline{5}$-point (middle), and 9-point (bottom). All values
      are shown as percentage of the central theory prediction.}
    \label{fig:diag_shift_validation_symmetric}
  \end{center}
  \begin{center}
    \includegraphics[width=14cm, height=4.4cm]{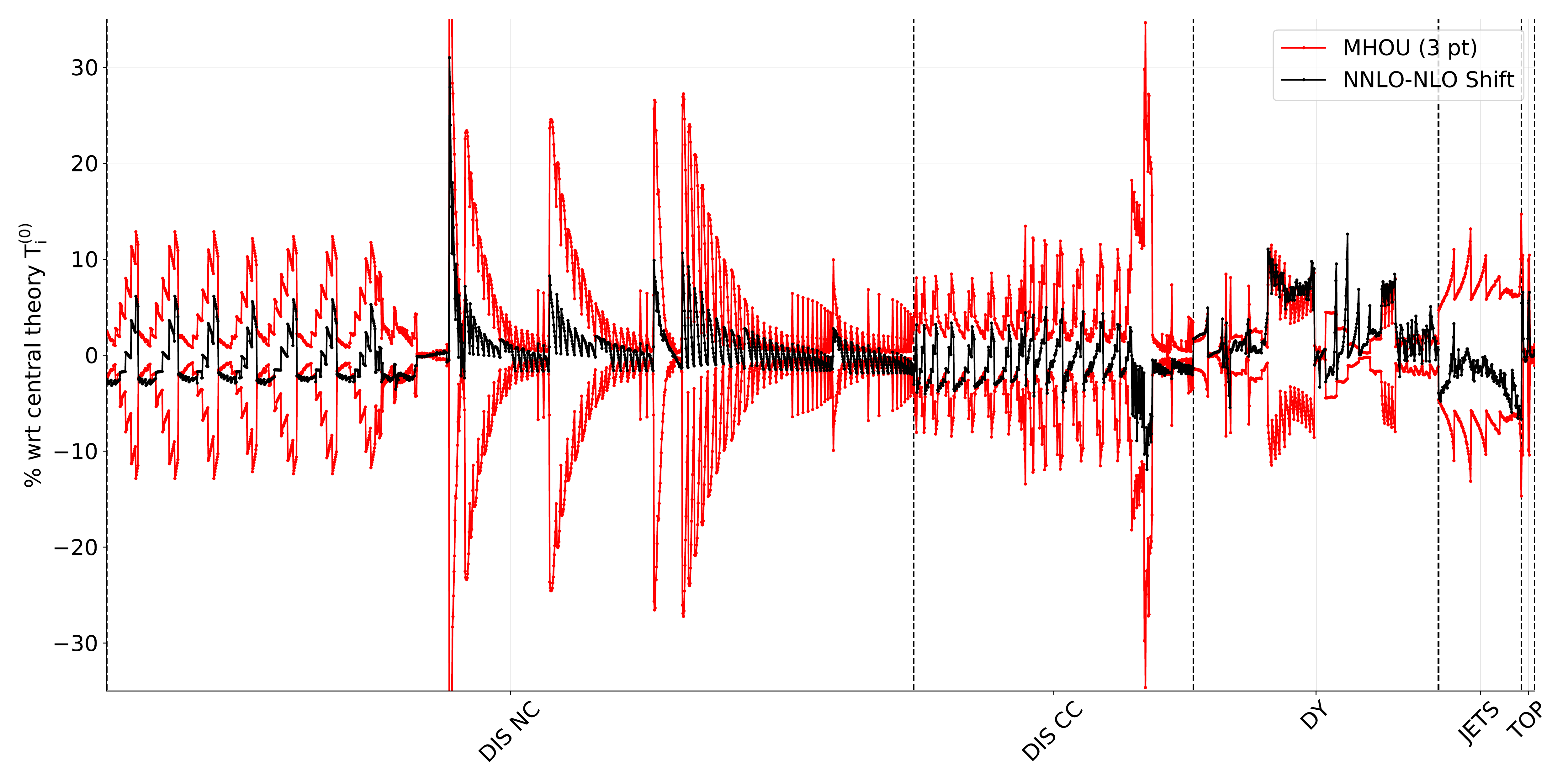}
    \includegraphics[width=14cm, height=4.4cm]{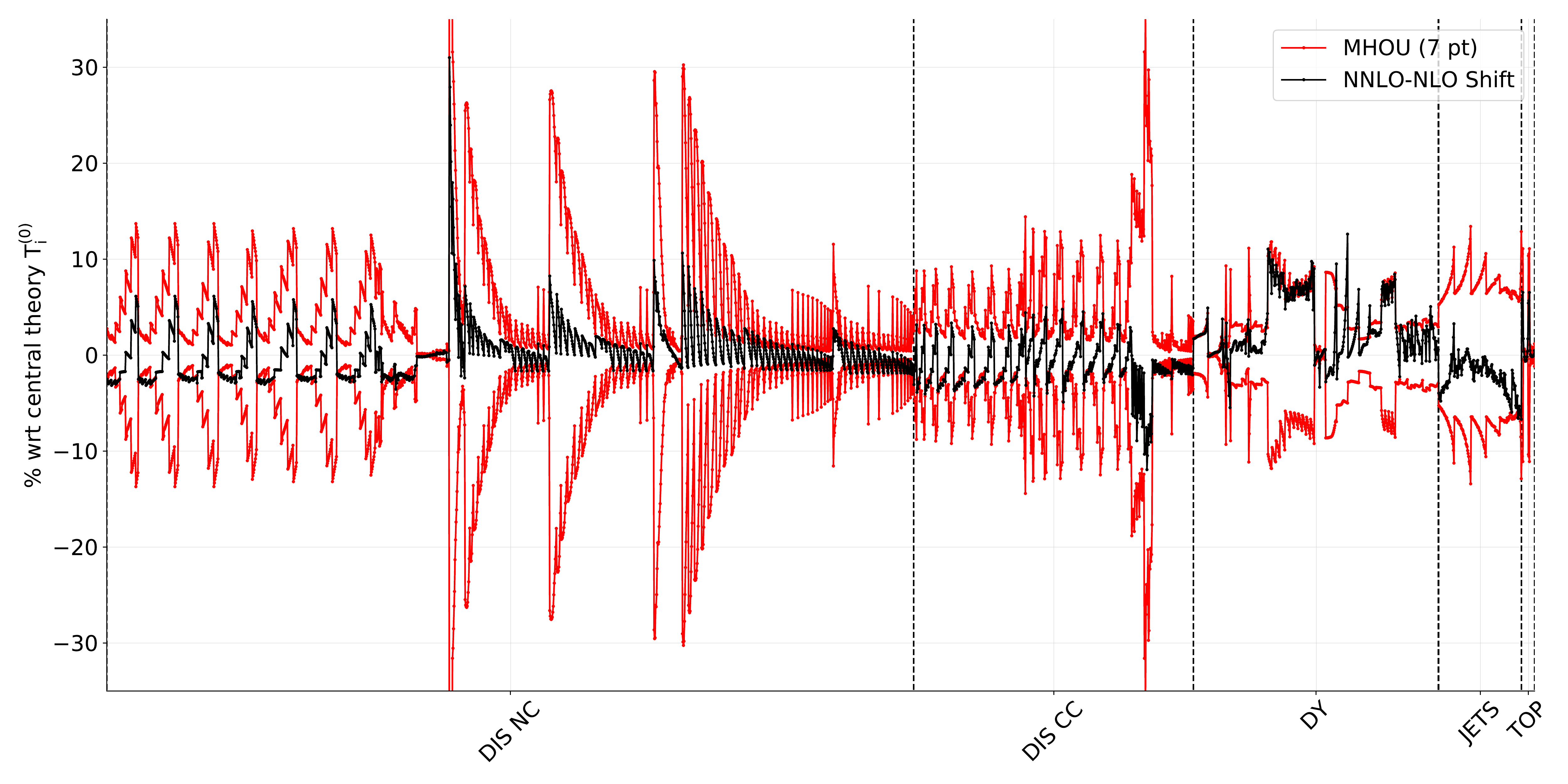}
    \caption{\small Same as Fig.~\ref{fig:diag_shift_validation_symmetric} but for
    the asymmetric prescriptions: 3-point (top) and 7-point (bottom).}
    \label{fig:diag_shift_validation_asymmetric}
  \end{center}
\end{figure}

We now explicitly perform the validation tests discussed in
Sect.~\ref{sec:validconstruction}, with the NLO theory covariance matrices 
$\widehat{S}_{ij}$ (normalized to NLO theory, as in
Eq.~(\ref{eq:Snorm})) constructed from scale variations for all data
points in Table~\ref{tab:datasets_process_categorisation}, and for
each prescription of
Sect.~\ref{sec:prescriptions}. These are then validated using the shift vector $\delta_i$ constructed as the difference of NNLO and NLO theory, normalized to the latter, as in Eq.~(\ref{eq:normshift}).

A very first comparison can be done at the level of diagonal elements $\sigma_i$, where $\widehat{S}_{ii} = (\sigma_i)^2$, 
by comparing them directly to the normalized shifts
${\delta}_i$ Eq.~(\ref{eq:normshift}).
This already tells us
whether the overall size of the scale variation 
is of the right order of magnitude: one expects
the shifts $\delta_i$ and the uncertainties $\sigma_i$ to be of roughly the
same order.

These comparisons are shown in
Figs.~\ref{fig:diag_shift_validation_symmetric}-\ref{fig:diag_shift_validation_asymmetric}.
In each plot the data points are presented sequentially on the horizontal axis, organized by process as in Table~\ref{tab:datasets_process_categorisation}.
The shape of the estimated MHOU imitates the 
shape of the true shift rather faithfully, for each of the five
processes, and for each prescription. This shows that the theory
covariance matrix gives a qualitatively reliable estimate of the true
MHOU, in the sense that the estimate is small when the MHOU is small,
large when it is large, and moreover correctly incorporates the
correlations in the HOU between nearby kinematic regions, responsible
for the shape. There is little discernible difference between all the
various point prescriptions, except in the overall size of the
estimates: for example comparing the symmetric prescriptions, we 
see that 5-point is the least conservative and $\overline{5}$-point is the most conservative, whilst 9-point lies somewhere between the two. This is particularly noticeable in the DY data. 

It is clear from these plots that the overall size of the estimated uncertainties, given by varying renormalization and factorization scales by a factor of two in either direction (i.e. as in 
Eq.~(\ref{eq:rangew}) with $w = \ln 4$) is, by and large, roughly
correct: if the range were significantly smaller, some of the
uncertainties would have been underestimated, whereas if it were
larger all uncertainties would have been overestimated. This said, for
several data points the MHOU at NLO is clearly overestimated by scale
variation: this is particularly true of the small-$x$ NC DIS data from
HERA in the center of the plot. 
  
Overall, these plots demonstrate that since there are only small differences in the diagonal elements of each prescription, it is in the detailed correlations between data points where the differences in
performance between the prescriptions lies. To expose this, we need to diagonalize the theory covariance matrix (using the procedure in Appendix~\ref{sec:diagonalization}), so that we can see in detail which components of the shift vector are correctly estimated, and which are missed, as explained in Sect.~\ref{sec:validconstruction}.

\begin{table}[ht!]
	\centering
\renewcommand*{\arraystretch}{1.30}
 \centering
\begin{tabular}{|c|c|C{30pt}|}
  \toprule
Prescription & $N_{\rm sub}$ & $\theta$ \\
\midrule
    5-pt & 8 & 33$^{\rm o}$ \\
    $\overline{5}$-pt & 12 & 31$^{\rm o}$ \\
    9-pt & 28 & 26$^{\rm o}$ \\\midrule
    3-pt & 6 & 52$^{\rm o}$ \\
    7-pt & 14 & 29$^{\rm o}$ \\
\bottomrule
\end{tabular}

        \vspace{5mm}
	\caption{\small The angle $\theta$  Eq.~(\ref{eq:theta}) between this
          shift and its component $\delta_i^S$ lying within the
          subspace $S$ (see Fig.~\ref{fig:subspace_diagram})
          spanned by the theory covariance matrix for  different
          prescriptions. The dimension of the subspace $S$ in each case
          is also given.}
	\label{tab:global_efficiencies}
\end{table}

\begin{table}[ht!]
	\centering
	\small
\renewcommand*{\arraystretch}{1.20}
\begin{tabular}{|c|c|c|c|c|c|c|c|}
 \toprule
Presc. & $N_{\rm sub}$ & DIS NC & DIS CC & DY & JET & TOP \\
\cline{3-7}
& & 1593 & 552 & 484 & 164 & 26 \\
\hline
 5-pt & 4 &39$^{\rm o}$ & 21$^{\rm o}$ & 25$^{\rm o}$ & 17$^{\rm o}$ & 11$^{\rm o}$	\\
$\overline{5}$-pt & 4 & 38$^{\rm o}$ & 17$^{\rm o}$ & 23$^{\rm o}$	& 22$^{\rm o}$ & 10$^{\rm o}$ \\
9-pt & 8 & 32$^{\rm o}$ & 16$^{\rm o}$ & 22$^{\rm o}$ & 14$^{\rm o}$ & 3$^{\rm o}$	\\
\hline
 3-pt & 2 &54$^{\rm o}$ & 36$^{\rm o}$ & 39$^{\rm o}$ & 24$^{\rm o}$ & 12$^{\rm o}$ \\
7-pt & 6 &35$^{\rm o}$ & 17$^{\rm o}$ & 22$^{\rm o}$ & 16$^{\rm o}$ & 3$^{\rm o}$	\\
    \bottomrule
\end{tabular}

        \vspace{3mm}
	\caption{Same as Table~\ref{tab:global_efficiencies}
          for each process of Table~\ref{tab:datasets_process_categorisation}. The number of data points in each process is given directly below the name of the process.}
	\label{tab:process_efficiencies}
\end{table}

As discussed in Sect.~\ref{sec:validconstruction}, once we have the eigenvectors corresponding to the nonzero eigenvalues of the theory covariance matrix, 
the first validation test consists of checking how much of the shift vector 
$\delta_i$ lies within the space spanned by these eigenvectors, $S$, and has thus been included in the estimation of MHOU provided by the theory 
covariance matrix.
The results of this test for the global dataset, described in Sect.~\ref{sec:inputdata}, are shown in  
Table~\ref{tab:global_efficiencies}: for each prescription we
give the dimension $N_{\rm sub}$
of $S$, i.e. the number of linearly independent eigenvectors
$e_i^\alpha$ of $S_{ij}$, and then the value of the angle $\theta$,
defined in Eq.~(\ref{eq:theta}),
between the shift $\delta_i$  and its component $\delta^S_i$, defined in Eq.~(\ref{eq:deltaS}), lying within 
the subspace $S$ spanned by $e_i^\alpha$. 
We note that all the angles are reasonably small, despite
the fact that $N_{\rm sub}$ is so much smaller that the dimension
$2819$ of the data space. 

The 9-point prescription performs best, with an angle of $\theta=26^{\rm o}$ between the shift $\delta_i$ and its projection $\delta_i^S$ in the subspace $S$: clearly the more complicated pattern of scale variations (compared to the other two symmetric prescriptions) improves the estimation of the MHOU.
The 3-point prescription performs worst, suggesting that lack of
correlation in the factorization scale between processes in this prescription 
means that much of the correlation in the MHOU due to universal PDF evolution has been missed.
The 7-point prescription is however only a little worse than 9-point, presumably due to the dilution of the correlation in factorization scale variation which is a feature of this prescription. Note that since these results for $\theta$ are geometrical, they are largely independent of the range of the scale variation Eq.~(\ref{eq:rangew}).

\begin{figure}[ht!]
  \begin{center}
    \includegraphics[width=0.99\textwidth]{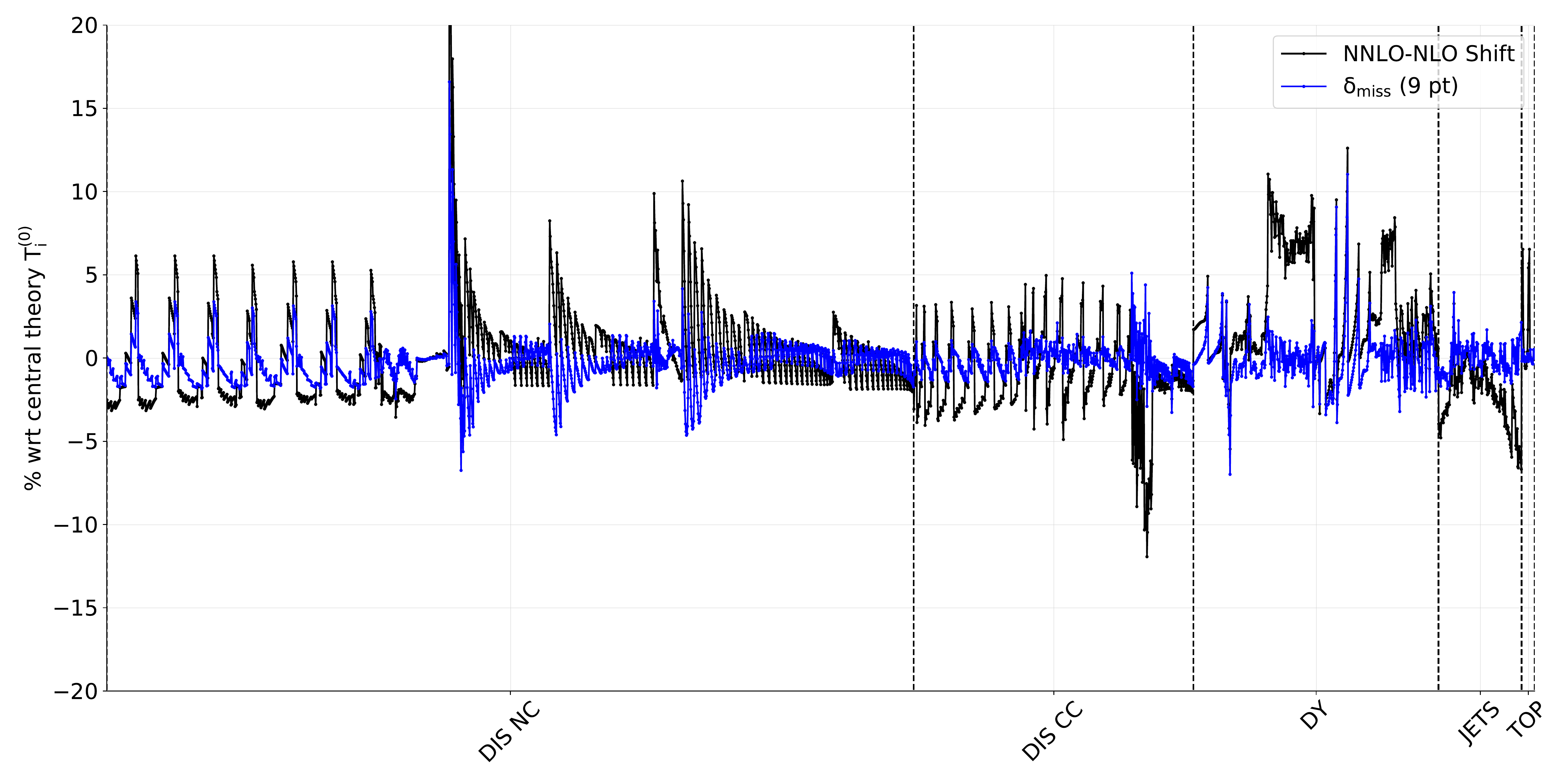}
    \caption{The NNLO-NLO shift $\delta_i$ (black) compared to its 
component $\delta_{\rm miss}$ (blue) which lies outside the subspace $S$, computed using the 9-point prescription.}
    \label{fig:deltamiss}
  \end{center}
\end{figure}

\begin{figure}[ht!]
  \begin{center}
    \includegraphics[scale=0.6]{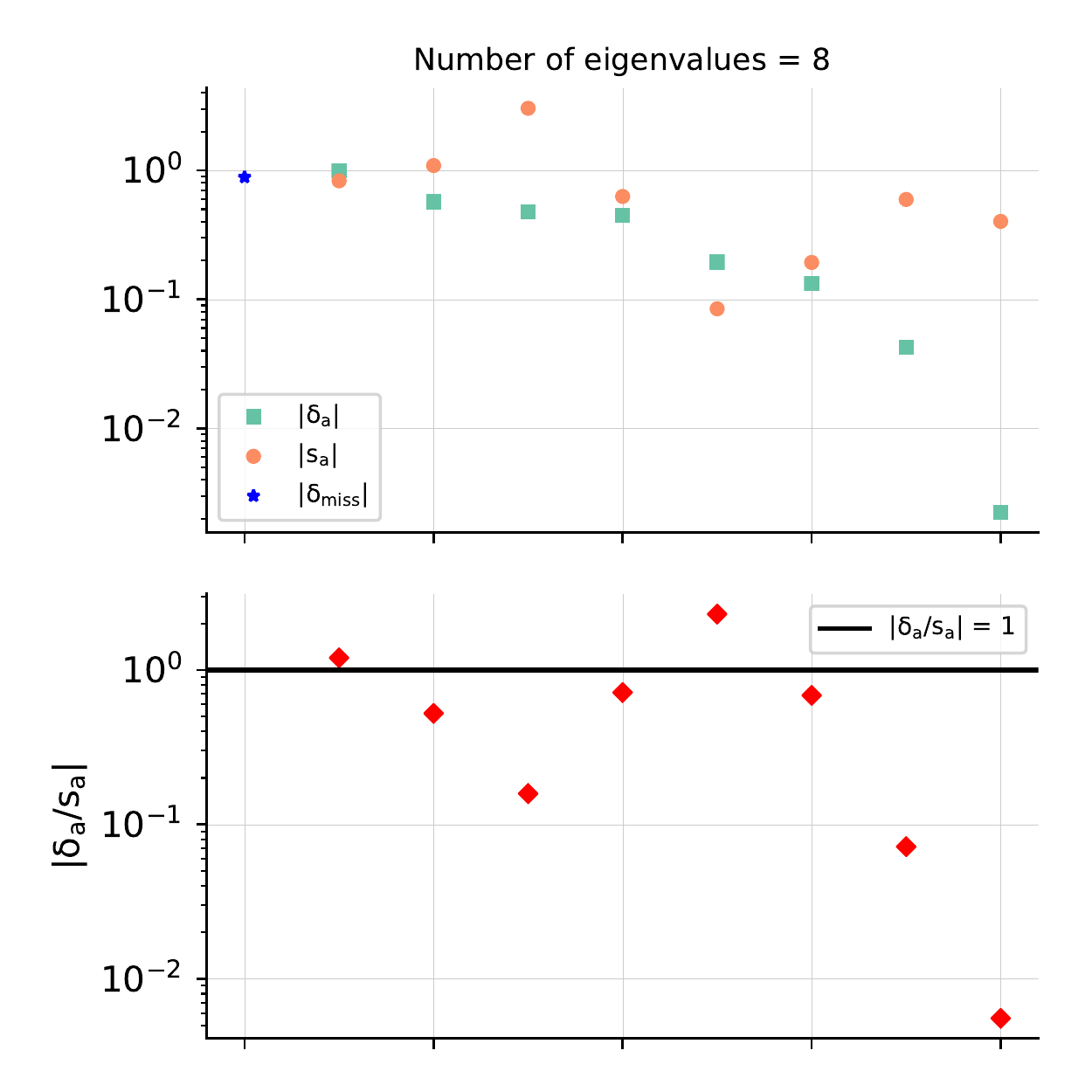}
    \includegraphics[scale=0.6]{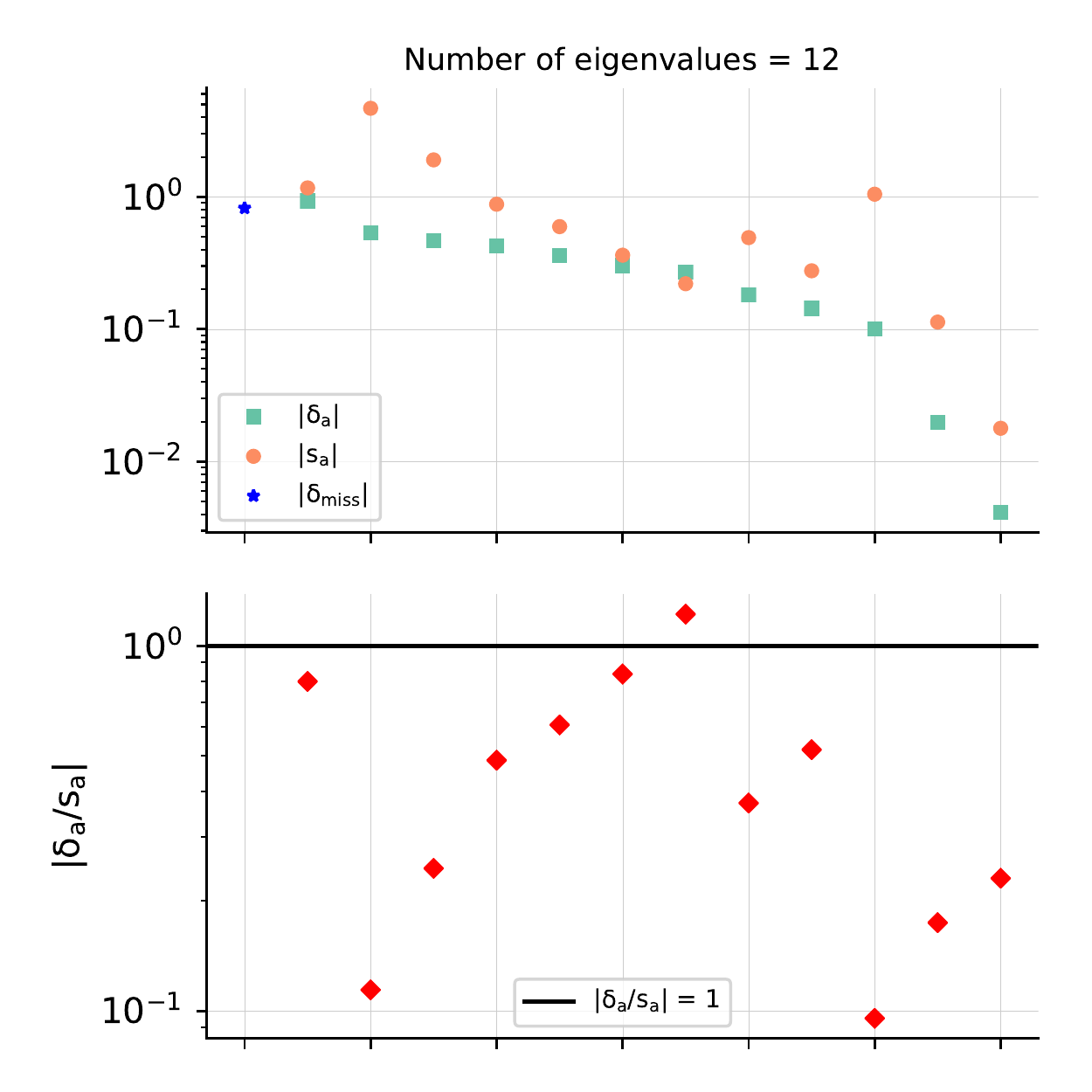}
    \includegraphics[scale=0.6]{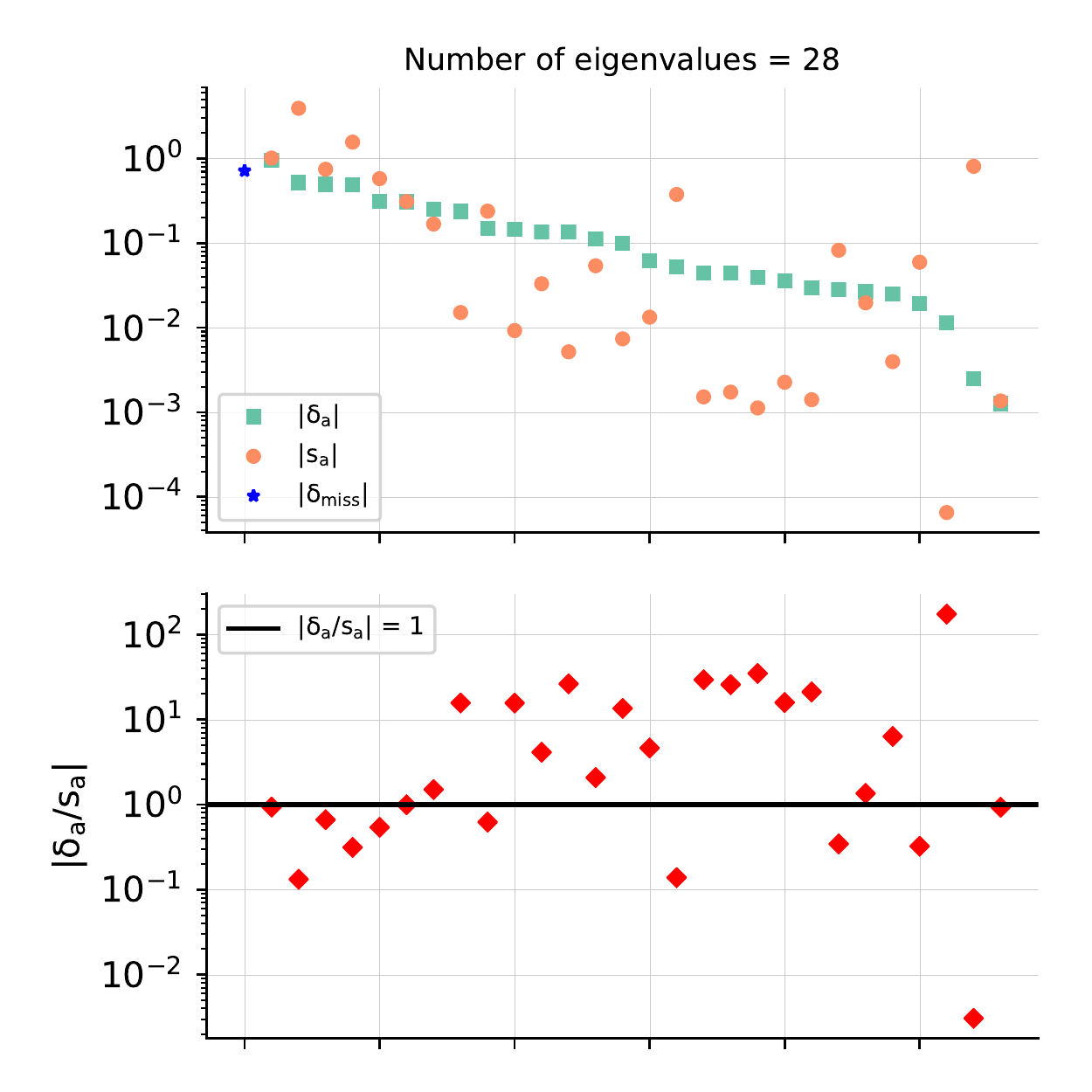}
    \caption{\small The projection $\delta^\alpha$ Eq.~(\ref{eq:deltaproj}) of the normalized shift vector
      $\delta_i$ Eq.~(\ref{eq:normshift}) along each eigenvector $e^\alpha_i$ of the normalized theory covariance
      matrix Eq.~(\ref{eq:Snorm}), compared to the corresponding eigenvalue 
    $s^\alpha$, ordered
      by the size of the projections (from largest to
      smallest). In each case results are shown as absolute (upper) and
      as ratios $\delta^\alpha/s^\alpha$ (lower), the horizontal line indicating when this ratio is one.  The length of the 
        component of 
       $\delta_i$ that is not captured at all by the theory covariance 
    matrix, $|\delta^{\rm miss}_i|$ is also shown (blue star).
      Results are shown for the symmetric prescriptions: 5-point (top left),
    $\overline{5}$-point (top right), and 9-point (bottom).}
    \label{fig:evals_all_prescriptions_symmetric}
  \end{center}
\end{figure}

It is interesting to ask whether all processes are equally well
described, and whether there are significant differences in
correlations between processes or within a process. 
To this purpose, in Table~\ref{tab:process_efficiencies} we list the angle 
$\theta$ computed for each individual process using the various prescriptions.
Three conclusions emerge from inspection of this table. First, when each process is taken individually, the results seen in Table~\ref{tab:global_efficiencies} for the relative merits of each prescription are replicated process by process: again 3-point is worst, and 9-point is best.
Secondly, processes with large numbers of data points are much harder to describe than those with only a few data points (i.e. $\theta$ is smallest for smaller datasets): this is hardly surprising, since the larger datasets cover a wider kinematic range and thus have more structure to predict. Finally, the quality of the description of the global
dataset for each prescription is in each case dominated by the process (DIS NC) which is described worst, however the global dataset is actually described a little better (for each prescription) than the dataset for this process, particularly for 9-point, less so for 3-point.
This suggests that correlations across 
processes are actually described reasonably well, and are anyway less critical than correlations within processes.

We next look in more detail at the part of $\delta_i$ which falls
outside the subspace $S$, $\delta^{\rm miss}_i =
\delta_i-\delta^S_i$. This is shown for the 9-point prescription in
Fig.~\ref{fig:deltamiss}. While this is generally uniformly small, of
order a few percent, across the full range of processes, it also has
nonzero components in all datasets, and all processes. Furthermore,
for most processes the shape of $\delta^{\rm miss}$ closely follows
that of the shift $\delta_i$. 
This may suggest that a significant fraction of  $\delta^{\rm miss}$
might be due to the fact that there is a component of $\delta_i$ which
is systematically missing for most or all processes. This in turn
suggests that a sizable part of $\delta^{\rm miss}$ might be due to
poor estimation of the MHOU
in PDF evolution, rather than poor estimation of MHOU in hard
cross-sections which can vary substantially between different
processes (and indeed different kinematics). Indeed, as already mentioned in
Sect.~\ref{multiple_var_sec}, our current treatment of factorization
scale variation is only approximate, and a more sophisticated
treatment would involve performing separate scale variation for each
eigenvalue of perturbative evolution.

\begin{figure}[ht!]
  \begin{center}
    \includegraphics[scale=0.6]{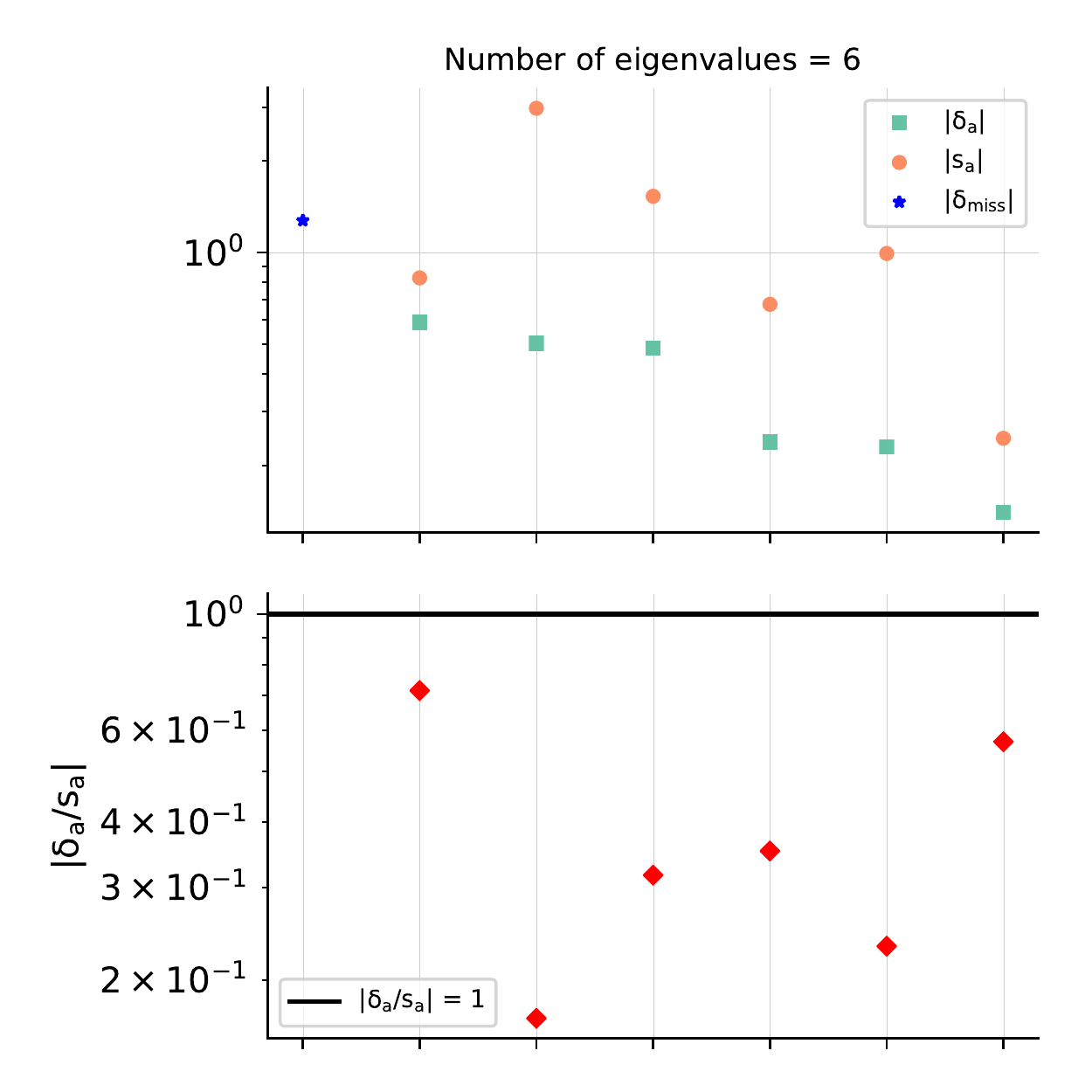}
    \includegraphics[scale=0.6]{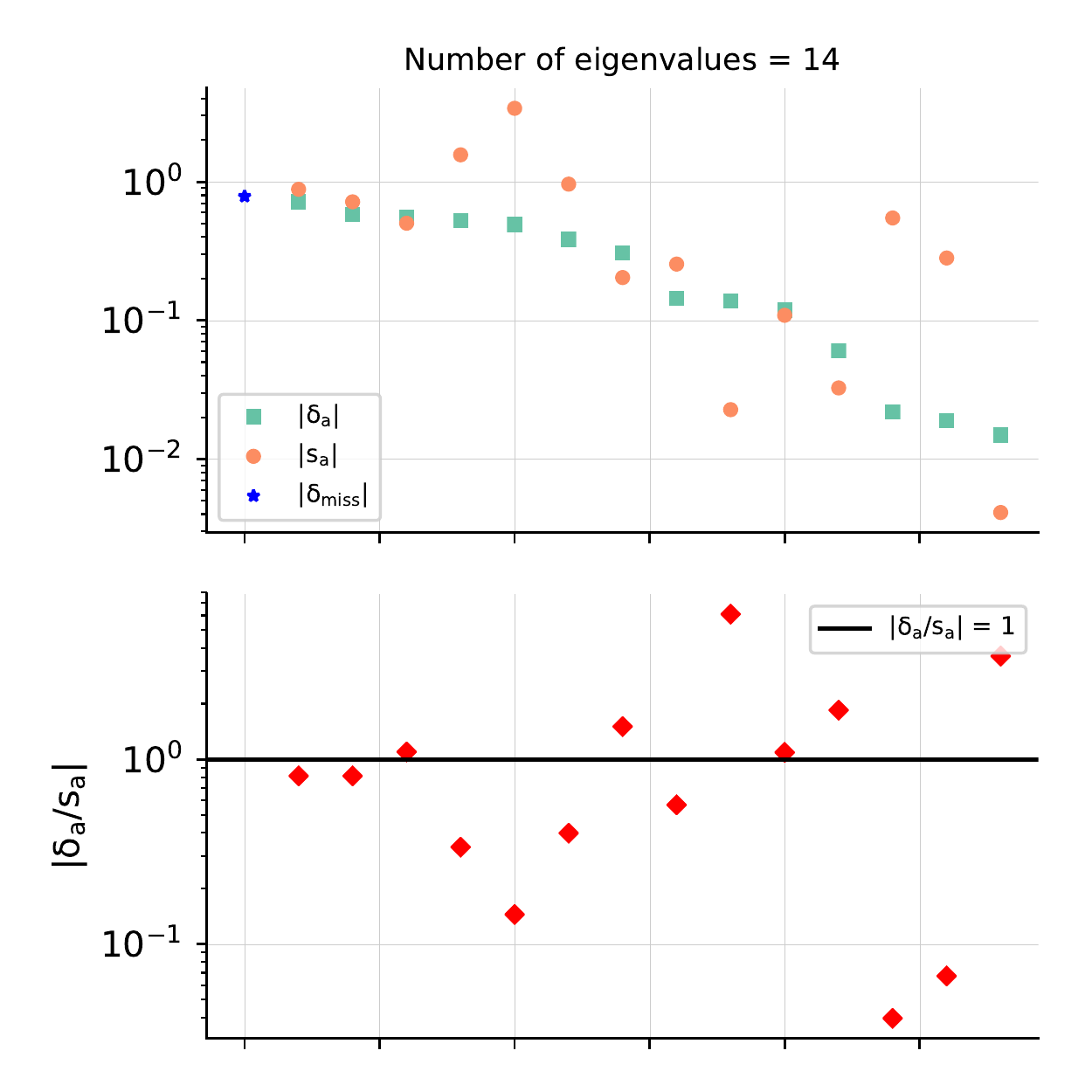}
    \caption{\small Same as Fig.~\ref{fig:evals_all_prescriptions_symmetric} but for
    the asymmetric prescriptions: 3-point (left) and 7-point (right).}
    \label{fig:evals_all_prescriptions_asymmetric}
  \end{center}
\end{figure}

Having established that most of the NNLO-NLO shift $\delta_i$ lies within $S$, we now proceed to examine what fraction of $\delta^S_i$ lies with the error ellipse $E$ specified by the theory covariance matrix. To that end, the 
eigenvalues $\lambda^\alpha = (s^\alpha)^2$ of the theory covariance 
matrix of the global dataset are shown in
Fig.~\ref{fig:evals_all_prescriptions_symmetric} for symmetric
prescriptions, and in
Fig.~\ref{fig:evals_all_prescriptions_asymmetric} for the asymmetric
ones: these define the length of the semi-axes of $E$. Since there are five distinct processes, there are $8$, $12$ and $28$ positive eigenvalues for the symmetric $5$-point, $\overline{5}$-point and $9$-point prescriptions respectively, and $6$, $14$ positive eigenvalues for the asymmetric $3$-point and $7$-point prescriptions, as explained in Appendix~\ref{sec:diagonalization}. Also shown are the projections $\delta^\alpha$ of the 
normalized shift vector $\delta$ Eq.~(\ref{eq:normshift}) along each 
corresponding eigenvector $e^\alpha_i$, Eq.~(\ref{eq:deltaproj}).

\begin{figure}[ht!]
  \begin{center}
    \includegraphics[width=0.95\textwidth]{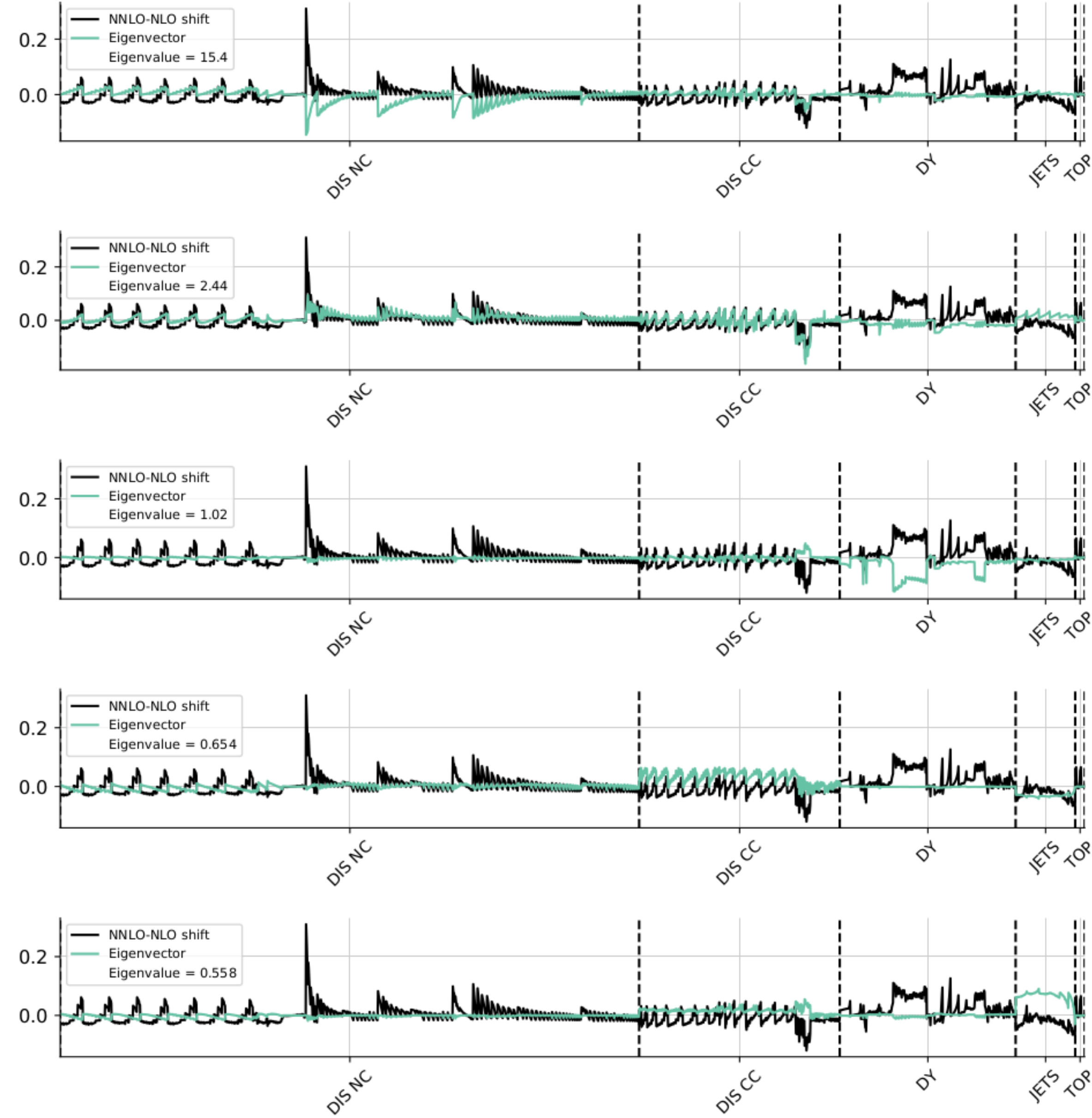}\\
    \caption{\small The components  $e^\alpha_i$ (green) of the eigenvectors, corresponding
      to the five largest eigenvalues for the 9-point theory covariance matrix, shown in the same format as Fig.~\ref{fig:diag_shift_validation_symmetric}. The NNLO-NLO shift, $\delta_i$ (black), is shown for comparison.
    \label{fig:evecs1} }
  \end{center}
\end{figure}

Inspection of these plots confirms that all the prescriptions seem to 
perform reasonably well. The largest eigenvalue is always very similar
in size to the shift, and  the size of the eigenvalues generally falls
as the projected shifts get smaller.  As expected, the 3-point
prescription clearly overestimates uncertainties, since $\delta^\alpha
< s^\alpha$ for all the eigenvalues. The same is true, but to a lesser
extent, for both 5-point and $\overline{5}$-point.  For the more complicated 7-point and 9-point 
prescriptions the largest projections (corresponding to the first seven or eight eigenvalues) are estimated rather well, though still perhaps a little conservatively, but for the smaller projections the scatter increases significantly, with some projected shifts hardly predicted at all. This is perhaps not surprising: when varying just six independent scales, we can only expect to obtain only a limited amount of information on the MHO terms. However the correct estimation of the largest projected shifts shows that the theory covariance matrix is giving a reasonable estimation of the MHOU, especially when implemented through the more complicated prescriptions.

On each of these plots, we also show the length of the component 
$\delta^{\rm miss}_i$ that is orthogonal to $S$, and thus completely outside $E$. For the symmetric prescriptions, $|\delta^{\rm miss}_i|$ is always less than the largest component of $\delta$ in $S$, while for the asymmetric prescriptions it is greater, very significantly so for the 3-point prescription. This is another indication that the symmetric prescriptions give a better account of the correlations in theoretical uncertainties. 

A more detailed understanding of the physical meaning of each
eigenvector can be acquired by inspecting its components $e_i^\alpha$ in the data space. These are shown in Fig.~\ref{fig:evecs1} for the
eigenvectors corresponding to the five largest eigenvalues in the
9-point prescription: the shift vector $\delta_i$ is also shown for comparison.
It is clear that there is a close correspondence between eigenvectors
and MHO contributions to individual processes. For instance the first
eigenvector contributes mostly to DIS NC, the second to both DIS NC and DIS CC, the third to DY, the fourth mainly to DIS CC, and the fifth mainly to JETS. Clearly the ordering of these larger eigenvalues is related to the number of data points for the respective processes: the more datapoints, the larger the eigenvalue of the (correlated) uncertainty estimate. Even relatively small eigenvalues can give an important contribution, though to processes with fewer datapoints: for example the ninth eigenvector (not shown) clearly dominates TOP.

In summary, from these validation tests it is apparent that the 9-point
prescription gives a reasonable estimate of most of the MHOU, both for individual processes and for the global dataset, with the 7-point being just slightly
worse. Based on this, we will therefore adopt 9-point as a default prescription
for the theory covariance matrix in the PDF determination to be discussed in 
the next section.

\section{PDFs with missing higher order uncertainties}
\label{sec:fitstherr}

We can now present the main results of this work: the
first determination of the parton distributions of the proton
which systematically accounts for the MHOUs affecting the theory
calculations of the input processes for the fit.
First  we present the results for PDFs obtained by 
fitting only DIS data. This provides us with an initial test case, which we
will study by comparing PDFs obtained including
the combined experimental and theoretical covariance matrix to
the corresponding baseline fit in which only experimental
uncertainties are included.

We then turn to the  global PDF determination,
which offers a nontrivial validation of our methodology,
specifically by comparing NLO PDFs, with and without MHOUs,
to NNLO PDFs. 
For global fits, we also study the stability of the results to
changes in the prescription used for the computation of the theory
covariance matrix: specifically, we compare
PDFs obtained with the  9-point
prescription  (which is our default)
to those based on the  7- and 3-point ones.
We  also study PDFs determined by only partially including the
theory covariance matrix, either only in the data generation or only in the
fitting. As discussed in the introduction, this provides us with a way of
disentangling the impact of the theory covariance matrix on the
central value of the PDFs or on the PDF uncertainty.

As discussed in Sect.~\ref{sec:thcovmat}, the theory uncertainties are included by simply replacing the experimental covariance
matrix $C_{ij}$ with the sum $(C+S)_{ij}$ of the experimental and theory
covariance matrices in the expression for the likelihood of the true
value given the data.
The NNPDF methodology, as used specifically in
the determination of the most recent NNPDF3.1 PDF
set~\cite{Ball:2017nwa}, is otherwise unchanged.
Within this methodology, the covariance matrix is used to generate
$ N_{\rm rep}$ pseudodata replicas $D^{(k)}_i$ for each datapoint $i$, with
$k=1,\dots, N_{\rm rep}$, whose distribution must reproduce the covariance
of any two data points. This means that with theory uncertainties
included,
\be
\label{eq:dgen}
\lim_{N_{\rm rep}\to\infty}\frac{1}{N_{\rm rep}(N_{\rm rep}-1)}\sum_{k=1}^{N_{\rm rep}} \left(D_i^{(k)}-\langle
  D_i\rangle \right)
  \left(D_j^{(k)}-\langle D_j\rangle \right)= C_{ij}+S_{ij},
  \ee
  with $\langle D_i\rangle= \frac{1}{N_{\rm rep}}\sum_{k=1}^{N_{\rm rep}}  D_i^{(k)}$ 
denoting the average over Monte Carlo replicas.

A PDF replica is then fitted to each pseudodata replica $D_i^{(k)}$ by
minimizing a figure of merit, which in the presence of theory
uncertainties becomes
\be
\label{eq:chi2_v3}
\chi^{2}=\frac{1}{N_{\rm dat}}\sum_{i,j=1}^{N_{\rm dat}}\lp  D_i-T_i\rp
\lp C+S \rp^{-1}_{ij} \lp D_j-T_j\rp ,
\ee
where $T_i$ is the theory prediction evaluated
with the central scale choice, and the 
theory covariance matrix $S_{ij}$ is computed using one of the prescriptions 
presented in Sect.~\ref{sec:prescriptions}. 

It is thus clear that the inclusion of a theory-induced contribution
in the covariance matrix affects only two steps of the procedure: the
pseudodata generation, and the minimization. Everything else is
unchanged, and is identical to the default NNPDF methodology.
Note
that in particular the experimental covariance matrix $C$ used in 
the fitting is determined, as in NNPDF3.1 and previous NNPDF releases using the
so-called $t_0$ method for the treatment of multiplicative
uncertainties, in order to 
avoid d'Agostini bias (see Refs.~\cite{Ball:2009qv,Ball:2012wy} for a
detailed discussion).
As in previous NNPDF releases, minimization is thus
performed using the $t_0$ definition of the $\chi^2$, but all $\chi^2$
values shown are computed using the covariance matrix as published by
the respective experiments.

In the sequel, in order to assess fit quality we will provide $\chi^2$
values, and also, we will study the estimator, defined in Ref.~\cite{Ball:2014uwa}
\begin{equation}
\phi= \sqrt{\langle \chi_{\rm exp}^2[T_i]\rangle -
  \chi^2_{\rm exp}[\langle T_i\rangle]}\, ,
\label{eq:frdbdef}
\end{equation}
where by $ \chi_{\rm exp}^2[T_i]$ we denote the value of the $\chi^2$
computed using the $i$-th PDF replica, and only including the
experimental covariance matrix (thus Eq.~(\ref{eq:chi2_v3}), but with $S_{ij}$ 
set to zero).
The average $\chi^2$ values which enter Eq.~(\ref{eq:frdbdef}) are then
$\langle
\chi^2[T_i]\rangle$,  the mean value of this $\chi^2$ averaged over
replicas, and $\chi^2[\langle T_i\rangle]$, the value of the
  $\chi^2$ computed using the ``central'' PDF set which is found
  by averaging over replicas.

  It was shown in Ref.\cite{Ball:2014uwa} that $\phi$ then gives the 
average over all datapoints of the ratio of 
  the uncertainties of the predictions to
  the uncertainties of the original experimental data, taking account of correlations:
\begin{equation}
\phi= \Big(\frac{1}{N_{\rm dat}}\sum_{i,j=1}^{N_{\rm dat}}(C)^{-1}_{ij} T_{ij}
\Big)^{1/2}\, ,
\label{eq:frdbrat}
\end{equation}
where $T_{ij}= \langle T_i T_j\rangle - \langle T_i\rangle\langle
T_j\rangle$ is the covariance matrix of the theoretical predictions.
For an uncorrelated
covariance matrix, this is just the ratio of the uncertainty in the
prediction using the output PDF to that of the original data.
Hence,  the value of $\phi$ provides an estimate
of the mutual theoretical consistency of the data which are being fitted:
consistent data are combined by the underlying theory and lead to an
uncertainty in the prediction which is significantly smaller than that
of the original data.
Note that $\phi$ is always defined so that the uncertainty in the prediction
is normalized to the original 
experimental uncertainty (rather than combined experimental and theory 
uncertainties). In particular, when considering PDFs determined
including a theory covariance matrix, this means that PDFs are
determined minimizing the $\chi^2$ Eq.~(\ref{eq:chi2_v3}), but
$\chi^2_{\rm exp}$ is instead used in the computation of $\phi$
Eq.~(\ref{eq:frdbdef}). 
 
When changing the covariance matrix from $C$ to $C'=C+S$ the
fluctuations of the replicas will change, according to
Eq.~(\ref{eq:dgen}), and if theoretical uncertainties change in the
same proportion one would expect the value of $\phi$ to become $\phi'=
r_\phi \phi$, with
\begin{equation}
r_\phi= \Big(1+\frac{1}{N_{\rm dat}}\sum_{i,j=1}^{N_{\rm dat}}(C)^{-1}_{ij} S_{ij}
\Big)^{1/2}.
\label{eq:frdbratnaive}
\end{equation}
Thus, when including MHOU, all else being equal, we would expect PDF
uncertainties to increase by a factor $r_\phi$. This
will provide us with a baseline to which we can compare the change in
uncertainty which is actually observed.

All the PDF sets which have been produced and which will be discussed in this
section are listed in Table~\ref{tab:thcovmatFits}.
For each of the fits, 
we indicate its label, the input dataset,
   the  perturbative order and the covariance matrix used.
   For the fits that include a theory covariance matrix, we also indicate the prescription
   with which it has been constructed. 
   In the remainder of this section we discuss the main features
   of these PDF sets.
  
\begin{table}[t]
  \centering
\footnotesize
  \renewcommand*{\arraystretch}{1.50}
  \begin{tabular}{lccccc}
    Label                   & Dataset  & $\quad$Order$\quad$  & Cov. Mat. &  Comments \\
    \toprule
        {\tt NNPDF31\_nlo\_as\_0118\_dis\_kF\_1\_kR\_1}  &  DIS   &   NLO  & $C$  & baseline DIS-only NLO  \\
        {\tt NNPDF31\_nlo\_as\_0118\_dis\_scalecov\_9pt} &  DIS   &   NLO  & $C+S^{(\rm 9pt)}$  &  \\
        \midrule
        {\tt NNPDF31\_nnlo\_as\_0118\_dis\_kF\_1\_kR\_1}  &  DIS   &   NNLO  & $C$  & baseline DIS-only NNLO  \\
        \midrule
        {\tt NNPDF31\_nlo\_as\_0118\_kF\_1\_kR\_1}  &  Global   &   NLO  & $C$  & baseline Global NLO  \\
        {\tt NNPDF31\_nlo\_as\_0118\_scalecov\_9pt}  &  Global   &   NLO  & $C+S^{(\rm 9pt)}$  &  \\
        {\tt NNPDF31\_nlo\_as\_0118\_scalecov\_7pt}  &  Global   &   NLO  & $C+S^{(\rm 7pt)}$  &  \\
        {\tt NNPDF31\_nlo\_as\_0118\_scalecov\_3pt}  &  Global   &   NLO  & $C+S^{(\rm 3pt)}$  &  \\
        \midrule
         {\tt NNPDF31\_nlo\_as\_0118\_scalecov\_9pt\_fit}  &  Global   &   NLO  & $C+S^{(\rm 9pt)}$  & $S$ only in $\chi^2$
            definition \\
            {\tt NNPDF31\_nlo\_as\_0118\_scalecov\_9pt\_sampl}  &  Global   &   NLO  & $C+S^{(\rm 9pt)}$  & $S$ only in sampling \\
            \midrule
        {\tt NNPDF31\_nnlo\_as\_0118\_kF\_1\_kR\_1}  &  Global   &  NNLO  & $C$  & baseline Global NNLO  \\
            \bottomrule
  \end{tabular}
  \vspace{0.3cm}
  \caption{\small Summary of the PDF sets discussed in this
    section. The dataset, perturbative order and nature of the
    treatment of uncertainties for each set are indicated.
    \label{tab:thcovmatFits}
  }
  \end{table}


\subsection{DIS-only PDFs}

We first discuss PDF sets based on DIS data only.
Fit quality indicators are collected in
Table~\ref{table:chi2table_covth_dis}. The theory covariance matrix is always
constructed using the 9-point prescription. 
We show the value of $\chi^2/N_{\rm dat}$ and of the $\phi$ estimator defined in
Eqs.~(\ref{eq:chi2_v3},\ref{eq:frdbdef})  respectively.
Results are shown
for both the total dataset and for the  individual DIS experiments of Table~\ref{tab:datasets_process_categorisation}.
Note that the total $\chi^2$ is no longer
just the weighted sum of the individual $\chi^2$s, because it now also includes
correlations between experiments.

\begin{table}[t]
\begin{center}
\renewcommand*{\arraystretch}{1.78}
\footnotesize
\begin{tabular}{|l|c|C{0.8cm}c|c|C{0.8cm}c|c|}
  \toprule
  \multicolumn{8}{|c|}{ NNPDF3.1 DIS-only fits}    \\
  &    & \multicolumn{3}{c|}{$\chi^2/N_{\rm dat}$}  & \multicolumn{3}{c|}{$\phi$}   \\
 Dataset & $n_{\rm dat}$ & \multicolumn{2}{c|}{NLO}  & NNLO  & \multicolumn{2}{c|}{NLO}  & NNLO  \\
 &  &  $C$ & $C+S^{(\rm 9pt)}$  &  $C$ &  $C$ & $C+S^{(\rm 9pt)}$  &  $C$   \\
\toprule
NMC         & 134 &  1.259  &  1.236     &   1.239        & 0.388   & 0.464    &  0.444      \\
SLAC        & 12  &  0.908 &  0.543     &   0.791         &  0.247  &  0.590   &  0.343            \\
BCDMS       & 530 &  1.046  &  1.017     &  1.047         &  0.339  &  0.505   &   0.389          \\
CHORUS      & 430 &  0.982   & 0.856      &   1.124        & 0.409   & 0.418    &  0.512           \\
NuTeV       & 41  & 0.628   &  0.491     &  0.872          & 0.940   & 0.994    &  1.35          \\
HERA incl   & 967 &  1.097  &   1.066    &   1.104         &  0.280  &  1.013   &  0.335         \\
HERA $F_2^c$& 31  &  1.047  &  0.997           &     1.033     &  0.526  &  1.097   &  0.631        \\
\midrule
Total       & 2145 &  1.061  &   1.032    &    1.095        & 0.358   & 0.780    &  0.441             \\
\bottomrule
\end{tabular}
\end{center}
\caption{The values of the $\chi^2/N_{\rm dat}$ and of the $\phi$ estimator
  in the NNPDF3.1 DIS-only fits
  with the theory covariance matrix $S^{\rm (9pt)}$, compared to the results based on including only
  the experimental covariance matrix $C$.
  \label{table:chi2table_covth_dis}
}
  \end{table}


It is apparent from Table~\ref{table:chi2table_covth_dis} that in
all cases the $\chi^2$ improves when including the theory covariance
matrix, both for individual experiments and for the total dataset.
Specifically, the $\chi^2$ decreases by about 2-3\%
when including theory a
covariance matrix $S^{\rm (9pt)}$ 
evaluated with the 9-point prescription.

The value of $\phi$ increases very substantially, suggesting a
significant increase in the PDF uncertainty. The expected increase
according to Eq.~(\ref{eq:frdbratnaive}) is  
$r_\phi=2.07$:  NLO MHOUs in DIS are much larger than
experimental uncertainties. The observed increase, by a factor of
$2.17$, is in good agreement with this expectation.
It is interesting to
observe that the NNLO value of $\phi$ is actually also rather larger than 
the NLO value, though not quite so much larger, suggesting that at NNLO 
the MHOUs in DIS might still be quite large.

\begin{figure}[t]
  \begin{center}
    \includegraphics[scale=0.39]{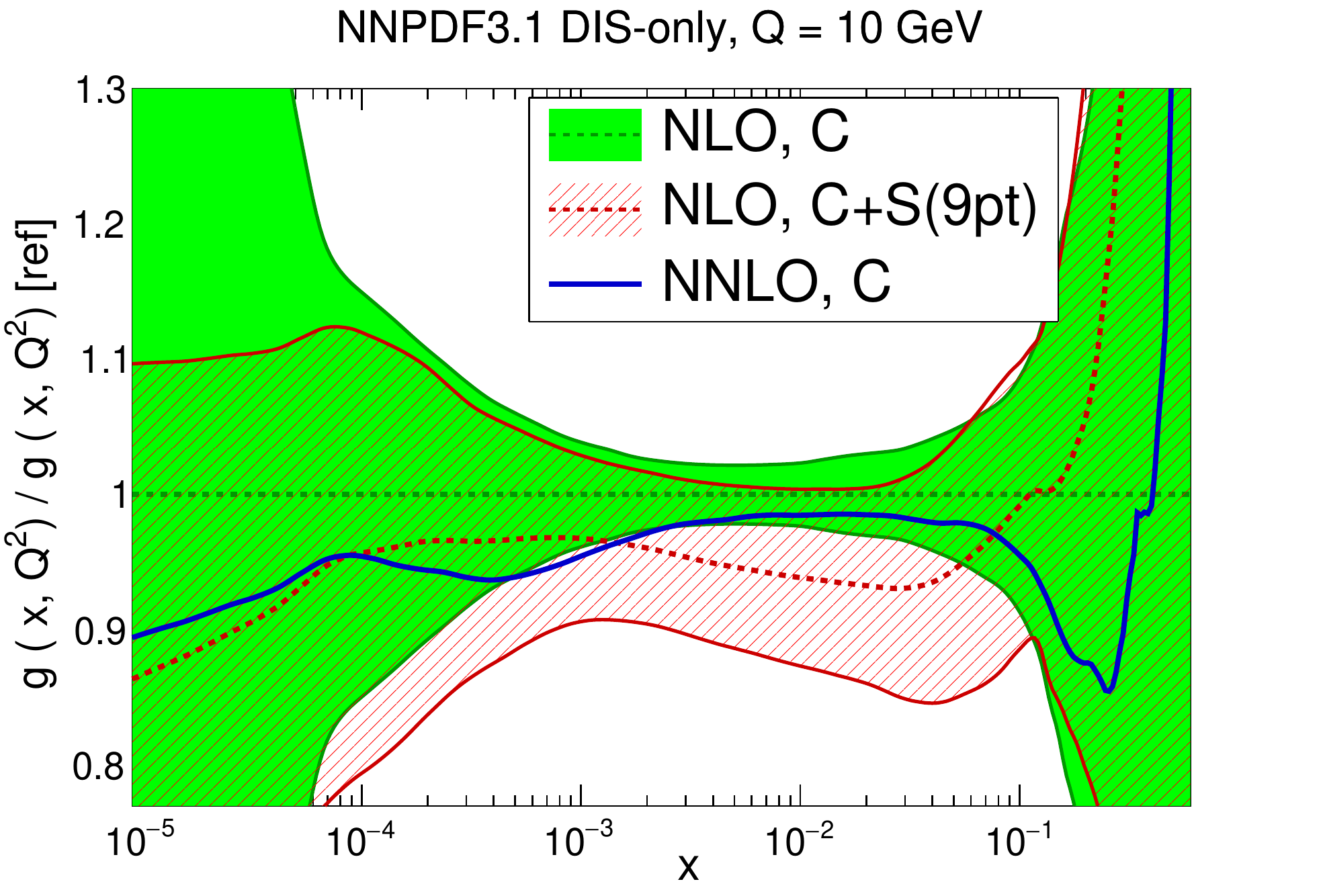}
    \includegraphics[scale=0.39]{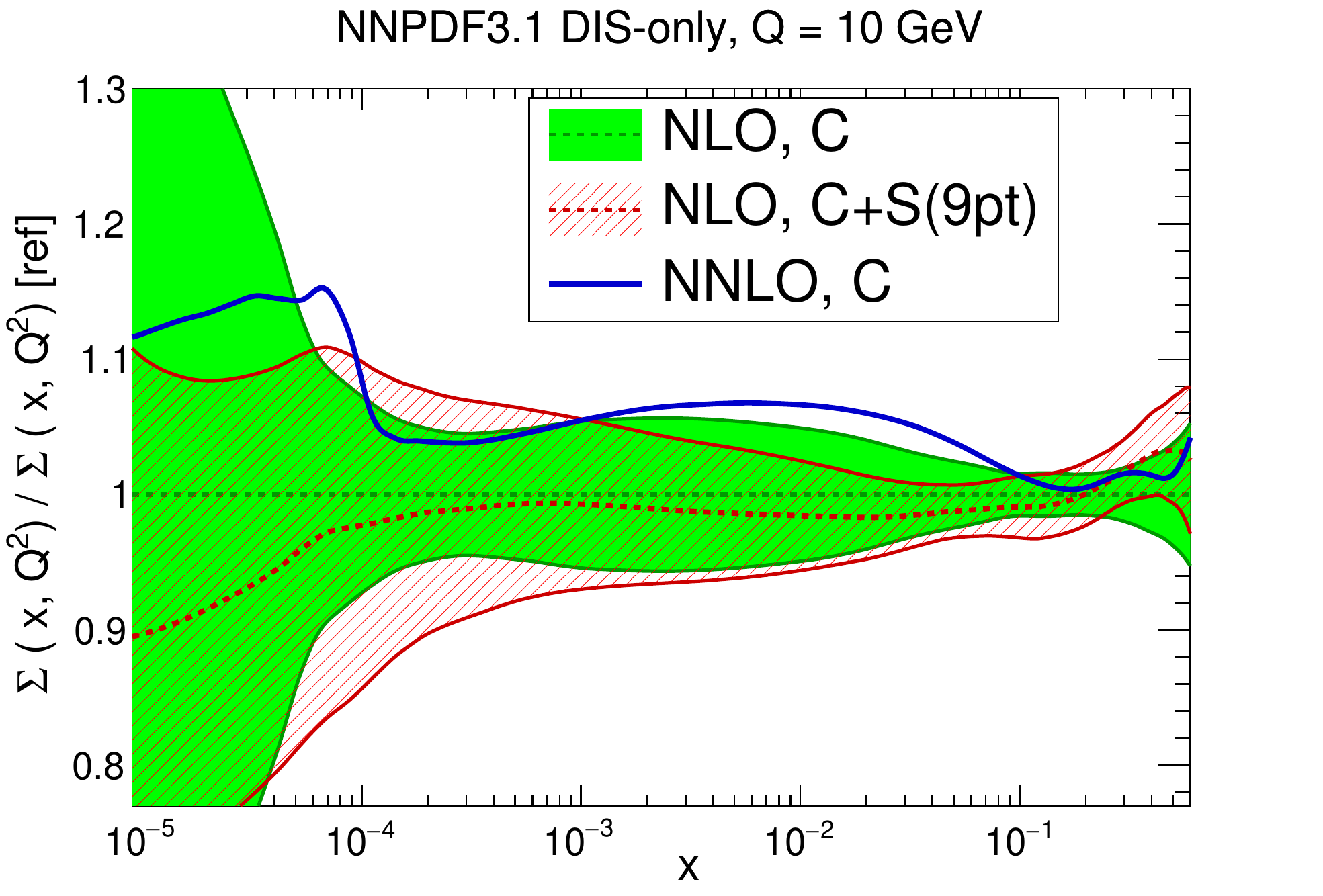}
     \caption{\small Comparison of DIS-only PDFs determined 
      with and without MHOUs in the covariance matrix.
      The gluon (left) and quark singlet (right) are shown at $Q=10$ GeV.
      The theory covariance matrix $S$ has been constructed using
      the 9-point prescription.
      The central value of the NNLO determined
      without MHOU is also shown. All results are shown as a ratio to
    the central value of the set with theory covariance
    matrix not included. Note that the uncertainty band has
      a different meaning according to whether the theory covariance
      matrix is included or not: if not it is the standard PDF 
      uncertainty coming from data, while if it is included, then it is the total uncertainty including the MHOU.
    \label{fig:DISonly-NLO-CovMatTH} }
  \end{center}
\end{figure}

Next we compare PDFs: in Fig.~\ref{fig:DISonly-NLO-CovMatTH} we compare
the gluon and the  total quark singlet PDF   at $Q=10$ GeV
with and without MHOUs in the covariance matrix, determined 
using the 9-point prescription.
      The NLO results are also compared with the central
      value of the NNLO fit based on the experimental covariance
      matrix only. Note that in these comparison plots the PDF
      uncertainty band is always computed using standard NNPDF
      methodology, i.e., as the standard deviation over the PDF
      replica sample. Therefore, this uncertainty band has a different
      meaning dependent on whether or not the theory covariance matrix is
      included: when it is not included, the band represents
       the conventional
      ``PDF uncertainty'', reflecting the uncertainties from the data
      (and methodology), while when it is included, the band provides the
      combined ``PDF'' and MHO uncertainty.

The comparison shows that for PDFs which are strongly constrained by
data, such as the quark singlet PDF for $x\gsim 10^{-3}$, the
uncertainty does not increase much upon inclusion
of the theory covariance matrix, and sometimes it even decreases.
However, for several PDFs, including the gluon PDF, which is only
loosely constrained by the 
DIS data, the uncertainty increases substantially with
MHOUs. This is of course consistent with the fact
that, in the absence of stringent experimental constraints, an extra
contribution to the covariance matrix will lead to increased
uncertainties in the best fit.


\subsection{Global PDFs}
\label{sec:globmhou}

We now discuss PDFs determined from the global dataset presented in
Sect.~\ref{sec:inputdata}. Only NLO PDFs will be discussed here, with 
global NNLO PDFs left for future work. %
The $\chi^2$ values and $\phi$ values are shown in
Tables~\ref{table:chi2table_covth_global_nlo}
and~\ref{table:phitable_covth_global_nlo} 
respectively, both for the total dataset and for the  individual
processes of Table~\ref{tab:datasets_process_categorisation}.
In comparison to the DIS-only case of
Table~\ref{table:chi2table_covth_dis} we now also show results
obtained using the 7-point and 3-point prescriptions, and also for the
default 9-point prescription but where the results were obtained by 
including the theory
covariance matrix either only in the $\chi^2$ definition
Eq.~(\ref{eq:chi2_v3}), or only in the data generation
Eq.~(\ref{eq:dgen}), in order to understand better the two distinct
effects. The baseline NLO and NNLO PDF sets (without theory covariance
matrix) are identical in all respects to the NNPDF3.1 PDF
sets~\cite{Ball:2018iqk}, except for the somewhat different dataset as
discussed in Sect.~\ref{sec:inputdata}.

As in the case of the DIS-only fit, upon adding the MHOU 
we find a reduction of $\chi^2$
both for the global fit and for individual datasets. Specifically, the
$\chi^2$ 
for the NLO global fit with theory covariance matrix computed
with the  9-point prescription decreases by about 3\%, and almost
coincides with the NNLO $\chi^2$, suggesting that indeed the theory
uncertainty is correctly accounting for the missing NNLO
correction. The pattern at the level of individual datasets is more
complex, due to a variety of reasons.
In particular, consider the CMS $Z$ $p_T$
distribution, where a very significant decrease in $\chi^2$
is observed when going from NLO to
NNLO, but not when adding the theory uncertainty to the
NLO. This turns out to be due to a sizable uncorrelated uncertainty which must
be added to the NNLO theory prediction in order to account for
numerical instabilities (see the discussion of Fig.~6 in Ref.~\cite{Ball:2018iqk}).

On the other
hand, the value of $\phi$ now increases much less than expected:
with our favorite
9-point prescription the increase is by about 30\%, while the expected 
$r_\phi=1.69$.  This is an indication 
that by accounting for the missing NNLO terms, 
the inclusion of MHOUs resolves some of the tensions in the fit with only 
the experimental uncertainties, thus reducing the overall effect of 
the MHOUs. The NNLO fit also 
shows an increase, to $0.36$, but the fact that this is already 
quite close to $0.41$ perhaps suggests that the effect of adding MHOUs 
to the NNLO global fit will be relatively modest.  

Comparing the different prescriptions, results are  reasonably
stable, even when comparing to the 3-point prescription which, as
discussed in Sects.~\ref{sec:asympres}-\ref{sec:validconstruction},
spans a much smaller subspace of theory variations. However, the
9-point prescription appears to perform best in terms of $\chi^2$
quality with very little difference in $\phi$, in agreement
with the results of Sect.~\ref{sec:validresults}.

We finally turn to fits in which the theory covariance matrix is
included either in the $\chi^2$ 
  definition Eq.~(\ref{eq:chi2_v3}) but not
  in the data generation Eq.~(\ref{eq:dgen}), 
  or in the data generation Eq.~(\ref{eq:dgen}) but not in the $\chi^2$ 
  definition Eq.~(\ref{eq:chi2_v3}).   In the former case, we 
expect the MHOUs to affect mostly the central value (since the relative weighting of different data points is altered during the fitting according to the relative size of their MHOUs), and to a lesser extent the
uncertainties (since the data
replicas only fluctuate according to the experimental uncertainties). 
The results show that indeed including the MHOU in the $\chi^2$
definition alone  leads to a 
$\chi^2$ value which is very close to that found when the MHOU is fully
included, consistent with the expectation that it is the inclusion of
the theory covariance matrix in the $\chi^2$ which mostly drives the best
fit, while the  $\phi$ value increases somewhat less.
In the latter case, we expect to obtain increased uncertainties but a worse fit,
since the data replica fluctuations are wider due to the MHOU, and this is 
not accounted for in the $\chi^2$. The results indeed 
show a significant deterioration of fit quality, as expected for an 
inconsistent fit: the $\chi^2$ goes up, and also the $\phi$ value 
goes up dramatically, showing the increase in
uncertainty due to the inclusion of MHOU in the sampling, now uncompensated 
by a rebalancing of the datasets through the inclusion of MHOU in the fit.

\begin{table}[!p]
\begin{center}
\renewcommand*{\arraystretch}{1.4}
\scriptsize
\begin{tabular}{|l|c|c|ccc|cc|c|}
  \toprule
  &    & \multicolumn{7}{c|}{$\chi^2/N_{\rm dat}$ in the NNPDF3.1 global fits}   \\
 Dataset & $n_{\rm dat}$ & \multicolumn{6}{c|}{NLO}  & NNLO  \\
 &  & $C$ & $C+S^{(\rm 9pt)}$   &  $C+S^{(\rm 7pt)}$  &  $C+S^{(\rm 3pt)}$ & $C+S^{(\rm 9pt)}_{\rm fit}$ &
  $C+S^{(\rm 9pt)}_{\rm samp}$   &  $C$ \\
\toprule
NMC                                     &  134 & 1.241 & 1.239 & 1.264 & 1.253 & 1.235 & 1.246 & 1.222 \\
SLAC                                    &   12 & 0.868 & 0.503 & 0.485 & 0.509 & 0.493 & 0.738 & 0.693 \\
BCDMS                                   &  530 & 1.040 & 1.029 & 1.046 & 1.062 & 1.033 & 1.042 & 1.062 \\
HERA $\sigma_{\rm NC}^p$                  &  886 & 1.086 & 1.044 & 1.046 & 1.079 & 1.044 & 1.190 & 1.098 \\
HERA $\sigma_{\rm NC}^c$                  &   31 & 1.395 & 1.037 & 1.082 & 1.172 & 1.055 & 1.563 & 1.163 \\
\midrule
\bf DIS NC                              & \bf 1593 & \bf 1.088 & \bf 1.079 & \bf 1.086 & \bf 1.095 & \bf 1.081 & \bf 1.227 & \bf 1.084 \\
\midrule
NuTeV dimuon                            &   41 & 0.474 & 0.388 & 0.355 & 0.359 & 0.421 & 0.406 & 0.470 \\
CHORUS                                  &  430 & 1.037 & 0.891 & 0.896 & 0.900 & 0.898 & 1.081 & 1.124 \\
HERA $\sigma_{\rm CC}^p$                  &   81 & 1.154 & 1.070 & 1.067 & 1.106 & 1.062 & 1.103 & 1.126 \\
\midrule
\bf DIS CC                              &  \bf 552 & \bf 1.012 & \bf 0.928 & \bf 0.933 & \bf 0.960 & \bf 0.929 & \bf 1.036 & \bf 1.079 \\
\midrule
ATLAS $W,Z$ 7 TeV 2010                  &   30 & 0.999 & 0.880 & 0.916 & 0.975 & 0.892 & 0.984 & 0.935 \\
ATLAS $W,Z$ 7 TeV 2011                  &   34 & 3.306 & 2.224 & 2.282 & 2.389 & 2.205 & 3.107 & 1.807 \\
ATLAS low-mass DY 7 TeV                 &    4 & 0.684 & 0.654 & 0.668 & 0.690 & 0.660 & 0.733 & 1.024 \\
ATLAS high-mass DY 7 TeV                &    5 & 1.677 & 1.736 & 1.700 & 1.660 & 1.667 & 1.577 & 1.498 \\
ATLAS $Z$ $p_T$ 8 TeV ($p_T^{ll},M_{ll}$) &   44 & 1.171 & 1.067 & 1.070 & 1.067 & 1.062 & 1.183 & 0.907 \\
ATLAS $Z$ $p_T$ 8 TeV ($p_T^{ll},y_{ll}$) &   48 & 1.666 & 1.583 & 1.614 & 1.688 & 1.638 & 1.641 & 0.865 \\
CMS Drell-Yan 2D 2011                   &   88 & 1.220 & 1.067 & 1.098 & 1.169 & 1.062 & 1.132 & 1.319 \\
CMS W asy 840 pb                        &   11 & 0.965 & 1.022 & 0.966 & 0.987 & 1.045 & 1.034 & 0.863 \\
CMS W asy 4.7 fb                        &   11 & 1.662 & 1.670 & 1.704 & 1.713 & 1.659 & 1.657 & 1.750 \\
CMS W rap 8 TeV                         &   22 & 0.955 & 0.611 & 0.609 & 0.587 & 0.627 & 0.665 & 0.826 \\
CMS $Z$ $p_T$ 8 TeV ($p_{T}^{ll},M_{ll}$)  &   28 & 3.895 & 3.745 & 3.712 & 3.836 & 3.706 & 3.905 & 1.339 \\
LHCb $Z$ 940 pb                         &    9 & 1.238 & 1.191 & 1.162 & 1.179 & 1.165 & 1.281 & 1.437 \\
LHCb $Z\to ee$ 2 fb                     &   17 & 1.305 & 1.303 & 1.305 & 1.313 & 1.334 & 1.250 & 1.203 \\
LHCb $W,Z\to\mu$ 7 TeV                  &   29 & 1.262 & 1.106 & 1.267 & 1.261 & 1.134 & 1.207 & 1.536 \\
LHCb $W,Z\to\mu$ 8 TeV                  &   30 & 1.194 & 1.027 & 1.125 & 1.154 & 1.054 & 1.152 & 1.438 \\
CDF $Z$ rap                             &   29 & 1.554 & 1.313 & 1.433 & 1.505 & 1.311 & 1.418 & 1.510 \\
D0 $Z$ rap                              &   28 & 0.649 & 0.601 & 0.626 & 0.640 & 0.597 & 0.618 & 0.604 \\
D0 $W\to e\nu$ asy                      &    8 & 1.176 & 1.066 & 1.055 & 1.083 & 1.029 & 1.200 & 2.558 \\
D0 $W\to \mu\nu$ asy                    &    9 & 1.400 & 1.450 & 1.372 & 1.361 & 1.439 & 1.395 & 1.374 \\
\midrule
\bf DY                                  &  \bf 484 & \bf 1.486 & \bf 1.447 & \bf 1.485 & \bf 1.483 & \bf 1.461 & \bf 1.434 & \bf 1.231 \\
\midrule
ATLAS jets 2011 7 TeV                   &   31 & 1.069 & 1.019 & 1.065 & 1.079 & 1.026 & 1.031 & 1.076 \\
CMS jets 7 TeV 2011                     &  133 & 0.869 & 0.786 & 0.790 & 0.830 & 0.795 & 0.883 & 0.921 \\
\midrule
\bf JETS                                &  \bf 164 & \bf 0.907 & \bf 0.839 & \bf 0.858 & \bf 0.901 & \bf 0.848 & \bf 0.911 & \bf 0.950 \\
\midrule
ATLAS $\sigma_{tt}^{\rm top}$              &    3 & 2.577 & 0.787 & 0.853 & 0.982 & 0.770 & 2.442 & 0.903 \\
ATLAS $t\bar{t}$ rap                    &   10 & 1.258 & 0.955 & 0.867 & 0.910 & 0.935 & 1.355 & 1.424 \\
CMS $\sigma_{tt}^{\rm top}$                &    3 & 0.984 & 0.170 & 0.234 & 0.333 & 0.158 & 0.859 & 0.140 \\
CMS $t\bar{t}$ rap                      &   10 & 0.950 & 0.910 & 0.923 & 0.933 & 0.916 & 0.942 & 1.039 \\ 
\midrule
\bf TOP                                 &  \bf  26 & \bf 1.260 & \bf 1.012 & \bf 1.016 & \bf 1.077 & \bf 1.001 & \bf 1.264 & \bf 1.068 \\ 
\midrule
\bf Total                               & \bf 2819 & \bf 1.139 & \bf 1.109 & \bf 1.129 & \bf 1.139 & \bf 1.113 & \bf 1.220 & \bf 1.105 \\
\bottomrule
\end{tabular}
\end{center}
\caption{The values of the $\chi^2/N_{\rm dat}$ in NLO global fits
  with the theory covariance matrix $S$, compared to the results based on including only
  the experimental covariance matrix $C$. Results are shown
  for  the 9-, 7-, and 3- point prescriptions.
  For the 9-point prescription we also show results obtained
  including the theory covariance matrix in the $\chi^2$ 
  definition Eq.~(\ref{eq:chi2_v3}) but not
  in the data generation Eq.~(\ref{eq:dgen}) (marked $S_{\rm fit}^{9pt}$) 
  and then in the data generation Eq.~(\ref{eq:dgen}) but not in the $\chi^2$ 
  definition Eq.~(\ref{eq:chi2_v3}) (marked $S_{\rm sampl}^{9pt}$).
  Values corresponding to the
  NNLO fit with  experimental covariance matrix $C$ only are also shown.
  \label{table:chi2table_covth_global_nlo}
}
  \end{table}


\begin{table}[t]
\begin{center}
\renewcommand*{\arraystretch}{1.78}
\footnotesize
\begin{tabular}{|l|c|ccc|cc|c|}
  \toprule
  & \multicolumn{7}{c|}{$\phi$ in the NNPDF3.1 global fits}   \\
 Process & \multicolumn{6}{c|}{NLO}  & NNLO  \\
 & $C$ & $C+S^{(\rm 9pt)}$   &  $C+S^{(\rm 7pt)}$  &  $C+S^{(\rm 3pt)}$ & $C+S^{(\rm 9pt)}_{\rm fit}$ &
  $C+S^{(\rm 9pt)}_{\rm sampl}$   &  $C$ \\
  \toprule
  DIS NC     & 0.266 & 0.412 & 0.393 & 0.384 & 0.414 & 1.137 & 0.305 \\
  DIS CC     & 0.389 & 0.408 & 0.427 & 0.442 & 0.388 & 0.502 & 0.471 \\
  \midrule
  DY         & 0.361 & 0.377 & 0.369 & 0.379 & 0.378 & 0.603 & 0.380 \\
  JETS       & 0.295 & 0.359 & 0.327 & 0.333 & 0.336 & 0.461 & 0.392 \\
  TOP        & 0.375 & 0.443 & 0.387 & 0.405 & 0.382 & 0.612 & 0.363 \\
  \midrule
  Total      & 0.314 & 0.405 & 0.394 & 0.394 & 0.400 & 0.932 & 0.362 \\
  \bottomrule
\end{tabular}
\end{center}
\caption{Same as Table~\ref{table:chi2table_covth_global_nlo}, but 
  for the values of the $\phi$ estimator.
  \label{table:phitable_covth_global_nlo}}
  \end{table}

\begin{figure}[t]
  \begin{center}
    \includegraphics[scale=0.39]{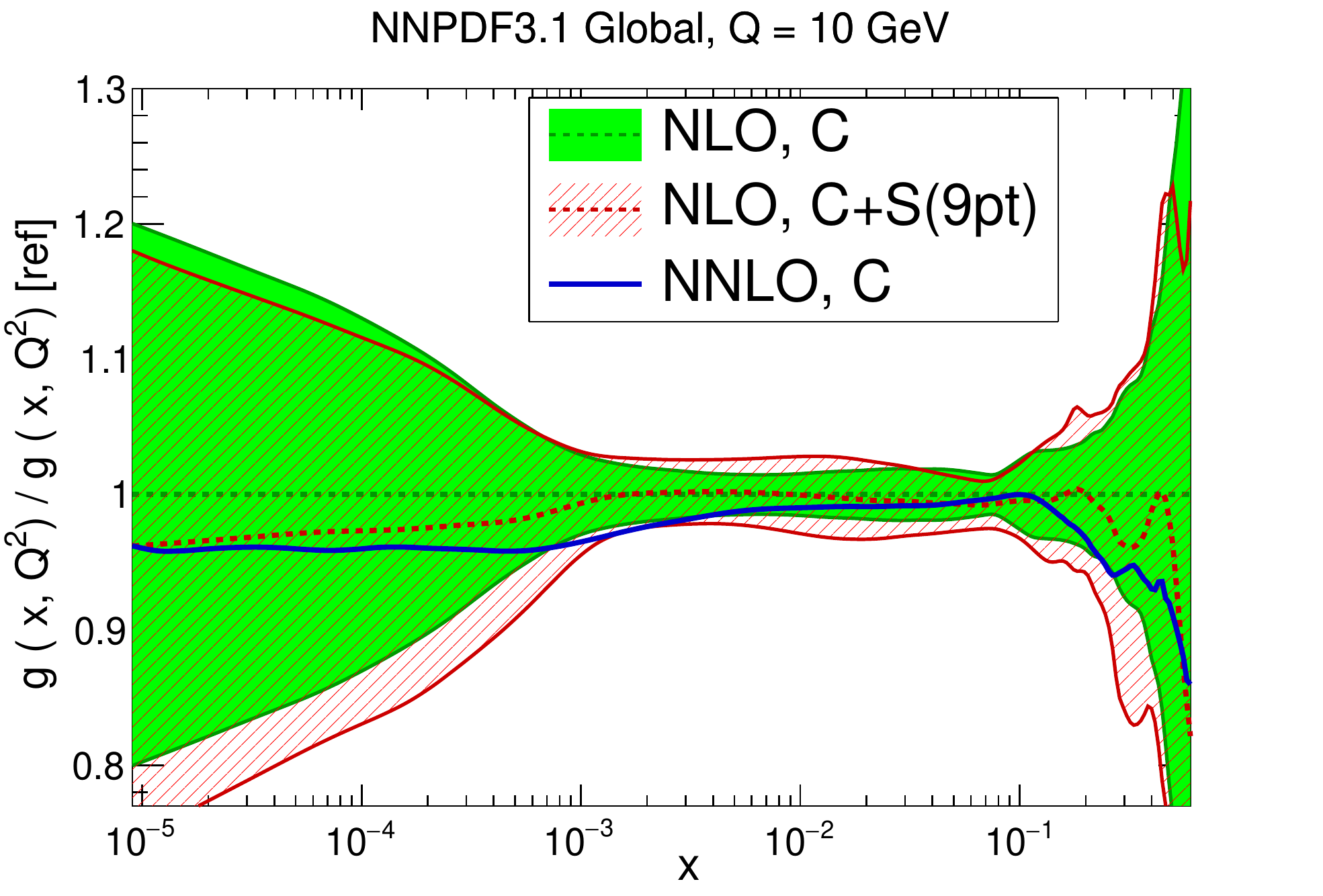}
    \includegraphics[scale=0.39]{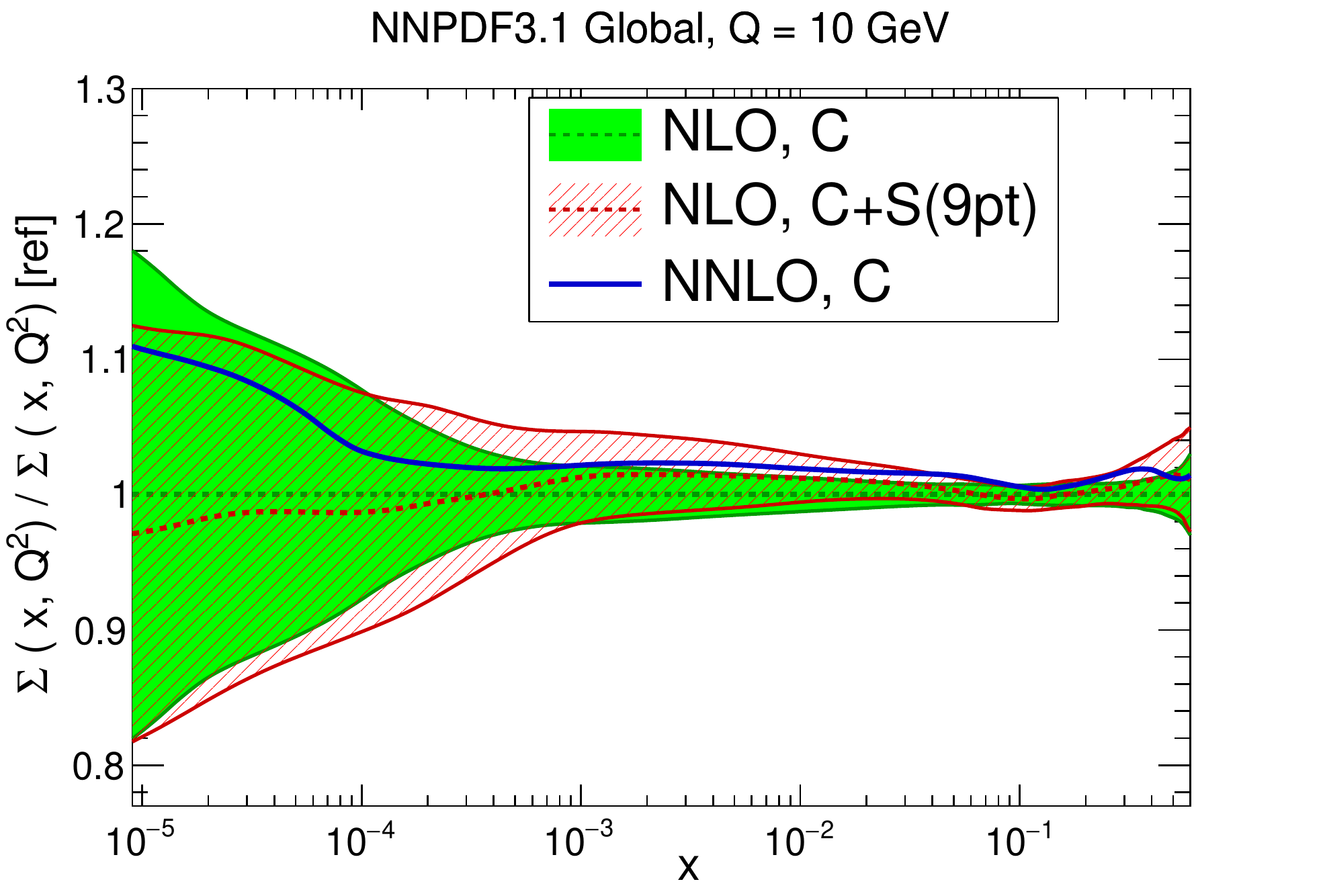}
    \includegraphics[scale=0.39]{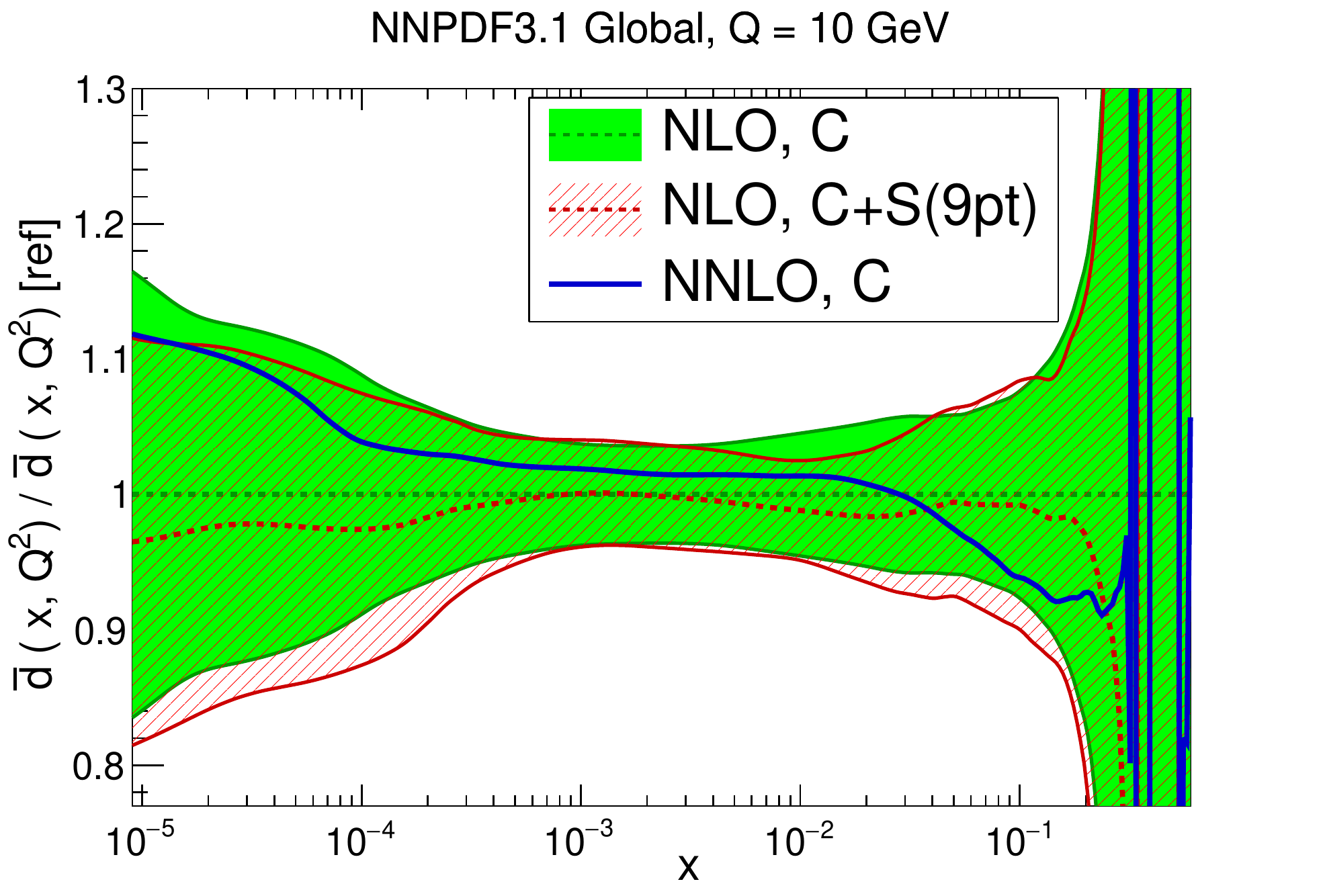}
   \includegraphics[scale=0.39]{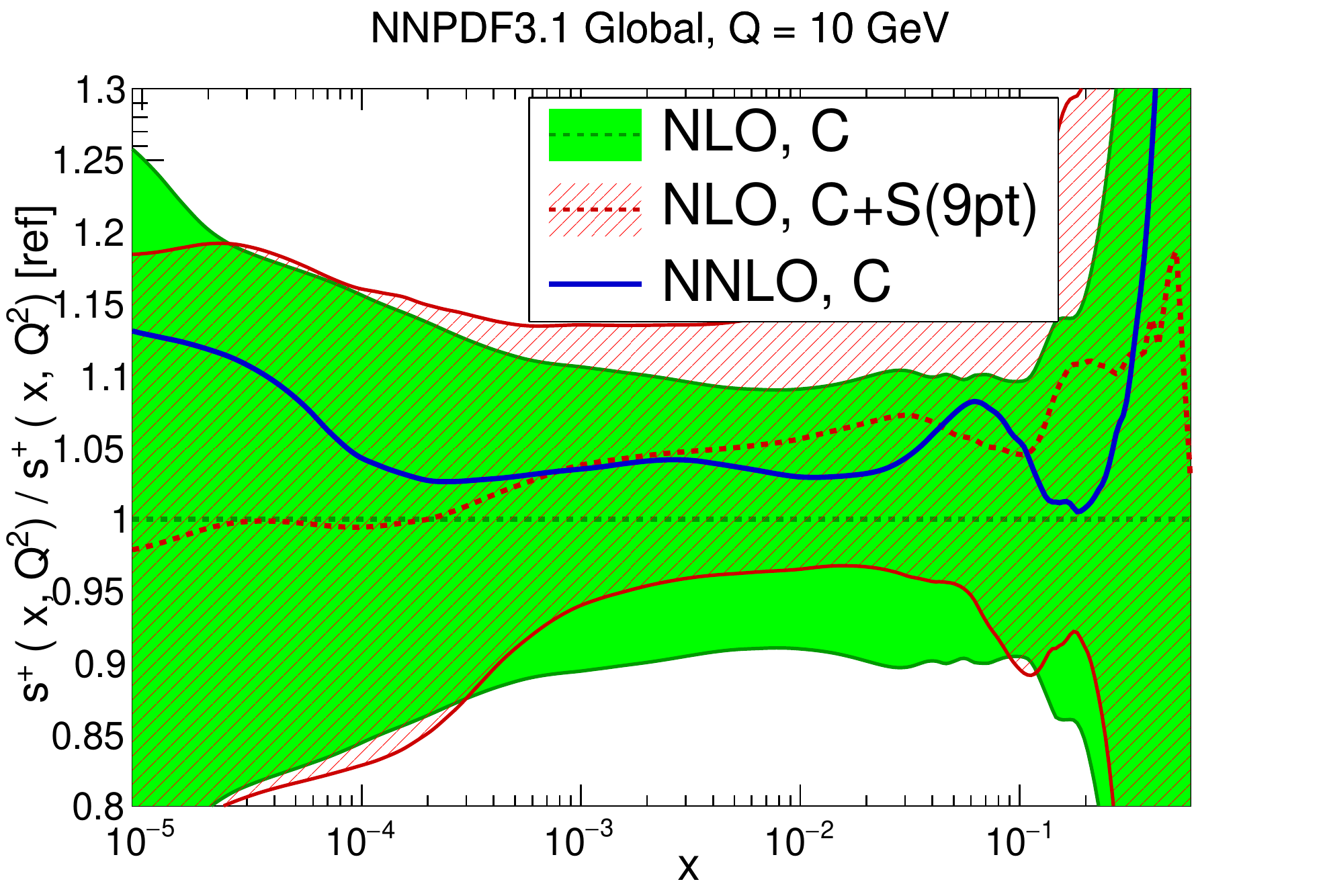}
   \caption{\small Same as Fig.~\ref{fig:DISonly-NLO-CovMatTH}
     now for the NNPDF3.1 global fits.
     We show the results of the NLO fits based on $C$ and $C+S^{\rm (9pt)}$ normalized
     to the former, as well as the central value of the NNLO fit based on $C$.
     for the gluon, the total quark singlet,
     the anti-down quark, and the total strangeness PDFs, all at $Q=10$ GeV.
    \label{fig:Global-NLO-CovMatTH} }
  \end{center}
\end{figure}

We now move on to discuss the corresponding
results at the PDF level, in analogy with
the comparisons presented for the DIS-only fits
in Fig.~\ref{fig:DISonly-NLO-CovMatTH}. Specifically, in 
Fig.~\ref{fig:Global-NLO-CovMatTH}.
we show the results of the NLO fits based on $C$ and $C+S^{\rm
(9pt)}$, as well as the central value of the NNLO fit based on $C$, 
for the gluon, the total quark singlet,
the anti-down quark, and the total strangeness PDFs, all at
$Q=10$~GeV. We also show in Fig.~\ref{fig:Global-NLO-CovMatTH} the
same PDFs but at the scale $Q=1.6$~GeV at which PDFs are
parametrized.

We find that in the data region the PDF uncertainty is only very moderately
increased by the inclusion of the theory covariance matrix, while central values
can shift significantly, by up to one sigma.
This is consistent with the observation that the $\phi$ values in
Table~\ref{table:phitable_covth_global_nlo} increase by only a
moderate amount upon
inclusion of the theory covariance matrix. This provides evidence that
in the data region the inclusion of the theory covariance matrix
resolves tensions which 
are otherwise present in the global dataset. In contrast, in regions where 
PDFs which are only loosely constrained by the data, and in particular in the extrapolation regions, the PDF uncertainty  increases significantly. 

\begin{figure}[t]
  \begin{center}
    \includegraphics[scale=0.39]{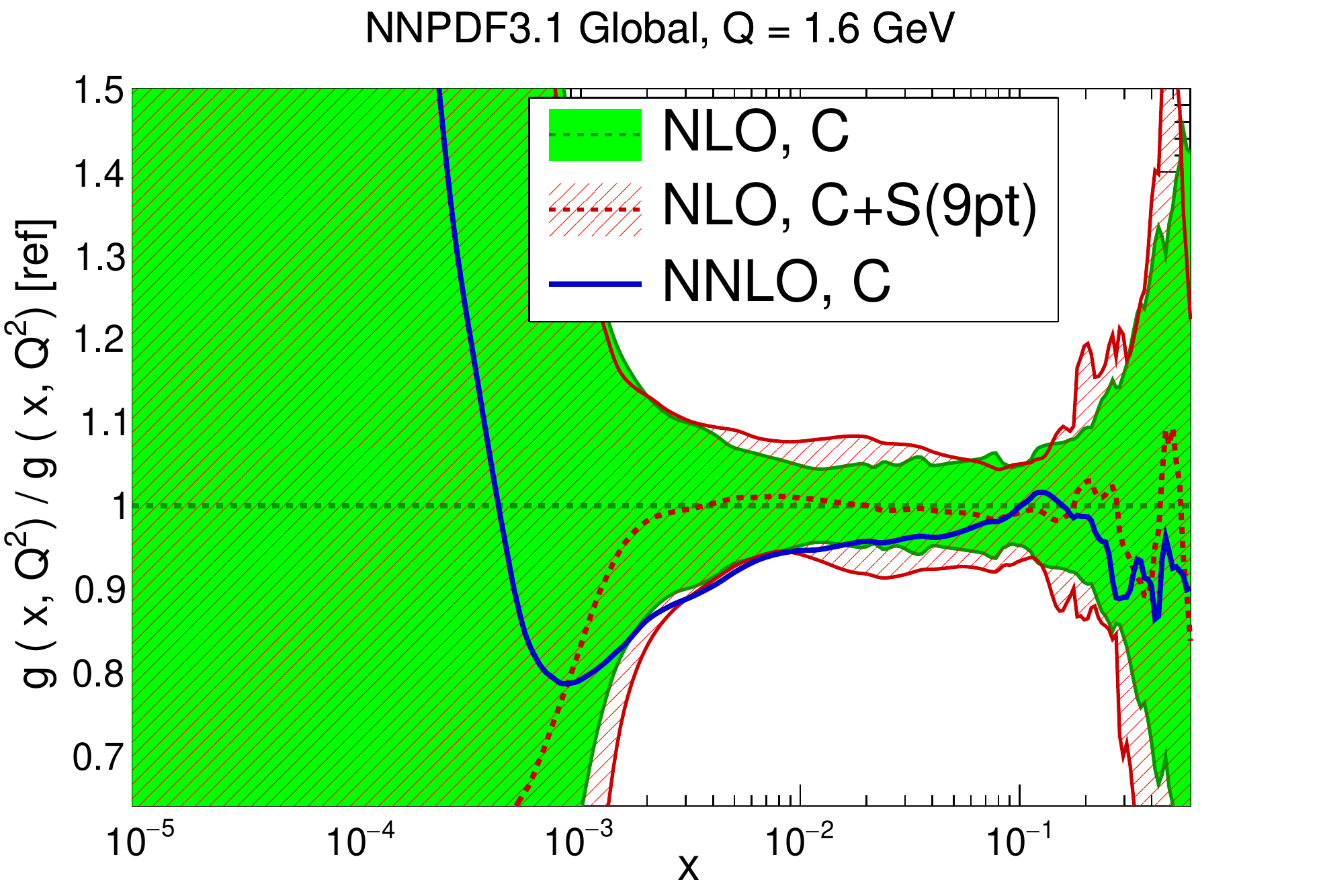}
    \includegraphics[scale=0.39]{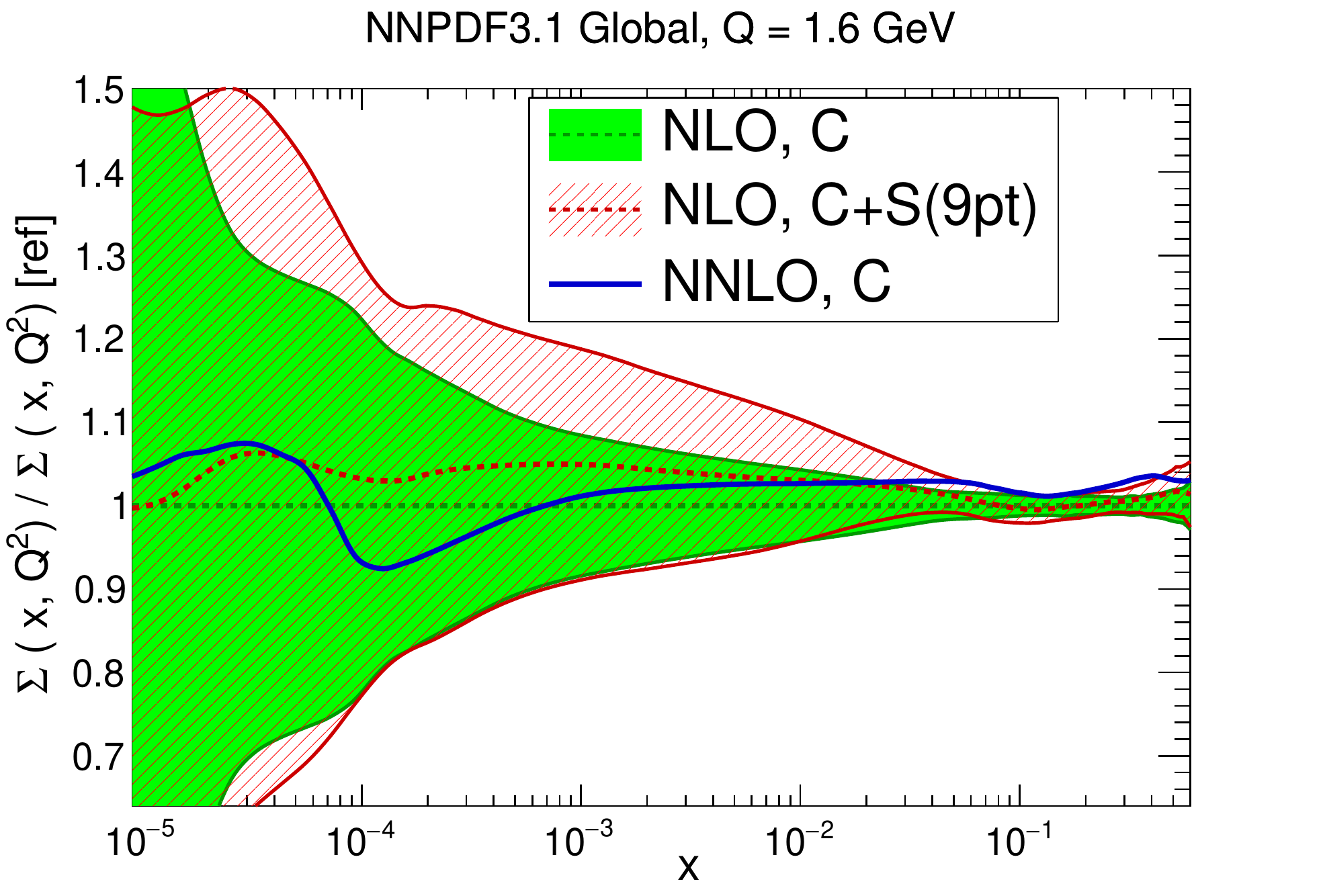}
    \includegraphics[scale=0.39]{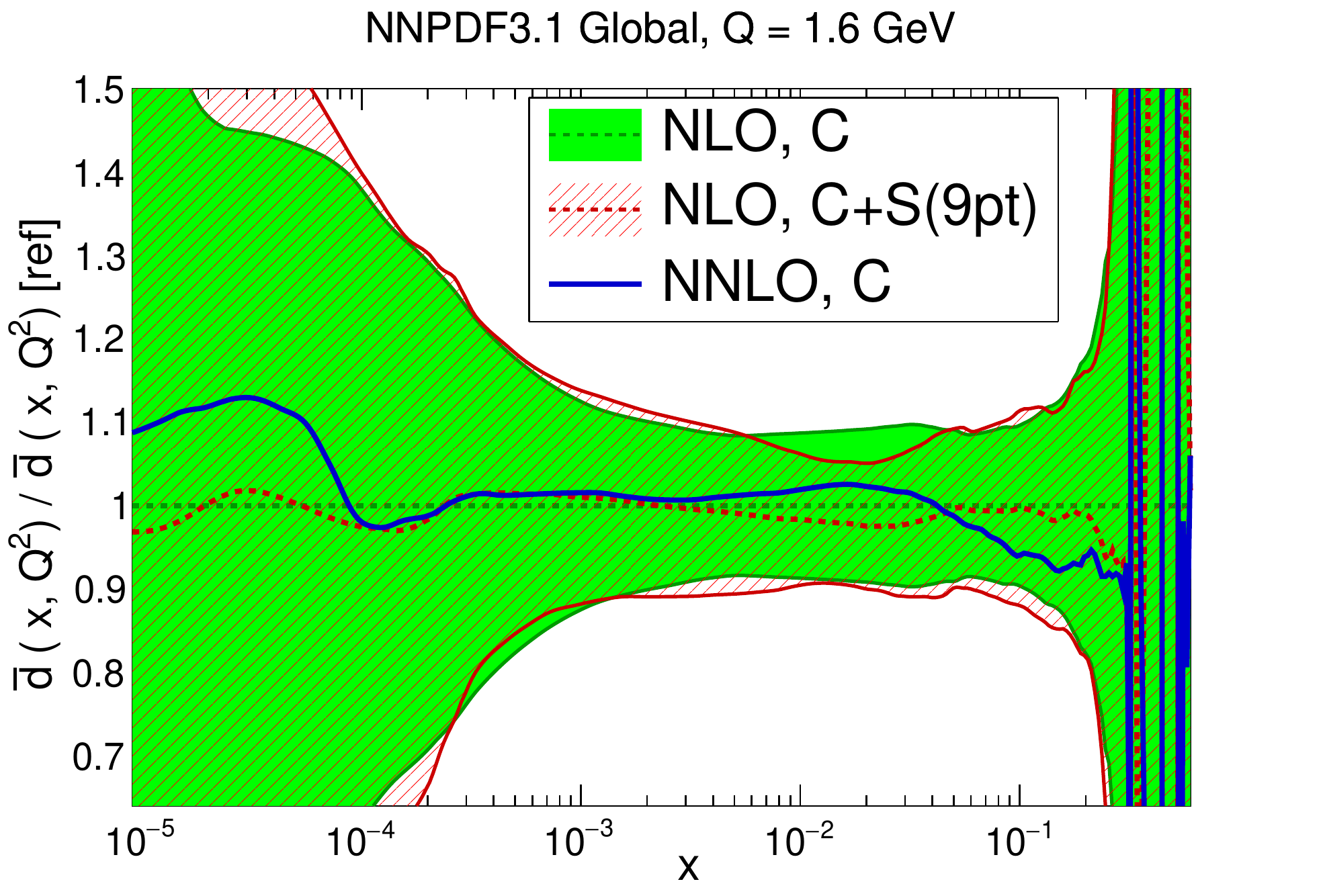}
   \includegraphics[scale=0.39]{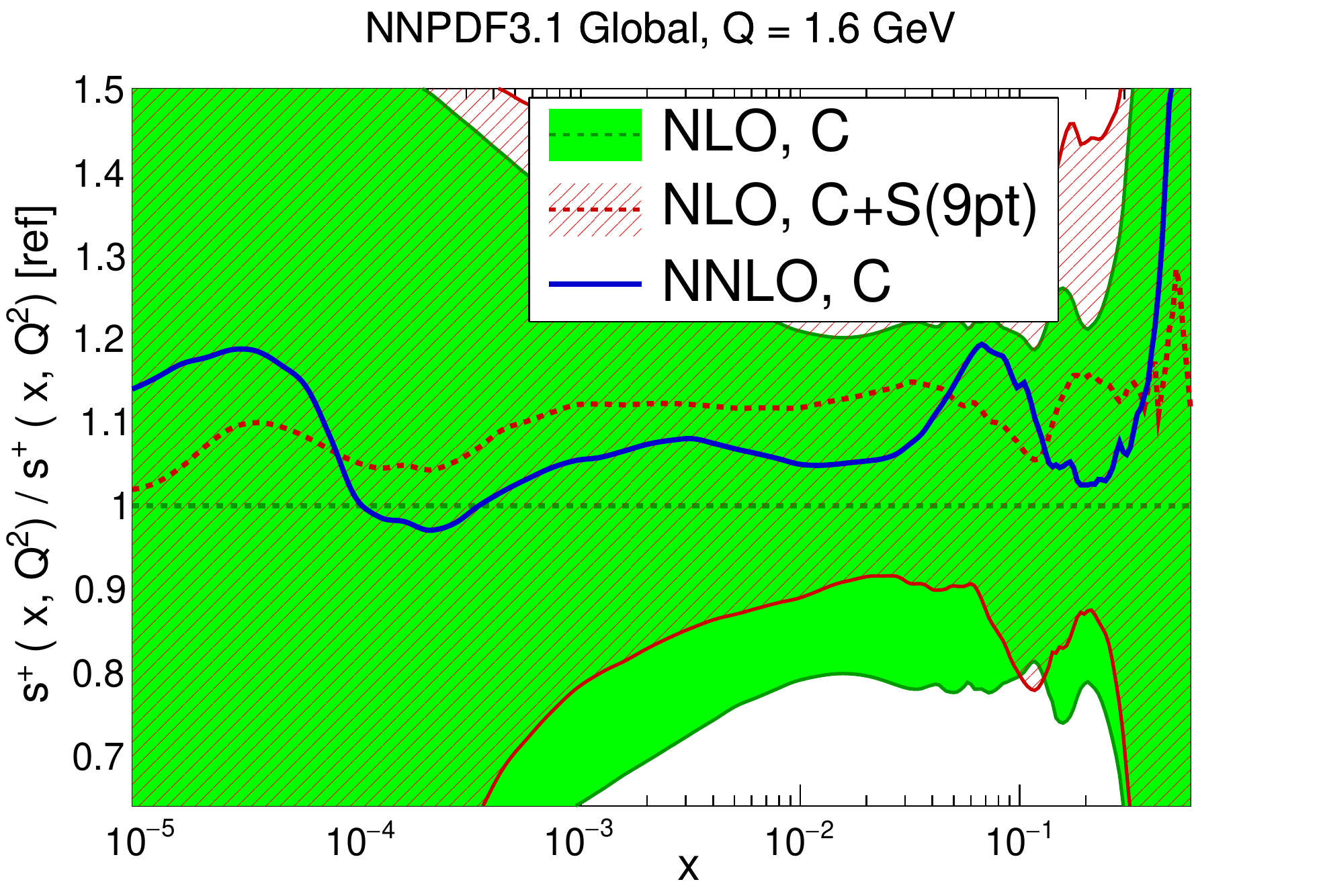}
   \caption{\small Same as Fig.~\ref{fig:Global-NLO-CovMatTH} but now
    with  results shown at the scale $Q=1.6$~GeV at which PDFs are
     parametrized.
    \label{fig:Global-NLO-CovMatTH16} }
  \end{center}
\end{figure}
When comparing PDFs at the parametrization scale in
Fig.~\ref{fig:Global-NLO-CovMatTH16},  an  especially interesting
comparison is with respect to the central NNLO 
value: not only is this quite compatible with the uncertainty band,
but there is now clear evidence that upon inclusion of the NLO MHOU 
the central best fit moves towards the correct NNLO
result. Of course, this improved agreement of the best-fit NLO and
NNLO PDFs is scale-dependent, since PDFs at NLO and NNLO evolve in
different ways, and the scale at which NLO and NNLO become closest
will depend on the scale of the data which dominate the determination
of each PDF combination. However, the agreement is seen in 
Fig.~\ref{fig:Global-NLO-CovMatTH} to persist by
and large also at high scale.
This is further evidence that indeed the theory covariance
matrix has resolved tensions due to MHOs.
This improved agreement of the central value of the NLO $C+S^{(\rm 9pt)}$
with the NNLO $C$ fits is  non-trivial: for instance, inclusion of the
theory covariance matrix leads to
a suppression of the gluon at large $x$ and an enhancement of
strangeness, both of which are indeed also observed at NNLO.

\begin{figure}[t]
  \begin{center}
    \includegraphics[scale=0.39]{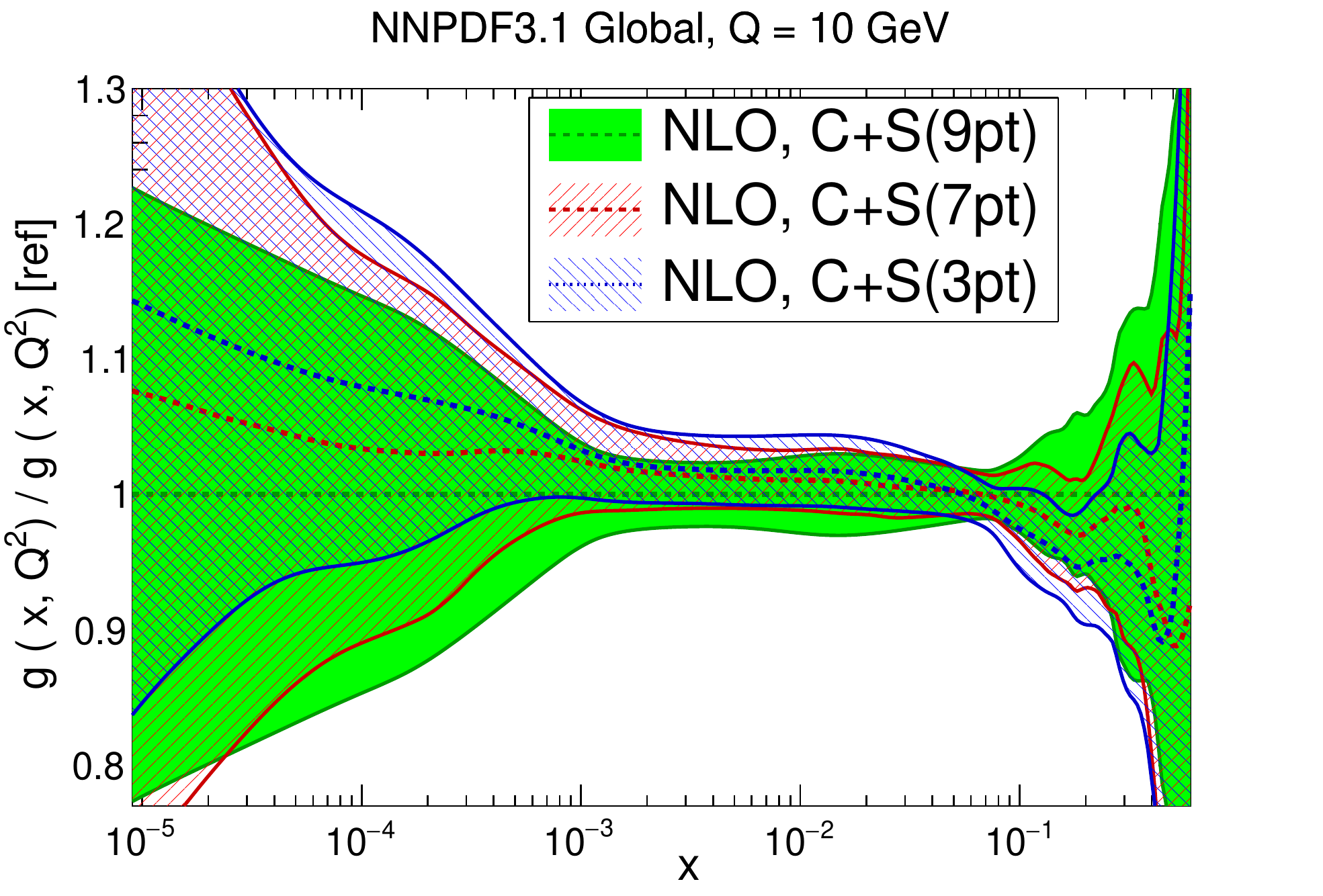}
    \includegraphics[scale=0.39]{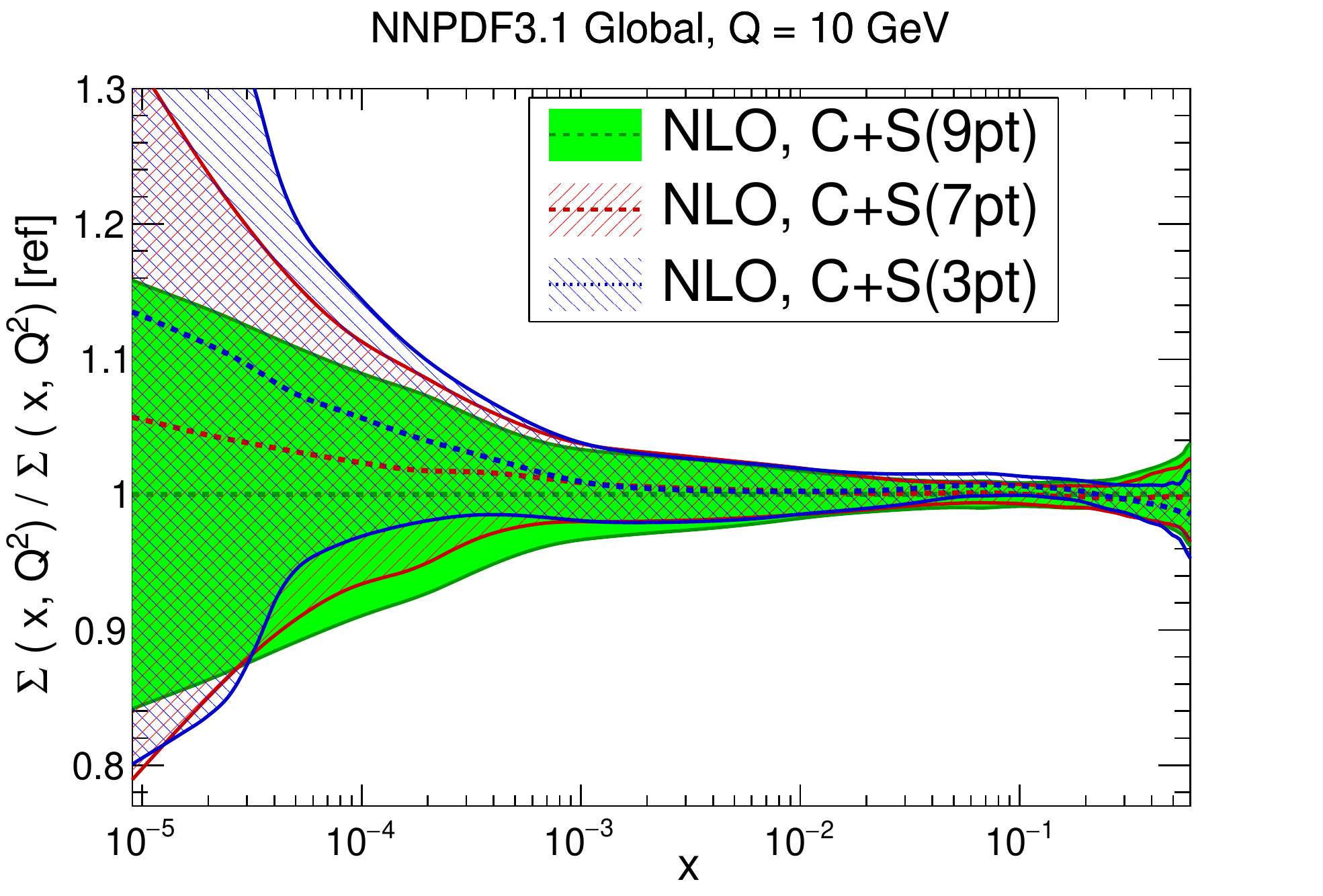}
    \includegraphics[scale=0.39]{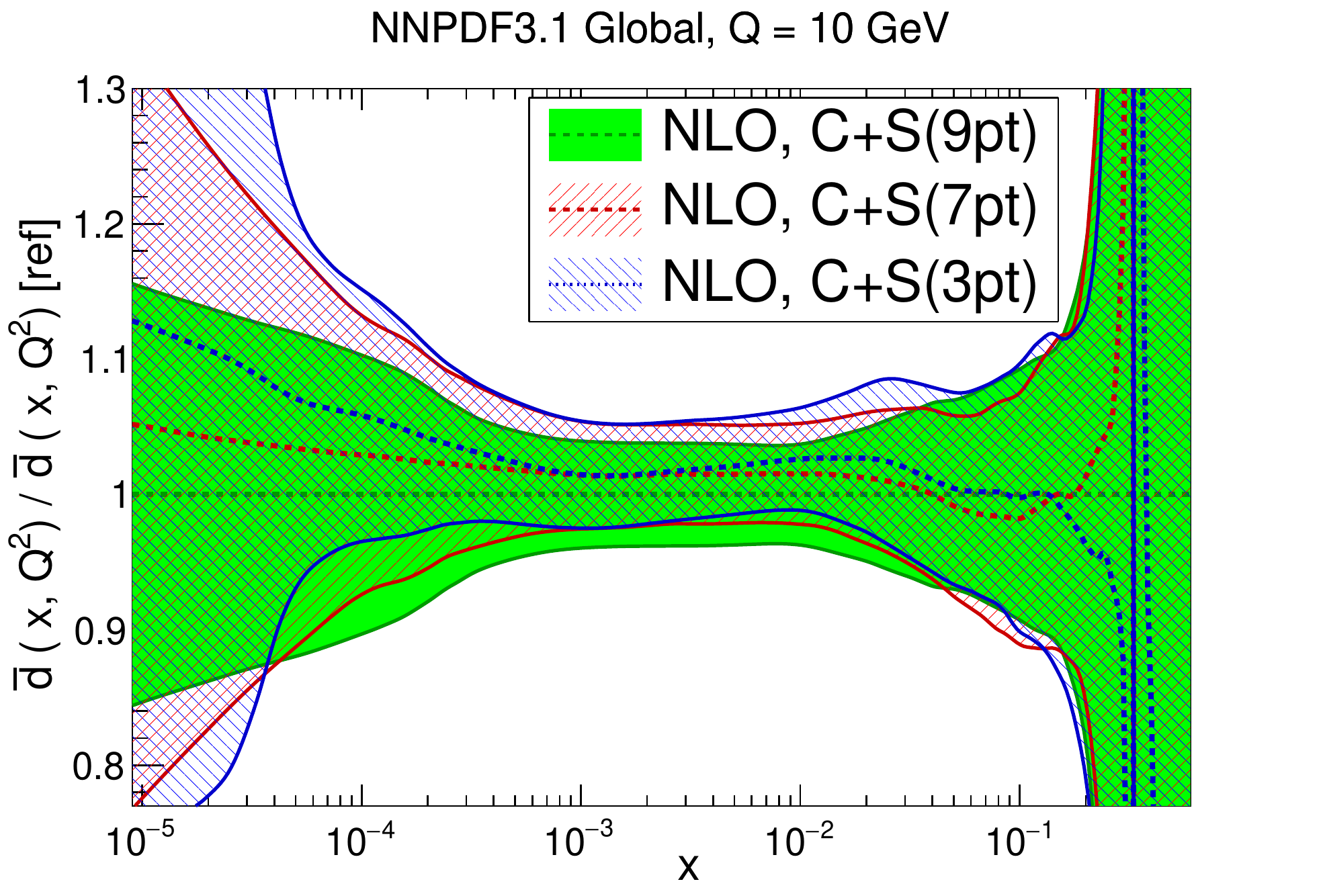}
   \includegraphics[scale=0.39]{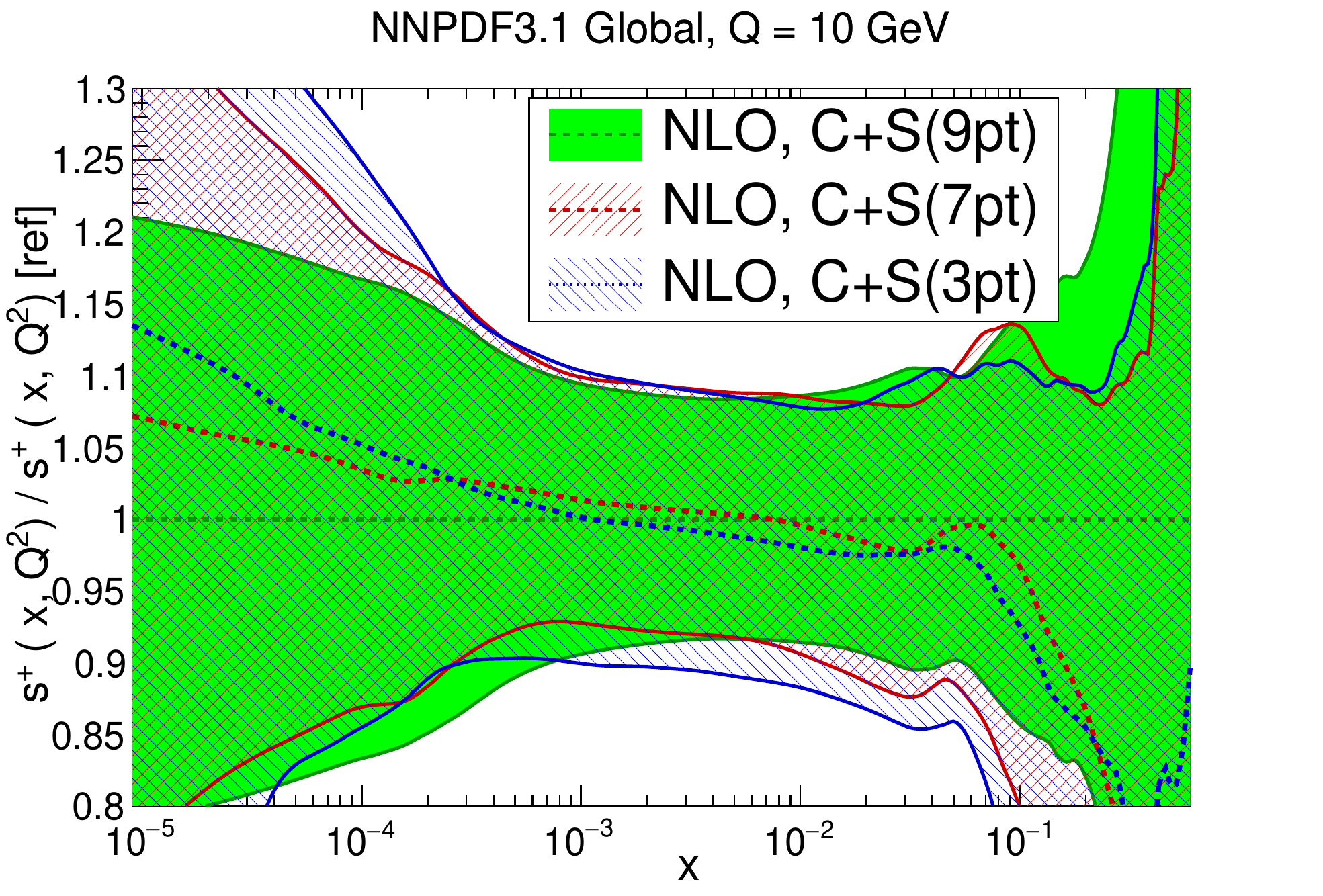}
   \caption{\small Same as Fig.~\ref{fig:Global-NLO-CovMatTH} now
     comparing the results of the NNPDF3.1 global fits with the theory covariance matrix
     constructed accordingly to the 3-, 7-, and 9-point prescriptions, normalized
     to the central value of the latter.
    \label{fig:Global-NLO-CovMatTH-prescriptions} }
  \end{center}
\end{figure}

\begin{figure}[t]
  \begin{center}
    \includegraphics[scale=0.39]{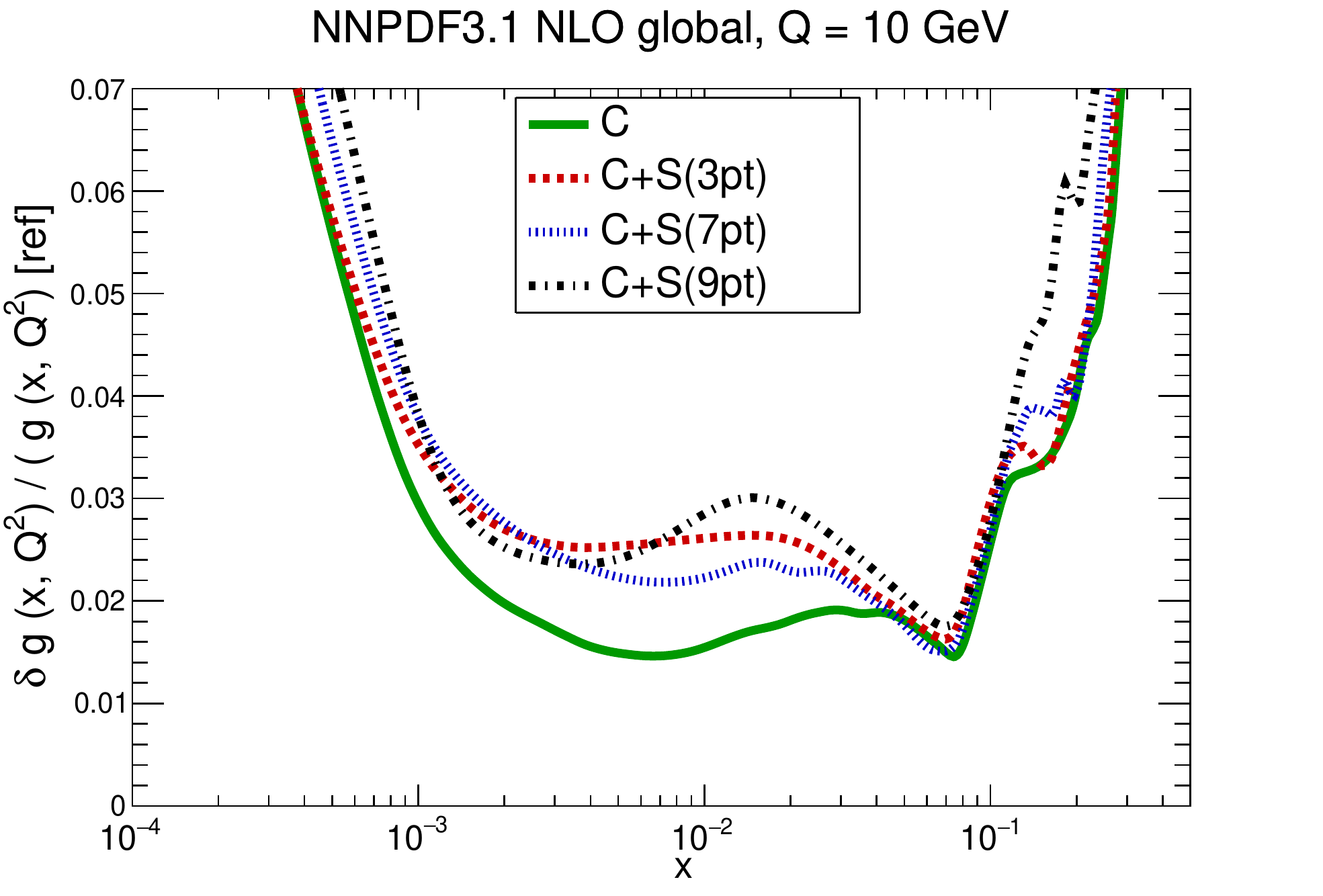}
    \includegraphics[scale=0.39]{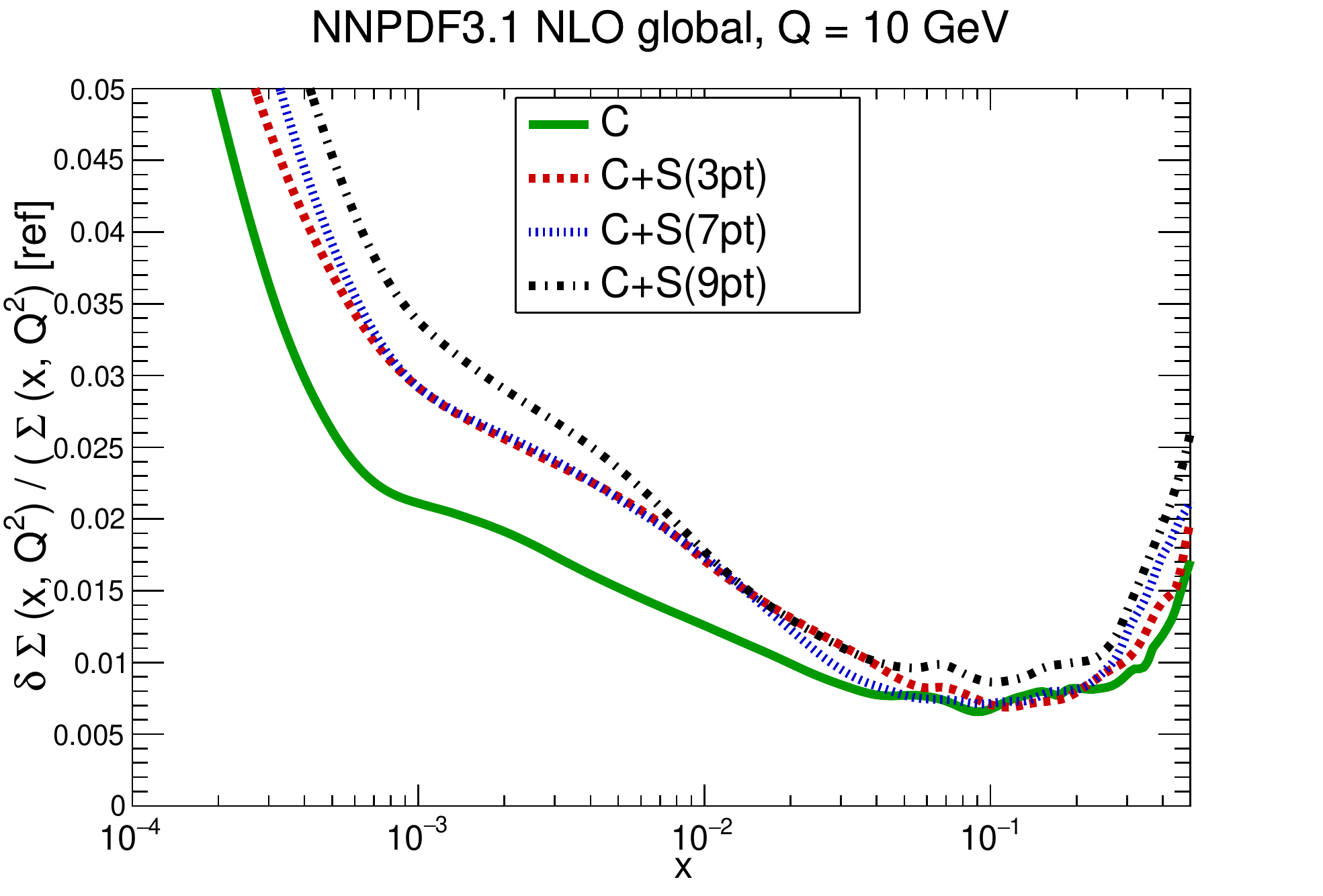}
    \includegraphics[scale=0.39]{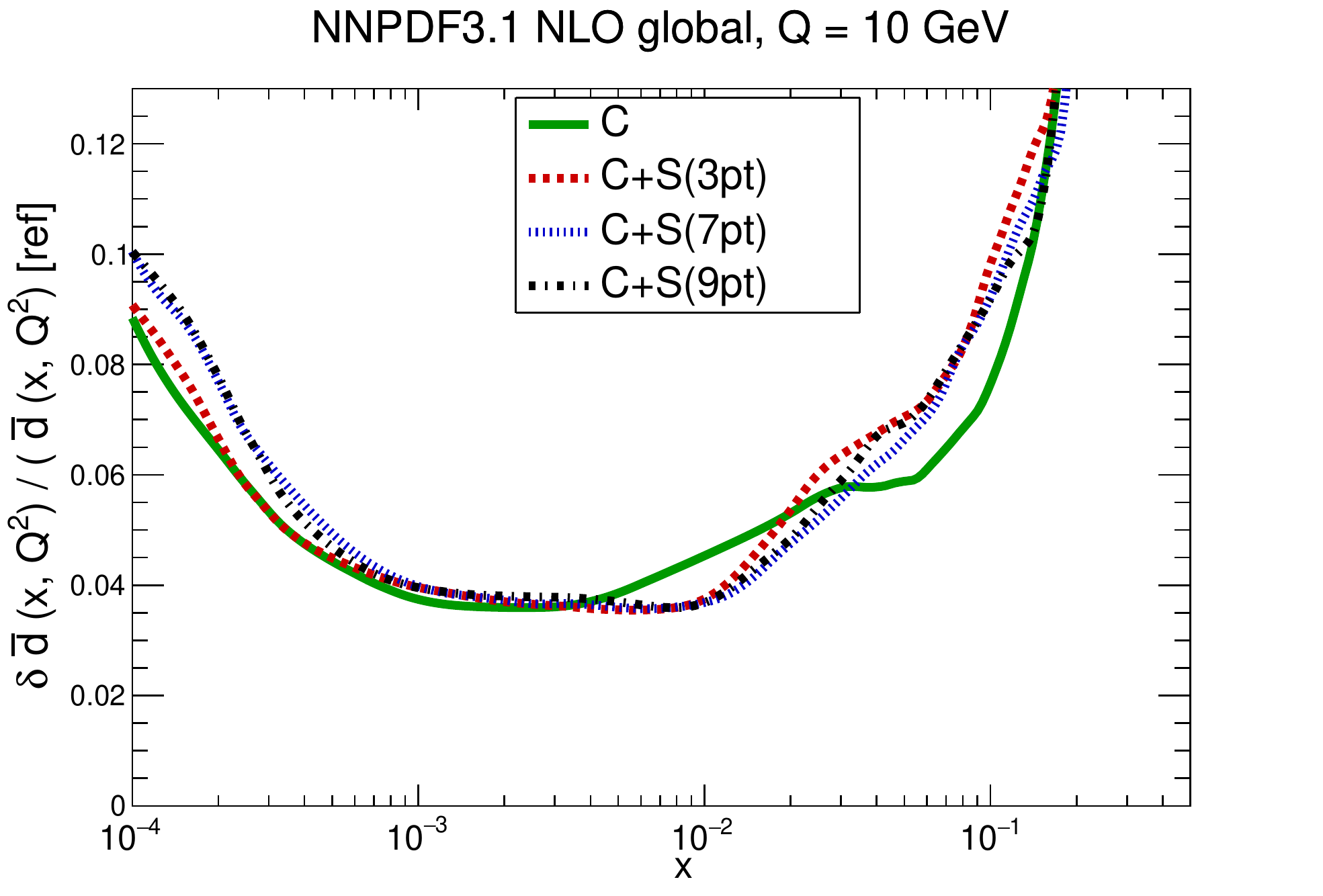}
   \includegraphics[scale=0.39]{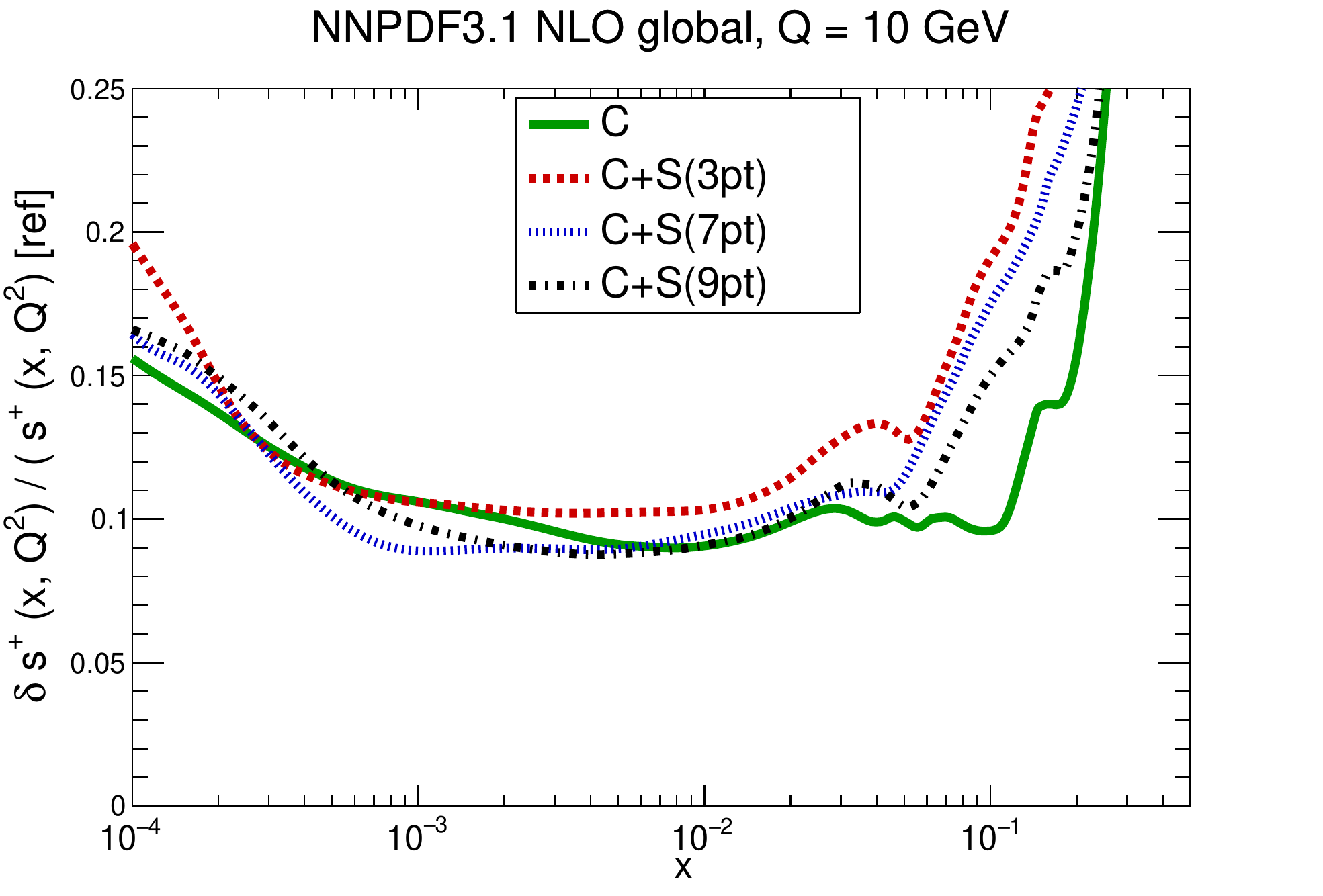}
   \caption{\small Same as
     Fig.~\ref{fig:Global-NLO-CovMatTH-prescriptions}, now showing
     relative PDF uncertainties, normalized to the central value of the baseline set.
     Note that the $y$-axes ranges are different
     for each PDF combination.
    \label{fig:Global-NLO-CovMatTH-prescriptions-uncertainties} }
  \end{center}
\end{figure}

Next, in  Fig. \ref{fig:Global-NLO-CovMatTH-prescriptions} we 
compare PDFs obtained using different prescriptions.
The corresponding relative PDF uncertainties are compared
in Fig.~\ref{fig:Global-NLO-CovMatTH-prescriptions-uncertainties}.
In agreement with what we saw for the $\chi^2$ and $\phi$ values in
Tables~\ref{table:chi2table_covth_global_nlo},~\ref{table:phitable_covth_global_nlo}
results are quite stable with respect to the choice of prescription, though in the most extreme case of the  3-point prescription, where factorization scale variations are entirely uncorrelated between different processes, we observe 
somewhat smaller uncertainties, and a central value
which is closer to that when the MHOU is not included.

%

\begin{figure}[h]
  \begin{center}
    \includegraphics[scale=0.39]{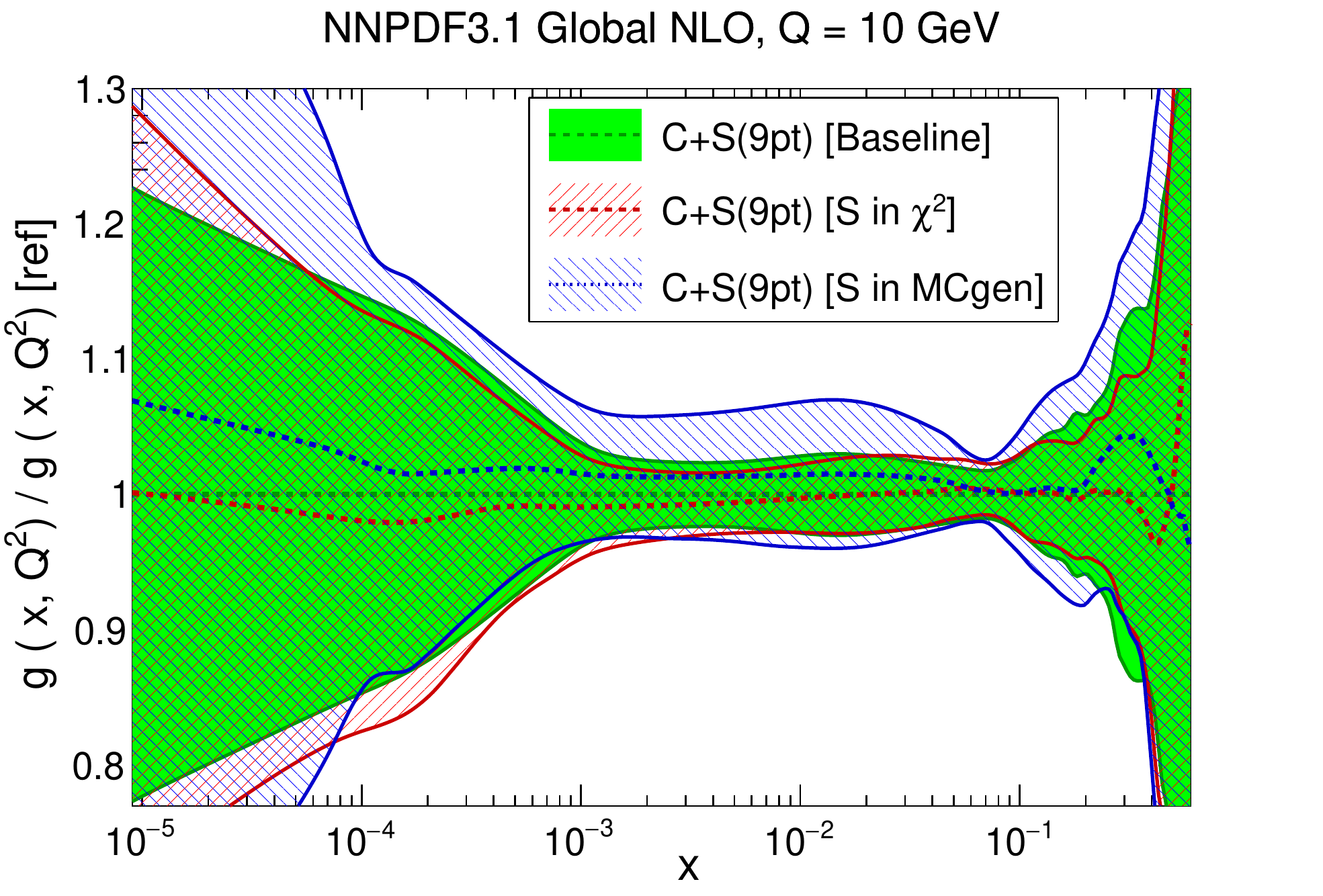}
    \includegraphics[scale=0.39]{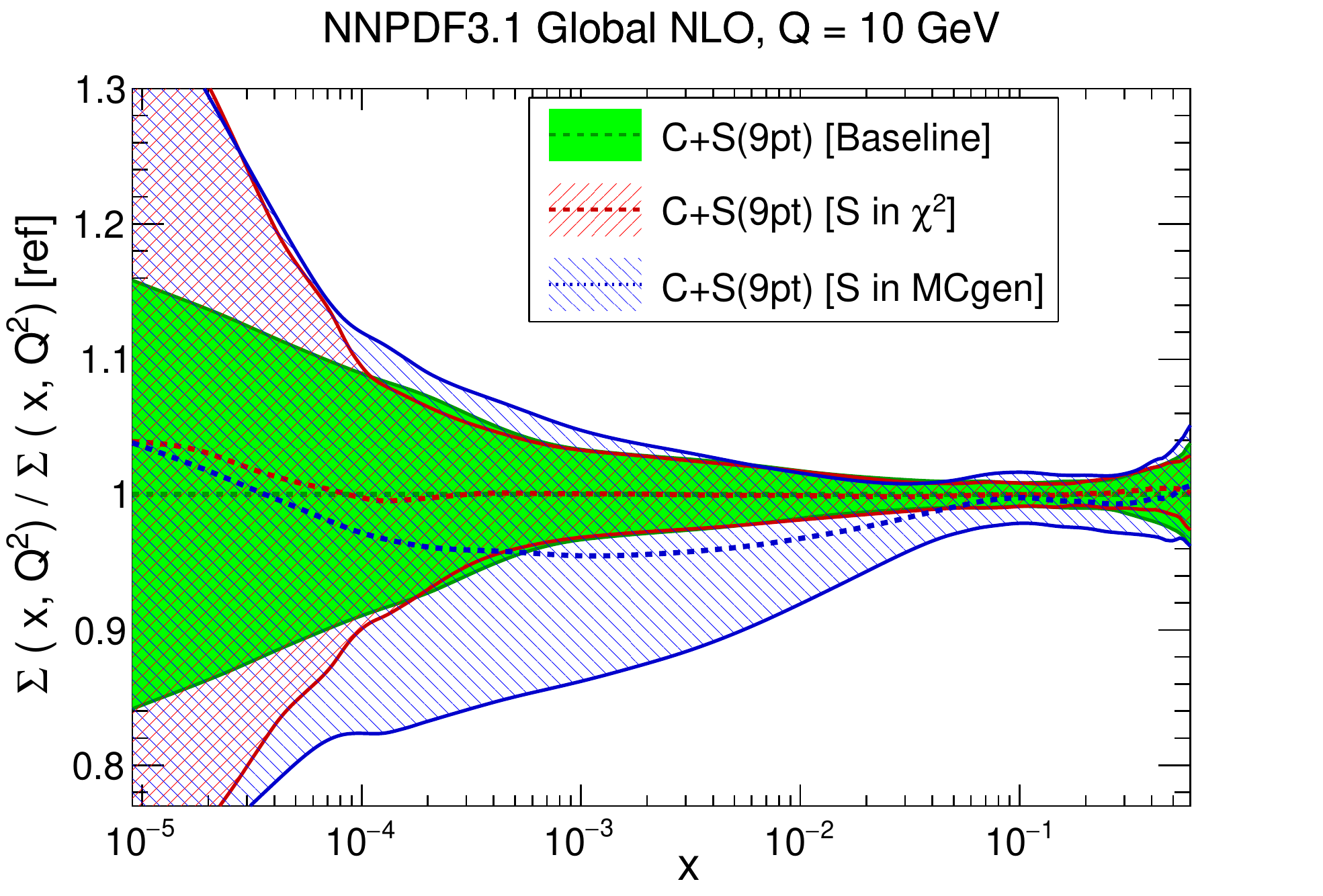}
    \includegraphics[scale=0.39]{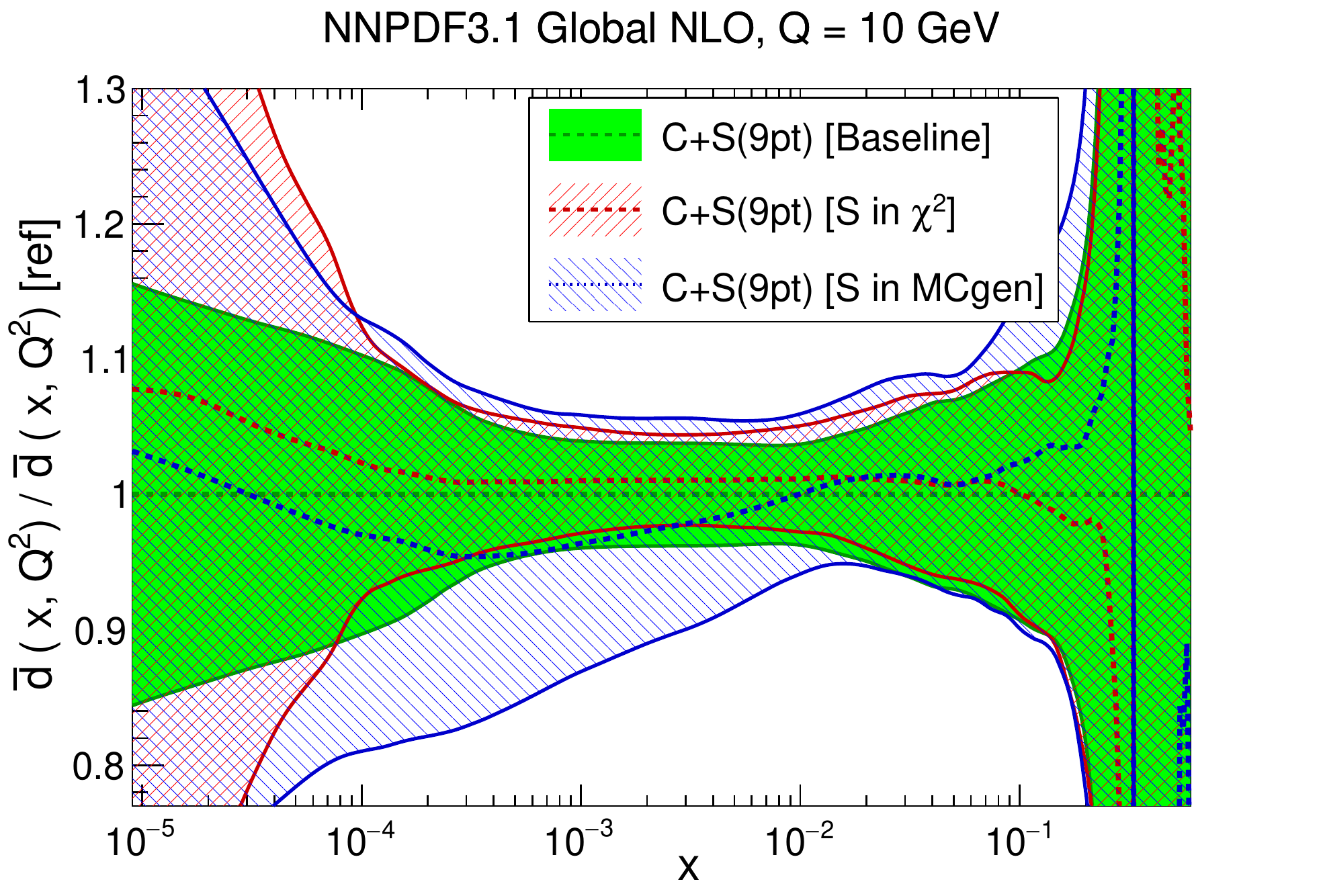}
   \includegraphics[scale=0.39]{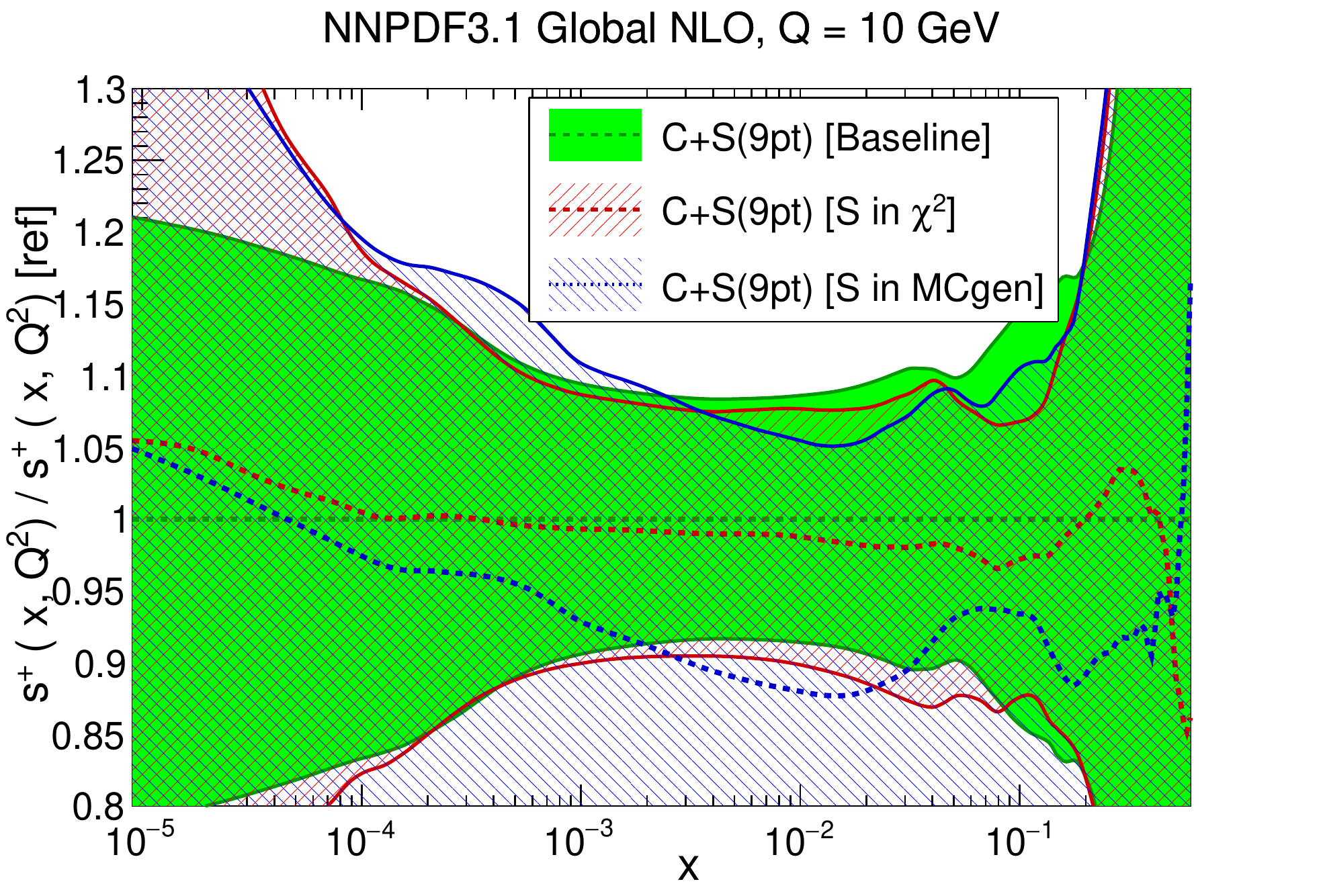}
   \caption{\small Same as Fig.~\ref{fig:Global-NLO-CovMatTH}, now comparing
     the results of the baseline $C+S^{(\rm 9pt)}$ fit with those in which
     the theory covariance matrix $S$ is included either in the $\chi^2$
     definition or in the generation of Monte Carlo replicas, but not on both.
    \label{fig:Global-NLO-CovMatTH-tests} }
  \end{center}
\end{figure}

Finally, in Fig.~\ref{fig:Global-NLO-CovMatTH-tests} we compare PDFs
obtained including the theory covariance matrix
only in the $\chi^2$ definition Eq.~(\ref{eq:chi2_v3}) but not
in the data generation Eq.~(\ref{eq:dgen}) and conversely.
We see that when the theory covariance matrix is included in the
replica generation but not in the $\chi^2$, uncertainties increase
very significantly.
This result is in agreement with the observation from
Table~\ref{table:chi2table_covth_global_nlo} that in this case the
fit quality significantly deteriorates, which is because the fit
becomes inconsistent due to the $\chi^2$ not matching the
wider fluctuations in the data.
The effect is particularly visible for the
quark distributions.
On the other hand, including the theory
covariance matrix only in the $\chi^2$  singles out the effect of
the theory covariance matrix on central values, due to rebalancing of 
datapoints in the fit according to their relative MHOU.
Indeed in this case
the central value is very close to that obtained when including the MHOU 
is both data generation and fit. We also see that the change in
uncertainties in the data region is now very small, consistent with
Table~\ref{table:chi2table_covth_global_nlo}.
These results confirm our 
expectation that in the full fit, while the MHOU results in a substantial 
increase in the fluctuations of data replicas, this is compensated by 
a relaxation of tensions due to the inclusion of MHOU the fit, with 
the net result that while central values shift, overall uncertainties 
do not increase much.

\FloatBarrier

\section{Implications for phenomenology}
\label{sec:pheno}

Whereas a  full assessment of the impact of the inclusion of MHOU in
PDFs will be possible only once we have global NNLO sets with MHOU, 
it is worth performing a first phenomenological
investigation, by computing reference LHC standard candles with the
NLO PDF sets which include MHOUs presented in Sect.~\ref{sec:fitstherr}, 
and comparing to results with the corresponding NLO PDF sets in which 
no MHOU is included.

In this section we will specifically consider
Higgs boson production in gluon-fusion and in vector-boson fusion, 
top quark pair and $Z$ and $W$ electroweak gauge boson
production. Note that the latter processes are among those which have
been used for PDF determination, see
Tab.~\ref{tab:datasets_process_categorisation}. This raises the issue
of possible double counting of uncertainties between the MHOU in the
PDF and in the hard matrix element. This will be addressed in
Sect.~\ref{sec:combmhou} below.

As discussed in Sect.~\ref{sec:fitstherr}, once the MHOU is included
in the covariance matrix, the standard NNPDF methodology can be used,
but with the PDF uncertainties now also including a theory-induced
contribution. Specifically,
PDF uncertainties (which now include the MHOU uncertainty) are
obtained as standard deviations over the replica sample. The total
uncertainty on a physical prediction is then obtained by combining this
uncertainty with that on the hard cross-section for the given
process. The latter is conventionally obtained as the envelope of a 7-point
scale variation, see e.g. Ref.~\cite{deFlorian:2016spz}. Of course, an
alternative possibility is to compute the theory uncertainty on the
hard cross-sections in exactly the same way as we compute it when
performing PDF determination, i.e. using the theory covariance
matrix. In this case, the MHOU on any measurement is found as the
diagonal element of the covariance matrix, evaluated for the given
measurement. Here we will compute the theory uncertainty both using the theory
covariance matrix (with the 9-point prescription,
given in Eq.~\eqref{9S}), and as a 7-point envelope.
The MHOU uncertainty
on the hard cross-section  can then be combined with the total
uncertainty on the PDF (which includes both MHOU and data
uncertainties) in quadrature. A more detailed discussion of
prescriptions for the computation of the total uncertainty on a
physical observable, including explicit formulae,
will be given in Sect.~\ref{sec:combmhou} below.

The current state of the art for precision phenomenology is NNLO, and
thus NNLO PDFs would be needed for accurate predictions. However, as
discussed in Sect.~\ref{sec:fitstherr}, at present only NLO global PDFs with
MHOU are available. In principle, NNLO PDFs from a DIS only fit are
also available. However, also as discussed in
Sect.~\ref{sec:fitstherr}, some of these PDFs (specifically the gluon)
are affected by large uncertainties due to the lack of
experimental constraints. The comparison of PDFs with and without MHOU
for such sets would thus be rather misleading. Therefore, in this
section we will focus on NLO PDFs. It should of course be kept in mind
that NNLO PDFs with MHOU are likely to have smaller uncertainties.

\subsection{Higgs production}
\label{sec:higgs}

We first discuss Higgs production in gluon 
fusion (ggF) and in vector boson fusion (VBF).
These two processes are of direct relevance for the characterization
of the Higgs sector and are both currently 
known at N$^3$LO accuracy~\cite{Anastasiou:2015ema,Anastasiou:2016cez,
Mistlberger:2018etf,Dreyer:2016oyx}.
Note that  the perturbative behavior and leading partonic channels
for these processes are quite different. %
Higgs production in gluon fusion is
driven by the gluon-gluon luminosity and its perturbative expansion
converges  slowly, with manifest
convergence reached only at N$^3$LO.
Vector boson fusion is driven by the quark-antiquark luminosity and
it exhibits fast perturbative
 convergence.

\begin{figure}[t]
  \begin{center}
   \includegraphics[width=0.49\textwidth]{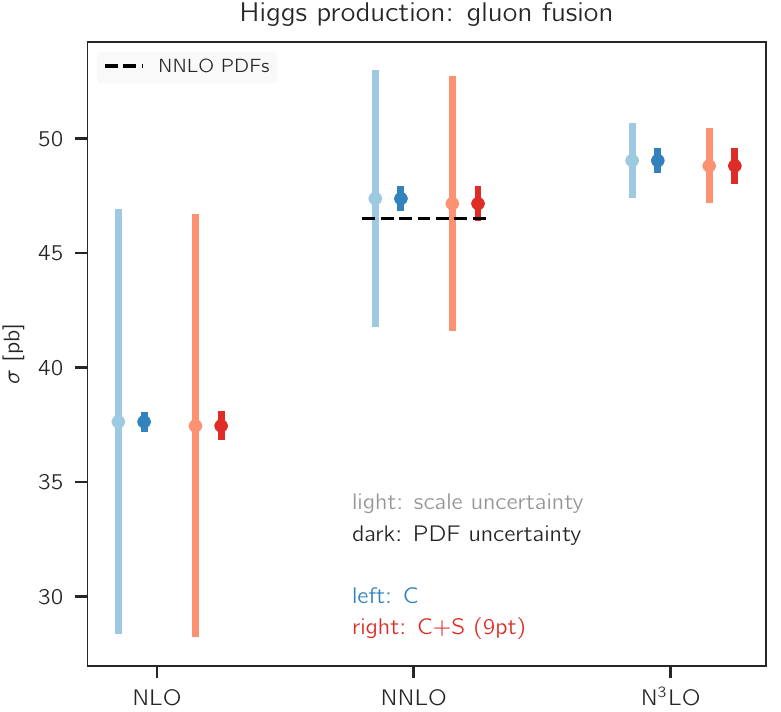}
  \includegraphics[width=0.49\textwidth]{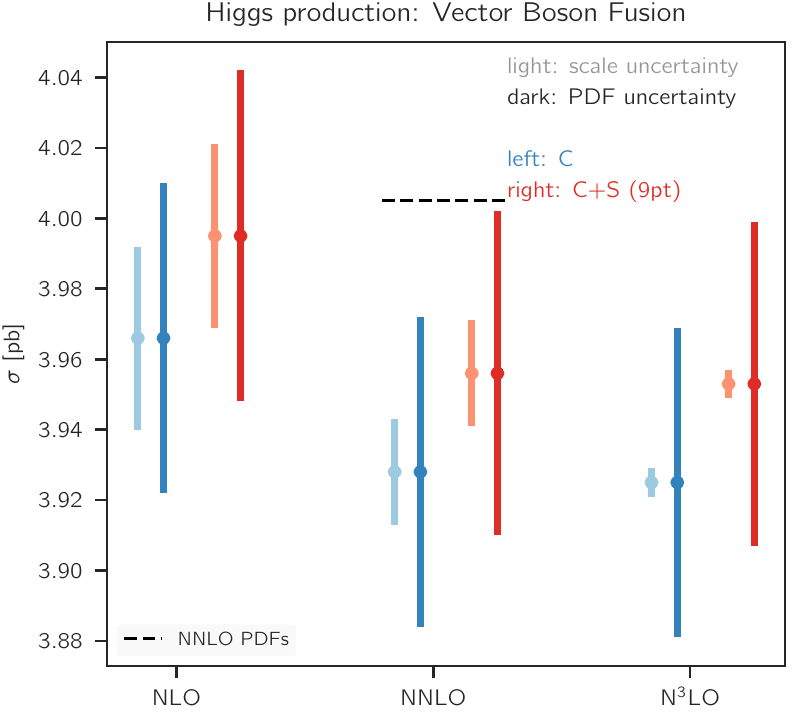}
\caption{\small Graphical
    representation of the results
    of Tables~\ref{tab:pheno-ggH} and~\ref{tab:pheno-VBF}.
At each perturbative order the pair of  uncertainty bands on the left
    (blue) is computed with PDFs
based on the  experimental covariance matrix $C$,
while the pair of  uncertainty
bands on the right (red) with  PDFs
based on the combined experimental and 
theoretical covariance matrix $C+S$ (9-point prescription).
The light-shaded bands represent the uncertainty on the hard cross-section 
(``scale uncertainty'') evaluated using the theory
covariance matrix (see text)
the dark bands represent the PDF uncertainty. For the NNLO result, we
also show the central value obtained using NNLO PDFs as a dashed horizontal
line.
    \label{fig:pheno-gghiggs} }
  \end{center}
\end{figure}

In Table~\ref{tab:pheno-ggH} 
we present predictions for Higgs production in gluon fusion
at the LHC for $\sqrt{s}=13$ TeV.
We perform the calculation at NLO, NNLO and N$^3$LO in the rescaled effective theory approximation
using {\tt  ggHiggs}~\cite{Ball:2013bra,Bonvini:2014jma,Bonvini:2016frm,Ahmed:2016otz,Bonvini:2018ixe,Bonvini:2018iwt} with
$\mu_f=\mu_r=m_H/2$ as central scale, with the
NLO global sets obtained in this paper, with and without MHOUs, as 
input PDFs at all orders.
The results are 
displayed graphically in Fig.~\ref{fig:pheno-gghiggs}, where, for the
NNLO computation, we also show the central value found using NNLO PDFs.

We find that for all perturbative orders the central values obtained with PDFs with and without MHOU are very similar, while the PDF uncertainty
is about 50\% larger when MHOU are included in the PDF fit.
This can be understood by noticing that for the intermediate values of
the momentum fraction,  $x\simeq 10^{-2}$, 
relevant for Higgs production in gluon fusion, the PDF uncertainty of
the gluon is increased in the $C+S^{(\rm 9pt)}$ fit 
as compared to the $C$-only fits, see
Fig.~\ref{fig:Global-NLO-CovMatTH}. Comparison to the result
obtained using NNLO PDFs  (for the NNLO computation) shows that
upon inclusion of the MHOU the PDF uncertainty band of the result
with NLO PDFs now includes the NNLO PDF result, while it would not in
the absence of MHOU, both because of the (small) shift in central
value and of the widening of the uncertainty band.

From Table~\ref{tab:pheno-ggH} one can also observe that the MHOU on
the hard matrix element uncertainty
$\sigma_{\mathcal{F}}^{\rm th}$ evaluated using the 9-point theory covariance matrix, Eq.~\eqref{9S},
is compatible with the canonical 7-point envelope
if the latter is symmetrized by taking the maximum value between the lower and upper uncertainties.
In particular, the theory covariance matrix estimate
is slightly larger than the  envelope prescription
at NLO and at NNLO, while it becomes a little smaller at N$^3$LO.
Even so, the NLO uncertainty band does not contain the NNLO
central value, which lies just above the edge of the band. 

\begin{table}[tbp]
	\centering
	\small
  \renewcommand*{\arraystretch}{1.7}
\begin{tabular}{l l l}
\multicolumn{3}{c}{Higgs production in gluon fusion at 13 TeV }\\
\toprule
           & \multicolumn{1}{c}{$C$} & \multicolumn{1}{c}{$C+S^{(\rm 9pt)}$} \\
\midrule
 NLO & 37.63 $\pm$  1.14\% $\pm$ 24.67 (22.69) \%  & 37.45 $\pm$  1.69\% $\pm$ 24.67 (22.69) \% \\
NNLO & 47.38 $\pm$  1.12\% $\pm$ 11.82 (10.09) \%
   & 47.16 $\pm$  1.65\% $\pm$ 11.83 (10.09) \% \\
N$^3$LO & 49.04 $\pm$  1.12\% $\pm$ 3.35 (3.85) \%
   & 48.81 $\pm$  1.65\% $\pm$ 3.35 (3.85) \% \\
\bottomrule
\end{tabular}

        \vspace*{3mm}
	\caption{\small The 
          total cross-sections for Higgs production in gluon fusion
	(in pb) obtained by  
          using NLO global PDFs based on either $C$ or
	$C+S^{{\rm (9pt)}}$, 
          see Table~\ref{tab:thcovmatFits}.
          We quote the central prediction, the total PDF uncertainty (first) and
          the MHOU  uncertainty on the hard cross-section
	(second) expressed as a percentage of the central value. The latter is evaluated both using the theory
	covariance matrix (9-point prescription) or, in parenthesis, a
         (symmetrized) envelope of the 7-point scale variations (see
	Sect.~\ref{sec:combmhou}), obtained by taking the maximum value between the lower and upper uncertainties. }
	\label{tab:pheno-ggH}
\end{table}

We conclude that using NLO PDFs in the N$^3$LO calculation, the
inclusion of MHOU in the PDFs translates into a few 
per-mille increase of the PDF uncertainty at the cross-section level.  
In Ref.~\cite{Anastasiou:2016cez} NNLO PDFs were used with the
N$^3$LO calculation in order to provide a state-of-the art result, and
a MHOU uncertainty on the NNLO PDF was estimated based on the
difference between results obtained using NLO and NNLO PDFs. Once NNLO
PDFs with MHOUs determined within our approach are available it
will be interesting to compare our results with this estimate.

We now turn  to Higgs production in vector boson fusion.
We perform the calculation at N$^3$LO accuracy
using {\tt proVBFH-inclusive}~\cite{Cacciari:2015jma,Dreyer:2016oyx}.
with central factorization and renormalization scales set equal
to the squared four-momentum of the vector boson. 
Results are collected in Table~\ref{tab:pheno-VBF} and shown
in Fig.~\ref{fig:pheno-gghiggs}.  
The  MHOU corrections to the PDFs are very small, so 
PDF uncertainties with or without theory covariance matrix are very similar.
Also in this case, like for gluon fusion, the uncertainty on the hard
matrix element computed with the
9-point theory covariance matrix
is similar to the one obtained by symmetrizing the 7-point envelope. 

The smallness of the MHOU in the PDF follows from the fact that VBF 
Higgs production
is driven by the quark-antiquark luminosity, which in turn is
dominated by the quark PDF in the data region, whose uncertainties,
as we have seen
in Sect.~\ref{sec:globmhou}, are almost unaffected by the inclusion of
MHOU.
Comparison to the result
obtained using NNLO PDFs  (for the NNLO computation) shows that
the NNLO PDF result is at the edge of the PDF uncertainty band of the result
with NLO PDFs if MHOU are included, while it is off by almost two $\sigma$
if they are not. This is  essentially due to the significant shift in
central value, in agreement with the observation made in
Sect.~\ref{sec:globmhou}, where we noticed that 
MHOUs have the effect of moving the central value of the PDFs in the data
region towards the NNLO result.
The shift in the central value of the VBF cross-section
due to the MHOU is in fact quite significant: its size is
comparable to the MHOU $\sigma_{\mathcal{F}}^{\rm th}$ on the
NLO matrix element, and indeed the shift when going from NLO to 
NNLO matrix elements, and thus much larger that the corresponding 
N$^3$LO correction.

We conclude that for VBF the main effect of including the MHOU in the PDF 
is a significant shift in the central value of the prediction.
Also in this case
estimates of the MHOU on the NNLO PDF were presented in
Ref.~\cite{Dreyer:2016oyx}, and it will be interesting to compare them
to our approach once NNLO PDFs with MHOU determined within our
approach are available.
%

\begin{table}[t]
	\centering
	\small
\renewcommand*{\arraystretch}{1.7}
\begin{tabular}{l l l}
\multicolumn{3}{c}{Higgs production in VBF at 13 TeV }\\
\toprule
           & \multicolumn{1}{c}{$C$} & \multicolumn{1}{c}{$C+S^{(\rm 9pt)}$} \\
\midrule
NLO & 3.966 $\pm$  1.12\% $\pm$ 0.66 (0.66) \% &  3.995 $\pm$  1.17\% $\pm$ 0.66 (0.65) \% \\
NNLO & 3.928 $\pm$  1.12\% $\pm$ 0.37 (0.42) \% &  3.956 $\pm$  1.17\% $\pm$ 0.37 (0.41) \%  \\
N$^3$LO & 3.925 $\pm$  1.12\% $\pm$ 0.11 (0.15)   &
3.953 $\pm$  1.17\% $\pm$ 0.11 (0.14) \%  \\
\bottomrule
\end{tabular}

        \vspace*{3mm}
	\caption{Same as Table~\ref{tab:pheno-ggH}, now for Higgs production in vector boson fusion.}
	\label{tab:pheno-VBF}
\end{table}

A common feature of gluon fusion and vector-boson fusion is that it is
only upon inclusion of the MHOU that the result found using NNLO PDFs
is within or at the edge of the PDF uncertainty band of the result
found with NLO PDFs.

\subsection{Top quark pair production}
\label{sec:top}

\begin{figure}[t]
  \begin{center}
    \includegraphics[width=0.49\textwidth]{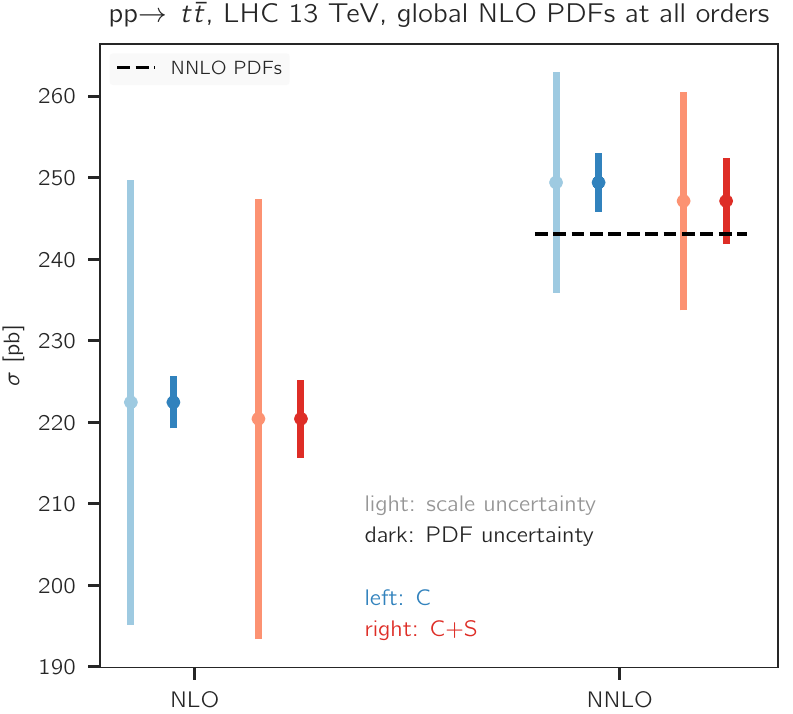}
    \includegraphics[width=0.49\textwidth]{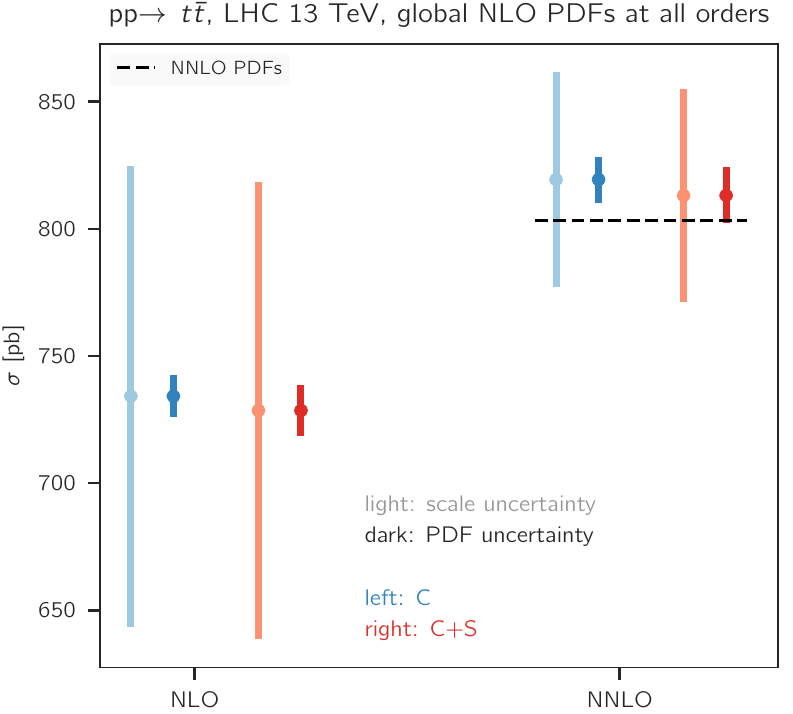}
    \caption{\small Same as Fig.~\ref{fig:pheno-gghiggs}
    for top-quark pair production at 8 and 13 TeV,
    see also Table~\ref{tab:pheno-ttbar}.
    \label{fig:pheno-ttbar} }
  \end{center}
\end{figure}

We now study the impact of the PDF-related MHOU on the
total top-quark pair production cross-section at the LHC for
different center-of-mass energies.
In Table~\ref{tab:pheno-ttbar} we collect, using the same
format as Table~\ref{tab:pheno-ggH}, the  predictions
for the top-quark pair-production cross-sections at $\sqrt{s}=7$, 8 and 13 TeV
obtained using the {\tt top++} code~\cite{Czakon:2011xx} and setting
the central scales to $\mu_f=\mu_r=m_t=172.5$ TeV.
The results in the case of 8 and 13 TeV are also displayed in
Fig.~\ref{fig:pheno-ttbar}, where again for at NNLO  we
also show the result obtained using NNLO PDFs.

\begin{table}[t]
	\centering
	\small
\renewcommand*{\arraystretch}{1.5}
\begin{tabular}{lll}
\multicolumn{3}{c}{ $t\bar{t}$ production at 7 TeV}\\
\toprule
     & \multicolumn{1}{c}{$C$} & \multicolumn{1}{c}{$C+S^{(\rm 9pt)}$} \\
\midrule
NLO   &  155.42 $\pm$ 1.57\% $\pm$ 12.2 (13.0) \%   & 153.94 $\pm$ 2.45\% $\pm$ 12.2 (13.0) \% \\
NNLO  &  174.48 $\pm$ 1.55\% $\pm$ 5.52 (6.46) \%   & 172.81 $\pm$ 2.42\% $\pm$ 5.52 (6.45) \% \\
\bottomrule
\\[-0.3cm]
\multicolumn{3}{c}{ $t\bar{t}$ production at 8 TeV}\\
\toprule
     & \multicolumn{1}{c}{$C$} & \multicolumn{1}{c}{$C+S^{(\rm 9pt)}$} \\
\midrule
NLO   &  222.45 $\pm$ 1.44\% $\pm$ 12.3 (12.8) \%   & 220.42 $\pm$ 2.17\% $\pm$ 12.3 (12.8) \% \\
NNLO  &  249.41 $\pm$ 1.43\% $\pm$ 5.43 (6.28) \%   & 247.14 $\pm$ 2.14\% $\pm$ 5.43 (6.27) \% \\
\bottomrule
\\[-0.3cm]
\multicolumn{3}{c}{ $t\bar{t}$ production at 13 TeV}\\
\toprule
     & \multicolumn{1}{c}{$C$} & \multicolumn{1}{c}{$C+S^{(\rm 9pt)}$} \\
\midrule
NLO   &  734.21 $\pm$ 1.11\% $\pm$ 12.4 (11.8) \%   & 728.57 $\pm$ 1.38\% $\pm$ 12.3 (11.8) \% \\
NNLO  &  819.43 $\pm$ 1.11\% $\pm$ 5.16 (5.64) \%   & 813.17 $\pm$ 1.35\% $\pm$ 5.16 (5.64) \% \\
\bottomrule
\end{tabular}

        \vspace*{3mm}
	\caption{\small
        Same as Table~\ref{tab:pheno-ggH}, now for
        top-quark pair-production at $\sqrt{s}=7,8$ and 13 TeV.}
	\label{tab:pheno-ttbar}
\end{table}

Just as in the case of Higgs production via gluon-gluon fusion,
we find  that for top-quark pair production the central values obtained with 
PDFs with and without MHOU are rather similar, and well within the one-$\sigma$
PDF uncertainty.
We also observe that the PDF uncertainty at $\sqrt{s}=7$ and 8 TeV (13 TeV) 
is about 50\% (20\%) larger once MHOU are included in the determination of the PDFs. 
This is again compatible with the  corresponding behavior of the gluon PDF shown in 
Fig.~\ref{fig:Global-NLO-CovMatTH}, where it can be observed that, for $x\simeq 0.1$, 
relevant for top pair production at $\sqrt{s}=7$ and 8 TeV, the PDF uncertainty is increased
in the $C+S^{\rm (9pt)}$ fit as compared to the $C$-only fit, while
this increase is less marked for $x\sim 0.3$, relevant for top pair
production at $\sqrt{s}=13$ TeV.
Also in this case, 
the NNLO prediction using NLO PDFs
is in better
agreement with the that using NNLO PDFs once MHOUs are included, and
in fact only in this case the latter is within the PDF error band of
the former.

In addition, we note once again that the uncertainty on the hard
cross-section
$\sigma_{\mathcal{F}}^{\rm th}$
evaluated using the 9-point covariance matrix is rather
similar to that obtained from the  symmetrized 7-point envelope.
 In particular, the 9-point result is slightly larger (smaller) than the 7-point envelope at NNLO (NLO).
Finally, from Fig.~\ref{fig:pheno-ttbar} we notice that for this process
the  MHOU on the hard cross-section  
dominates the PDF uncertainty (with or without MHOU included), 
even with NLO PDFs.

\subsection{$Z$ and $W$ gauge boson production}
\label{sec:zw}
We finally turn to gauge boson production, for which we obtain predictions
using  the computational framework {\tt Matrix}~\cite{Grazzini:2017mhc}.
In this formalism, all tree-level and one-loop amplitudes are 
obtained from {\tt
OpenLoops}~\cite{Cascioli:2011va,Matsuura:1988sm,Denner:2016kdg}. 
For these theoretical predictions for  inclusive $W$ and $Z$ production cross sections at
$\sqrt{s}$ = 13 TeV, we adopt realistic kinematic cuts similar to those
applied by ATLAS and CMS.
The fiducial phase space for the $W^{\pm}$ cross-section is defined
by requiring $p_{l,T}\ge 25$ GeV and $\eta_{l} \le$ 2.5 for the charged lepton
transverse momentum and pseudo-rapidity and a missing energy
from the neutrino of $p_{\nu,T}\ge 25$ GeV.
In the case of $Z$ production, we require 
$p_{l,T}\ge$ 25 GeV and $|\eta_l|\le$ 2.5 for the charged leptons
transverse momentum and rapidity and 66 $\le m_{ll} \le$ 116 GeV for 
the di-lepton invariant mass.

\begin{figure}[t]
  \begin{center}
    \includegraphics[width=0.48\textwidth]{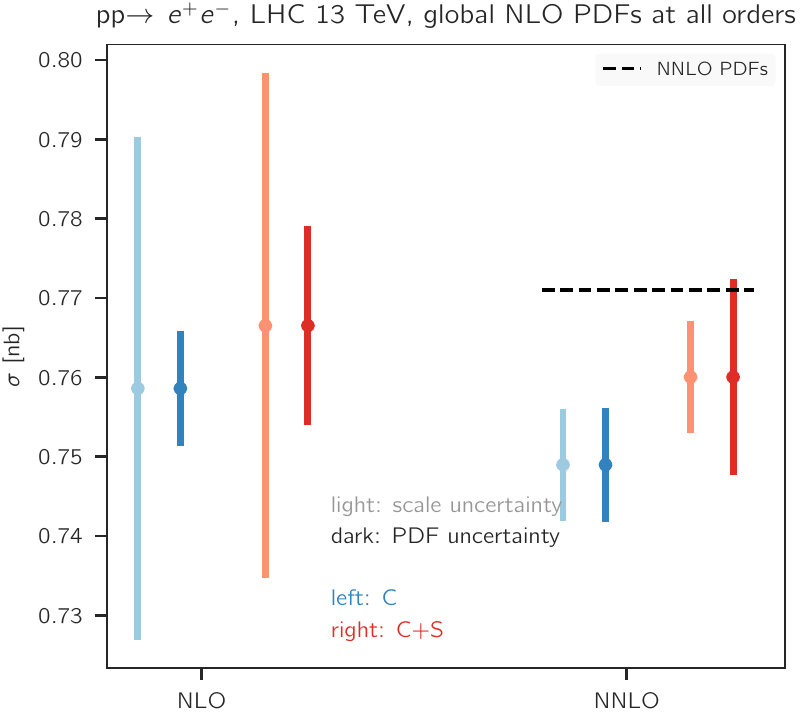}
    \includegraphics[width=0.48\textwidth]{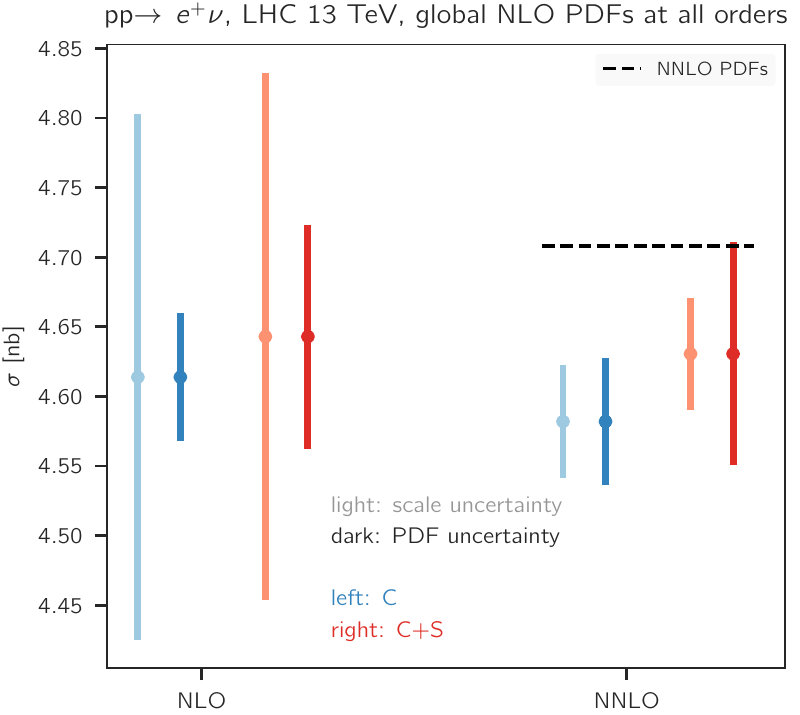}
    \includegraphics[width=0.48\textwidth]{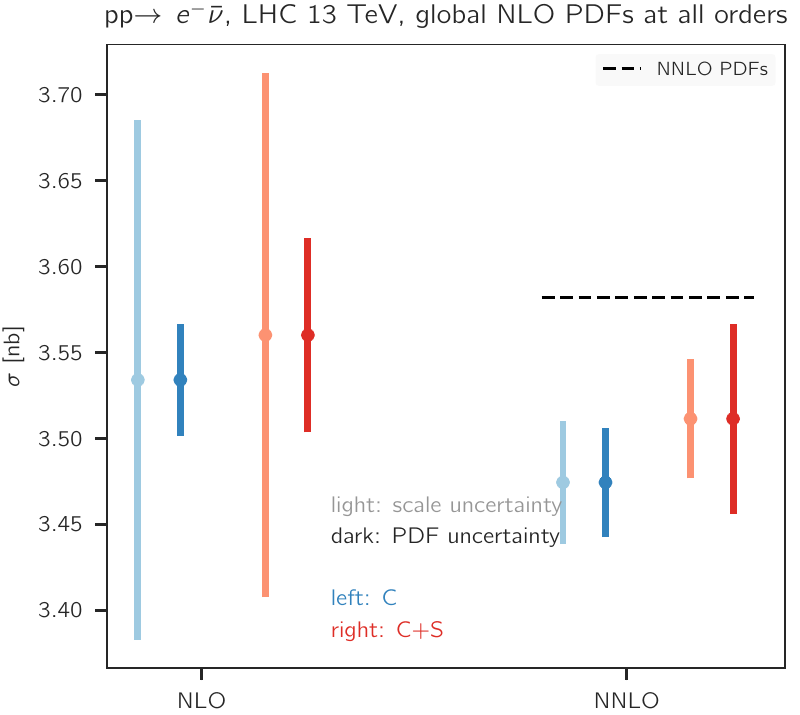}
    \caption{\small Same as Fig.~\ref{fig:pheno-gghiggs}
    for $W^\pm$ and $Z$ gauge boson production
    at $\sqrt{s}=13$ TeV,
    see also Table~\ref{tab:pheno-ZW}.
    \label{fig:pheno-zw} }
  \end{center}
\end{figure}

In Table~\ref{tab:pheno-ZW} we display a similar
comparison as in Table~\ref{tab:pheno-ggH} now for
        $W$ and $Z$ gauge boson production at $\sqrt{s}=13$ TeV.
The corresponding graphical representation of the results is provided in
Fig.~\ref{fig:pheno-zw}, again using the same conventions as in
Fig.~\ref{fig:pheno-gghiggs} and again also showing the NNLO result
with NNLO PDFs.

We find that when including the
 MHOU the PDF uncertainty
is increased by
$\simeq 70\%, 30\%$ and $75\%$ for $Z$, $W^+$, and $W^-$ production
 respectively. 
Given that $W$ and $Z$ production at ATLAS and CMS at $\sqrt{s}=13$ TeV
is sensitive to the light sea quarks down to $x\simeq 10^{-3}$, this increase
in the PDF uncertainty once MHOU are accounted for is consistent
with the corresponding increase reported in the case
of the singlet PDF in Fig.~\ref{fig:Global-NLO-CovMatTH-prescriptions-uncertainties}.

\begin{table}[t]
  \centering
  \small
\renewcommand*{\arraystretch}{1.5}
\begin{tabular}{lll}
\multicolumn{3}{c}{$Z$ production at 13 TeV}\\
\toprule
     & \multicolumn{1}{c}{$C$} & \multicolumn{1}{c}{$C+S^{(\rm 9pt)}$} \\
\midrule
NLO   &  0.759 $\pm$ 0.96\% $\pm$ 4.18 (4.18) \%   & 0.767 $\pm$ 1.63\% $\pm$ 4.16 (4.15) \% \\
NNLO  &  0.749 $\pm$ 0.97\% $\pm$ 0.94 (0.63) \%   & 0.760 $\pm$ 1.64\% $\pm$ 0.93 (0.66) \% \\
\bottomrule
\\[-0.3cm]
\multicolumn{3}{c}{$W^-$ production at 13 TeV}\\
\toprule
     & \multicolumn{1}{c}{$C$} & \multicolumn{1}{c}{$C+S^{(\rm 9pt)}$} \\
\midrule
NLO   &  3.534 $\pm$ 0.92\% $\pm$ 4.28 (4.34) \%   & 3.560 $\pm$ 1.58\% $\pm$ 4.28 (4.34) \% \\
NNLO  &  3.474 $\pm$ 0.92\% $\pm$ 1.03 (0.64) \%   & 3.511 $\pm$ 1.59\% $\pm$ 0.99 (0.63) \% \\
\bottomrule
\\[-0.3cm]
\multicolumn{3}{c}{$W^+$ production at 13 TeV}\\
\toprule
     & \multicolumn{1}{c}{$C$} & \multicolumn{1}{c}{$C+S^{(\rm 9pt)}$} \\
\midrule
NLO   &  4.614 $\pm$ 1.00\% $\pm$ 4.09 (4.15) \%   & 4.643 $\pm$ 1.73\% $\pm$ 4.08 (4.14) \% \\
NNLO  &  4.582 $\pm$ 0.99\% $\pm$ 0.88 (0.58) \%   & 4.631 $\pm$ 1.72\% $\pm$ 0.87 (0.62) \% \\
\end{tabular}

  \vspace{2mm}
  \caption{Same as Table~\ref{tab:pheno-ggH}, now for
        $W$ and $Z$ gauge boson production at $\sqrt{s}=13$ TeV. The cross-section 
         is given in nb.}
  \label{tab:pheno-ZW}
\end{table}

Similarly to  Higgs production in vector-boson-fusion, we find that
the inclusion of MHOU in the PDF shifts the central value of the prediction, 
by an amount
which is comparable to or larger than the data-driven PDF
uncertainty. Yet again,  the agreement of
the NNLO prediction with NLO PDFs with that which is obtained when
NNLO PDFs are used is significantly improved: for $Z$ production
within the PDF error band and for $W$ production just barely outside it.
We  conclude that for weak gauge boson production at
the LHC the impact of the MHOU associated to the PDFs is twofold: on
the one hand an overall 
increase in the PDF uncertainties that ranges between 30\% and 70\% depending on the process,
and on the other hand a shift in the central values which is comparable to that
of the PDF uncertainties of the fit without MHOU.

\section{Usage and delivery}
\label{sec:usage}

As mentioned previously, the PDF sets with MHOU presented in
Sect.~\ref{sec:fitstherr} 
can be used in essentially the same way as the standard NNPDF
sets.
In this section we discuss how MHOUs included in PDF sets should
be combined with those in hard matrix elements, specifically
addressing some conceptual issues, and we then
provide detailed instructions for their use.
We then discuss the delivery of the
PDF sets presented in
this work, and provide a list of the sets which are being made publicly
available by means of the {\tt LHAPDF} interface.

\FloatBarrier

\subsection{Combining MHOUs in PDFs and hard matrix elements}
\label{sec:combmhou}

As discussed in the introduction, the MHOU on PDFs discussed in this
paper arises due to the fact that PDFs are determined using perturbative
computations performed at a finite order in the perturbative expansion,
and it manifests itself in the fact that PDFs change when varying the order 
at which they are
determined: NLO and NNLO PDFs differ.
We have further seen in
Sect.~\ref{sec:scalevarn} that there exist two distinct 
sources of MHOU in the PDF: that related to MHOs in the computation of
the hard cross-sections for those processes used for PDF determination, and
that coming from MHOs in the anomalous dimensions.
These two sources of MHOU in the PDFs
are respectively associated with
renormalization and factorization scale variation and can be treated
as independent of each other, at least
with the definition given here and summarized in
Table~\ref{tab:scale_nomenclature}.

On top of this MHOU on the PDF, when computing a factorized
prediction for a PDF-dependent hard process not used in the 
determination of the PDFs, 
but rather predicted using a given PDF set, there is then the usual
MHOU on the hard process itself. This, in turn, just like the MHOU on
the PDF, comes from two separate sources: the MHOU on the hard
cross-section for the given process, and the MHOU on
the evolution of the PDF from the initial scale to the scale of the 
process. This has been
seen explicitly in the phenomenological results presented in
Sect.~\ref{sec:pheno}, Tables~\ref{tab:pheno-ggH}-\ref{tab:pheno-ZW}
and Figs.~\ref{fig:pheno-gghiggs}-\ref{fig:pheno-zw}. So
each prediction carries two uncertainties, a PDF uncertainty, which
includes the MHOU in the determination of the PDFs (shown as
a dark band in the plots, and given as the first uncertainty in the
tables), and a ``scale'' uncertainty in the prediction (shown as
a light band in the plots, and given as the second uncertainty in the
tables). Note that in all these plots and tables the PDF uncertainty
(when including the theory covariance matrix) includes both the MHOU,
and the standard PDF uncertainty due to the uncertainties in the experimental 
data, while the ``scale'' uncertainty is just the usual MHOU in the prediction.

In summary, a factorized prediction is affected by
two different sources of MHOU:
the MHOU in the PDF determination, included in the PDF uncertainty, and then 
the MHOU in the calculation of the prediction itself. Each in turn
receives contributions from both
renormalization and factorization scale variation.
This immediately raises the question as to whether some of these 
uncertainties are correlated, and --- if this is the case --- 
whether this correlation can be easily accounted for.

A first obvious source of correlation arises when producing a
prediction for a process which is among those included for the PDF
determination.
Examples of this category of processes are top quark pair and gauge
 boson production, discussed
in Sect.~\ref{sec:pheno}. They are already included among the
 processes of
Table~\ref{tab:datasets_process_categorisation}.
The MHOU coming from
renormalization scale variation is then correlated.
Indeed, we know from Fig.~\ref{fig:corrmats} that any two
predictions for the same physical process are highly correlated, particularly
at points which are kinematically close. 
One might choose to ignore this problem, on the grounds that the main 
purpose of 
PDF determinations is to predict new processes, such as Higgs production, 
or BSM processes: after all, if there is new data for an existing process, it 
can be included in the PDF fit, and then all correlations would be retained.
However this (partial) solution is not available for factorization scale 
variations, which are used to estimate the MHOU in the evolution between 
different scales: since 
the PDFs are universal, these MHOUs are correlated across all processes, both 
within the fit and also in any predictions made subsequently using the PDFs. 

The existence of correlations between MHOU in the fitted process and MHOU 
in the predicted process can be demonstrated rather 
clearly~\cite{Harland-Lang:2018bxd} by 
noting that PDFs are merely a tool to express a physical observable 
in terms of other physical observables. In particular QCD predicts 
the cross-section for one
observable in terms of measurements of cross-sections for the same or other
observables. Normally to do this one first 
extracts the PDF from the cross-section data at a range of scales, 
and then computes cross-sections at some other scale 
using the extracted PDFs. However in the case of nonsinglet structure 
functions (discussed in Ref.~\cite{Harland-Lang:2018bxd} as a simple 
paradigm), where the relation between structure function and PDF is 
straightforward and linear, one can eliminate the nonsinglet PDF altogether: 
given the structure function at one scale, QCD then predicts the 
structure function at a different scale, with no reference to any PDF. 

Now, it is clear that when expressing one process in terms of another
process directly, without any PDFs, there is a significant cancellation of MHOU,
specifically that related to perturbative evolution, estimated by
means of factorization
scale variation. In the example of the nonsinglet structure function,
if the structure function at one scale is predicted from its value at some
different scale, the factorization
scale uncertainty will only depend on the evolution between the
two scales involved. Hence there is only one source of MHOU in the prediction.
On the other hand, when using a PDF, there are, as  explained above, two 
sources of MHOU estimated through factorization scale variation: that from 
evolving the initial PDF up to the scale of the data used in fit, and that from evolving the initial PDF up to the scale of the prediction. Hence, one has 
in effect two sources of MHOU, and if these are assumed to be uncorrelated, and thus added in quadrature, any cancellations are lost and the result will inevitably be an over-estimate of the uncertainty. 

If PDFs are to be delivered in the usual way as a universal ({\it
  i.e.} process independent) PDF set, much of the detailed information
about the specific data, their uncertainties, and the theoretical
calculations, and in particular their MHOUs that have gone into
determining the PDFs is lost:  all that remains are the process
independent PDFs.  Given only the PDFs, it is clearly impossible 
to reconstruct the original  data, or the MHOUs specific to calculations 
at each data point, since many different data sets, 
from different processes, can yield the same PDFs. Consequently, when 
using PDFs to make a prediction, the correlation between the MHOU in the prediction and that in the calculations used to determine the PDFs cannot be 
computed, even in principle: with only the universal PDFs as input, the correlation it is no longer available. The loss of this correlation is the
inevitable price to pay for PDF universality. 

Having understood that neglecting such correlations is
inevitable, at least without extending the range of deliverables, one may ask how serious the issue is. The total MHOU in 
the determination of the PDF arises
from the combination of the MHOU of theoretical predictions made for  
a large number of datapoints. The
correlations between all these are automatically kept into account by
the fitting procedure. Inevitably the fit adjusts to take the MHOU into 
account: datapoints associated with large MHOU (compared to the experimental 
uncertainty) will be deweighted in the fit, while the effect of data with 
small MHOU (compared to their experimental uncertainty) will be relatively 
unchanged. This rebalancing of the fit is one of the main consequences
of including the MHOU.

Hence, as we saw in our global fit results, the MHOUs 
have only a relatively small impact on the overall PDF uncertainty: rather 
by resolving tensions in the fit due to MHOs in the theoretical predictions, 
they lead to significant shifts in the central value. However when making a 
prediction, the uncertainty due to MHOU in the hard process can be large: in 
fact in many cases as large or even larger than the total PDF uncertainty 
(including its MHOU). Neglecting the correlation between the MHOU in the 
prediction (which might be large) and the MHOU in the PDFs (which is relatively 
small) by adding them in quadrature is then likely to be a small effect. 
Note that this does not mean that
the MHOU on the PDF was negligible in the first place: and indeed as
we have seen it may significantly affect the central value of the
prediction. Rather, it is its effect on the overall PDF uncertainty
which, at least in the data region that we are discussing here, is
relatively small. Furthermore,
because what is being neglected is a correlation which would lead to a
cancellation of uncertainties, it can at worst lead to a small overestimate 
of uncertainties. 

We conclude that the while there is clearly a correlation between the MHOU in 
the determination of the PDFs and the MHOU of the hard matrix element of the predicted process, ignoring this correlation, and thus adding the two sources of MHOU in quadrature, will give a result which is at worst a little conservative.
Given all the well known uncertainties intrinsic to 
the estimate of MHOUs through scale variation, we consider such an approach both pragmatic and realistic. 

\subsection{Computation of the total uncertainty}
\label{eq:unccomp}

Having concluded that uncorrelated combination of the MHOU on the PDF
and on the hard matrix element is justified, we summarize
our procedure for computing uncertainties in practice.

To begin with, the PDF uncertainty $\sigma^{\rm PDF}_{\mathcal{F}}$
associated with a given cross-section $\mathcal{F}$ is  evaluated
as usual in the NNPDF methodology as the standard deviation over the
replica set:
 \be
\sigma^{\rm PDF}_{\mathcal{F}} =
\left( \frac{1}{N_{\rm rep}-1}
\sum_{k=1}^{N_{\rm rep}}   
\lp \mathcal{F} [ \{  q^{(k)} \}] 
-   \la \mathcal{F} [ \{  q \}] \ra\rp^2 
 \right)^{1/2}.
 \label{eq:mastersig}
 \ee
If this prescription is applied to a PDF set with ``standard'' PDF uncertainty (such as the published
NNPDF3.1~\cite{Ball:2017nwa}) set, the resulting uncertainty only includes
the correlated statistical and systematic uncertainties from the data,
and the methodological uncertainty intrinsic to any PDF fit.
If the PDF sets including MHOU
presented in Sect.~\ref{sec:results}
of this paper are used instead, the resulting uncertainty obtained
from Eq.~(\ref{eq:mastersig}) accounts for both the data-driven
and MHOU on the PDF, with all correlations taken into account.

Because the MHOU on the hard matrix element is treated as uncorrelated
to the PDF uncertainty, it can in principle be computed with any
prescription preferred by the end-user.
A commonly used
prescription is 7-point scale variation~\cite{deFlorian:2016spz}.
Our preferred prescription is instead to use the same methodology as used for
the computation of the theory covariance matrix.
In this case, the uncertainty on
the cross-section $\mathcal{F}$ is then simply the corresponding diagonal 
entry of the covariance matrix element, namely
\begin{equation}\label{eq:thuncmh}
\sigma_{\mathcal{F}}^{\rm th} = \left[S^{(\rm 9pt)}_{{\mathcal{F}}{\mathcal{F}}}\right]^{1/2},
\end{equation}
where $S^{(\rm 9pt)}_{{\mathcal{F}}{\mathcal{F}}}$ is evaluated using  our default
9-point prescription defined by
Eq.~\eqref{9S}, with $\Delta_{ij}$ computed for $i=j=\mathcal{F}$,
i.e. the theory prediction for the given observable.
We showed in 
Sect.~\ref{sec:pheno} that for various standard candles our 9-point theory 
covariance matrix prescription and the 7-point envelope prescription give 
very similar results, provided the envelope prescription is symmetrized.

The PDF uncertainty Eq.~(\ref{eq:mastersig}) and the uncertainty on
the hard matrix element Eq.~(\ref{eq:thuncmh}) can then be treated as
uncorrelated uncertainties.
It is then appropriate to combine them in quadrature, so the total
uncertainty on the cross-section $\mathcal{F}$ is simply
\be
\sigma_{\mathcal{F}}^{\rm tot} = \lp \lp \sigma_{\mathcal{F}}^{\rm th} \rp^2 + \lp \sigma^{\rm PDF}_{\mathcal{F}}
\rp^2\rp^{1/2} \, .
\ee
We believe that this prescription provides a conservative estimate of the combined MHOU on the predicted cross-section.

Note that when using a $\chi^2$ to assess the quality of the agreement between 
experimental data and the associated theory predictions for a PDF set which includes MHOUs, the MHOU must be always be included in the definition of the 
$\chi^2$ estimator, ideally (though not necessarily) by means of the 
theory covariance matrix.
This is because, as seen in Sect.~\ref{sec:globmhou}, the inclusion of 
MHOU modifies the best-fit central value, and thus if the MHOU were 
not included in the $\chi^2$, these PDFs would not provide the best fit, and 
the results might be misleading.
Because  the theory covariance matrix has been included in the fitting
(based on the argument of Sect.~\ref{sec:thcovmat}) as uncorrelated to the experimental covariance matrix, when assessing
fit quality
it should  be regarded as an 
additional systematic uncertainty, specific to the determination of PDFs 
from the data, to be added in quadrature to the usual experimental systematics.

\subsection{Delivery}
\label{sec:delivery}

The variants of the NNPDF3.1 NLO global sets presented in this work are publicly available
in the {\tt LHAPDF} format~\cite{Buckley:2014ana} from the NNPDF website:

\begin{center}
 \href{http://nnpdf.mi.infn.it/nnpdf3-1th/}{\tt http://nnpdf.mi.infn.it/nnpdf3-1th/}
\end{center}

\noindent
In the following, we list the PDF sets that are made available.
The NLO sets based on the theory covariance matrix are: 
\begin{center}
  \tt NNPDF31\_nlo\_as\_0118\_scalecov\_9pt \\
  \tt NNPDF31\_nlo\_as\_0118\_scalecov\_7pt \\
  \tt NNPDF31\_nlo\_as\_0118\_scalecov\_3pt
\end{center}
which correspond to the fits based on Eq.~(\ref{eq:chi2_v3})
in the cases in which the theory covariance matrix $S_{ij}$ has been evaluated with
the 9-, 7-, and 3-point prescriptions, respectively.

We have also constructed NLO PDF sets based on 
scale-varied theories, to be discussed in
Appendix~\ref{sec:fitsscalesvar} below. These are  determined
using Eq.~(\ref{eq:chi2_v2}), and they are
\begin{center}
  \tt NNPDF31\_nlo\_as\_0118\_kF\_1\_kR\_1 \\
  \tt NNPDF31\_nlo\_as\_0118\_kF\_2\_kR\_2 \\
  \tt NNPDF31\_nlo\_as\_0118\_kF\_0p5\_kR\_0p5 \\
  \tt NNPDF31\_nlo\_as\_0118\_kF\_2\_kR\_1 \\
  \tt NNPDF31\_nlo\_as\_0118\_kF\_1\_kR\_2 \\
  \tt NNPDF31\_nlo\_as\_0118\_kF\_0p5\_kR\_1 \\
  \tt NNPDF31\_nlo\_as\_0118\_kF\_1\_kR\_0p5 
\end{center}
where the naming convention indicates the values of the scale ratios $k_f$ and $k_r$.
Note that the {\tt NNPDF31\_nlo\_as\_0118\_kF\_1\_kR\_1} set is also the baseline (central scales and
experimental covariance matrix only) to be used in the comparisons with the fits based
on the theory covariance matrix listed above.
Finally, we also provide the set
\begin{center}
  \tt NNPDF31\_nnlo\_as\_0118\_kF\_1\_kR\_1
\end{center}
which corresponds to the NNLO fit with central scales and
experimental covariance matrix only, that has been produced for validation purposes.

It is  important to bear in mind
that the variants of the NNPDF3.1 fits presented in this work
are based on a somewhat different dataset to that used in the default 
NNPDF3.1 analysis.
Therefore, when using these sets it is important to be consistent:
for example by
comparing fits with and without MHOU that are based on a common input dataset.

In addition to the sets listed above, the other PDF sets presented in this paper, such as the DIS-only fits based on scale-varied calculations and on the theory covariance matrix, are available from the authors upon request.

\section{Summary and outlook}
\label{sec:summary}

In this work we have presented the first PDF determination that includes 
MHOU as part of the PDF uncertainty.
This is in principle required for consistency,
given that MHOU are routinely part of the theoretical predictions for hadron
collider processes, and likely to become a requirement for precision collider
phenomenology as other sources of uncertainties decrease.

The bulk of our work amounted to establishing a general
language and formalism for the inclusion of MHOU when multiple
processes  are considered at once in the global PDF fit, constructing 
prescriptions for
estimating these MHOU by means of scale variation, and for validating them
in cases in which the higher order corrections are known.
The formalism presented here is sufficiently flexible that it can also be
applied to different sources of theoretical uncertainty, such as nuclear
corrections or higher twists, and could also be used in
conjunction with alternative ways of estimating MHOU, such as for example 
the Cacciari-Houdeau method.

The validation studies presented here suggest however that the conventional 
scale variation method to estimate the MHOU works remarkably well.
Indeed, when coupled to the
theory covariance matrix formalism that we introduced, this method turns 
out to be free of
the instabilities that plague envelope techniques, and it leads to results
which appear to be reasonably stable and thus insensitive to the arbitrary
choices that are inherent to its implementation.
The reason for these properties is
essentially that, within a covariance matrix approach, possible
directions which do not correspond to actual MHO have no impact on the fitting.

Our results however also suggest that even more realistic estimates of
MHOU might be obtained through more complex patterns of scale variation 
than those considered here.
Specifically, a more refined treatment of factorization scale
variation is likely to be advantageous, in which independent variation
is performed for each eigenvalue of of the anomalous dimension
matrix. Also,  it might be advantageous to vary 
independently the renormalization scales in different partonic
sub-channels.
Indeed, we have observed from the validation of our
estimate of MHOU, while always reasonably successful for the datasets
considered here, deteriorates as the size of the dataset increases,
which suggests that more complex structures  might be required.
Here we have performed a first investigation, and the
exploration of these more complex patterns of scale variation will be 
left for future work.

On the phenomenological side, our results show that at least at NLO the
main effect of the inclusion of MHOU in PDF determination is to
improve the accuracy of the result, while not significantly reducing
its precision.
Indeed, whenever experimental information is abundant,
in particular for a global dataset, we have found that the
total PDF uncertainty is only moderately affected by the inclusion 
of MHOU --- in fact, for the datapoints included in PDF determination 
it even decreases --- but the central value moves closer to the true
result.
Moreover, the fit quality improves, thereby showing that the main effect 
of the inclusion of MHOU is in reducing tensions between datasets due to 
imperfections in their theoretical description.

The most interesting future phenomenological development will be of
course the extension of our methodology to the determination of MHOU
in a state-of-the-art global NNLO PDF set.
It will be interesting to
assess to what extent the behaviour observed at NLO persists there.
More generally, the inclusion of MHOU at NNLO is expected to lead to 
the most precise and accurate PDF sets that can be determined with 
currently available theoretical and experimental information.

\subsection*{Acknowledgments}
We are grateful to V.~Bertone and N.~P.~Hartland for collaboration
in the early stages of this work.
We would also like to acknowledge useful discussions with
J.~Bendavid, M.~Bonvini, F.~Caola, M.~Duehrssen, L.~Harland-Lang, P.F.~Monni, G.~Salam,
and R.~Thorne on the topic of theory uncertainties
and PDFs.

R.D.B. is supported by the UK
Science and Technology Facility Council through grant
ST/P000630/1.
S.F. is supported by the European
Research Council under the European Union's Horizon
2020 research and innovation Programme (grant agreement n.740006).
T.G. is supported by The Scottish
Funding Council, grant H14027.
Z.K. is supported
by the European Research Council Consolidator Grant
``NNLOforLHC2''.
E.R.N. is supported by the European Commission through the Marie Sklodowska-Curie
Action ParDHonS FFs.TMDs (grant number 752748).
R.L.P. and M.W. are supported by the STFC grant
ST/R504737/1.
J.R. is supported by the European
Research Council Starting Grant ``PDF4BSM'' and by
the Netherlands Organization for Scientific Research
(NWO).
L.R. is supported by the European Research
Council Starting Grant ``REINVENT'' (grant number
714788).
M.U. is partially supported by the STFC grant
ST/L000385/1 and funded by the Royal Society grants
DH150088 and RGF/EA/180148.
C.V. is supported by
the STFC grant ST/R504671/1.

\appendix
\section{Diagonalisation of the theory covariance matrix}
\label{sec:diagonalization}

To carry out the validation described in 
Sect.~\ref{sec:validconstruction} and thus
compute the angles $\theta$ defined Eq.~(\ref{eq:theta}), 
we must first diagonalise $\widehat{S}_{ij}$ for the various
prescriptions.
In  this appendix we provide  details concerning
this diagonalization process.

The diagonalisation
of the theory covariance matrix $\widehat{S}_{ij}$
is difficult due to the very large number of zero eigenvalues.
To get around this problem we first project $\widehat{S}_{ij}$ onto $S$, and then perform the diagonalization in this subspace (in which all eigenvalues are positive, by construction).
The projection is easily achieved, since $S$ is spanned by the vectors $\{\Delta_i(\kappa_f, \kappa_{r_a}): \kappa_f, \kappa_{r_a} \in V_m\}$ defined in Eq.~(\ref{P.1}), used to construct $S_{ij}$ in Eq.~(\ref{P.2}).
Similarly, $\widehat{S}_{ij}$ is constructed from normalized vectors
$\{\widehat{\Delta}_i(\kappa_f, \kappa_{r_a}): \kappa_f, \kappa_{r_a} \in V_m\}$,
where $\widehat{\Delta}_i = \Delta_i/T^{\rm NLO}_i$.
However these vectors
are not all linearly independent, and a linearly independent set is 
best constructed on a case by case basis. This construction also gives us $N_{\rm sub}$, the dimension of $S$, for each of the prescriptions. 

\begin{itemize}

\item \textbf{5-point}: when there are $p$ processes, $V_4$ has $2+2^p$ distinct elements \\$\{(\pm;0,0,0,\ldots), (0;\pm,\pm,\pm,\ldots)\}$, so there are $2+2^p$ different vectors, $\widehat{\Delta}_i^{\pm 0}$ and $\widehat{\Delta}_i^{0\pm}$: if $i_n\in \pi_n$,with $n=1,\ldots,p$ labeling the different processes, then
\begin{equation}
\widehat{\Delta}_i^{+0}\equiv \left( 
\widehat{\Delta}_{i_1}^{+0},
\widehat{\Delta}_{i_2}^{+0},
\ldots,\widehat{\Delta}_{i_p}^{+0}
\right),
\end{equation}
and similarly for $\widehat{\Delta}_i^{-0}$, while
\begin{equation}
\widehat{\Delta}_i^{0\pm}\equiv \left( 
\widehat{\Delta}_{i_1}^{0\pm},
\widehat{\Delta}_{i_2}^{0\pm},
\ldots,\widehat{\Delta}_{i_p}^{0\pm}
\right),
\end{equation}
where for each process the renormalization scale is varied independently.
Not all of the second class of vectors, $\widehat{\Delta}_i^{0\pm}$, are linearly independent: for example, when $p=2$, there exists the linear relation
\begin{equation}
\left(\widehat{\Delta}_{i_1}^{0+},\widehat{\Delta}_{i_2}^{0+}\right)-\left(\widehat{\Delta}_{i_1}^{0-},\widehat{\Delta}_{i_2}^{0+}\right) = \left(\widehat{\Delta}_{i_1}^{0+},\widehat{\Delta}_{i_2}^{0-}\right)-\left(\widehat{\Delta}_{i_1}^{0-},\widehat{\Delta}_{i_2}^{0-}\right),
\end{equation}
so there are five rather than six linearly independent vectors. The number of linear relations for general $p$ can be deduced inductively: if we have in total $n_p$ linearly independent vectors $v_i^a$ for $p$ processes, where $a=1,\ldots,p$, $i=(i_1,\ldots,i_p)$, then when there are $p+1$ processes, we have $2n_p$ distinct vectors $\left(v_i^a,\widehat{\Delta}_{i_{p+1}}^{0\pm}\right)$, but with linear relations
 \begin{equation}
\left(v_i^a,\widehat{\Delta}_{i_{p+1}}^{0+}\right)-\left(v_i^{a+1},\widehat{\Delta}_{i_{p+1}}^{0+}\right) = \left(v_i^a,\widehat{\Delta}_{i_{p+1}}^{0-}\right)-\left(v_i^{a+1},\widehat{\Delta}_{i_{p+1}}^{0-}\right).
\end{equation}
There are $n_p(n_p-1)$ of these relations, but of these only $n_p-1$ are linearly independent. So $n_{p+1} = 2n_p -(n_p-1)=n_p+1$. For $p=2$, $n_p=5$, so in general we must have $n_p = p+3$, i.e. the dimension of the subspace $S$ is $N_{\rm sub} = p+3$ for the 5-point prescription.

\item \textbf{$\overline{5}$-point}: when there are $p$ processes, $\overline{V}_4$ has $2^{p+1}$ distinct elements $\{(\pm;\pm,\pm,\pm,\ldots)\}$, so there are $2^{p+1}$ different vectors $\widehat{\Delta}_i^{+\pm}$ and $\widehat{\Delta}_i^{-\pm}$, where   
\begin{equation}
\widehat{\Delta}_i^{+\pm}\equiv \left( 
\widehat{\Delta}_{i_1}^{+\pm},
\widehat{\Delta}_{i_2}^{+\pm},
\ldots,\widehat{\Delta}_{i_p}^{+\pm}\right),
\end{equation}
and similarly for $\widehat{\Delta}_i^{-\pm}$: the scale variation of $\ln k_f$ is fully correlated across all processes, but all the renormalization scales for the processes $\pi_r$ are varied independently from it and the others. Again not all of these $2^{p+1}$ vectors are linearly independent, but it can be shown using a similar inductive argument as for 5-point that the number of independent vectors is $N_{\rm sub} = 2p+2$.

\item \textbf{9-point}: when there are $p$ processes, $V_8$ has $2^p+2\cdot3^p$ distinct elements\\ $V_8= \{(0;\pm,\pm,\pm,\ldots), (\pm;\pmz,\pmz,\pmz,\ldots)\}$, so the corresponding vectors are $\widehat{\Delta}_i^{0\pm}$, $\widehat{\Delta}_i^{+\pm}$, $\widehat{\Delta}_i^{+0}$, $\widehat{\Delta}_i^{-\pm}$ and $\widehat{\Delta}_i^{-0}$ where
\begin{equation}
\widehat{\Delta}_i^{0\pm}\equiv \left(
\widehat{\Delta}_{i_1}^{0\pm},
\widehat{\Delta}_{i_2}^{0\pm},
\ldots,\widehat{\Delta}_{i_p}^{0\pm}\right),
\end{equation}
while
\begin{equation}\begin{split}
\widehat{\Delta}_i^{+\pm}&\equiv \left( 
\widehat{\Delta}_{i_1}^{+\pm},
\widehat{\Delta}_{i_2}^{+\pm},
\ldots,\widehat{\Delta}_{i_p}^{+\pm}
\right),\\
\widehat{\Delta}_i^{+0}&\equiv \left( 
\widehat{\Delta}_{i_1}^{+0},
\widehat{\Delta}_{i_2}^{+0},
\ldots,\widehat{\Delta}_{i_p}^{+0}
\right),
\end{split}
\end{equation}
and similarly for $\widehat{\Delta}_i^{-\pm}$ and $\widehat{\Delta}_i^{-0}$. Again there are many linear relations between these vectors: if for $p$ processes there are $n^0_p$ independent vectors of class $\widehat{\Delta}_i^{0\pm}$, and $n^\pm_p$ independent vectors of classes $\widehat{\Delta}_i^{+\pm}$, $\widehat{\Delta}_i^{+0}$ and $\widehat{\Delta}_i^{-\pm}$, $\widehat{\Delta}_i^{-0}$ then while for $p+1$ processes $n^0_{p+1} = 2n^0_p-(n_p^0-1)$ (i.e. $n_p^0-1$ linearly independent linear relations), $n^\pm_{p+1} = 3n^\pm_p-2(n_p^\pm-1)$ (i.e. $2(n_p^\pm-1)$ linearly independent linear relations). So we now find $n^0_p= p+1$, $n^\pm_p= 2p+1$, and $N_{\rm sub} = n^0_p+n_p^++n_p^- = 5p+3$.  

\item \textbf{3-point}: when there are $p$ processes, $V_2$ has $2^p$ distinct elements $V_2= \{(\pm,\pm,\pm,\ldots)\}$, so the independent vectors are $\widehat{\Delta}_i^{++}$ and $\widehat{\Delta}_i^{--}$, where
\begin{equation}
\widehat{\Delta}_i^{++}\equiv \left(
\widehat{\Delta}_{i_1}^{++},
\widehat{\Delta}_{i_2}^{++},
\ldots,\widehat{\Delta}_{i_p}^{++}\right),
\end{equation}
and similarly for $\widehat{\Delta}_i^{--}$. The number of linearly independent vectors is $N_{\rm sub} = p+1$.

\item \textbf{7-point}: since $V_6 = V_4\oplus V_2$, the $2+2^{p+1}$ independent vectors are simply those for 5-point and those for 3-point together, i.e. $\widehat{\Delta}_i^{\pm 0}$, $\widehat{\Delta}_i^{0\pm}$, $\widehat{\Delta}_i^{++}$ and $\widehat{\Delta}_i^{--}$. The number of linearly independent vectors is thus $N_{\rm sub} = 2p+4$.

\end{itemize}

Once we have a set of linearly independent vectors spanning the space, 
we can use them to construct an orthonormal basis $v_i^a$, such that $\sum_i v_i^a v_i^b = \delta^{ab}$.
Then the projection of $\widehat{S}_{ij}$ into the subspace $S$ will be given by 
\begin{equation}   
\widehat{S}^{ab} = \sum_{i,j} v_i^a v_j^b \widehat{S}_{ij},
\end{equation} 
and the diagonalization of $\widehat{S}^{ab}$ gives the positive eigenvalues $\lambda_\alpha=(s^\alpha)^2$. The eigenvectors $e_i^\alpha$ can then be constructed from the basis vectors $v_i^{a}$: if $e_a^\alpha$ is the eigenvector of $s^{ab}$ corresponding to eigenvalue $\lambda_\alpha$, then $e_i^\alpha = \sum_a v_i^a e_a^\alpha$.

\section{PDF sets with different scale choices}
\label{sec:fitsscalesvar}

The approach that we have pursued in this work for the determination of
MHOUs in PDFs is based on the idea of utilising scale
variation of the theory prediction to produce an estimate of
the MHOU, and then using this information to construct a
theory covariance matrix to be used in PDF fitting.
Results from this approach have been presented in
Sect.~\ref{sec:fitstherr}.
An alternative, and perhaps more naive, option would be that of simply
performing PDF fits in which different choices are made for the
factorization and renormalization scales used in the fit.
One may then take the envelope of the resulting fits, for some set of scale
choices, as an estimate of the MHOU.

In this appendix, we will construct PDF sets based on varying the
renormalization and factorization scales in the PDF fit.
These PDFs are obtained from the
minimization of the usual figure of merit
\be
\label{eq:chi2_v2}
\chi^{2(s)}=\frac{1}{N_{\rm dat}}\sum_{i,j=1}^{N_{\rm dat}}(D_i-T_i^{(s)})
(C_0^{-1})_{ij} (D_j-T_j^{(s)}),
\ee
where  $T_i^{(s)}=T_i ( \kappa_f^{(s)},\kappa_r^{(s)})$, $s$ labels the
scale choices used for the determination of each PDF set, and $C_0$ is the experimental covariance matrix evaluated using the usual $t_0$-prescription.
We will then
study the resulting PDFs.

As we shall see, whereas this approach
provides an independent way of assessing the dependence of PDFs on scale
choice, it does not provide a stable way of estimating
MHOUs. Although these PDF sets do not appear to be
advantageous for MHOU, we present them here because they are nevertheless interesting for their own sake. This is especially true in view of the fact that PDF
sets based on systematic scale variation of the underlying theory have
never been presented before.

Based on the experimental and theoretical settings described
in  Sect.~\ref{sec:inputdata}, we have produced a number of PDF sets with
different choices for $k_r$ and $k_f$, input dataset, and perturbative
order, which are summarized in
Table~\ref{tab:scalevarFits}.
The PDF sets corresponding to the central
scale choices are the same as discussed in Sect.~\ref{sec:fitstherr}.
In the same way as in Sect.~\ref{sec:fitstherr}, 
we determine PDFs at NLO both  from a
DIS-only dataset and a global dataset, with NNLO PDFs determined with
central scale choices as a reference.

In all of these PDF determinations, the factorization and
renormalization scale are varied in a fully correlated way between all
datasets.
So for example, if $k_r=2$, then the renormalization scale is taken
to be twice its default value for all processes. This immediately exposes 
a defect in this method: in principle, the MHOU in the hard cross-sections of 
different processes are uncorrelated. However, uncorrelated variations of
$k_r$ across the five processes that we consider would require $3^6$ fits of
each type, or ${\mathcal O}(70,000)$ replicas, which is of course impractical.

\begin{table}[t]
  \centering
\footnotesize
  \renewcommand*{\arraystretch}{1.25}
  \begin{tabular}{lcccc}
    Label                   & $\quad$Dataset$\quad$  & $\quad$Order$\quad$  &
    $k_f=\mu_r/Q$
    & $k_r=\mu_f/Q$  \\
 \toprule   
     {\tt NNPDF31\_dis\_nlo\_as\_0118\_kF\_1\_kR\_1}  & DIS  & NLO  &  1  &  1  \\
     {\tt NNPDF31\_dis\_nlo\_as\_0118\_kF\_2\_kR\_2}   & DIS  & NLO  &  2  &  2  \\
     {\tt NNPDF31\_dis\_nlo\_as\_0118\_kF\_0p5\_kR\_0p5}   & DIS  & NLO  &  $\half$  &  $\half$  \\
     {\tt NNPDF31\_dis\_nlo\_as\_0118\_kF\_2\_kR\_1}    & DIS  & NLO  &  2  &  1  \\
     {\tt NNPDF31\_dis\_nlo\_as\_0118\_kF\_1\_kR\_2}  & DIS  & NLO  &  1  &  2  \\
     {\tt NNPDF31\_dis\_nlo\_as\_0118\_kF\_0p5\_kR\_1}  & DIS  & NLO  &  $\half$  &  1  \\
     {\tt NNPDF31\_dis\_nlo\_as\_0118\_kF\_1\_kR\_0p5}   & DIS  & NLO  &  1  &  $\half$ \\
     {\tt NNPDF31\_dis\_nlo\_as\_0118\_kF\_2\_kR\_0p5}   & DIS  & NLO  &  2  &  $\half$  \\
     {\tt NNPDF31\_dis\_nlo\_as\_0118\_kF\_0p5\_kR\_2}  & DIS  & NLO  &  $\half$  &  2  \\
     \midrule
    {\tt NNPDF31\_nlo\_as\_0118\_kF\_1\_kR\_1}  & Global  & NLO  &  1  &  1  \\
    {\tt NNPDF31\_nlo\_as\_0118\_kF\_2\_kR\_2}      & Global  & NLO  &  2  &  2  \\
     {\tt NNPDF31\_nlo\_as\_0118\_kF\_0p5\_kR\_0p5}   & Global  & NLO  &  $\half$  &  $\half$  \\
     {\tt NNPDF31\_nlo\_as\_0118\_kF\_2\_kR\_1}    & Global  & NLO  &  2  &  1  \\
     {\tt NNPDF31\_nlo\_as\_0118\_kF\_1\_kR\_2}    & Global  & NLO  &  1  &  2  \\
     {\tt NNPDF31\_nlo\_as\_0118\_kF\_0p5\_kR\_1}    & Global  & NLO  &  $\half$  &  1  \\
     {\tt NNPDF31\_nlo\_as\_0118\_kF\_1\_kR\_0p5}      & Global  & NLO  &  1  &  $\half$ \\
     {\tt NNPDF31\_nlo\_as\_0118\_kF\_2\_kR\_0p5}      & Global  & NLO  &  2  &  $\half$  \\
     {\tt NNPDF31\_nlo\_as\_0118\_kF\_0p5\_kR\_2}     & Global  & NLO  &  $\half$  &  2 \\
         \midrule
     {\tt NNPDF31\_dis\_nnlo\_as\_0118\_kF\_1\_kR\_1}  & DIS  & NNLO  &  1  &  1  \\
     \midrule
      {\tt NNPDF31\_nnlo\_as\_0118\_kF\_1\_kR\_1}     & Global  & NNLO  &  1  &  1 \\
       \bottomrule
  \end{tabular}
  \vspace{0.3cm}
  \caption{\small List of PDF sets  with different choices of the
    renormalization $\mu_f$ and factorization $\mu_f$.
    For each set, we indicate its label, the input dataset,
    the perturbative order, and the ratios $k_r=\mu_r/Q$
    and $k_f=\mu_f/Q$ to the central scale $Q$.
    \label{tab:scalevarFits}
  }
  \end{table}


\subsection{DIS-only PDFs}
\begin{table}[t]
\begin{center}
\renewcommand*{\arraystretch}{1.7}
\footnotesize
\begin{tabular}{|l|c|ccccccccc|c|}
  \toprule
Dataset   & $N_{\rm dat}$ & \multicolumn{9}{c|}{$\chi^2/N_{\rm dat}$ NNPDF3.1 DIS-only NLO} & NNLO  \\
&  & (1,1) & (2,2)  &  (\half,\half)  &  (2,1) & (1,2) &  (1,\half) & (\half,1) & $(2,\half)$ & $(\half,2)$  & (1,1) \\
\toprule
NMC  & 134         & 1.259   &  1.274  & 1.272  & 1.263   & 1.268  &
1.247   & 1.286  & 1.277  &  1.301  & 1.244  \\
SLAC  & 12         & 0.908   &  1.144  & 0.941  & 0.881   & 0.922  &
0.709   & 0.7651  & 0.668  & 0.745  & 0.794      \\
BCDMS  &  530      & 1.046   &  1.047  & 1.057  & 1.046   & 1.040  &
1.050   & 1.059  & 1.093  &  1.053  & 1.046    \\
CHORUS  &  430     & 0.982   &  1.024  & 1.069  & 1.031   & 1.018  &
1.030   & 1.024  & 1.055  &  1.038   & 1.093    \\
NuTeV  &   41      & 0.628   &  0.564  & 0.712  & 0.711   & 0.642  &
0.693   & 0.634  & 0.725  &  0.736  & 0.892     \\
HERA incl  & 967   & 1.097   &  1.126  & 1.136  & 1.167   & 1.091  & 1.152   & 1.131  & 1.357  &  1.122 &  1.103    \\
HERA $F_2^c$  & 31 & 1.047   &  0.983  & 1.153  & 1.058   & 1.012  & 1.257   & 2.122  & 1.868  &  2.137  &1.055    \\  
\midrule
Total  &  2145     & 1.061  & 1.083    & 1.103  & 1.104   & 1.064  &
1.098   & 1.104  &1.218  &   1.104 & 1.089      \\
\bottomrule
\end{tabular}
\end{center}

  \caption{\small The values of $\chi^2/N_{\rm dat}$ for the DIS-only
     PDF sets
     based on scale-varied theories.
    We display the values of the $\chi^2/N_{\rm dat}$ for the nine combination
    of scale variations, $(k_f,k_r)$,
    listed in Table~\ref{tab:scalevarFits}.
    For each dataset, we also indicate the number of data points after cuts,
    see also Table~\ref{tab:datasets_process_categorisation}.
  \label{tab:chi2table_dis_nlo}}
\end{table}
    
In Table~\ref{tab:chi2table_dis_nlo} we collect the
values of $\chi^2/N_{\rm dat}$ for the PDFs determined
from a DIS-only dataset with various choices of renormalization and
factorization scale.
We note that
the central scale choice
leads to the lowest value of $\chi^2$.
The scale choice  $(2,\half)$
leads to a much larger $\chi^2$ than any other choice. This choice,
which involves a large scale ratio, is typically omitted when
estimating scale uncertainties 
from an envelope prescription. Note however that the reciprocal choice
$(\half,2)$, which is also usually discarded for the same reason,
leads to a $\chi^2$ which is not particularly large.
 Variation of  $\chi^2$ values
with the scales is more marked for  HERA experiments than
for fixed target, consistent with the observation that scale variation is larger in the small $x$ region covered by the HERA data.

\begin{figure}[t]
  \begin{center}
  \includegraphics[scale=0.37]{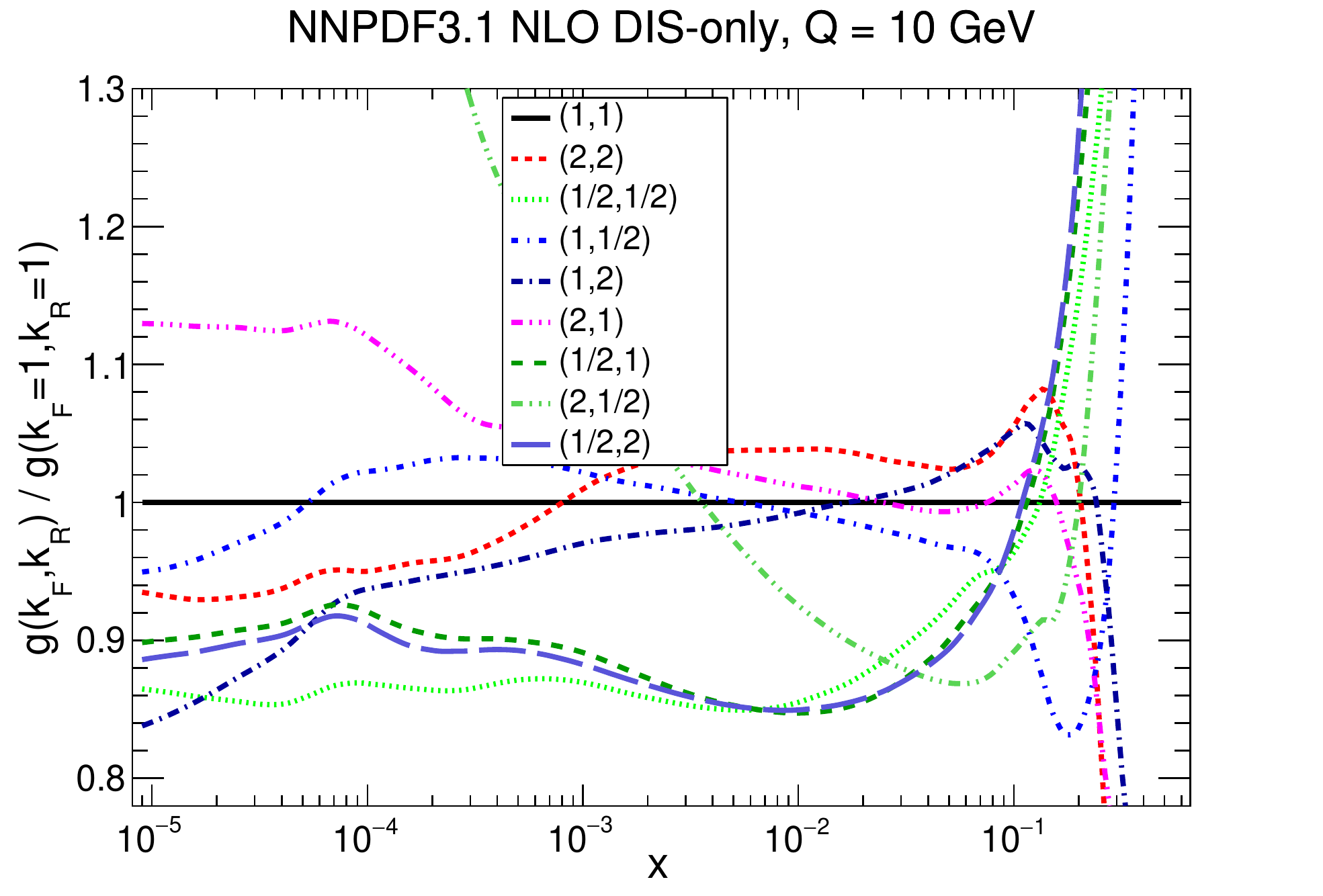}
    \includegraphics[scale=0.37]{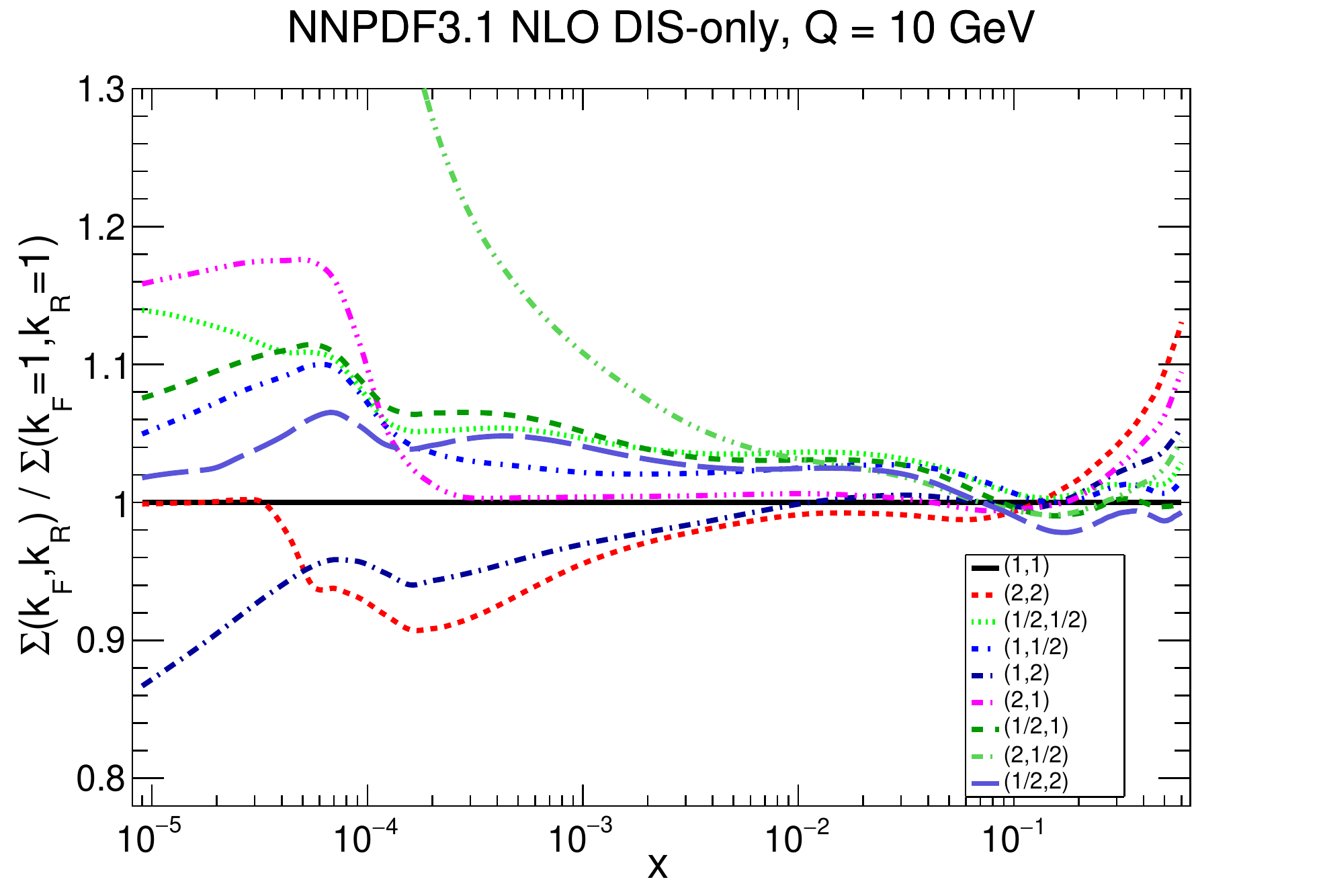}
 \includegraphics[scale=0.37]{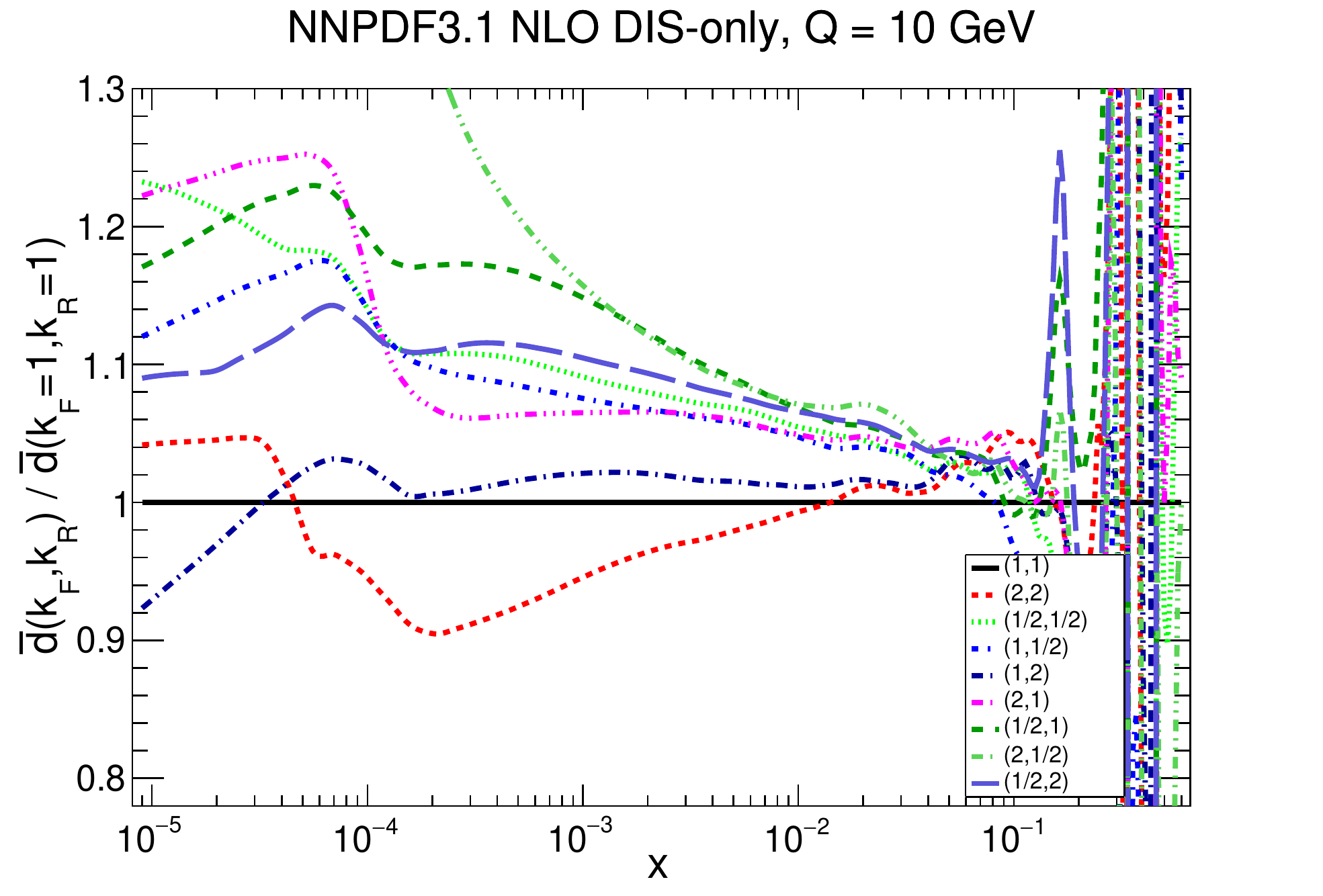}
    \caption{\small Comparison of the central values of the  DIS-only
      PDF with different values of $(k_f,k_r)$. All results are
      normalized to the baseline $(k_f,k_r)=(1,1)$. 
      The gluon,
      the total quark singlet and  down antiquark (top to bottom)  are
      shown at  $Q=10$~GeV.
    \label{fig:NLODIS-scalevars} }
  \end{center}
\end{figure}

We next  assess the impact of scale variation on the PDFs. The
impact on PDF uncertainties turns out to be moderate, and thus we concentrate
on central values.
In Fig.~\ref{fig:NLODIS-scalevars} we compare the central values of
the DIS-only
PDFs obtained  with the values of $(k_f,k_r)$ listed in
Table~\ref{tab:scalevarFits}, normalized to
the $(k_f,k_r)=(1,1)$ baseline. The gluon, total quark singlet and
down antiquark are shown at $Q=10$~GeV. The behaviour for other quark flavors is similar.

Scale variation for the gluon turns out to be reasonably asymmetric, with
all scale choices leading to a gluon which is below the central
scale choice for $x\lsim 10^{-2}$.
The scale choice
$(k_f,k_r)=(2,\half)$ which, as already noted, leads to a much
worse $\chi^2$ value appears to lead to a rather unstable PDF.
In the singlet case the spread of scale variations about the central
choice is more symmetric. Whereas for the
gluon the spread is considerable for all $x$ values, for the quark
singlet the spread becomes quite small at large 
$x\gsim 0.1$. The behavior of the down antiquark is similar to that
of the singlet.

In general, the sensitivity of results to scale variation
appears to be directly linked to the data-driven PDF uncertainty,
with a much wider spread observed whenever  the information coming
from data is reduced and PDF uncertainties are large.

\subsection{Global PDFs}\label{sect:globalNLOfits}

\begin{table}[t]
\begin{center}
\renewcommand*{\arraystretch}{1.7}
\scriptsize
\begin{tabular}{|l|c|ccccccccc|c|}
  \toprule
  Process  
& $N_{\rm dat}$ 
& \multicolumn{9}{c|}{$\chi^2/N_{\rm dat}$ NNPDF3.1 global NLO}  
& NNLO 
 \\
&  
& (1,1) & (2,2) & $(\half,\half)$ & (2,1) & (1,2) & $(1,\half)$ & $(\half,1)$ & $(2,\half)$ & $(\half,2)$   
& (1,1) 
 \\
 \toprule
DIS NC & 1593 & 1.088 & 1.182 & 1.209 & 1.191 & 1.103 & 1.144 & 1.188 & 1.394 & 1.200 & 1.084 \\
DIS CC &  552 & 1.012 & 1.014 & 1.045 & 1.018 & 1.020 & 1.042 & 1.089 & 1.065 & 1.079 & 1.079 \\
\midrule
DY     &  484 & 1.486 & 1.500 & 1.437 & 1.439 & 1.461 & 1.347 & 1.441 & 1.772 & 1.664 & 1.231 \\
JETS   &  164 & 0.907 & 0.875 & 0.947 & 0.911 & 0.874 & 0.914 & 0.938 & 1.023 & 0.945 & 0.950 \\
TOP    &   26 & 1.260 & 2.542 & 1.390 & 1.143 & 2.352 & 1.277 & 1.121 & 1.493 & 1.756 & 1.068 \\
\midrule
Total  & 2819 & 1.139 & 1.200 & 1.256 & 1.214 & 1.153 & 1.190 & 1.240 & 1.405 & 1.253 & 1.105  \\
\bottomrule
\end{tabular}
\end{center}

  \caption{\small Same as Table~\ref{tab:chi2table_dis_nlo} for global PDFs.
  \label{tab:chi2table_global_nlo}}
\end{table}

We now turn to  PDF fits based on the global dataset.
As we will show, the use of a global dataset reduces not only the PDF 
uncertainties
but also the relative impact of varying the scales, compared to the
DIS-only fits. This is consistent with previous results~\cite{nnpdflh}
showing that the perturbative stability of PDFs improves as the size
of the dataset used for their determination grows.

In Table~\ref{tab:chi2table_dis_nlo} we collect the
values of $\chi^2/N_{\rm dat}$ for the NNPDF3.1 NLO global PDF sets
for all scale choices, with the NNLO value for the central scale
choice also shown.
As in the DIS-only case, the best fit is found for the central scale
choice, with all others leading to a worse
$\chi^2$ value, and the scale choice  $(2,\half)$ leading to much
worse fit quality.
All scale choices at NLO give a worse fit than the NNLO fit, due to 
the fact that NNLO corrections are needed for a good description of several
high-precision LHC data~\cite{Ball:2017nwa}.

\begin{figure}[t]
  \begin{center}
 \includegraphics[scale=0.37]{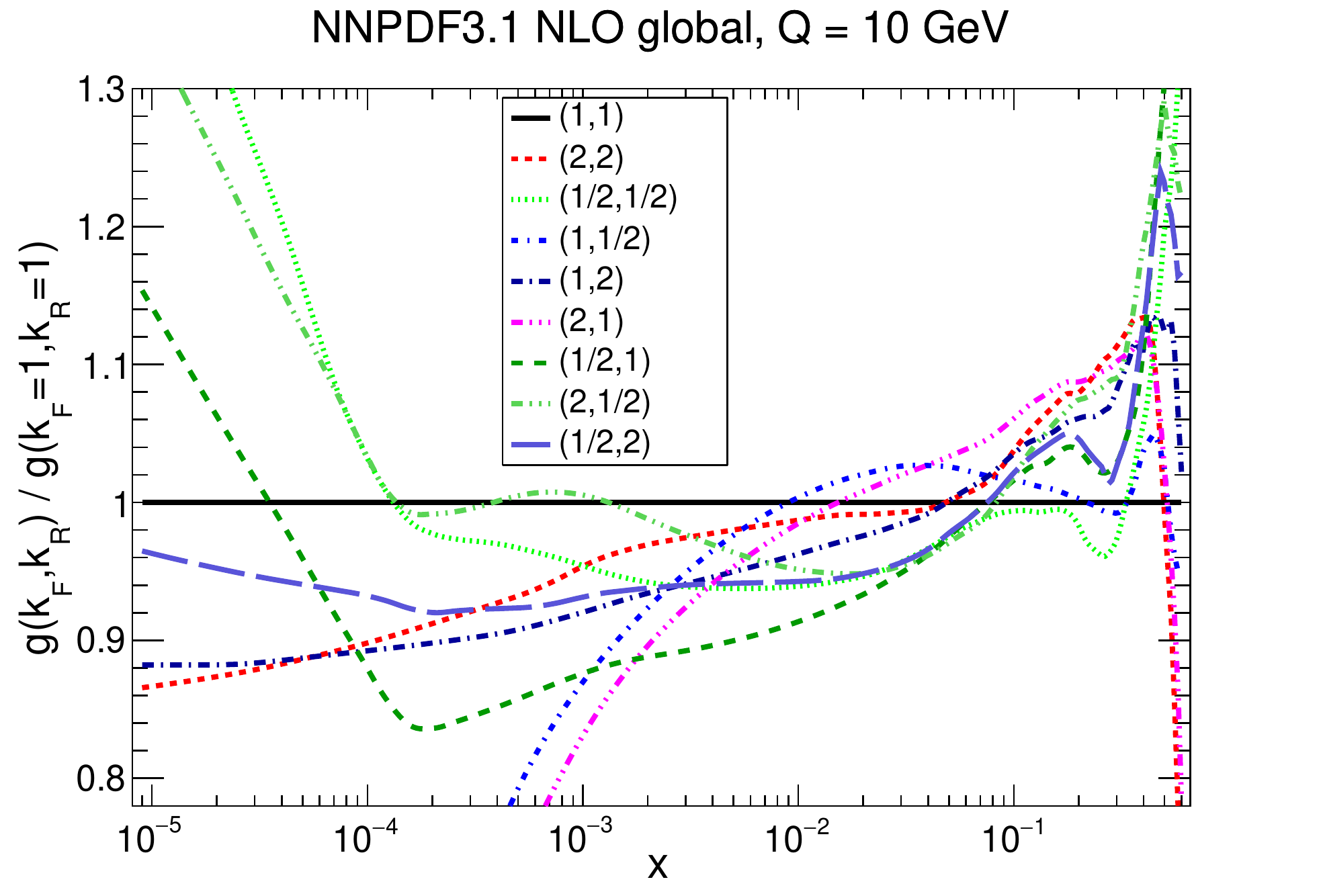}
 \includegraphics[scale=0.37]{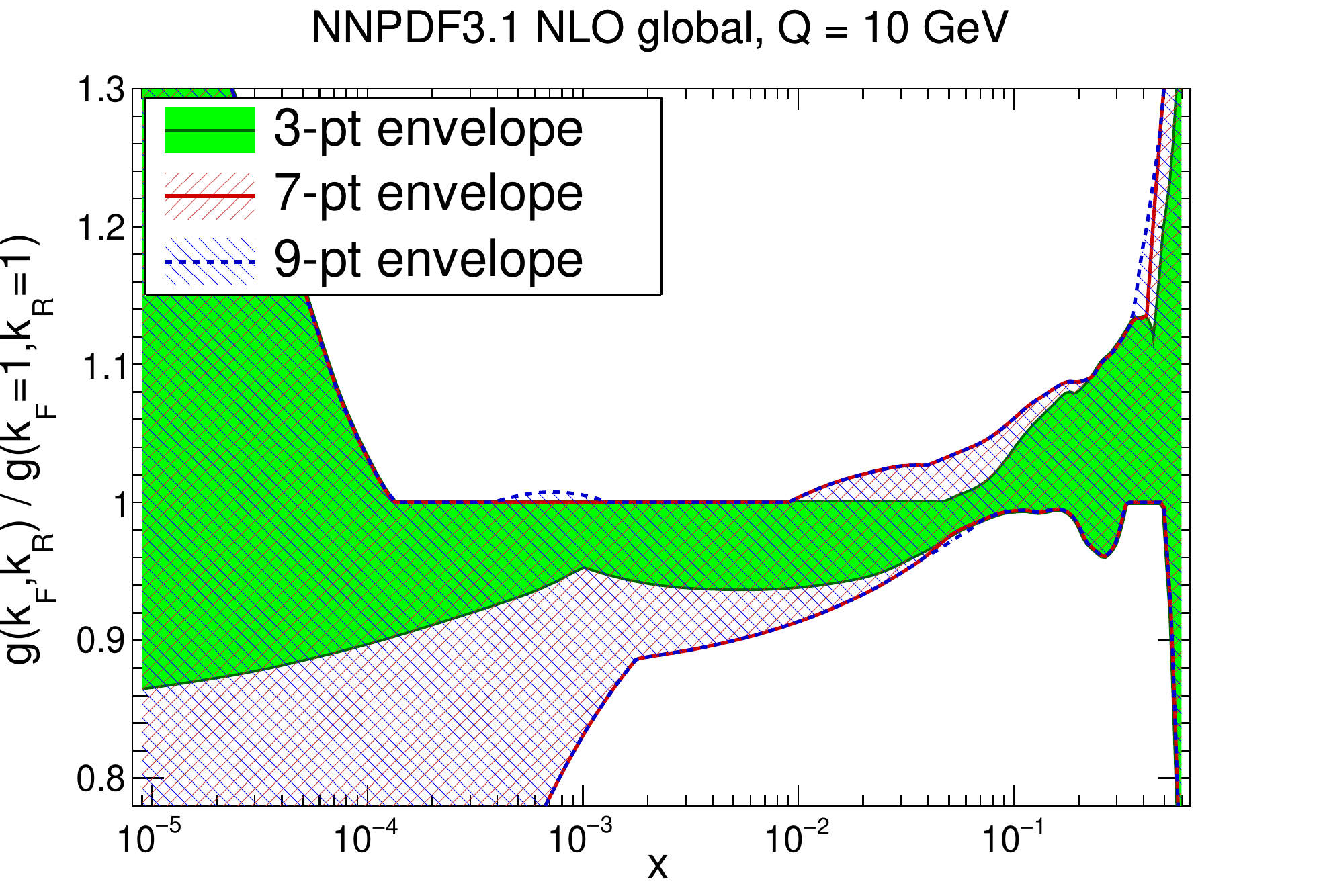}
 \includegraphics[scale=0.37]{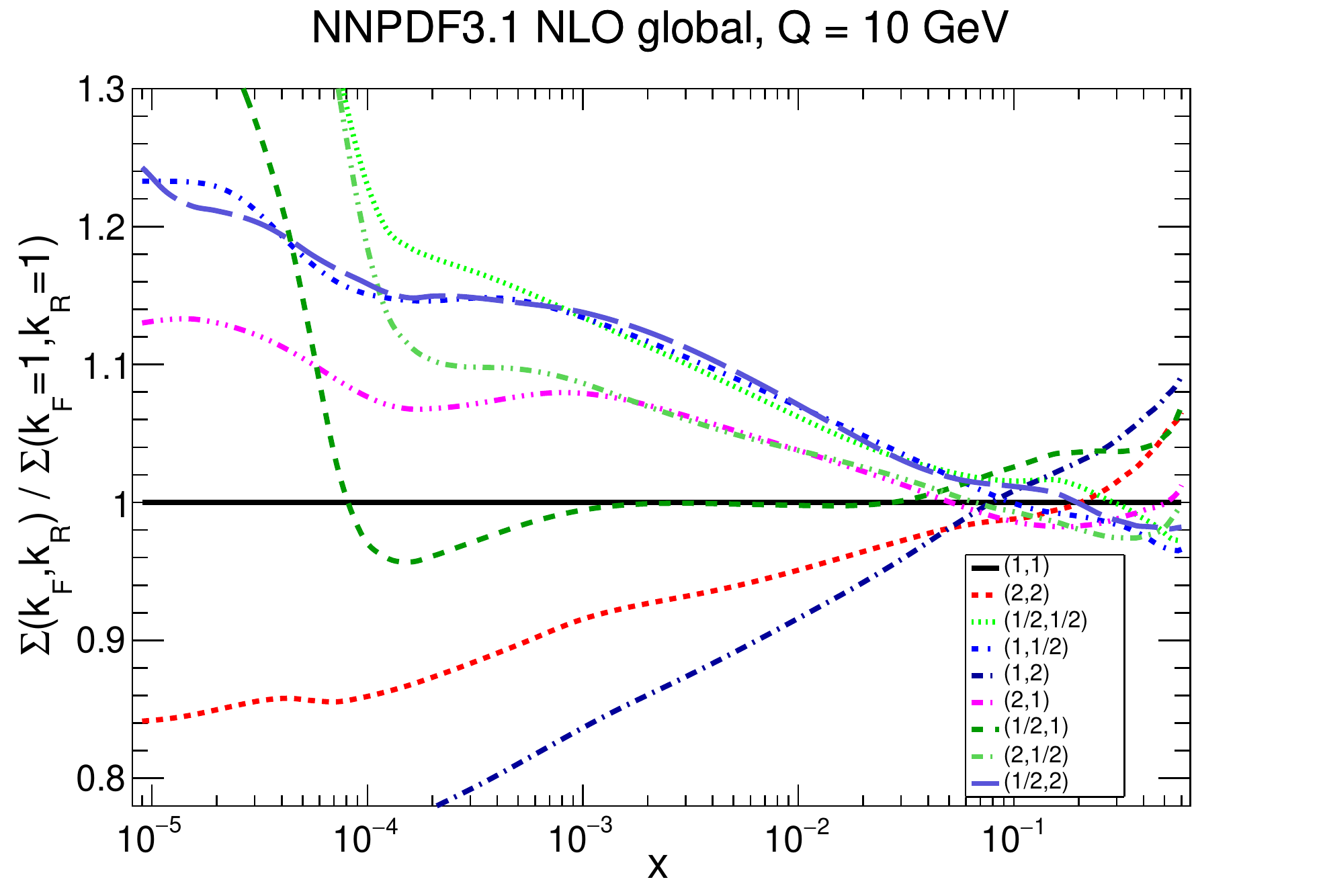}
 \includegraphics[scale=0.37]{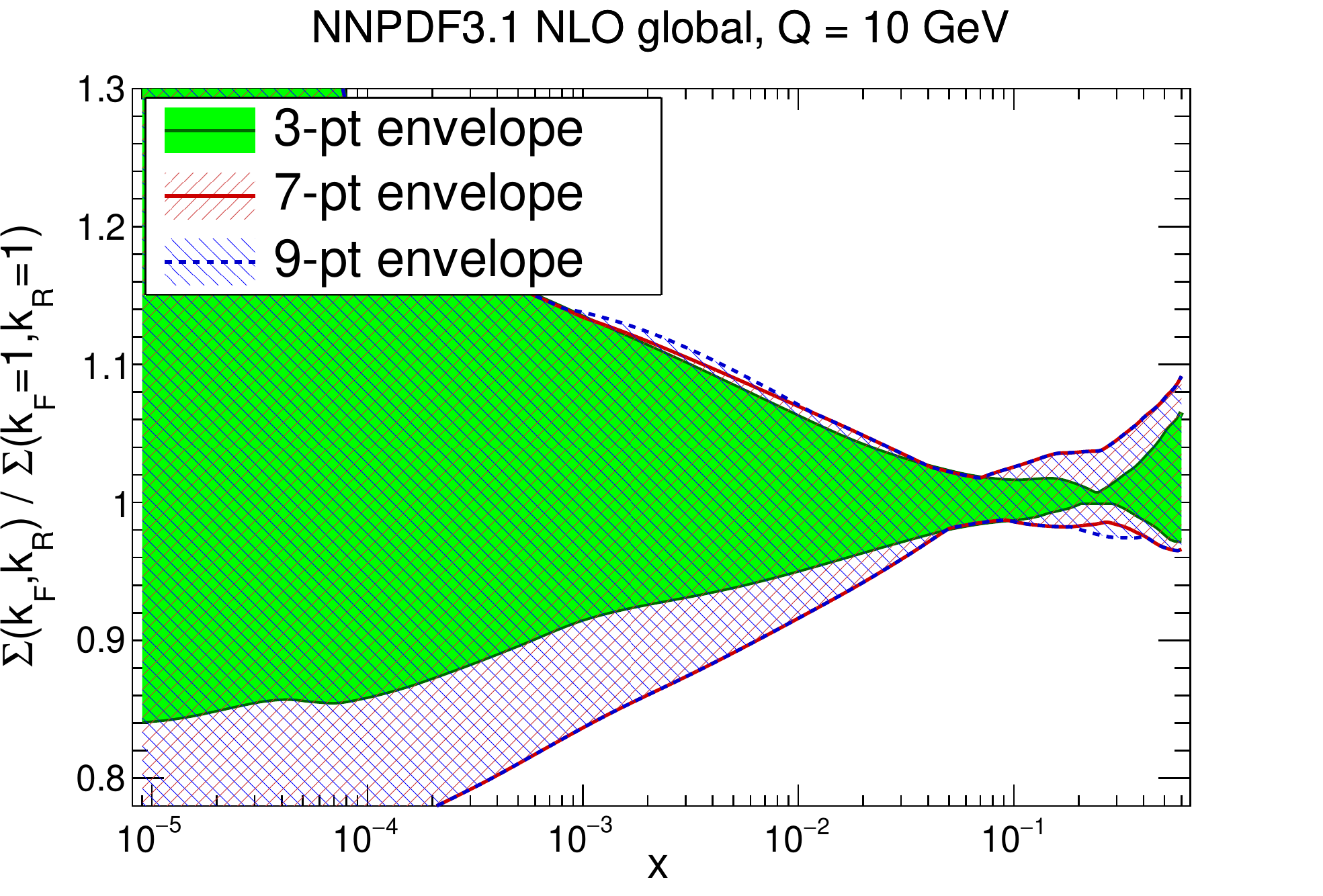}
 \includegraphics[scale=0.37]{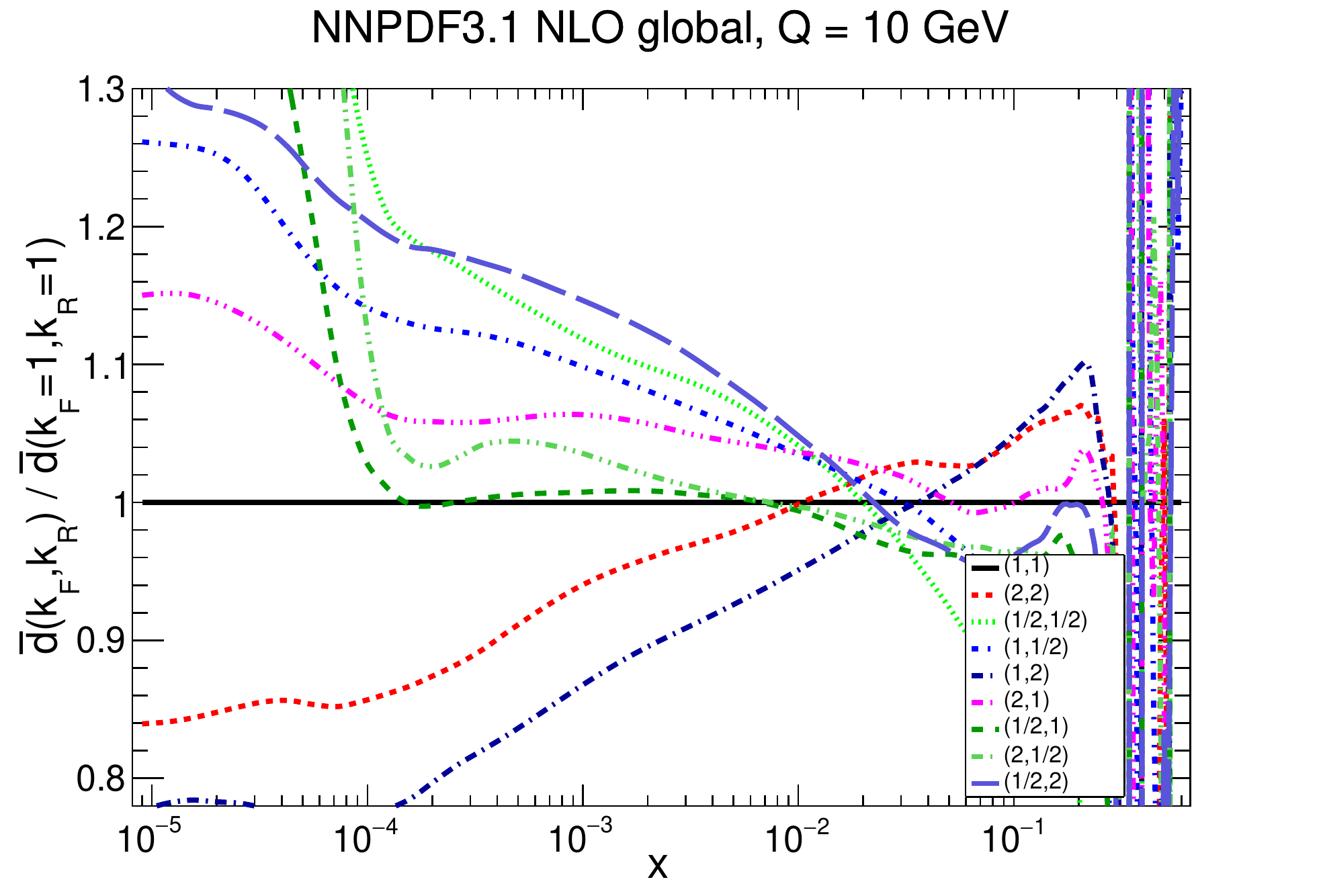}
 \includegraphics[scale=0.37]{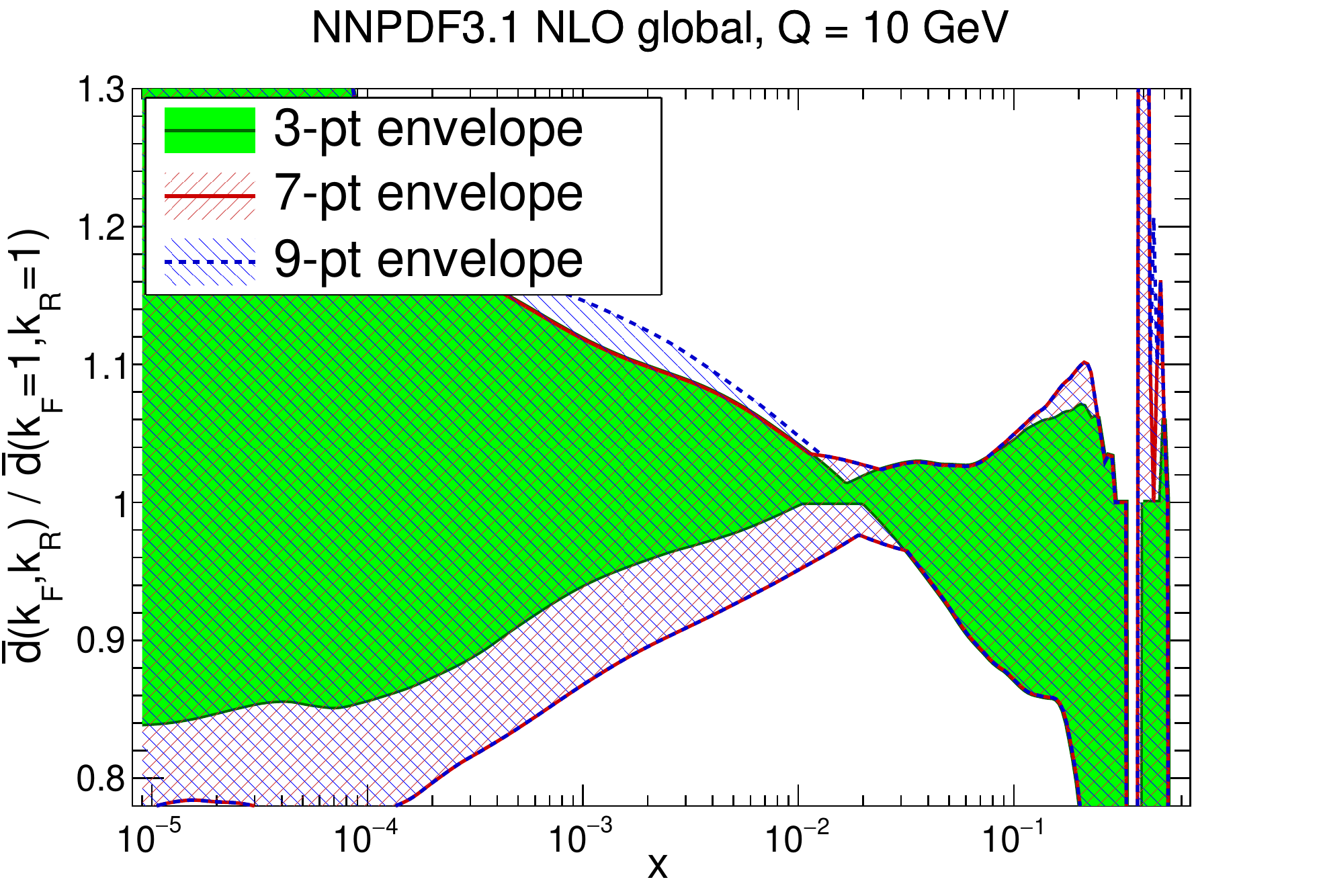}
 \caption{\small Left: same as Fig.~\ref{fig:NLODIS-scalevars} (left)
 for global PDF sets.
  Right:  Envelope of scale variations computed using various
  prescriptions (see text).
    \label{fig:NLOGlobal-scalevars} }
  \end{center}
\end{figure}

The corresponding PDFs are shown 
in Fig.~\ref{fig:NLOGlobal-scalevars}, where we also display various
envelopes of scale variations which will be discussed in Sect.~\ref{sec:envelope} below.
The general features are  similar to those of the
DIS-only fits shown in Fig.~\ref{fig:NLODIS-scalevars}.
In the case of the gluon, varying the scales with respect to
the central choice leads in general to a suppression for $x\lsim 10^{-2}$
and an enhancement for larger values.
For the singlet and down quark PDFs, one can observe a large spread
as the scales are varied at small-$x$, which is approximately symmetric,
a point of minimum sensitivity around $x \simeq 0.1$, and then a further
increase of the spread of central values at large-$x$ especially
for the poorly constrained down antiquark.
The scale combination $(k_f,k_r)=(2,\half)$ also leads to
a large distortion 
of the PDF central values in this case, as in the DIS-only fits.

\subsection{The envelope prescription for MHOU}
\label{sec:envelope}
Now we would like to assess the possibility of using PDF sets obtained
with different choices of renormalization and factorization scales as a
means to estimate MHOUs. We first systematically compare
fit quality as a function of the scale choice.
In order to facilitate this comparison
we evaluate
\be
\label{eq:deltachi2}
\Delta_{\chi^2}^+ \equiv \lp \chi^2_{\rm max}-
    \chi^2_{\rm central} \rp \, ; 
 \, \qquad
\Delta_{\chi^2}^- \equiv \lp  \chi^2_{\rm min}-
 \chi^2_{\rm central}\rp \, ,
 \ee
where $\chi^2_{\rm max}$~($\chi^2_{\rm min}$) denotes the largest (smallest) value
of the $\chi^2$, for either the total dataset or for individual
experiments,  among all the NLO fits based on scale-varied theories
listed in Tables~\ref{tab:chi2table_dis_nlo} 
and~\ref{tab:chi2table_global_nlo}, and $\chi^2_{\rm central}$ indicates
the values from the baseline fit with central scales.

The values of these quantities for both the DIS-only fits
and the global fits are collected
in Table~\ref{tab:chi2_scales_shift}, both for 
individual experiments and  for the total dataset.
From this comparison, see that for
DIS-only PDFs the HERA data drive the differences in  $\chi^2$ values,
with rather smaller contributions
from the fixed-target experiments.
%
\begin{table}[t]
\begin{center}
\renewcommand*{\arraystretch}{1.7}
\footnotesize
\begin{tabular}{|l|c|l|}
  \toprule
  Dataset  &   $N_{\rm dat}$ & {$\lc
     \Delta_{\chi^2}^-, \Delta_{\chi^2}^+\rc$}\\
    &   &    DIS-only NLO     \\
\toprule
NMC         & 134 &    $\lc    -2  , +6  \rc$        \\ 
SLAC        & 12  & $\lc    -3  ,  +3  \rc$    \\ 
BCDMS       & 530 &  $\lc    -3  ,    +25  \rc$    \\ 
CHORUS      & 430 &  $\lc     0  ,    +38  \rc$   \\ 
NuTeV       & 41  & $\lc    -3  ,   +4  \rc$ \\ 
HERA incl   & 967 & $\lc    -6  ,   +251  \rc$ \\ 
HERA $F_2^c$& 31  & $\lc    -3  ,    +34  \rc$  \\ 
\midrule
Total       &2145 & $\lc     0  ,   +337  \rc$  \\ 
\bottomrule
\end{tabular}
\qquad
\begin{tabular}{|l|c|c|}
\toprule
Process  &   $N_{\rm dat}$ & $\lc\Delta_{\chi^2}^-, \Delta_{\chi^2}^+\rc$\\
         &               & global NLO \\
\midrule
DIS NC & 1593 & $\lc   0, +487\rc$\\
DIS CC &  552 & $\lc   0,  +42\rc$\\
\midrule
DY     &  484 & $\lc -67, +138\rc$\\
JETS   &  164 & $\lc  -5,  +19\rc$\\
TOP    &   26 & $\lc  -4,  +33\rc$\\
\midrule
Total  & 2819 & $\lc   0, +750\rc$\\
\bottomrule
\end{tabular}
\end{center}

\caption{\small The values of $\Delta_{\chi^2}^\pm$
  Eq.~(\ref{eq:deltachi2}) for  DIS-only and  global
  PDFs. 
  The maximum and minimum of the $\chi^2$
  are evaluated for the PDF sets listed in Table~(\ref{tab:scalevarFits}).
  \label{tab:chi2_scales_shift}}
\end{table}

The global fits behave in a similar way, with the HERA data still
dominating $\chi^2$ differences. However, a marked $\chi^2$ spread is
now also seen for the LHC experiments. This shows that the precise
collider (HERA and LHC) data are most sensitive to higher order corrections.
As in the case of DIS-only fits,  for almost all experiments
the central scale choice $\lp k_f,k_r\rp=(1,1)$ provides
the best overall description of the various datasets. A notable
exception is DY, for which the scale choice $\left(1,\half\right)$
leads to an improved fit (see Tab.~\ref{tab:chi2table_global_nlo}),
in agreement with the argument (often used for Higgs production in
gluon fusion) that the natural renormalization scale for inclusive
production of a 
colorless object is half its mass.


The fact that, with the exception of the combination
$(k_f,k_r)=(2,\half)$ all scale choices lead to PDFs in reasonable
agreement with the data, suggests that an estimate of MHOU might be
obtained by taking an envelope of PDFs determined with different scale
choices. We consider in particular: the 9-point envelope, in which all
combinations of scales of Table.~(\ref{tab:scalevarFits}) are included
in the envelope; the 7-point envelope, in which the two
choices $(k_f,k_r)=(2,\half)$ and 
$(k_f,k_r)=(\half,2)$ are removed from the 9-point envelope; and the
3-point envelope, in which only the two choices $(k_f,k_r)=(2,2)$ and
$(k_f,k_r)=(\half,\half)$ are considered together with the central
scale choices.
These envelopes are shown in
Fig.~\ref{fig:NLOGlobal-scalevars}. It is clear that the size of the
envelope is extremely sensitive to the choice of scales to be
included. Of course, by construction, an  envelope including more
scale choices always leads to a wider band than envelopes with fewer
choice, and indeed the 7-point envelope leads to significantly larger
uncertainties than the 3-point envelope, though the 7-point and
9-point envelopes essentially coincide.

\begin{figure}[t]
  \begin{center}
  \includegraphics[width=0.48\textwidth]{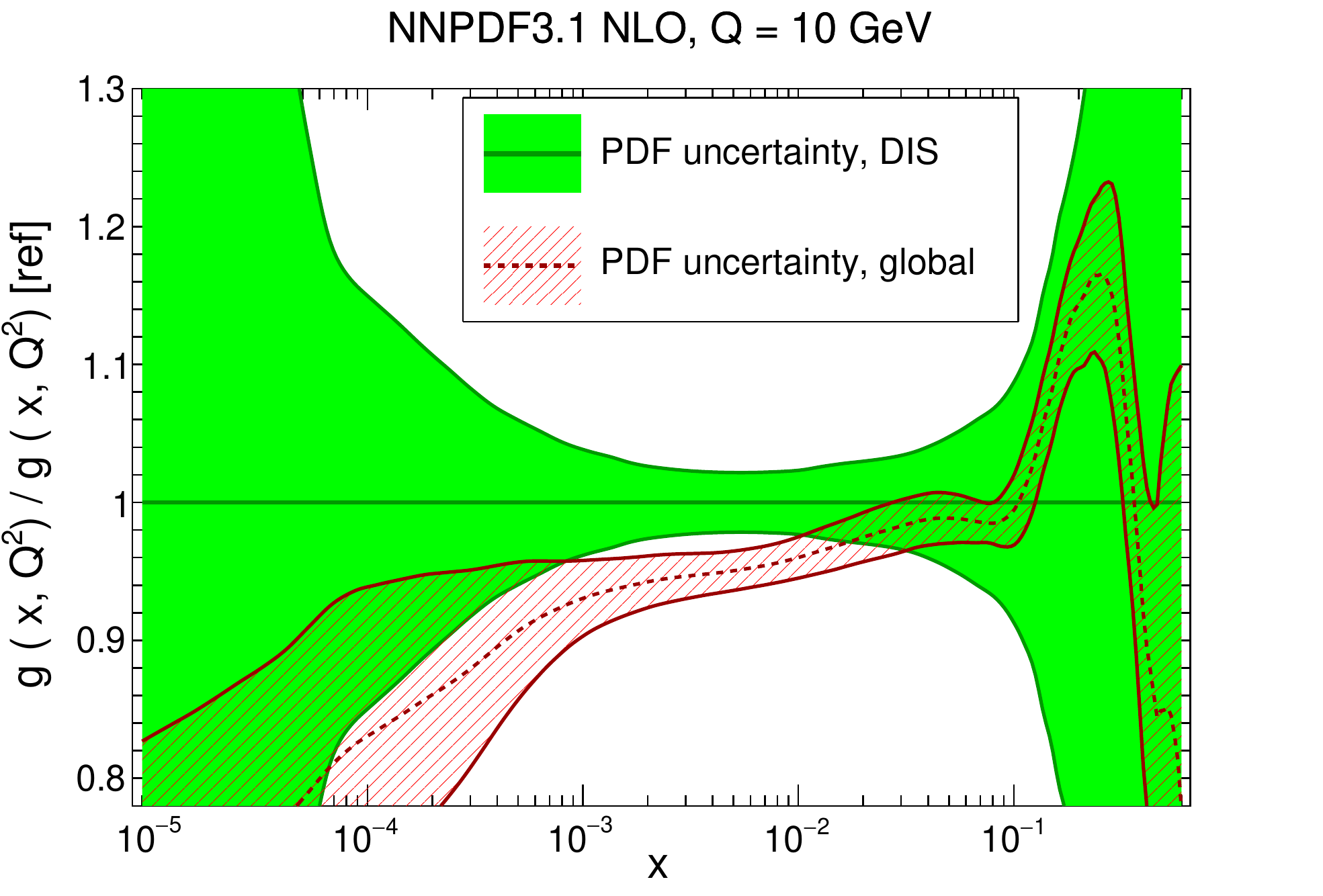}
  \includegraphics[width=0.48\textwidth]{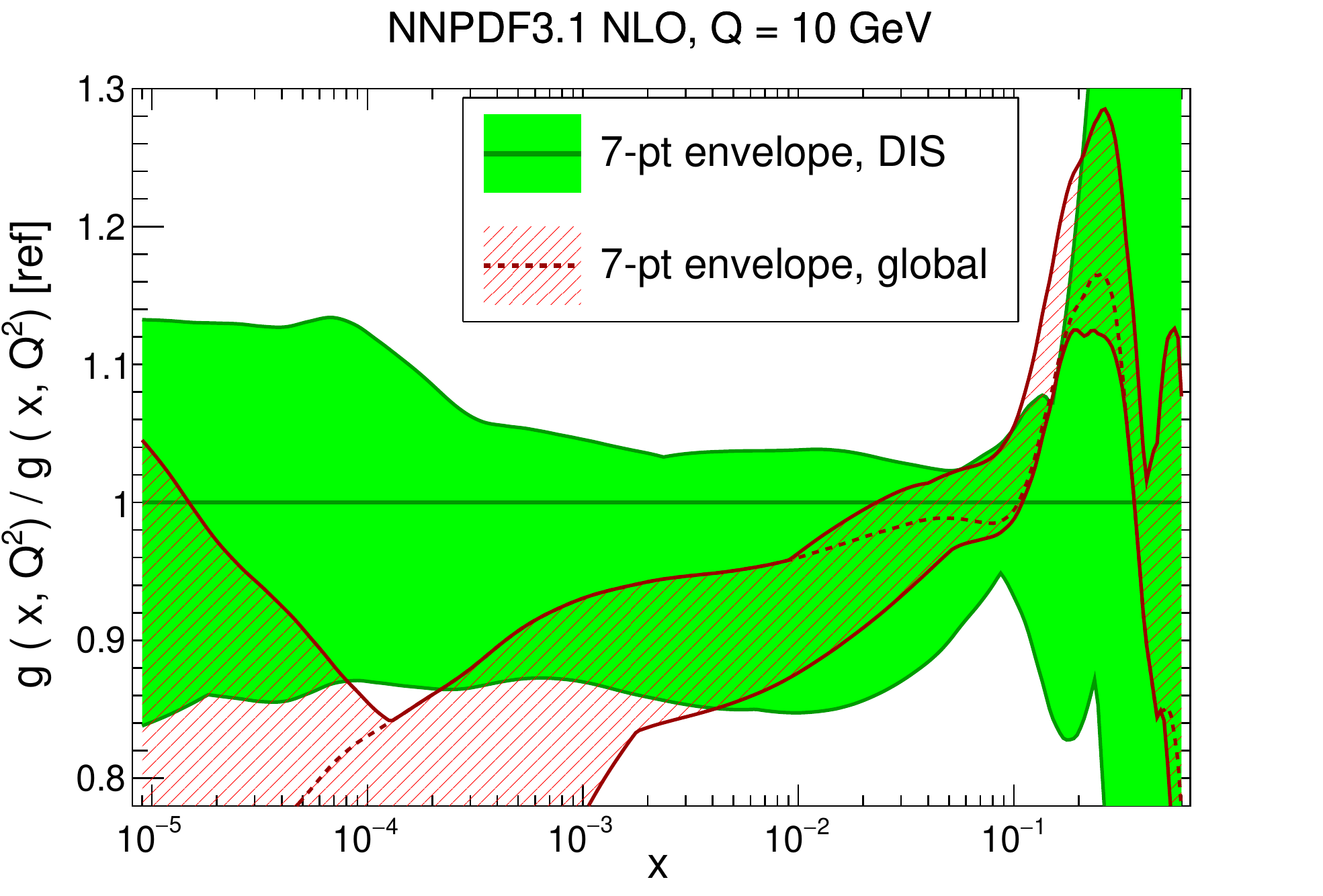}
  \includegraphics[width=0.48\textwidth]{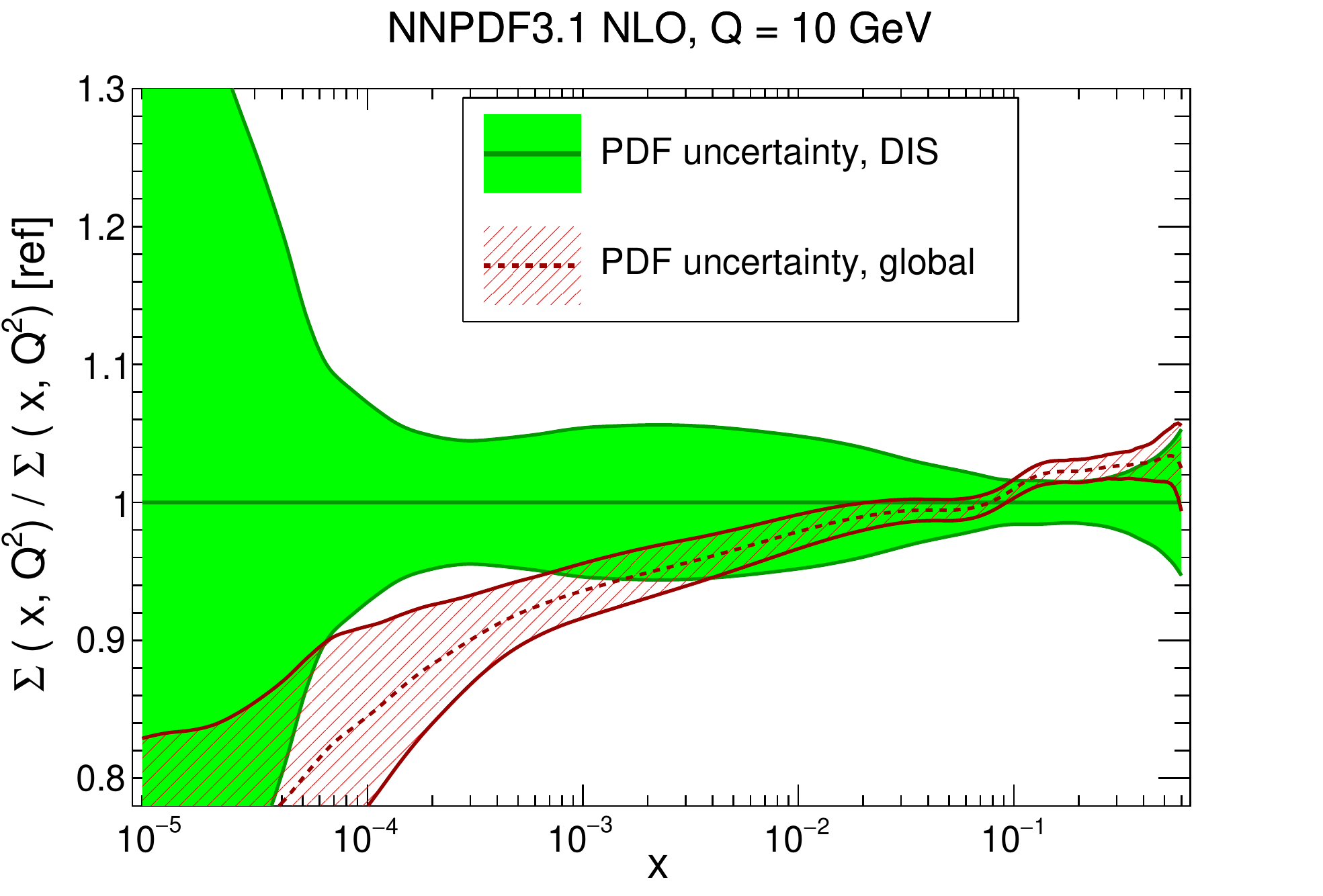}
  \includegraphics[width=0.48\textwidth]{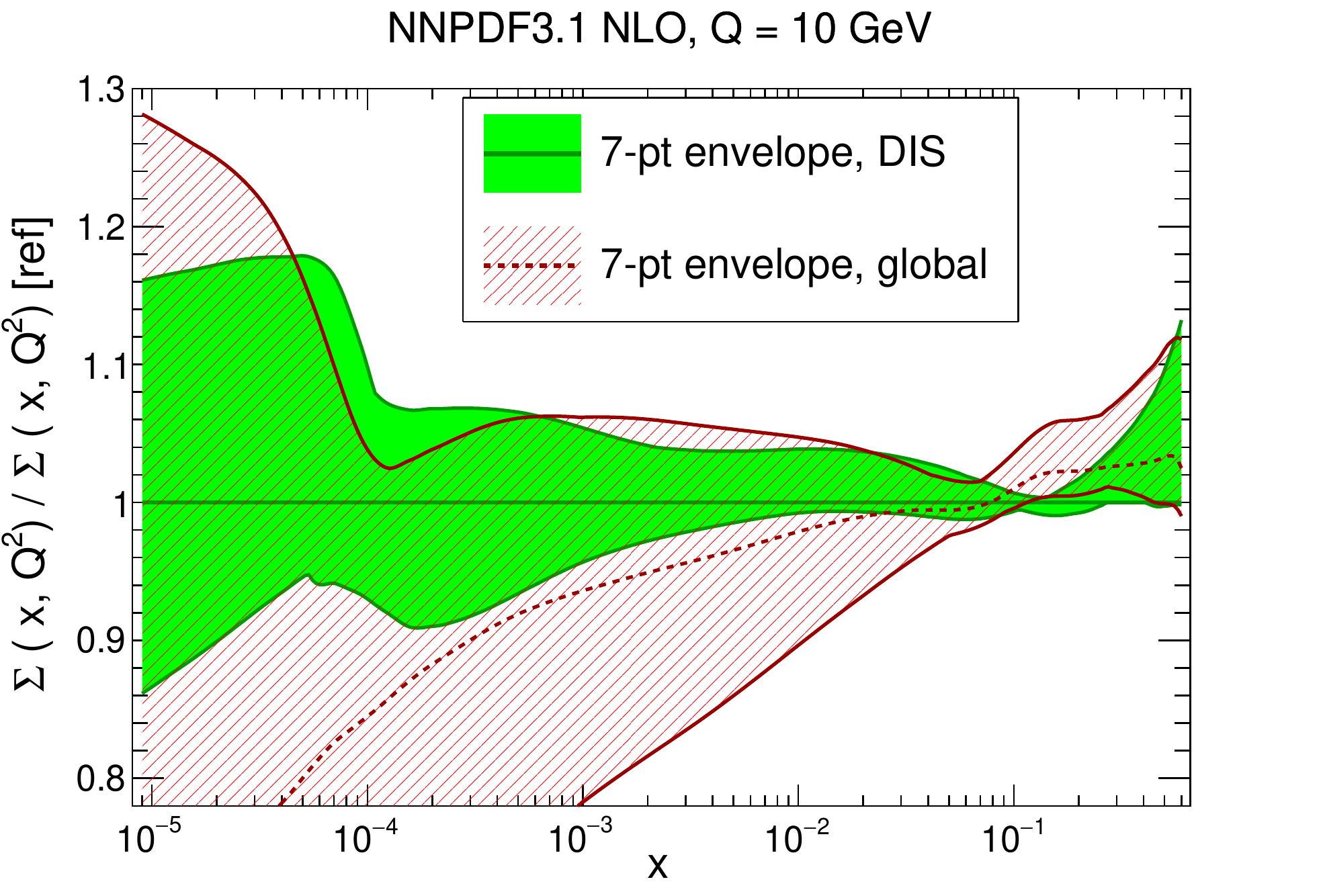}
  \includegraphics[width=0.48\textwidth]{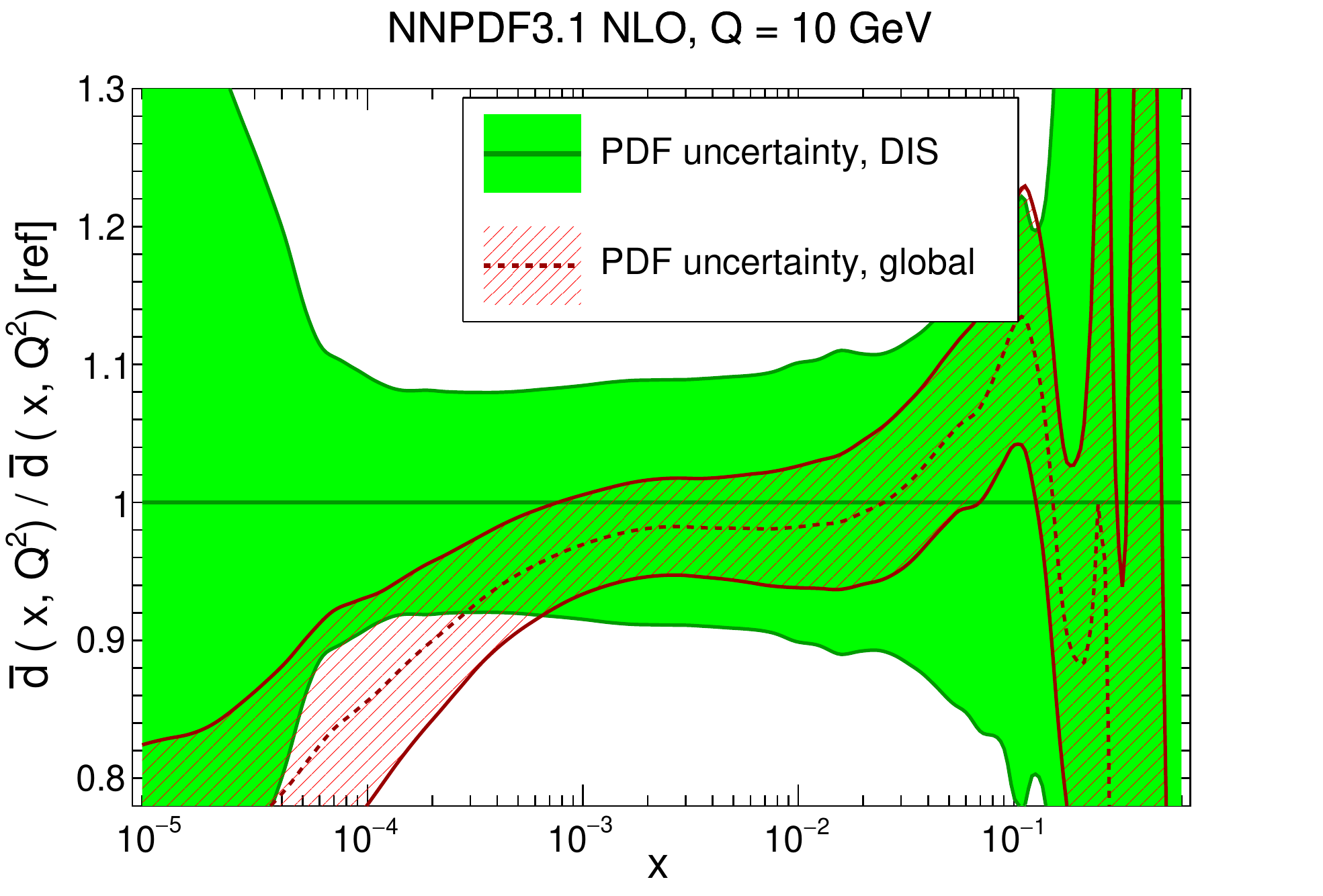}
  \includegraphics[width=0.48\textwidth]{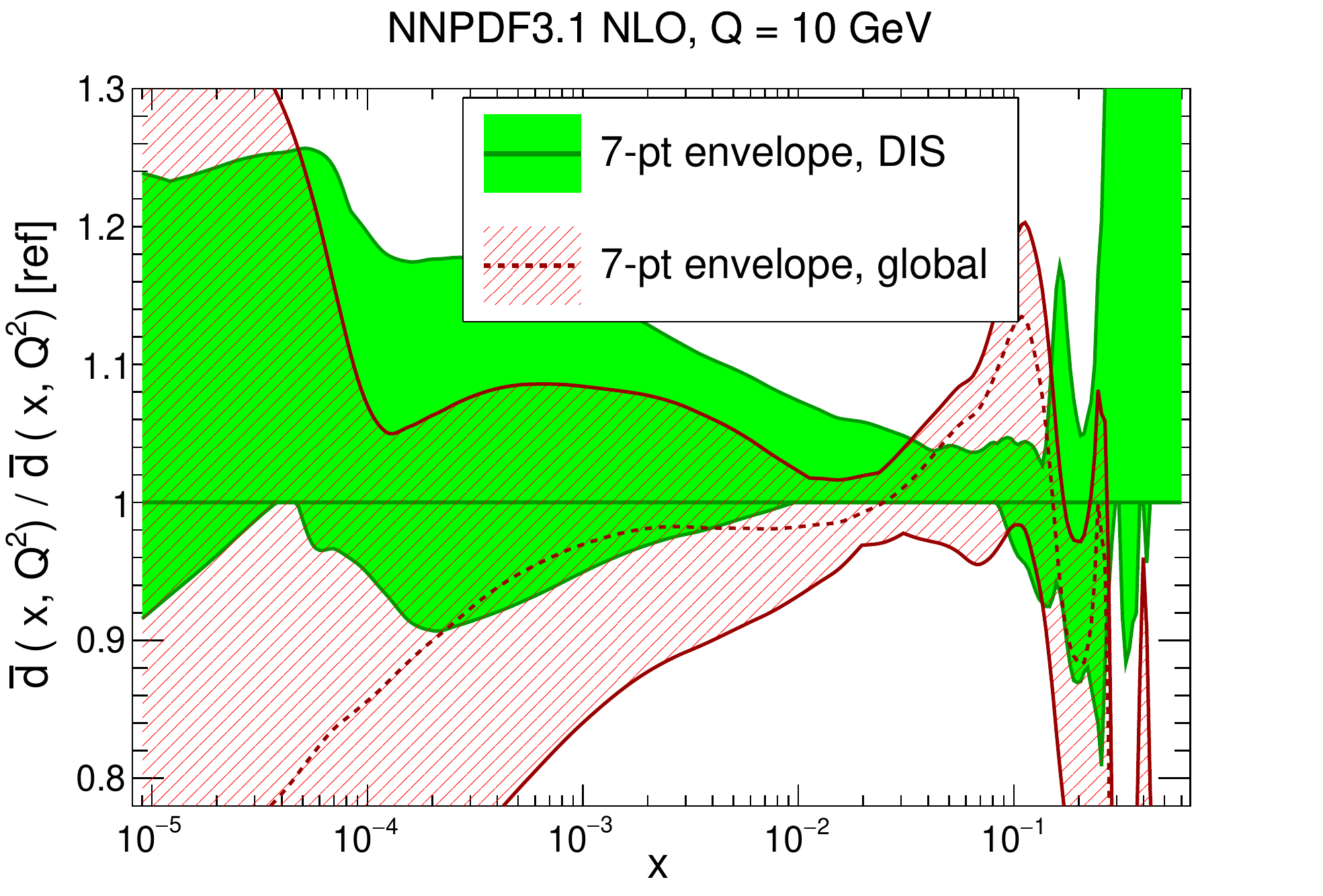}
  \caption{\small Comparison between PDF uncertainties (left) and
    7-point envelopes of scale variations (right) for the DIS and
    global PDFs. The gluon, quark singlet and antidown at $Q=10$~GeV
    are shown. 
      \label{fig:PDFuncertainties-DIS-vs-Global} }
  \end{center}
\end{figure}

Further insight on the envelope method can be obtained by comparing
the envelope of scale variations, taken as a candidate MHOU, for
DIS-only and global PDF sets.
This is done in
Fig.~\ref{fig:PDFuncertainties-DIS-vs-Global}, where the standard,
data-driven PDF uncertainties are also shown for reference. Whereas
the PDF uncertainties always decrease when from DIS-only to global
PDFs (and so did the theory uncertainties when determined using the
theory covaraince matrix in Sect.~\ref{sec:fitstherr}) the theory
uncertainties estimated from the envelope prescription behave more
erratically, with the envelope for the global fit leading to a wider
band in the case of the singlet distribution. Quite in general,
the envelope estimates of MHOUs appear to be rather large 
in comparison to PDF uncertainties, and unstable upon changes in dataset.

Finally, we can ask how MHOUs estimated from an envelope could be
combined with the data-driven PDF uncertainties.
In the case of
the covariance matrix approach discussed in Sect.~\ref{sec:fitstherr}, results are found using default NNPDF methodology including an extra contribution to the covariance matrix. Here, however, we need a
prescription for the combination of MHOUs (obtained from the envelope of scale variations) with the data-driven PDF uncertainties (obtained using
default NNPDF methodology with a purely experimental covariance matrix).

A possible prescription would be to calculate the total uncertainty on PDFs as a sum
in quadrature of the envelope MHOU and the standard PDF uncertainty. 
 Taking into account that these envelopes are asymmetric the
 prescription is then
  \be
 \sigma^{\rm tot,\pm} = \lp (\sigma^{\rm mho,\pm})^2+(\sigma^{\rm PDF})^2\rp^{1/2} \,
 \label{eq:ScaleVarUncComb},
 \ee
 where $\sigma^{\rm mho,+(-)}_{q}$ indicates the upper (lower) limit
 of the envelope, and $\sigma^{\rm PDF}_{q}$ is the standard PDF
 uncertainty.

\begin{figure}[t]
  \begin{center}
 \includegraphics[width=0.49\textwidth]{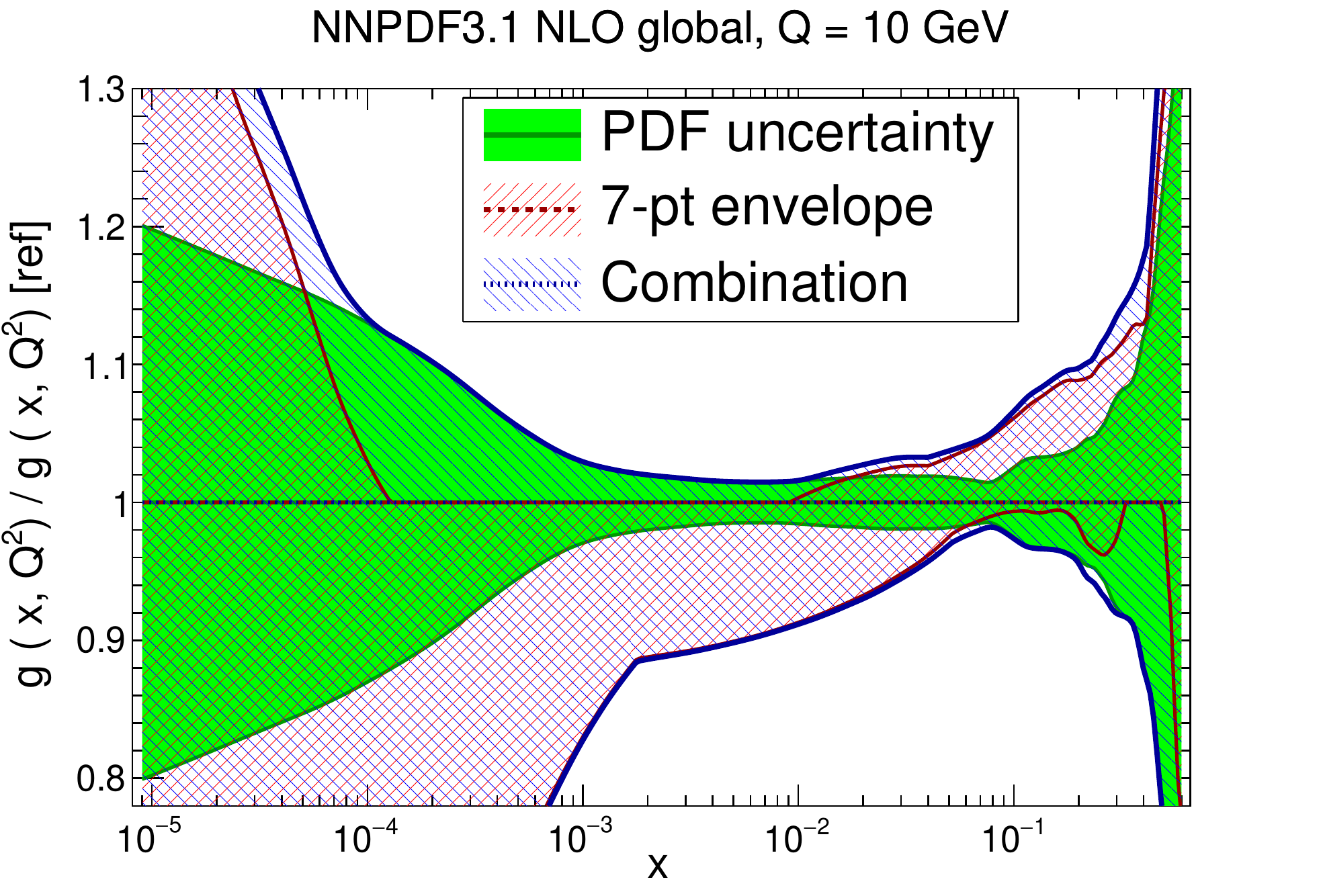}
 \includegraphics[width=0.49\textwidth]{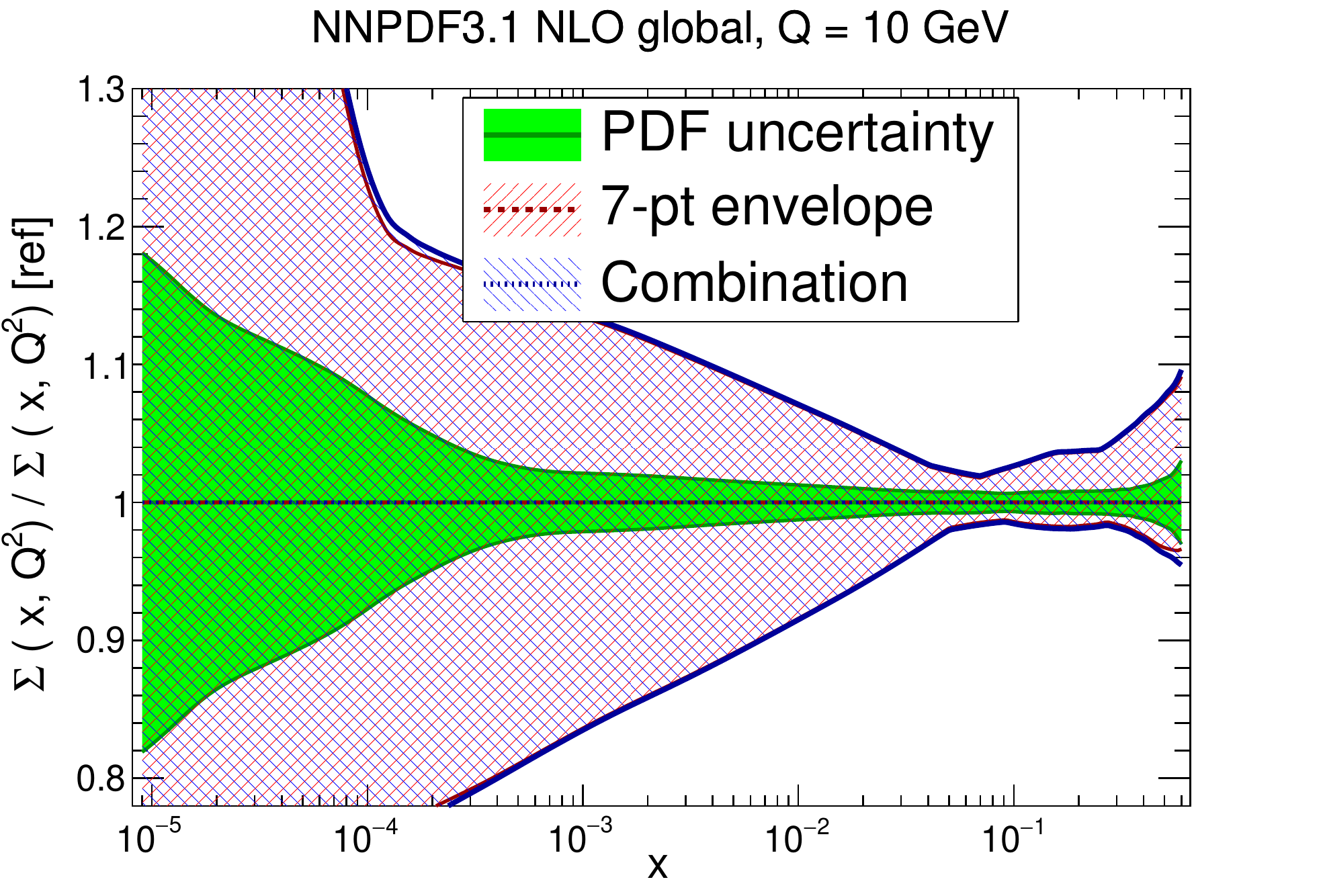}
 \includegraphics[width=0.49\textwidth]{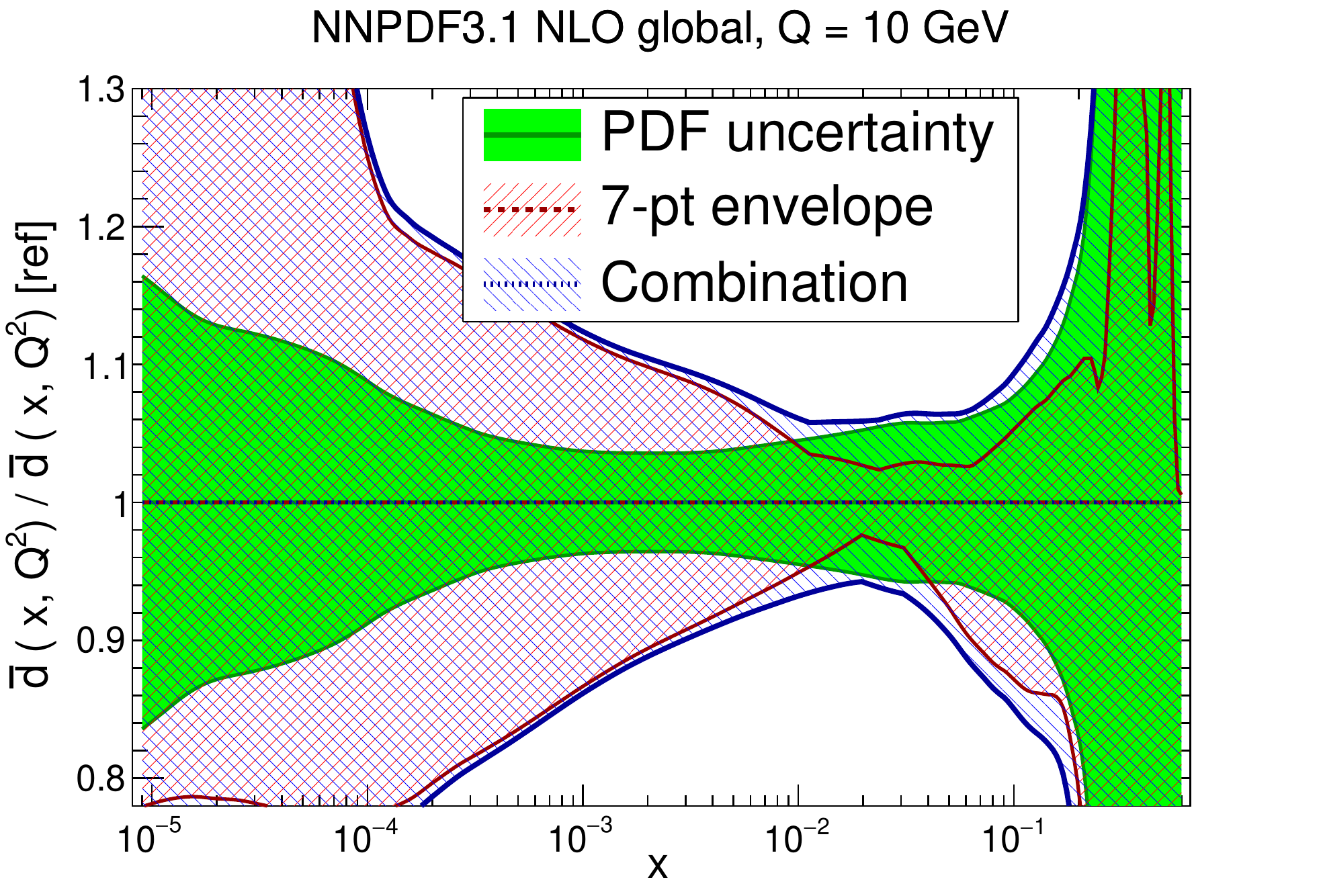}
 \includegraphics[width=0.49\textwidth]{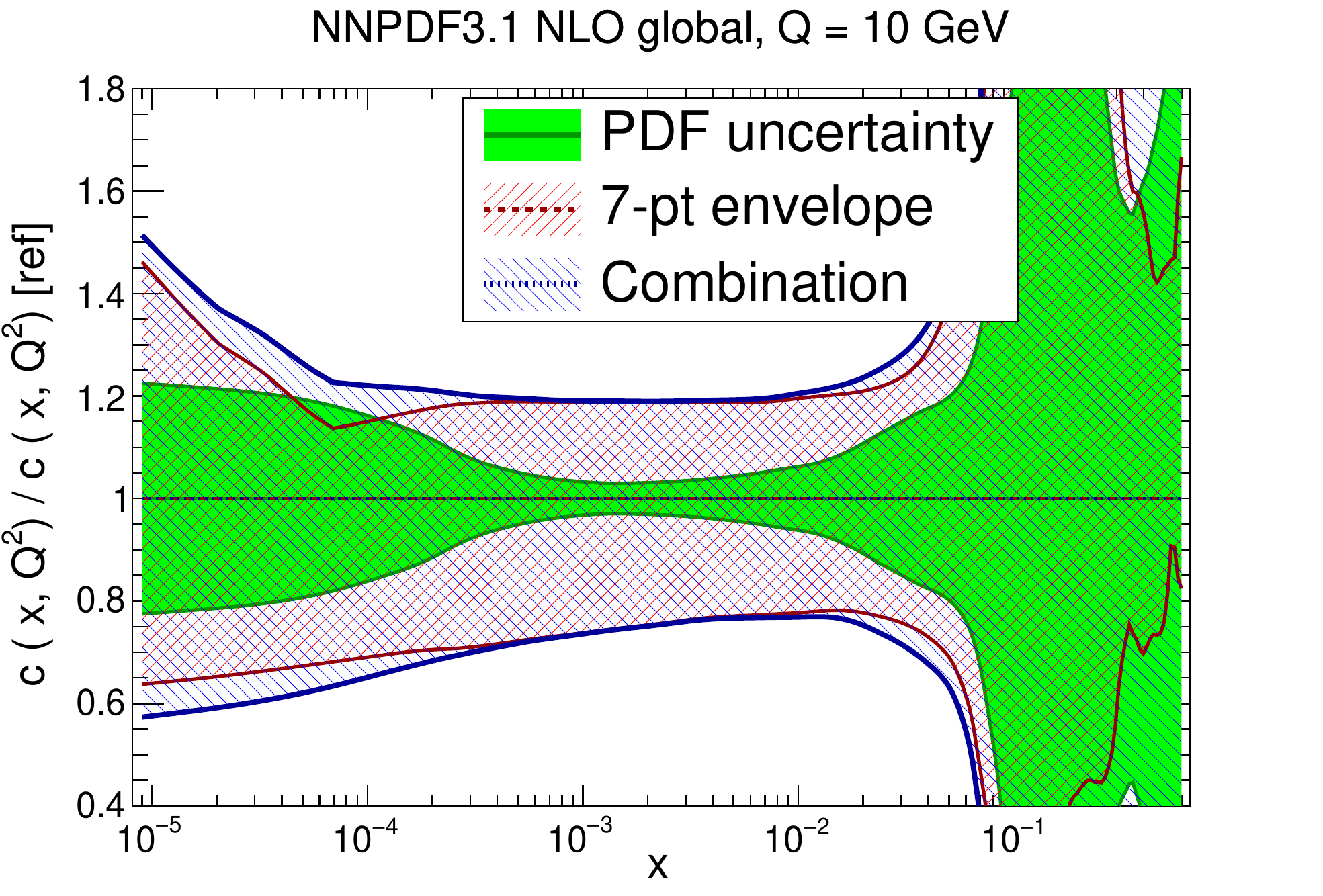}
 \caption{\small The  PDF uncertainties
   Eq.~(\ref{eq:ScaleVarUncComb}):  the PDF uncertainty  $\sigma^{\rm
     PDF}$ computed using standard NNPDF methodology, the 7-point
   envelope estimate of the MHO uncertainty  $\sigma^{\rm mho,\pm}$
   and the total combined uncertainty   $\sigma^{\rm tot,\pm}$. 
   Results are shown at $Q=10$ GeV for the gluon (top left), quark
   singlet (top right), down
   antiquark (bottom left), and  charm PDFs (bottom right)
    normalized to the central value.
    \label{fig:MHOU-envelope-combination} }
  \end{center}
\end{figure}

\begin{figure}[t]
  \begin{center}
    \includegraphics[scale=0.39]{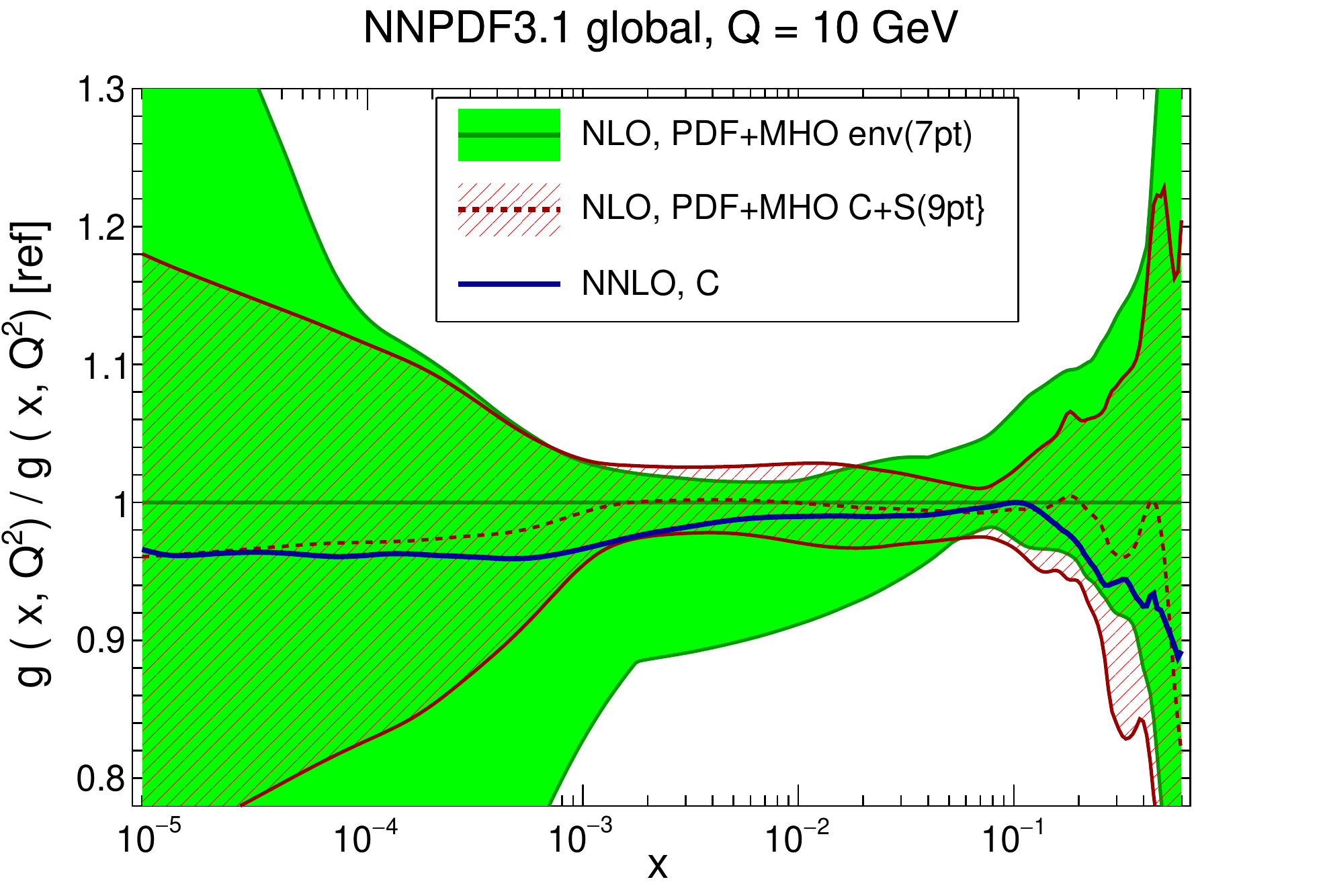}
    \includegraphics[scale=0.39]{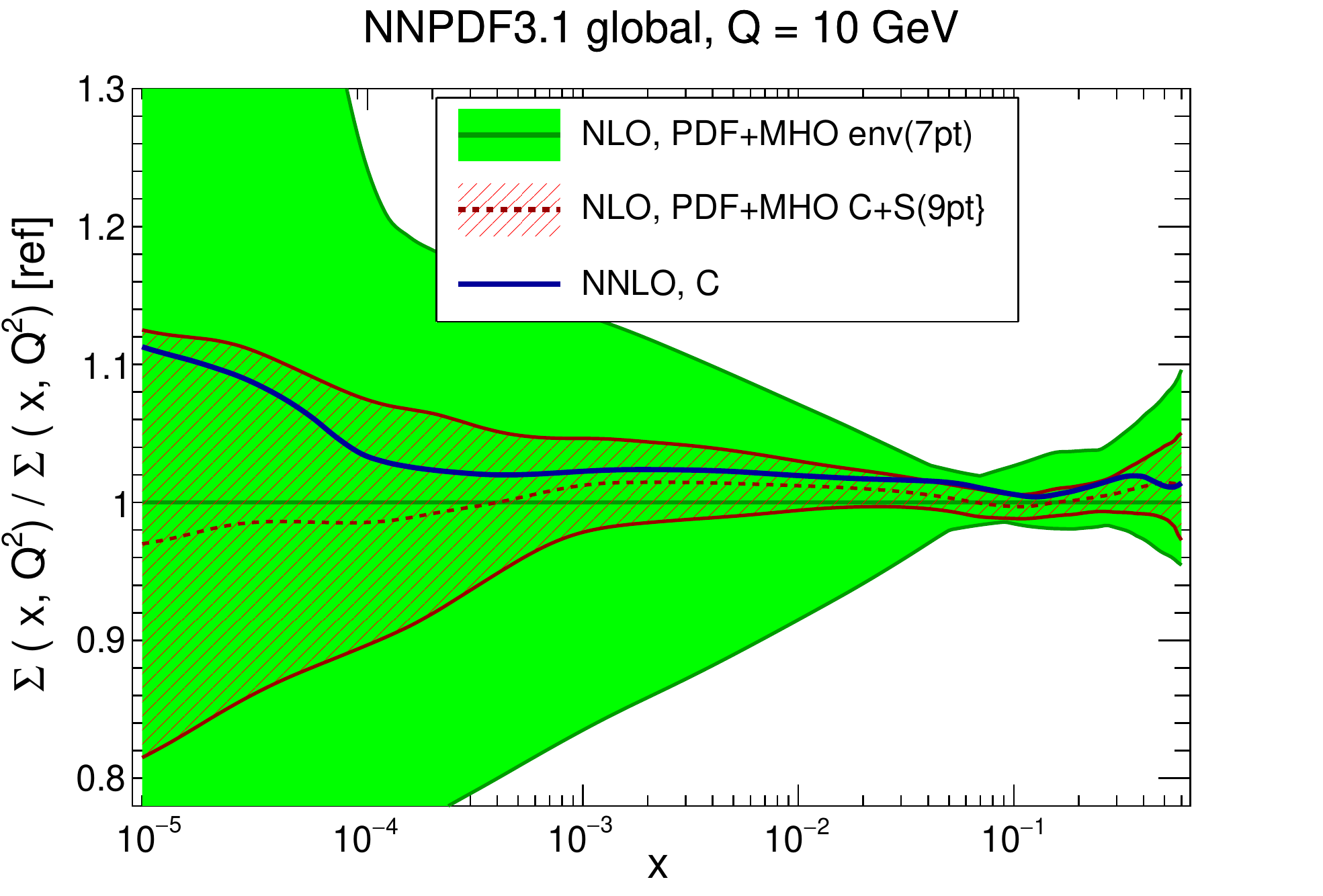}
    \includegraphics[scale=0.39]{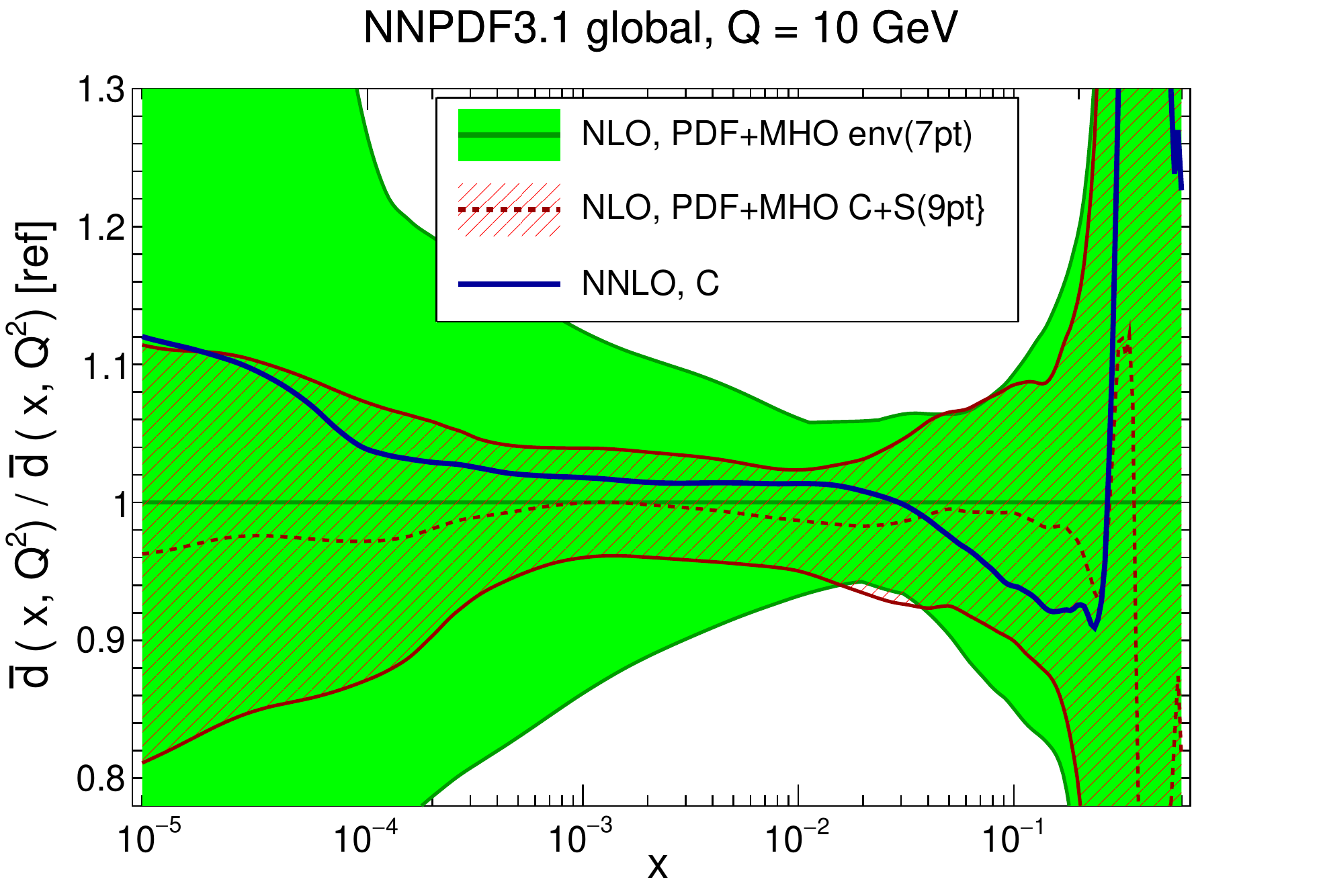}
    \includegraphics[scale=0.39]{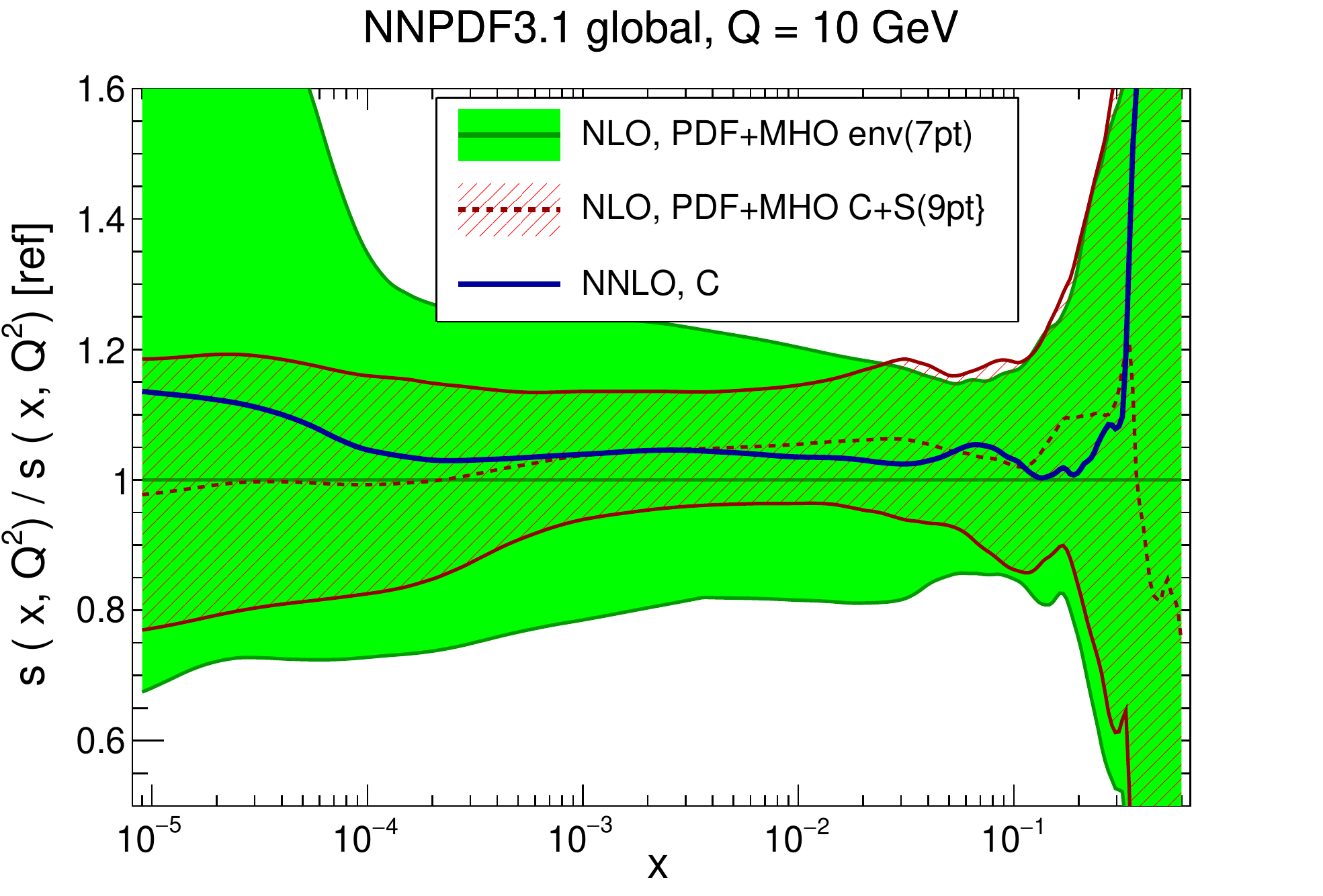}
    \caption{\small Same as Fig.~\ref{fig:MHOU-envelope-combination}
      comparing the total uncertainty obtained from the 7-point
      estimate of the MHOU using
      Eq.~(\ref{eq:ScaleVarUncComb}) to the total uncertainty obtained
      from the theory covariance matrix [same  as
      Fig.~\ref{fig:Global-NLO-CovMatTH})] computed using the 9-point
      prescription. The central NNLO value obtained using the
      experimental covariance matrix is also shown. All results are
      normalized to the central NLO value with experimental covariance
      matrix.
    \label{fig:finalcomparison} }
  \end{center}
\end{figure}

In Fig.~\ref{fig:MHOU-envelope-combination} we show the uncertainties
 $\sigma^{\rm tot,\pm}$, $\sigma^{\rm mho,\pm}$  and $\sigma^{\rm PDF}$
Eq.~(\ref{eq:ScaleVarUncComb}) using the 7-point envelope
for the gluon, the quark singlet, the down
  antiquark, and the charm PDFs, all normalized to the central value.
  In Fig.~\ref{fig:finalcomparison} we further compare, for the same
  PDF combinations, the total 
  uncertainties obtained with the envelope method (shown in
  Fig.~\ref{fig:MHOU-envelope-combination}) with the total uncertainties
  obtained with our theory covariance matrix methodology (shown
in  Fig.~\ref{fig:Global-NLO-CovMatTH}), all normalized to the central
curve of the envelope method. The NNLO central curve (with
experimental covariance matrix only) is also shown.
      Results are obtained using the baseline settings:
      the 7-point prescription for the envelope method
      and the 9-point prescription for the theory covariance matrix.

It is clear that some qualitative features are common to both
uncertainty estimates; in particular, the asymmetry of the envelope
prescription  favors variations which go towards the direction of the
true NNLO result. However, it is also clear that the envelope prescription
has a number of shortcomings: it leads to discontinuous and asymmetric
uncertainties, which are difficult to accommodate in a Gaussian
framework; it is very unstable and strongly dependent on arbitrary
choices for the set of the scale variations over which the envelope should
be taken; it is quite cumbersome and again arbitrary in requiring one
to postulate a specific way of combining MHOU and data-induced PDF
uncertainties; it leads to very large MHOUs which appear to be
overestimated in comparison to the known shift to the NNLO result if a
7-point prescription is used.

The reason for the much greater stability of MHOU estimated using the
covariance matrix prescription should be clear: when using an envelope
prescription, any large deviation in a given direction leads to
large uncertainties in that direction, regardless of whether indeed
there are large MHOU or not. In a covariance matrix approach, a large
eigenvalue in any given direction will allow the fit to move in that
direction. This, however, at least in the presence of abundant
experimental information, will actually happen only if the data pull in
that direction due to MHOU, and otherwise it will have little
effect.
Note also that, by construction, in an envelope approach the best fit
will be the same as that in which MHOU are not included. So it is possible to have a more conservative estimate of the overall uncertainty, but not a more accurate result.

We conclude that, whereas results for MHOU based on scale varied fits and 
an envelope prescription are by and large
consistent with those obtained with a covariance matrix approach, they
are less stable, less reliable, and less accurate.

\clearpage
\bibliography{thcovmat}

\end{document}